\documentclass[12pt,a4paper,english]{report}
\usepackage[T1]{fontenc}
\usepackage[utf-8]{inputenc}
\usepackage[polish,english]{babel}
\usepackage{amsmath}
\usepackage{amsfonts}
\usepackage{indentfirst}
\usepackage[dvips]{graphicx}
\usepackage{youngtab}
\usepackage{epsfig}
\usepackage{rotating}

\newcommand{\bea}{\begin{eqnarray}}
\newcommand{\eea}{\end{eqnarray}}
\newcommand{\be}{\begin{equation}}
\newcommand{\ee}{\end{equation}}

\newcommand{\nn}{\nonumber\\}

\newcommand{\hf}{\frac{1}{2}}

\newcommand{\Z}{{\mathbb Z}}
\newcommand{\R}{{\mathbb R}}
\newcommand{\C}{{\mathbb C}}

\def\Tr{{\rm Tr}}

\def\Gp{\Gamma_{+}}
\def\Gm{\Gamma_{-}}

\def\G{\Gamma}

\def\tz{\tilde{Z}}

\newcommand{\cF}{{\cal F }}

\newcommand{\cY}{{\cal Y }}

\begin{document}

\sloppy

\newpage
\thispagestyle{empty}

\begin{center}
 \large \textsc{Uniwersytet Warszawski} \\
\textsc{Wydział   Fizyki}\\
\textsc{ Instytut Fizyki Teoretycznej}
\vskip 2 cm

%{\bf \Huge{{ Piotr Sułkowski }}}\\
{\bf \Huge{\textsc{piotr sułkowski}} } \\
\vskip 2cm

\begin{center}
\epsfig{file=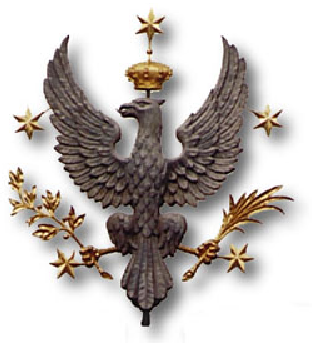,width=5cm}
\end{center}
\vspace*{2cm}

{\bf\Huge \textsc{calabi-yau crystals} \\
\vskip 10pt
\bf\Huge \textsc{in topological string theory} }

\vskip 15pt

%{\bf \large{\textsc{KRYSZTAŁY CALABI-YAU W TOPOLOGICZNEJ TEORII STRUN}}}\\
{\bf \large{\textsc{kryształy calabi-yau w topologicznej teorii strun}}}\\
\vskip 15 pt

{\large \textsc{rozprawa doktorska}} \\

%\vskip 0.8 cm
%\begin{center}
%Zak"lad Teorii Cz"astek Elementarnych \\ i Oddzia"lywa"n Fundamentalnych
%\end{center}

\vskip 2 cm
\begin{center}
{\large \textsc{Warszawa, 2007}}
\end{center}
\end{center}
\newpage 
\thispagestyle{empty}
{\tiny{{.}}}
\vskip 15cm

\begin{tabular}{l}

{\large \textsc{Promotorzy:}} \\

\\

%{{\bf Prof. dr Robbert Dijkgraaf,} $\quad$  {\bf dr hab. Jacek Pawe\l czyk, prof UW}}
{\bf \large{Prof. dr Robbert Dijkgraaf}} \\

{\bf \large{Universiteit van Amsterdam}} \\

\\

%\vskip 1cm
{\bf \large{dr hab. Jacek Pawełczyk, prof. UW }} \\

{\bf \large{Uniwersytet Warszawski}} \\

\end{tabular}

\newpage
\thispagestyle{empty}
{\tiny{{.}}}
\vskip 18.5cm

\begin{flushright}
{\bf \huge{\emph{Całej mojej rodzinie  }}}\\
\end{flushright}

%***********************************************************

%\begin{center}

%{\LARGE Calabi-Yau crystals}\\
%{\LARGE in topological string theory}\\
%{\Large -   -   -}\\
%{\Large Piotr Su\l kowski}

%\end{center}

%***********************************************************

\selectlanguage{english}

\tableofcontents

\newpage

\newpage

%***********************************************************
%***********************************************************
\selectlanguage{polish}
\chapter*{Wstęp i streszczenie}
\addcontentsline{toc}{chapter}{Wstęp i streszczenie}

Mimo odwiecznych poszukiwań wciąż nie znamy pełnej odpowiedzi na pytanie jak brzmią podstawowe prawa Przyrody. W wyniku ogromnego rozwoju nauki w minionym stuleciu stwierdzono, że prawa te z pewnością obejmują teorię względności i mechanikę kwantową. Teoria względności wyjaśnia zachowanie materii na dużych skalach, sięgających nawet rozmiarów Wszechświata. Z kolei mechanika kwantowa, a w szczególności kwantowa teoria pola, opisują  świat na bardzo małych odległościach. Wedle obecnego stanu wiedzy, większość zjawisk na poziomie kwantowym jest opisywana przez Model Standardowy, który jest pewną szczególną realizacją kwantowej teorii pola z grupą cechowania $SU(3)\times SU(2)\times U(1)$.

Tym niemniej, dotychczas nie udało się dokonać unifikacji mechaniki kwantowej z teorią względności, i mimo wielu spektakularnych sukcesów teorie te nie wystarczają do wyjaśnienia wszystkich obserwowanych zjawisk. Przykładami takich niewyjaśnionych zjawisk są łamanie symetrii elektrosłabej, niskoenergetyczne zachowanie chromodynamiki kwantowej oraz przebieg wysokoenergetycznych procesów w teorii względności. Symetria elektrosłaba związana jest z podgrupą $SU(2)\times U(1)$ Modelu Standardowego i jest ona spontanicznie złamana poprzez tzw. mechanizm Higgsa. Mechanizm ten wymaga istnienia dodatkowej cząstki skalarnej, zwanej bozonem Higgsa, która jako jedyna w Modelu Standardowym nie została dotychczas zaobserwowana doświadczalnie. W istocie nie wiadomo nawet czy bozon Higgsa jest cząstką elementarną czy też jakimś stanem związanym, a jego niezwykle mała masa w stosunku do skali Plancka jest źródłem tzw. problemu hierarchii. Chromodynamika kwantowa jest częścią Modelu Standardowego związaną z grupą $SU(3)$. Jedną z jej najistotniejszych cech jest nieperturbacyjny charakter stałej sprzężenia dla małych energii, co uniemożliwia otrzymanie ilościowych wyników w tym zakresie przy pomocy rozwinięcia perturbacyjnego. Z kolei teoria grawitacji Einsteina nie jest renormalizowalna, co uniemożliwia przeprowadzenie jej kwantyzacji jakąkolwiek standardową metodą. Tym niemniej, kwantowe efekty grawitacyjne bez wątpienia muszą odgrywać istotną rolę w warunkach które panowały we wczesnych stadiach rozwoju Wszechświata, lub które występują we wnętrzu czarnych dziur.

W celu rozwiązania powyższych problemów zaproponowano kilka niezwykle interesujących idei teoretycznych. Niestety, dotychczas nie zostały one potwierdzone doświadczalnie, aczkolwiek wiele różnorodnych argumentów przemawia za tym, że mogą być one istotne przy opisie Przyrody na najbardziej podstawowym poziomie. Mechanika kwantowa i teoria względności mogłyby być unifikowane w ramach teorii strun, która opisuje wszystkie cząstki elementarne i ich oddziaływania jako stany jedynego fundamentalnego obiektu, czyli \emph{struny}. Z fenomenologicznego punktu widzenia najbardziej obiecujące wydają się być teorie \emph{superstrun}, które posiadają pewną dodatkową symetrię, tzw. \emph{supersymetrię}. Supersymetria ma istotne konsekwencje także jeśli zapostuluje się ją w teoriach pola, a w szczególności w Modelu Standardowym. Przewiduje ona między innymi, że każda cząstka powinna mieć swojego \emph{superpartnera} o innym spinie. Mimo, że superpartnerzy znanych cząstek elementarnych dotychczas nie zostali zaobserwowani, supersymetryczny Model Standardowy ma wiele zalet: na przykład mógłby on pozwolić na rozwiązanie problemu hierarchii, jak też przewiduje on unifikację stałych sprzężenia przy dużych energiach. W związku z tym istnieją pewne przypuszczenia, iż supersymetria mogłaby być odkryta nawet w najnowszym akceleratorze LHC w niedalekiej przyszłości. 

Innym fascynującym pomysłem teoretycznym jest idea \emph{dualności}, zgodnie z którą zjawiska zachodzące poza zakresem stosowalności danej teorii mogą być opisane przez teorię do niej \emph{dualną} o zupełnie odmiennych stopniach swobody. Jedną z propozycji realizacji tej idei jest dualność teorii z cechowaniem i teorii grawitacyjnych, która może być zrealizowana w teorii strun jako tzw. \emph{dualność otwarto-zamknięta}.

Supersymetria oraz dualność są interesujące także dlatego, że uproszczone teorie związane z tymi ideami są często ściśle rozwiązywalne. Bardzo duże nadzieje wiąże się z faktem, iż szczegółowe rozwiązanie takich teorii pozwoli także na zrozumienie najistotniejszych aspektów teorii opisujących realny świat, i rozwiązanie ich w następnej kolejności już bez stosowania uproszczeń. Takie właśnie podejście prezentujemy w niniejszej pracy: jest ona poświęcona ścisłym rozwiązaniom tzw. teorii \emph{topologicznych}, w których wiele skomplikowanych zjawisk kwantowych może być jawnie zanalizowanych w pewien uproszczony sposób.

Istnieją zarówno topologiczne teorie pola jak i strun. Uproszczenia w nich występujące związane są z faktem, iż zazwyczaj teorie te posiadają dużo mniej fizycznych stopni swobody niż teorie określane mianem \emph{fizycznych}, które opisują pewne aspekty Przyrody (jak np. teorie z cechowaniem) lub też wiąże się z nimi takie nadzieje (jak np. teorie superstrun). Tym niemniej, istnieje wiele motywacji dla studiowania teorii topologicznych. Po pierwsze, jak wyżej wspomniano, w wielu przypadkach są one ściśle rozwiązywalne. Po drugie, amplitudy teorii topologicznych są często podzbiorem amplitud odpowiednich teorii fizycznych, a zatem ich rozwiązanie jest automatycznie częściowym rozwiązaniem odpowiednich teorii fizycznych. Ponadto istnieją coraz bardziej przekonujące argumenty na to, iż opis pewnych rzeczywistych zjawisk, na przykład związanych z zachowaniem się materii skondensowanej w bardzo niskich temperaturach, rzeczywiście wymaga zastosowania teorii topologicznych. Rozwój tych teorii ma także niebywały wkład w wiele dziedzin matematyki, i niejednokrotnie prowadził do niespodziewanych i zaskakujących rozwiązań skomplikowanych zagadnień matematycznych. 

W niniejszej rozprawie szczególną uwagę poświęcamy topologicznym teoriom strun. Istnieją dodatkowe motywacje, oprócz tych wymienionych wyżej, by zrozumieć tę właśnie klasę teorii. W ich ramach można ściśle zrozumieć pewne modele dualności otwarto-zamkniętej. Są one głęboko związane z różnymi czterowymiarowymi teoriami pola i pozwalają na wyliczenie wielkości takich jak prepotencjały w teoriach supersymetrycznych typu $\mathcal{N}=2$, superpotencjały w teoriach typu $\mathcal{N}=1$, czy też entropii pewnej klasy czarnych dziur. Z matematycznego punktu widzenia topologiczne teorie strun łączą na pozór bardzo odległe zagadnienia, na przykład teorię Kodairy-Spencera z teoriami niezmienników Gromova-Wittena, Gopakumara-Vafy, oraz Donaldsona-Thomasa. 

Niniejsza praca poświęcona jest pewnej niezwykłej właściwości teorii topologicznych: w niektórych sytuacjach teorie te okazują się być związane z prostymi modelami kryształów znanymi z fizyki statystycznej. Modele takie, odpowiadające topologicznym teoriom pola oraz strun, zdefiniowane są odpowiednio poprzez pewne zespoły dwu- i trójwymiarowych partycji, co pozwala konkretnie zinterpretować amplitudy topologiczne. Ponadto, sumy statystyczne związane z takimi modelami kryształów mogą być ściśle wyznaczone, i są one równe pewnym amplitudom topologicznym. Teorie pola oraz strun dla których znane są modele kryształu zdefiniowane są na tzw. rozmaitościach \emph{Calabi-Yau} i w związku z tym modele te zwane są \emph{kryształami Calabi-Yau}.

Reasumując, niniejsza rozprawa poświęcona jest badaniu kryształów Calabi-Yau w teoriach topologicznych, ze szczególnym uwzględnieniem topologicznych teorii strun. Poniżej zwięźle charakteryzujemy wiele niezależnych idei związanych z opisaną wyżej tematyką rozprawy z nadzieją, że ułatwi to Czytelnikowi lekturę pozostałych rozdziałów. Poza ogólnymi uwagami dotyczącymi teorii strun, wszystkie pozostałe aspekty poruszone w niniejszym wstępnym rozdziale są dużo bardziej szczegółowo opisane w głównej części pracy.

%****************************************************

\section*{Teoria strun}
\addcontentsline{toc}{section}{Teoria strun}

Podstawowe informacje dotyczące teorii strun znaleźć można w podręcznikach \cite{gsw,polchinski}. Najważniejszym jej założeniem jest to, iż źródłem wszystkich cząstek elementarnych oraz ich oddziaływań są drgania jednowymiarowych strun. Poruszając się w czasoprzestrzeni $M$ struny zakreślają dwuwymiarową \emph{powierzchnię świata struny} $\Sigma$. Rozróżniamy dwa rodzaje strun: \emph{zamknięte} i \emph{otwarte}. Powierzchniami świata strun zamkniętych są powierzchnie Riemanna o dowolnym genusie $g$, ewentualnie z rozciągającymi się asymptotycznie końcami reprezentującymi cząstki nadchodzące z nieskończoności. Z kolei powierzchnie świata strun otwartych charakteryzowane są zarówno przez genus $g$ jak też liczbę spójnych składowych brzegu $h$. Współrzędne oraz metrykę na powierzchni świata oznaczamy odpowiednio jako $\sigma^a$ oraz $g_{ab}$ dla $a,b=1,2$. Współrzędne czasoprzestrzenne $X^{\mu}$ definiują odwzorowanie
\be
X^{\mu}:\quad \Sigma \longrightarrow M,   \label{pl-sigma-model}
\ee
natomiast metrykę czasoprzestrzenną oznaczamy $G_{\mu\nu}(X)$. W ogólności teoriom, które opisują odwzorowania w pewną nietrywialną zakrzywioną przestrzeń, nadaje się miano \emph{nieliniowych modeli sigma}, lub po prostu \emph{modeli sigma}. Klasyczne działanie dla strun zamkniętych jest przykładem modelu sigma i można je zapisać jako 
\be
S_{struna} = \frac{1}{4\pi\alpha'} \int_{\Sigma} d^2\sigma\,g^{1/2} g^{ab}G_{\mu\nu}\partial_a X^{\mu} \partial_b X^{\nu}.  \label{pl-S-Polyakov}
\ee
W przypadku strun otwartych należy dodatkowo uwzględnić warunki brzegowe związane z istnieniem końców takich strun. Parametr $\alpha'$ o wymiarze (\emph{długość czasoprzestrzenna})$^2$ jest jedyną arbitralną stałą w teorii strun. Jest on związany z napięciem struny $T$ poprzez $T=(2\pi\alpha')^{-1}$ i przyjmuje się, iż długość struny jest rzędu $\sim\alpha'^{1/2}$. 

\begin{figure}[htb]
\begin{center}
\includegraphics[width=0.3\textwidth]{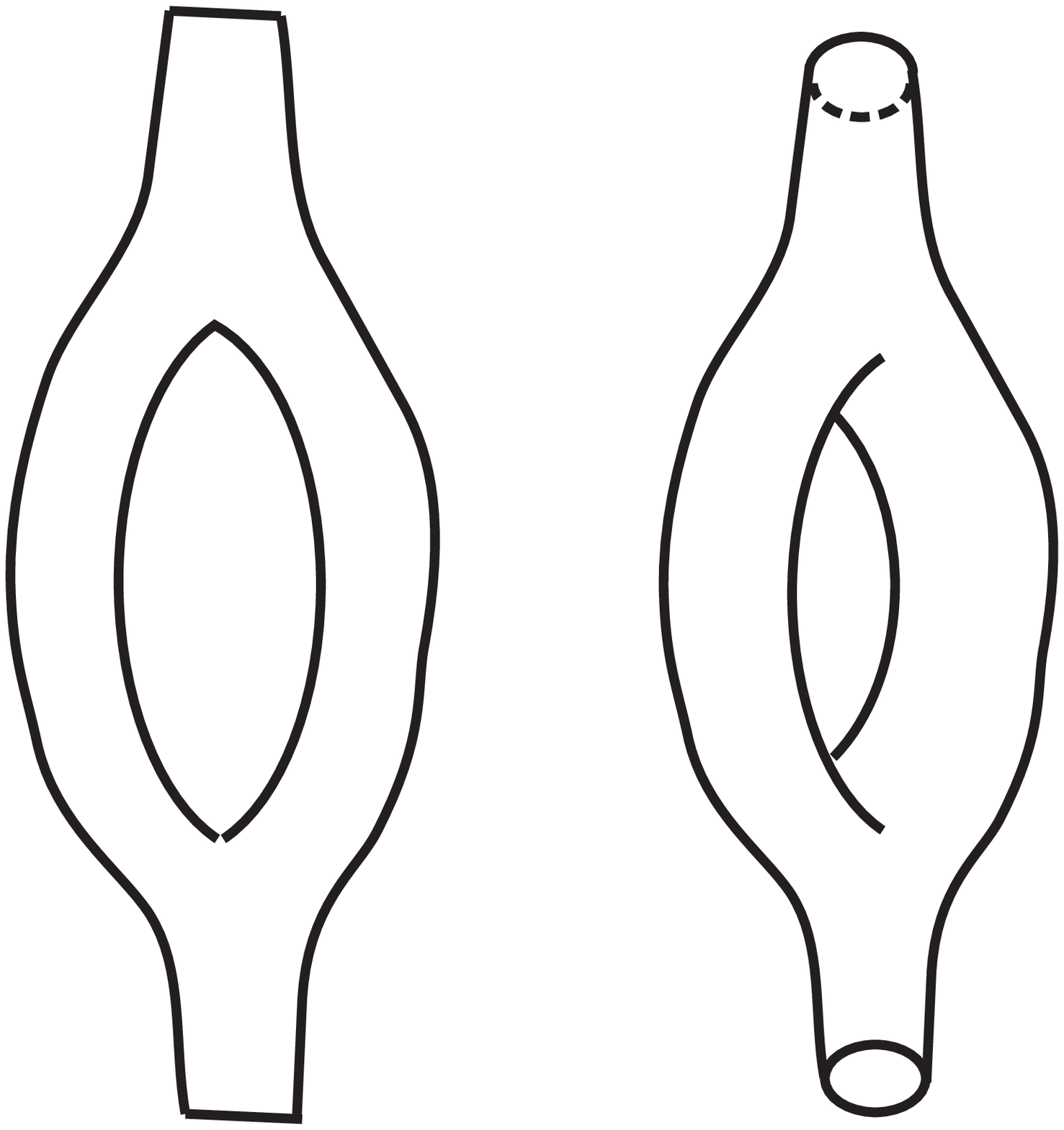}
\caption{Powierzchnie świata strun otwartych (po lewej) oraz zamkniętych (po prawej).} \label{pl-fig-worlds}
\end{center}
\end{figure}

Jednym ze sposobów kwantyzacji teorii strun jest wprowadzenie całki po trajektoriach Polyakova, która w Euklidesowej sygnaturze przyjmuje postać
$$
Z = \int \mathcal{D}X\,\mathcal{D}g\,e^{-S_{struna}}.
$$
Klasyczne działanie $S_{struna}$ jest konforemnie niezmiennicze, w związku z czym teoria kwantowa jest dobrze określona wtedy i tylko wtedy jeśli także jest konforemnie niezmiennicza. Narzuca to bardzo silne warunki na postać czasoprzestrzeni $M$: jej wymiar musi być równy 26 oraz musi ona spełniać próżniowe równania Einsteina
$$
R_{\mu\nu}+\mathcal{O}(\alpha') = 0.
$$

Niemniej jednak, spełnienie powyższych warunków nie zapewnia jeszcze poprawności teorii, gdyż jej najlżejszy stan jest tachionem, którego kwadrat masy jest ujemny. Okazuje się, że zapostulowanie supersymetrii w teorii strun eliminuje z niej tachiony i zapewnia konsystentncję. W tak zwanym formalizmie Ramonda-Neveu-Schwarza wiąże się to z uwzględnieniem superpartnerów $X^{\mu}$ oznaczanych jako  $\psi^{\mu}$, oraz rozpatrywaniem rozszerzonego działania Polyakova z supersymetrią typu $\mathcal{N}=(2,2)$. Uważna analiza takiej teorii pokazuje, że oprócz braku tachionów charakteryzuje się ona supersymetrycznym spektrum cząstek w czasoprzestrzeni, co ma duże znaczenie z fenomenologicznego punktu widzenia. W teorii superstrun znikanie anomalii konforemnej także implikuje konieczność spełnienia próżniowych równań Einsteina w czasoprzestrzeni, aczkolwiek w tym wypadku jej wymiar musi być równy $\textrm{dim}\,M = 10$. W związku tym, aby otrzymać fenomenologiczny model z czterowymiarową czasoprzestrzenią reprezentującą realny Wszechświat, rozpatruje się teorie z 6 wymiarami skompaktyfikowanymi na pewnej rozmaitości, na tyle małej by nie była obserwowalna w rzeczywistości. Warunki zachowania supersymetrii implikują, iż musi to być \emph{rozmaitość Calabi-Yau}, czyli rozmaitość o zespolonym wymiarze 3 i znikającym tensorze Ricciego.

Po wyeliminowaniu tachionów spektrum strunowe składa się z pewnej ilości stanów bezmasowych oraz nieskończonej wieży stanów masywnych o masach rzędu  $\sim\alpha'^{-1/2}$. W granicy punktowej $\alpha'\to 0$ wszystkie masywne stany odprzęgają się i teoria strun redukuje się do tzw. \emph{teorii efektywnej} dla stanów bezmasowych. W teorii strun zamkniętych spektrum stanów bezmasowych składa się ze stanu o spinie 2, który można zidentyfikować z grawitonem odpowiadającym metryce $G_{\mu\nu}$ w działaniu (\ref{pl-S-Polyakov}), antysymetrycznego pola Kalba-Ramonda  $B_{\mu\nu}$, oraz dylatonu $\Phi$, który zadaje strunową stałą sprzężenia $g_s=\exp\,\Phi$. W teorii superstrun występują ponadto tzw. pola R-R o całkowitym spinie oraz pola R-NS o połówkowym spinie. Spektrum strun otwartych jest odmienne z powodu warunków związanych z obecnością ich końców, które pozwalają na wprowadzenie dodatkowych stopni swobody, tzw. \emph{czynników Chana-Patona} związanych z pewną grupą symetrii Liego. Bezmasowy stan w bozonowej teorii strun otwartych ma spin 1 i jego oddziaływania są odtwarzane przez efektywne działanie Yanga-Millsa z symetrią cechowania związaną z powyższą grupą Liego. Stan ten może być zatem zidentyfikowany z polem cechowania  $A^{\mu}$. W teorii supersymetrycznej pojawiają się także odpowiedni partnerzy supersymetryczni.

Jeszcze jednym istotnym składnikiem teorii strun są tzw. \emph{D-brany}. Mogą być one opisane jako podrozmaitości do których zaczepione muszą być końce strun otwartych. Obiekt taki nazywamy $Dp$-braną, jeśli rozciąga się w wymiarze czasowym oraz $p$ wymiarach przestrzennych. $D$-brany są w istocie nieodłącznym składnikiem teorii strun: po pierwsze okazują się one być źródłami dla pól typu R-R, a po drugie pewne symetrie teorii strun wiążą je ze zwykłymi strunami fundamentalnymi. 

Z uważnej analizy wszystkich wspomnianych wyżej warunków spójności wynika, że istnieje 5 niezależnych teorii superstrun. Z punktu widzenia niniejszej pracy najistotniejsze są teorie strun zamkniętych typu IIA i IIB. Teorie te zawierają pola R-R o szczególnych spinach, co związane jest z istnieniem $D$-bran będących ich źródłami o odpowiednich wymiarach. W szczególności, w teorii IIA i IIB $Dp$-brany charakteryzują się odpowiednio przestrzennym wymiarem $p$ parzystym i nieparzystym. Pozostałe konsystentne teorie superstrun to teoria strun otwartych typu I z symetrią  $SO(32)$ oraz dwie tzw. teorie heterotyczne z symetriami  $SO(32)$ i $E_8\times E_8$. Ponadto okazuje się, że wszystkie te teorie powiązane się różnorodnymi dualnościami i wedle pewnej śmiałej hipotezy reprezentują one zaledwie kilka szczególnych granic jednej fundamentalnej teorii, tzw. $M$-teorii.

%****************************************************

\subsubsection*{Kompaktyfikacje Calabi-Yau w teoriach typu II}

Jak wspomnieliśmy powyżej, efektywną, supersymetryczną, czterowymiarową teorię można otrzymać poprzez kompaktyfikację teorii superstrun na rozmaitości Calabi-Yau. Własności teorii efektywnej, takie jak skład cząstek i sposób ich oddziaływania, zdeterminowane są przez geometrię tej rozmaitości. Idea zawarcia infomacji dotyczącej cząstek elementarnych we własnościach pewnego obiektu geometrycznego jest niezwykle interesująca. Ilość rozmaitości Calabi-Yau jest bardzo duża, toteż identyfikacja jednej z nich jako hipotetycznie odpowiadającej stanowi Wszechświata nie jest łatwa. Niemniej jednak, w ostatnich latach prowadzone są intensywne badania w tej dziedzinie \cite{susskind,M-landscape}.

Aby zrozumieć ideę kompaktyfikacji, rozważmy przypadek teorii superstrun typu II na przestrzeni Calabi-Yau $M$. W teorii IIA redukuje się ona do czterowymiarowej teorii supergrawitacji typu $\mathcal{N}=2$ z $h^{1,1}(M)$ multipletami wektorowymi oraz $(h^{2,1}(M)+1)$ hipermultipletami. W teorii IIB ilości tych multipletów są przyporządkowane przeciwnie. Multiplety te zawierają pola skalarne odpowiadające tak zwanym \emph{modułom} przestrzeni $M$, które determinują jej geometryczne własności, takie jak kształt i rozmiar. Nietrywialna konfiguracja pól skalarnych w czterowymiarowej czasoprzestrzeni odpowiada nietrywialnemu rozwłóknieniu nad nią rozmaitości Calabi-Yau. W ogólności zachowanie tych pól zadawane jest poprzez różne potencjały. Na przykład, dla jednego modułu $\phi$ odpowiadającego całkowitemu rozmiarowi przestrzni $M$, z teorii strun otrzymujemy konkretną postać tzw. \emph{potencjału K\"ahlera} dla związanego z nim multipletu
$$
K = -3\,\textrm{ln}(\phi + \phi^*).
$$
Mimo iż pola skalarne rzadko występują w Przyrodzie, istotnymi ich przykładami które mogłyby wynikać z kompaktyfikacji strunowych są pole Higgsa oraz hipotetyczny \emph{inflaton} związany z teorią inflacji kosmicznej. Konkretnych modele fenomenologiczne wymagają oczywiście bardziej skomplikowanej postaci potencjałów dla pól skalarnych. Mechanizmy pozwalające generować takie potencjały są w ostatnich latach szczegółowo badane; tego typu przykład dyskutowany jest w pracy \cite{F-theory}.

Podjęto również wiele prób odtworzenia całego spektrum Modelu Standardowego poprzez kompaktyfikacje strunowe. Szczególnie obiecującą klasą takich modeli są tzw. \emph{scenariusze branowe}. Mają one kilka charakterystycznych właściwości. Aby uwzględnić bozony cechowania, zakłada się, że nasza czasoprzestrzeń pokrywa się z $D3$-braną. Skład pól materii zadawany jest poprzez nawinięcie --- także nawzajem przecinających się --- bran na cykle przestrzni Calabi-Yau. Mimo iż teorie typu II prowadzą do supersymetrii typu  $\mathcal{N}=2$, bardziej prawdopodobny fenomenologicznie przypadek typu  $\mathcal{N}=1$ można uzyskać poprzez kompaktyfikacje na przestrzeniach z osobliwościami, na przykład tzw. \emph{orbifoldach}. Interesująca próba odtworzenia Modelu Standardowego przedstawiona jest w pracy \cite{SM-brane}.

%****************************************************

\subsubsection*{Dualność teorii z cechowaniem i teorii grawitacji}

Jak wyżej wspomniano, bozony cechowania związane są z bezmasowymi stanami strun otwartych, natomiast grawitony z bezmasowymi stanami strun zamkniętych. W teorii strun znana jest odpowiedniość pomiędzy opisem w języku strun otwartych i zamkniętych, z której zatem wynikają dualności pomiędzy teoriami z cechowaniem i teoriami grawitacji. Po raz pierwszy tego typu dualność rozważał 't Hooft już w 1974 roku \cite{thooft}, znajdując ogólne argumenty za tym, iż każdej teorii z cechowaniem powinna odpowiadać pewna teoria strun zamkniętych. Jego oryginalną motywacją było znalezienie teorii grawitacyjnej dualnej do chromodynamiki kwantowej, która pozwoliłaby zrozumieć naturę oddziaływań silnych w niskich energiach. Mimo że problem ten wciąż nie jest rozwiązany, idea takiej dualności stała się w ostatnich latach szerokim polem badań, przede wszystkim dzię odkryciu Maldaceny, iż teorią dualną do supersymetrycznej teorii z cechowaniem typu $\mathcal{N}=4$ jest teoria grawitacji na przestrzeni AdS (anty de Sittera) \cite{adscft}. Wkrótce znaleziono wiele podobnych związków dla całej klasy tego typu teorii, które określa się mianem odpowiedniości AdS/CFT. Jako że bozonowa część supersymetrycznej teorii typu $\mathcal{N}=4$ jest zwykłą teorią Yanga-Millsa, taką samą jak w przypadku chromodynamiki kwantowej, odkrycie Maldaceny ożywiło nadzieję na rozwiązanie zagadki oddziaływań silnych zgodnie z oryginalną ideą 't Hoofta. Tym niemniej, duża ilość supersymetrii w teorii typu $\mathcal{N}=4$ powoduje, że jest ona konforemnie niezmiennicza. Oznacza to w szczególności, że jej stała sprzężenia nie zależy od skali, co jest zachowaniem zupełnie odmiennym niż wyjściowy problem nieperturbacyjnej stałej sprzężenia dla oddziaływań silnych. Jednakże istnieją pewne zjawiska związane z chromodynamiką kwantową --- jak na przykład zachowanie się plazmy gluonowo-kwarkowej --- dla których nawet przewidywania teorii supersymetrycznej wydają się mieć pewne znaczenie. Można mieć zatem nadzieję, iż idea dualności między teoriami z cechowaniem oraz grawitacją okaże się słyszna przy opisie niektórych aspektów Przyrody. 

Istnieje wiele innych modeli strunowych oprócz teoriami opisanych powyżej. Szczególnie interesującym przykładem są przedstawione w dalszej częsci topologiczne teorie strun, będące głównym tematem niniejszej rozprawy. W teoriach tych znany jest pewien szczególny przykład dualności otwarto-zamkniętej, zwanej dualnością Gopakumara-Vafy, którą można analizować w ścisły i ilościowy sposób. Zrozumienie istotnych aspektów dualności otwarto-zamkniętej dla tego przypadku może okazać się cenne przy analizie bardziej skomplikowanych modeli, takich jak hipoteza AdS/CFT, jak też natury oddziaływań silnych i kwantowej grawitacji. 

%****************************************************

\subsubsection*{Podsumowanie}

W tej części podsumowaliśmy wyłaniający się z teorii strun obraz zjawisk fundamentalnych. Niestety, w chwili obecnej nie jest możliwe dokonanie jego bezpośredniej eksperymentalnej weryfikacji. Niemniej jednak, powyżej zwróciliśmy uwagę na wiele silnych ograniczeń, które każda teoria pretendująca do miana fundamentalnej powinna spełniać. Po pierwsze, w odpowiednich granicach teoria taka powinna odtwarzać znane prawa teorii względności i kwantowej teorii pola. Po drugie, sama jej struktura matematyczna musi być spójna, co narzuca niebanalne warunki z uwagi na skomplikowany aparat matematyczny stosowany w omawianych teoriach. Teoria strun spełnia te wszystkie ograniczenia przy stosunkowo niewielkiej liczbie eleganckich założeń. Ponadto, w jej ramach można zrealizować wiele fenomenologicznie atrakcyjnych modeli. Pozwala to żywić nadzieję, iż rzeczywiście jest ona dobrym kandydatem na teorię fundamentalną.

%****************************************************

\section*{Topologiczne teorie pola}
\addcontentsline{toc}{section}{Topologiczne teorie pola}

Kwantową teorię pola określoną na rozmaitości $M$ nazywamy \emph{topologiczną}, jeśli jej obserwable nie zależą od ciągłych deformacji metryki przestrzeni $M$, i w związku z tym są topologicznymi niezmiennikami $M$ lub pewnych obiektów z tą przestrzenią związanych. Istnieją zarówno topologiczne teorie pól jak i strun, przy czym znane są dwa odmienne sposoby konstrukcji tych pierwszych, które zwięźle przestawiamy poniżej.

Jedna klasa topologicznych teorii pola to teorie takie, których lagranżjany można zapisać całkowicie bez użycia metryki przestrzeni na której są one określone. Taka sytuacja zachodzi na przykład wtedy, gdy lagrażjan danej teorii można zapisać jedynie przy pomocy form różniczkowych, i ponadto miara w całce po trajektoriach nie wprowadza żadnej dodatkowej zależności od metryki. Przykładem takiej teorii jest teoria Cherna-Simonsa, której zaskakujące rozwiązanie znalezione przez Wittena \cite{witten-cs} miało ogromny wpływ na tę gałąź fizyki matematycznej. Działanie Cherna-Simonsa dla koneksji cechowania $A$ ma postać 
\be
S_{CS} = \frac{k}{4\pi} \int_M \textrm{Tr}\,\big(A\wedge dA + \frac{2}{3}A\wedge A\wedge A \big),   \label{pl-CS}
\ee
przy czym warunek niezmienniczości względem cechowania w teorii kwantowej implikuje iż $k$ musi być liczbą całkowitą.

\begin{figure}[htb]
\begin{center}
\includegraphics[width=0.4\textwidth]{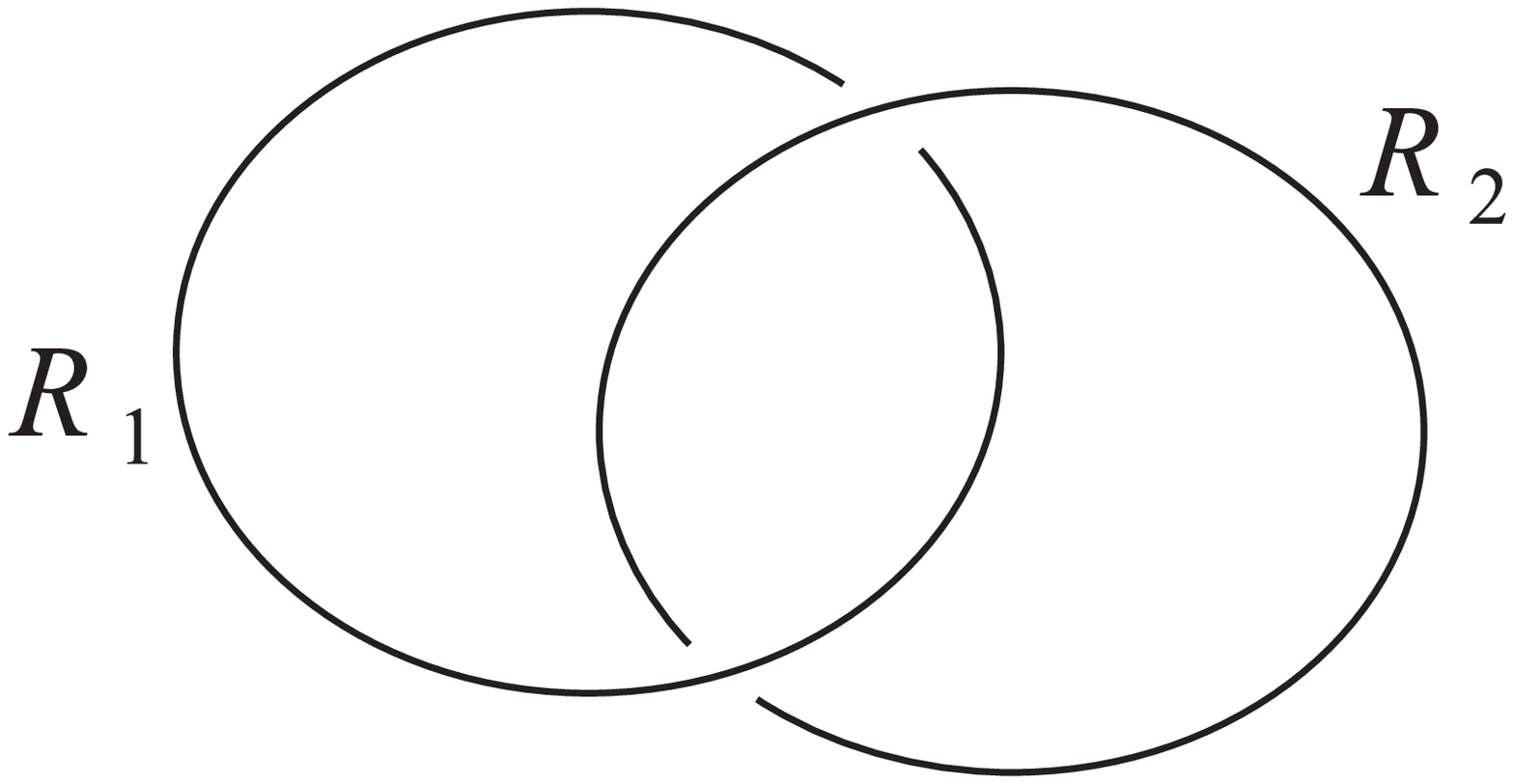}
\caption{Splot Hopfa.} \label{pl-fig-hopf-intro}
\end{center}
\end{figure}

Ważną klasą obserwabli w teorii Cherna-Simonsa są wartości oczekiwane pętli Wilsona
$$
\mathcal{W}^K_R = \langle \textrm{Tr}_R P e^{\oint_K A} \rangle = \int \mathcal{D}A \, e^{i S_{CS}}\,\textrm{Tr}_R P e^{\oint_K A},
$$
obliczane dla pewnej reprezentacji $R$ grupy cechowania i dla zamkniętej pętli $K$. Z matematycznego punktu widzenia pętla taka definiuje węzeł, a niezmienniczość powyższej amplitudy ze względu na ciągłe deformacje metryki oznacza, iż amplituda ta jest pewnym niezmiennikiem węzła $K$. Najprostszym przykładem węzła jest tak zwany \emph{węzeł trywialny}, czyli zwykła niezawęźlona pętla homeomorficzna z okręgiem. Z kolei jako \emph{splot} definiuje się obiekt złożony z kilku węzłów i analogicznie wartości oczekiwane iloczynu kilku pętli Wilsona w teorii Cherna-Simonsa są niezmiennikami splotów. Prostym przykładem splotu jest tzw. \emph{splot Hopfa} pokazany na rysunku \ref{pl-fig-hopf-intro}, złożony z dwóch  zawęźlonych węzłów trywialnych, którego niezmienniki można obliczyć w teorii Cherna-Simonsa jako
\be
\mathcal{W}_{R_1  R_2} = \langle \textrm{Tr}_{R_1} P e^{\oint_{K_1} A}\, \textrm{Tr}_{R_2} P e^{\oint_{K_2} A} \rangle. \label{pl-CS-hopflink}
\ee
Jeśli jedna z reprezentacji $R_1,R_2$ jest trywialna, niezmiennik ten redukuje się do niezmiennika węzła trywialnego  $\mathcal{W}_{R \bullet}$.

%This connection between quantum Chern-Simons theory and knot invariants was made precise by E. Witten in 1989, and it had enormous impact both on physics and mathematics. 

Druga klasa topologicznych teorii pola to tzw. \emph{teorie z twistem}. Konstruowane są one poprzez zmieszanie symetrii czasoprzestrzennych z symetrią $R$ w teoriach supersymetrycznych w taki sposób, że obserwable nowej teorii można utożsamić z klasami kohomologii pewnego operatora $Q$ typu BRST, a obserwable związane z metryką są kohomologicznie trywialne. Część wyników przedstawionych w niniejszej rozprawie odnosi się do supersymetrycznych teorii typu  $\mathcal{N}=4$ z twistem wprowadzonych w pracy \cite{V-W}.

%****************************************************

\section*{Topologiczne teorie strun}
\addcontentsline{toc}{section}{Topologiczne teorie strun}

W supersymetrycznych modelach sigma typu $\mathcal{N}=(2,2)$, będących uogólnieniami teorii odwzorowań typu (\ref{pl-sigma-model}), także można wprowadzić powyżej opisany twist \cite{top-sigma,witten-mirror}. Topologiczne teorie strun są zdefiniowane poprzez rozwinięcie w genusie $g$, tak że amplituda  $F_g(t^I)$ dana jest przez całkę po trajektoriach analogiczną do  (\ref{pl-S-Polyakov}), lecz zawierającą działanie dla supersymetrycznego modelu sigma z twistem. Całkowita energia swobodna strun topologicznych zdefiniowana jest poprzez szereg
\be
F_{top}(t^I) = \sum_{g=0} F_g(t^I) \, g_s^{2g-2},  \label{pl-F-top-g}
\ee
gdzie parametr $g_s$ mierzy wkłady odpowiadające poszczególnym genusom, i jest interpretowany jako strunowa stała sprzężenia. Suma statystyczna strun topologicznych dana jest jako
\be
Z_{top} = e^{F_{top}(t^I)}.          \label{pl-Z-top-g}
\ee
Konstrukcja topologicznych teorii strun otwartych jest analogiczna, aczkolwiek należy uwzględnić w niej powierzchnie Riemanna z brzegiem. W powyższych wzorach $t^I$ są tymi samymi modułami rozmaitości Calabi-Yau, które wprowadziliśmy w dyskusji kompaktyfikacji teorii superstrun powyżej. 

W konstrukcji strun topologicznych także pojawiają się różnorakie warunki konsystencji. Wynika z nich między innymi, że struny te mogą się propagować na rozmaitościach Calabi-Yau o zespolonym wymiarze 3. W ogólności przestrzeń modułów rozmaitości Calabi-Yau składa się z dwóch niezależnych części, odpowiadającym tzw. deformacjom K\"ahlera oraz deformacjom zespolonym. Deformacje te związane są odpowiednio z modyfikacją rozmiarów oraz kształtu cyklów homologii danej rozmaitości. Jak się okazuje, amplitudy w teoriach strun topologicznych zależą od deformacji tylko jednej z tych klas. W związku z tym istnieją dwa typy topologicznych strun, tzw. \emph{model A} oraz \emph{model B}, takie że:
\begin{itemize}
\item W modelu A moduły $t^I$ odpowiadają jedynie deformacjom K\"ahlera, a amplitudy obliczane w tych teoriach nie zależą od deformacji zespolonych.
\item W modelu B moduły $t^I$ odpowiadają jedynie deformacjom zespolonym, a amplitudy obliczane w tych teoriach nie zależą od deformacji K\"ahlera.
\end{itemize}
Analogicznie jak w zwykłych teoriach strun omawianych wcześniej, w teoriach otwartych strun topologicznych końce strun muszą być przymocowane do tzw. bran. W modelu A brany takie nazywane są A-branami i okazuje się, że muszą one być nawinięte na trójwymiarowe (rzeczywiste) podrozmaitości lagranżowskie przestrzeni Calabi-Yau. Brany w modelu B nazywane są B-branami i muszą być nawinięte na zespolone podrozmaitości przestrzeni Calabi-Yau, więc w szczególności są one parzystego wymiaru rzeczywistego. Ponadto, teorie modelu A oraz B są z sobą związane tzw. \emph{symetrią lustrzaną}, wedle której wszystkie rozmaitości Calabi-Yau można pogrupować w pary, takie że moduły K\"ahlera jednej z rozmaitości w danej parze odwzorowywane są w moduły zespolone drugiej rozmaitości i \emph{vice versa}. Manifestacją symetrii lustrzanej w topologicznych teoriach strun jest równość odpowiednich amplitud modelu A dla jednej rozmaitości oraz amplitud modelu B dla drugiej rozmaitości w danej parze. Ponadto symetria lustrzana odwzorowuje A-brany w B-brany i \emph{vice versa}.

%****************************************************

\subsubsection*{Związek z teoriami superstrun}

Fakt, iż czasoprzestrzeń w teorii strun topologicznych musi być rozmaitością Calabi-Yau o zespolonym wymiarze 3, jest odpowiednikiem dyskutowanego wcześniej warunku istnienia 10-wymiarowej czasoprzestrzeni w teorii superstrun. W istocie 6 rzeczywistych wymiarów skompaktyfikowanych na rozmaitości Calabi-Yau w teorii superstrun w celu uzyskania efektywnej czterowymiarowej teorii można w ścisły sposób utożsamić z trzema zespolonymi wymiarami rozmaitości Calabi-Yau w teorii strun topologicznych. Zbieżność ta jest źródłem głębokiego związku pomiędzy teorią superstrun i strun topologicznych, i w konsekwencji teorie strun topologicznych zawierają wiele interesujących informacji o efektywnych teoriach w 4 wymiarach ortogonalnych do rozmaitości Calabi-Yau w kompaktyfikacjach superstrunowych. Liczne związki z teoriami czterowymiarowymi są także silną motywacją dla dogłębnego studiowania strun topologicznych, niezależnie od ich pozostałych właściwości. Oto kilka przykładów takich właśnie związków:
\begin{itemize}
\item Moduły K\"ahlera i zespolone, od których zależą amplitudy w modelu A i B zdefiniowane na rozmaitości Calabi-Yau, są związane odpowiednio z multipletami wektorowymi oraz hipermultipletami w kompaktyfikacji teorii IIA na tej samej rozmaitości. W teorii IIB multiplety te są zamienione rolami. 
\item Amplitudy $F_g(t^I)$ wprowadzone we wzorze (\ref{pl-F-top-g}) okazują się być podzbiorem amplitud superstrunowych związanych z rozpraszaniem grawitonów i ich supersymetrycznych partnerów (tzw. grawifotonów) w efektywnej teorii supersymetrycznej teorii pola typu $\mathcal{N}=2$ \cite{agnt,bcov}. 
\item Energia swobodna strun topologicznych $F_{top}$ na odpowiednio dobranej rozmaitości Calabi-Yau odtwarza tzw. prepotencjał, który zawiera ścisłą nieperturbacyjną informację o rozwiązaniach instantonowych w odpowiedniej supersymetrycznej teorii z cechowaniem typu $\mathcal{N}=2$ \cite{Nek,Nek-Ok}. 
\item Energia swobodna $F_{top}$ wyznacza także wkład do superpotencjału dla multipletów związanych z modułami $t^I$ rozmaitości Calabi-Yau $M$ w teorii z cechowaniem typu $\mathcal{N}=1$ w obecności strumieni pewnych pól przechodzących przez cykle $M$ \cite{gvw}. Ponadto, w wyniku szczególnej realizacji dualności 't Hoofta która odwzorowuje strumienie pól w $D$-brany \cite{largeN}, superpotencjał ten można zinterpretować jako superpotencjał dla stanu związanego pól cechowania w odpowiedniej teorii typu $\mathcal{N}=1$. 
\item W teoriach supergrawitacji możliwe jest opisanie mikroskopowych stopni swobody pewnej klasy czarnych dziur jako zespołów $D$-bran. W szczególności zostało zapostulowane, że suma statystyczna dla tych czarnych dziur $Z_{czarna\ dziura}$ związana jest z suma statystyczną odpowiedniej topologicznej teorii strun \cite{OSV}
\be
Z_{czarna\ dziura} = |Z_{top}|^2.     \label{pl-Z-bh}
\ee
\end{itemize}

Mimo iż analiza powyższych związków nie jest głównym tematem niniejszej rozprawy, wyniki tu przedstawione zawierają szereg amplitud dla strun topologicznych, które w efekcie mogą mieć znaczenie także w pewnych teoriach czterowymiarowych.

%****************************************************

\subsubsection*{Dualność Gopakumara-Vafy}
%\addcontentsline{toc}{section}{Dualność Gopakumara-Vafy}

Podobnie jak w zwykłych teoriach strun, teorie efektywne związane z bezmasowymi modami strun otwartych i zamkniętych istnieją także dla strun topologicznych. W istocie wszystkie stany masywne w teoriach strun topologicznych odprzęgają się od fizycznych amplitud, w związku z czym odpowiednie teorie efektywne są jednocześnie teoriami \emph{ścisłymi}. Takie teorie efektywne mają także topologiczny charakter i często są o wiele prostsze do rozwiązania niż związane z nimi teorie strun. Fakt ten leży u podstaw wielu wyników uzyskanych w niniejszej pracy.

Jak zostało wykazane przez Wittena w pracy \cite{cs-string}, dyskutowana wcześniej teoria Cherna-Simonsa jest ścisłą teorią efektywną dla otwartych strun topologicznych. W przypadku modelu A --- który jest głównym obiektem naszych zainteresowań --- tę odpowiedniość można podsumować jako następującą równoważność
\be
\begin{array}{|ccc|} \hline
\begin{array}{c}
\textrm{otwarte struny topologiczne } \\ 
\textrm{określone na}\  T^* M
\end{array}
& \iff  &  
\begin{array}{c}
\textrm{teoria Cherna-Simonsa } \\ 
\textrm{określona na}\ M  
\end{array}   
 \\ \hline            
\end{array}               \label{pl-cs-effective}
\ee
gdzie $M$ jest lagranżowską podrozmaitością rzeczywistego wymiaru 3 pełnej wiązki kostycznej $T^*M$ (będącej rozmaitością Calabi-Yau o wymiarze zespolonym 3), przy czym na cykl $M$ jest nawiniętych $N$ A-bran w teorii strun otwartych. W teorii Cherna-Simonsa $N$ odpowiada rzędowi grupy cechowania.

\begin{figure}[htb]
\begin{center}
\includegraphics[width=0.5\textwidth]{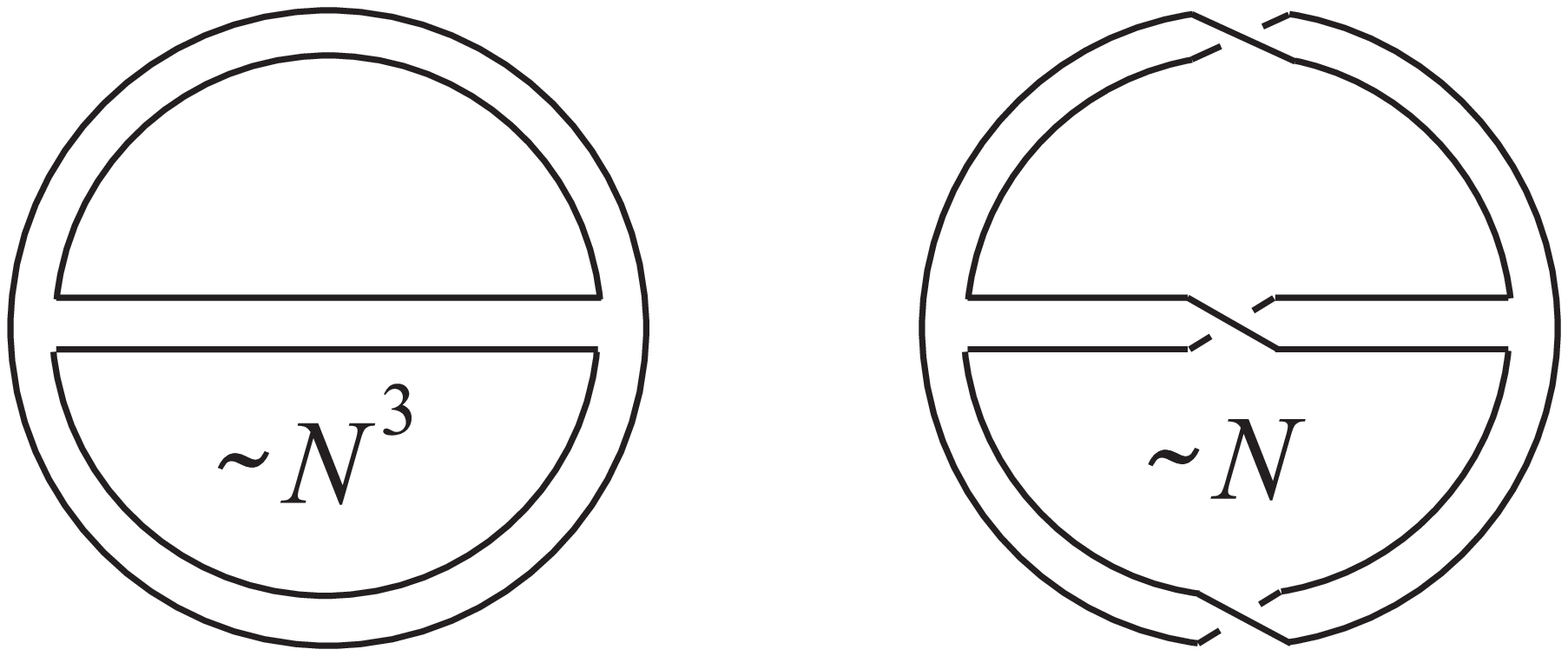}
\caption{Diagram planarny w notacji 't Hoofta po lewej stronie można narysować na sferze, czyli powierzchni o genusie $g=0$. Diagramy które nie są planarne można narysować na powierzchni o większym genusie; w przykładzie po prawej stronie powierzchnią tą jest torus o genusie $g=1$. Wkład od ustalonego diagramu w notacji 't Hoofta jest proporcjonalny do $\lambda^{E-V} N^h$, gdzie $E,V$ i $h$ są odpowiednio liczbą propagatorów, wierzchołków i spójnych składowych brzegu.} \label{pl-fig-diagrams-intro}
\end{center}
\end{figure}

Aby kontynuować nasze rozważania dotyczące topologicznych teorii strun musimy przypomnieć słynną dualność 't Hoofta pomiędzy teorią z cechowaniem a pewną teorią strun zamkniętych \cite{thooft}, w granicy dużego rzędu grupy cechowania $N$. Związek ten jest oczywisty jeśli dokona się rozwinięcia amplitud teorii z cechowaniem w potęgach $N$. Rozwinięcie takie jest szczególnie łatwo uzyskać w tzw. notacji 't Hoofta, która różni się od notacji zwykłych diagramów Feynmana jedynie poszerzeniem linii reprezentujących pola cechowania. W efekcie takiego poszerzenia powstają dwie krawędzie, które reprezentują dwa indeksy reprezentacji dołączonej. Przykłady diagramów w notacji 't Hoofta przedstawione są na rysunku \ref{pl-fig-diagrams-intro}. Najistotniejszą obserwacją jest to, że amplituda dla każdego takiego diagramu jest proporcjonalna do ustalonej potęgi $N$ oraz stałej sprzężenia $\lambda$ i może być narysowana na powierzchni Riemanna o genusie $g$ i $h$ spójnych składowych brzegu. Energia swobodna teorii jest sumą takich właśnie próżniowych diagramów oznaczanych przez $F_{g,h}$
$$
F^{pert} = \sum_{g,h}  F_{g,h} \,\lambda^{2g-2} t^h,
$$
gdzie $t= N\lambda$ jest tzw. \emph{stałą sprzężenia 't Hoofta}. Istotą pomysłu 't Hoofta jest obserwacja, iż w granicy dużego $N$ wykonanie w powyższej amplitudzie sumy po $h$ może być zinterpretowane jako wypełnienie wszystkich dziur w odpowiednich powierzchniach Riemanna, co prowadzi do rozwinięcia w genusie dla powierzchni bez brzegu. Dla teorii Cherna-Simonsa można zapisać
\be
\quad F^{CS}_g(t)=F^{non-pert}_g + \sum_{h=0}^{\infty} F_{g,h}t^h,  \label{pl-F-closed-intro}
\ee
gdzie $F^{non-pert}_g$ są wyrazami nieperturbacyjnymi w rozwinięciu 't Hoofta. W ogólności takie rozwinięcie w genusie jest charakterystyczne dla teorii strun zamkniętych, jeśli $F_g(t)$ zinterpretuje się jako amplitudę strun zamkniętych dla genusu $g$, natomiast parametr $t$ utożsami się z pewnym modułem czasoprzestrzennej rozmaitości tej teorii strun. W szczególności, identyfikując stany strun zamkniętych ze wzbudzeniami grawitacyjnymi, prowadzi to do dualności pomiędzy daną teorią cechowania i teorią grawitacyjną.

Duża część wyników niniejszej pracy związana jest z niezwykle interesującym przykładem dualności 't Hoofta znalezionym przez Gopakumara i Vafę dla teorii topologicznych \cite{G-V-transition}. Zidentyfikowali oni teorię dualną do teorii Cherna-Simonsa na $S^3$ jako model A topologicznych strun zamkniętych na rozmaitości zwanej \emph{rozwiązanym konifoldem}. Rozwiązany konifold jest prostym, aczkolwiek nietrywialnym przykładem rozmaitości Calabi-Yau zawierającym nietrywialną sferę $S^2$. W tej odpowiedniości stałe sprzężenia teorii Cherna-Simonsa oraz strun są z sobą identyfikowane, natomiast stała sprzężenia 't Hoofta $t$ jest identyfikowana z rozmiarem sfery $S^2$ w rozwiązanym konifoldzie. W szczególności, Gopakumar i Vafa pokazali \emph{explicite} równość rozwinięcia energii swobodnej dla strun topologicznych (\ref{pl-F-top-g}) oraz zsumowanego rozwinięcia Cherna-Simonsa (\ref{pl-F-closed-intro}) 
$$
F^{CS}_g(t) = F_{g}^{konifold}(t).
$$
Podsumowując, w modelu A strun topologicznych 
\be
\begin{array}{|ccc|} \hline
\begin{array}{c}
\textrm{teoria Cherna-Simonsa} \\ 
\textrm{określona na}\  S^3
\end{array}
& \iff  &  
\begin{array}{c}
\textrm{zamknięte struny topologiczne} \\ 
\textrm{na rozwiązanym konifoldzie} 
\end{array}   
 \\ \hline            
\end{array}               \label{pl-cs-closed}
\ee

\begin{figure}[htb]
\begin{center}
\includegraphics[width=0.8\textwidth]{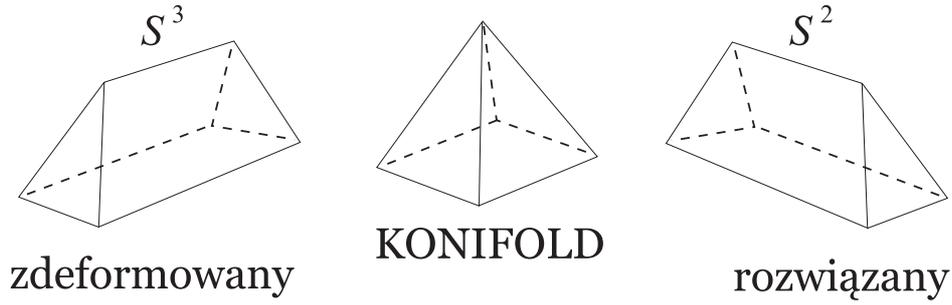}
\caption{Przemiana Gopakumara-Vafy pomiędzy modelem A otwartych strun topologicznych na zdeformowanym konifoldzie oraz zamkniętymi strunami topologicznymi na rozwiązanym konifoldzie.} \label{pl-fig-transition}
\end{center}
\end{figure}

Powyższe obserwacje można zinterpretować w następujący sposób. Ze związku (\ref{pl-cs-effective}) dla $M=S^3$ oraz (\ref{pl-cs-closed}) wynika, iż w granicy dużych $N$ model A topologicznych strun \emph{otwartych} na $T^*S^3$ jest równoważny modelowi A topologicznych strun \emph{zamkniętych} na rozwiązanym konifoldzie. Jest to przykład dualności \emph{otwarto-zamkniętej} pomiędzy dwoma teoriami strun. Ponadto, $T^*S^3$ jest rozmaitością Calabi-Yau zwaną \emph{zdeformowanym konifoldem}, zawierającą nietrywialną trójsferę $S^3$. Zarówno konifold rozwiązany jak i zdeformowany są dwoma wygładzeniami osobliwego konifoldu, który jest opisywany następującym równaniem w $\C^4$ ze współrzędnymi $(z_1,\ldots,z_4)$
$$
z_1^2 + z_2^2 + z_3^2 + z_4^2 = 0.
$$
Powyższa dualność może być zobrazowana jako ciągły proces, w którym trójsfera $S^3$ w zdeformowanym konifoldzie kurczy się całkowicie, po czym następuje rozszerzanie sfery $S^2$ odpowiadającej konifoldowi rozwiązanemu. Równocześnie znika $N$ A-bran które były nawinięte na $S^3$, czemu towarzyszy pojawienie się stopni swobody strun zamkniętych. Proces ten, zwany przemianą geometryczną Gopakumara-Vafy, pokazany jest schematycznie na rysunku \ref{pl-fig-transition}.

%*********************************************************

\subsubsection*{Wierzchołek topologiczny}
%\addcontentsline{toc}{section}{Wierzchołek topologiczny}

\begin{figure}[htb]
\begin{center}
\includegraphics[width=0.3\textwidth]{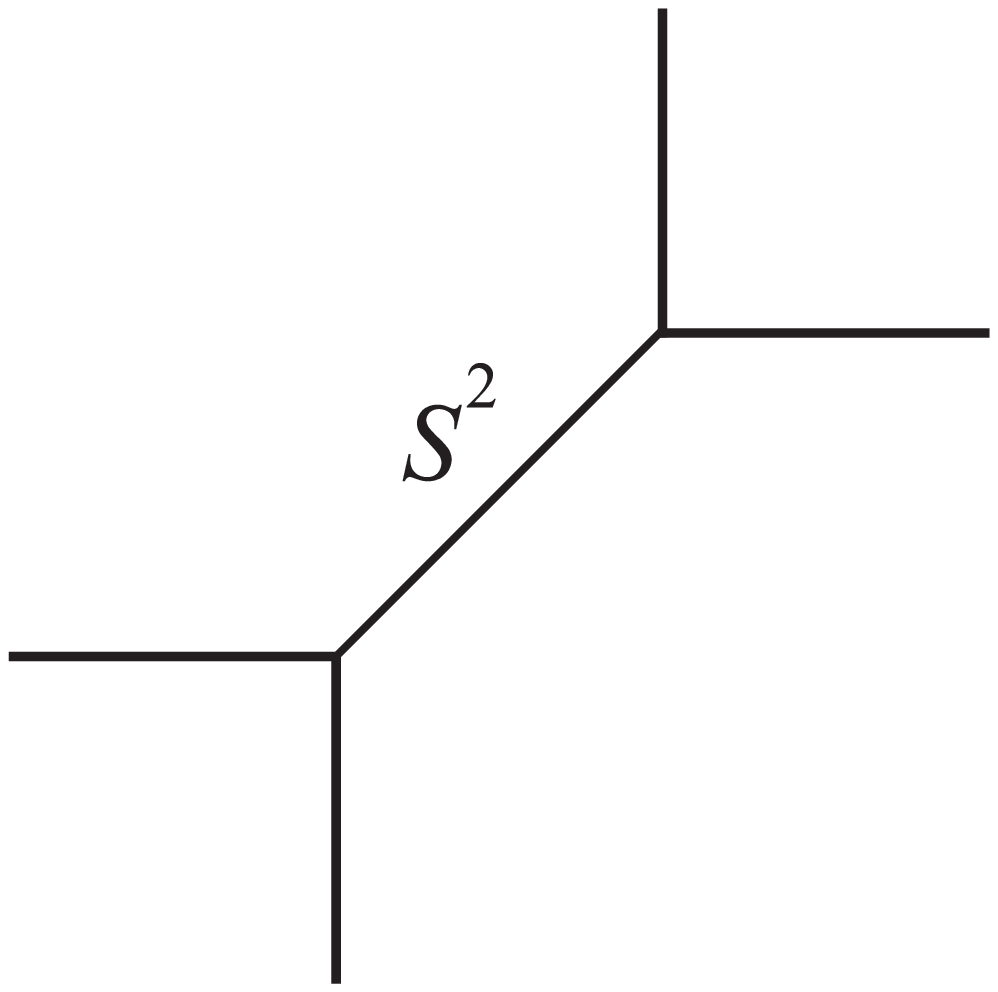}
\caption{Rozwiązany konifold jest reprezentowany przez diagram złożony z dwóch wierzchołków reprezentujących lokalne układy współrzędnych $\C^3$ połączonych wzdłuż wspólnego odcinka odpowiadającego nietrywialnej sferze $S^2$.} \label{pl-fig-coni}
\end{center}
\end{figure}

Z dualności Gopakumara-Vafy wynika równość sum statystycznych dla teorii zamkniętych strun topologicznych na rozwiązanym konifoldzie oraz teorii Cherna-Simonsa na $S^3$. Okazuje się, że związek ten można uogólnić w sposób, który pozwala na obliczanie amplitud strun topologicznych w modelu A dla całej klasy niezwartych, \emph{torycznych} rozmaitości Calabi-Yau. Z definicji, rozmaitość toryczną można przedstawić jako rozwłóknienie torusów nad pewną bazą. Pełna informacja o takim rozwłóknieniu może być zawarta w postaci dwuwymiarowego diagramu, którego krawędzie reprezentują podrozmaitości przestrzeni bazowej, wzdłuż których jedno z działań $S^1$ jest zdegenerowane. Podstawowym elementem takiego diagramu jest wierzchołek z trzema odnogami, który reprezentuje pojedynczy lokalny układ współrzędnych $\C^3$. Wzajemne sklejenie takich lokalnych układów współrzędnych jest reprezentowane na diagramie jako sklejenie dwóch wierzchołków wzdłuż wspólnego odcinka reprezentującego nietrywialną sferę $S^2$ zawartą w rozmaitości. Cały dwuwymiarowy diagram zawiera informację, jak lokalne układy współrzędnych $\C^3$ są sklejone w pełną toryczną rozmaitość Calabi-Yau. Przykład takiego diagramu dla rozwiązanego konifoldu pokazany jest na rysunku \ref{pl-fig-coni}.

Diagramatyczny sposób przedstawienia rozmaitości Calabi-Yau można również uogólnić do obliczania różnorakich amplitud strun topologicznych, poprzez przypisanie pewnej konkretnej amplitudy do pojedynczego wierzchołka. Amplituda taka jest funkcją $q=e^{-g_s}$ gdzie $g_s$ jest strunową stałą sprzężenia. Jest ona oznaczana jako $C_{R_1 R_2 R_3}$ i reprezentuje konfiguracje strun otwartych o końcach umieszczonych na trzech układach bran, których konfiguracje są zakodowane w diagramach Younga $R_i,\  i=1,2,3$. Amplituda ta nazywana jest \emph{wierzchołkiem topologicznym} i została wyprowadzona w pracy \cite{vertex}. Wyprowadzenie to w całości opiera się na dualności Gopakumara-Vafy, a wierzchołek topologiczny wyrażony jest poprzez niezmienniki splotu Hopfa (\ref{pl-CS-hopflink}) w granicy dużego $N$, co uwidacznia związek strun topologicznych z teorią Cherna-Simonsa. W istocie obliczanie amplitud strun topologicznych w formalizmie wierzchołka topologicznego przypomina reguły Feynmana, przy czym pojedynczy diagram reprezentuje całą rozmaitość Calabi-Yau, a jego wierzchołki i propagatory odpowiadają strunom propagującym się w lokalnych układach współrzędnych oraz nawiniętym na nietrywialne sfery wewnątrz rozmaitości. Formalizm wierzchołka topologicznego pozwala również na wyznaczenie amplitud topologicznych w obecności bran.

Wiele wyników otrzymanych przy użyciu wierzchołka topologicznego zostało porównanych z bezpośrednimi rachunkami w topologicznej teorii strun, oraz sprawdzonych innymi nietrywialnymi metodami. Zgodność uzyskana we wszystkich tych przypadkach jest bardzo silnym potwierdzeniem dualności otwarto-zamkniętej. Poza niebywałą możliwością ilościowego sprawdzenia przewidywań tej dualności, formalizm wierzchołka topologicznego niezmiernie upraszcza rachunki amplitud strun topologicznych, z czego wielokrotnie będziemy korzystać w niniejszej pracy. 

%*********************************************************

\section*{Kryształy Calabi-Yau}
\addcontentsline{toc}{section}{Kryształy Calabi-Yau}

Powyżej podsumowaliśmy konstrukcję topologicznych teorii strun, jak też różnorodne motywacje fizyczne i matematyczne stojące za wyznaczeniem i analizą ich amplitud. Opisaliśmy również jak dualność otwarto-zamknięta jest realizowana w topologicznych teoriach strun i w jaki sposób prowadzi ona do sformułowania wierzchołka topologicznego, który pozwala na obliczenie amplitud strun topologicznych w modelu A na szerokiej klasie niezwartych torycznych rozmaitości Calabi-Yau. W pierwszej chwili wydawać by się mogło, że struktura tych amplitud jest niezmiernie skomplikowana --- sama amplituda wierzchołka topologicznego $C_{R_1 R_2 R_3}$ jest wysoce nietrywialną funkcją $q=e^{-g_s}$. Jednakże, w topologicznych teoriach strun modelu A ukryta jest niezwykła prostota i elegancja --- okazuje się, że są one ściśle związane ze statystycznymi modelami topnienia kryształów. W szczególności, ich amplitudy są równe sumom statystycznym po wszystkich konfiguracjach topniejącego kryształu. Obserwacja ta pozwala zinterpretować rozmaitości Calabi-Yau jako \emph{rozmaitości kwantowe} w języku odpowiednio rozumianej \emph{kwantowej geometrii}. Każda konfiguracja kryształu może być zinterpretowana jako kwantowa fluktuacja zmieniająca strukturę czasoprzestrzeni i można ją także opisać w języku pewnej efektywnej teorii grawitacji w tej czasoprzestrzeni \cite{foam}. Suma po fluktuujących topologiach czasoprzestrzeni jest też nazywana \emph{pianą kwantową}, a ta teoria grawitacyjna, jako że jest efektywną teorią strun zamkniętych z rysunku \ref{pl-fig-transition}, jest równocześnie teorią dualną do teorii Cherna-Simonsa. W ten sposób otrzymujemy niezwykle interesujący przykład odpowiedniości teorii z cechowaniem i teorii grawitacyjnej, którego ścisła ilościowa analiza może być przeprowadzona. Aby podkreślić związek modeli statystycznych oraz topologicznych teorii strun, te pierwsze zwane są w tym kontekście \emph{kryształami Calabi-Yau} --- i właśnie one są głównym tematem niniejszej rozprawy.

\begin{figure}[htb]
\begin{center}
\includegraphics[width=0.4\textwidth]{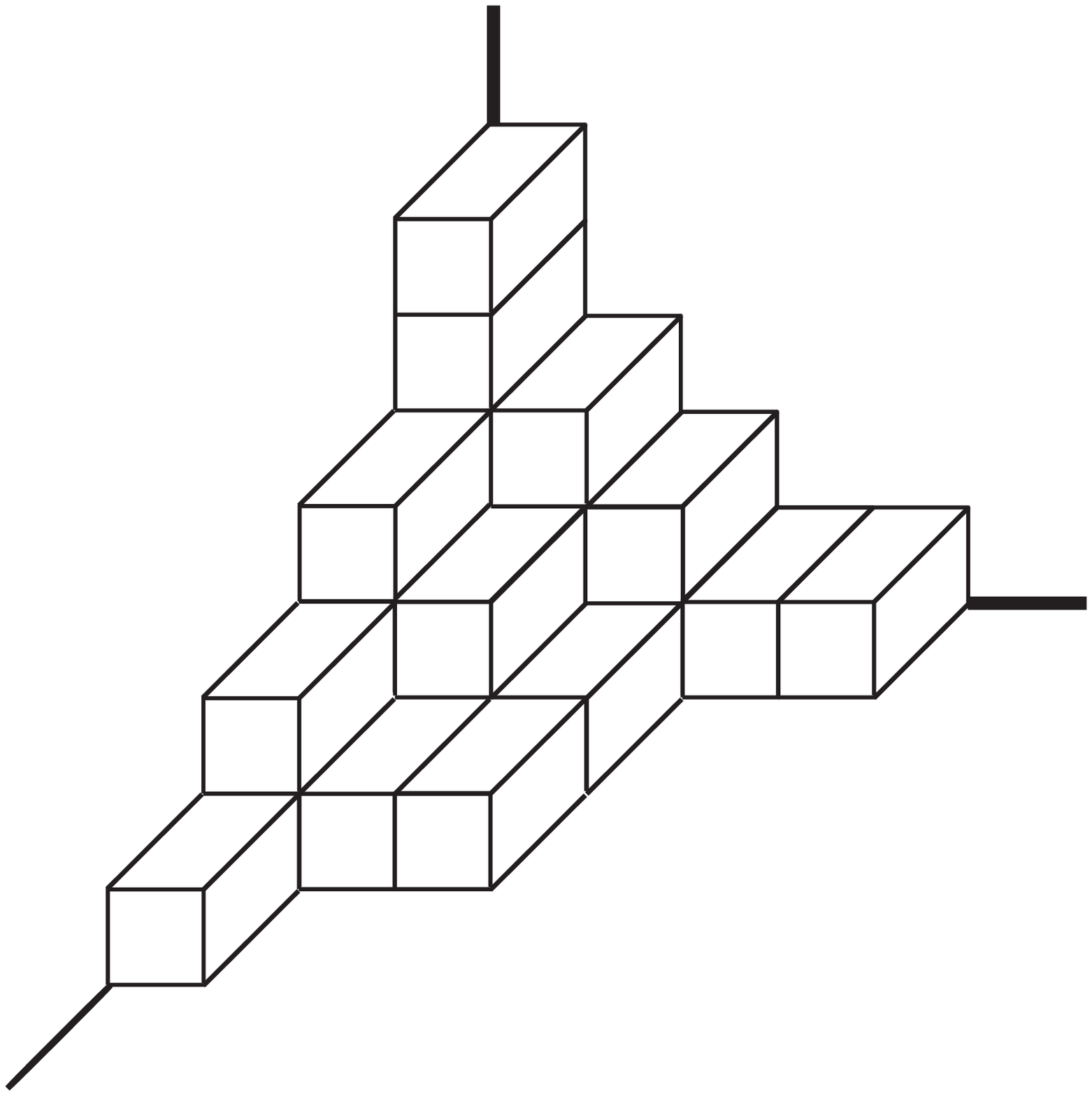}
\caption{Przykład trójwymiarowej partycji odpowiadającej pewnej konfiguracji kryształu Calabi-Yau.} \label{pl-fig-crystal-3d}
\end{center}
\end{figure}

Kryształy Calabi-Yau związane z topologicznymi teoriami strun  są zdefiniowane w trzech wymiarach rzeczywistych, co jest związane z obecnością trzech wymiarów zespolonych przestrzeni Calabi-Yau. Najprostszy przykład takiego kryształu jest zdefiniowany poprzez \emph{trójwymiarowe partycje}, które wypełniają dodatni oktant $\mathbb{Z}^3$. Trójwymiarowa partycja jest --- w ściśle określonym sensie --- maksymalnie upakowaną konfiguracją złożoną z elementarnych kostek zlokalizowanych wokół narożnika kryształu, jak na rysunku \ref{pl-fig-crystal-3d}. Każda taka konfiguracja $\pi$ reprezentuje jeden stan topniejącego kryształu, a jej wkład do sumy statystycznej kryształu zadany jest przez liczbę kostek $|\pi|$ ważoną pewnym parametrem $q$. W przypadku zwykłych opisanych tutaj trójwymiarowych partycji, ich suma statystyczna zadana jest tzw. funkcją McMahona
$$
M(q) = \sum_{\pi} q^{|\pi|} = \prod_{n=1}^{\infty} \frac{1}{(1-q^n)^n}.
$$
Okazuje się, iż suma statystyczna (\ref{pl-Z-top-g}) topologicznych strun zamkniętych modelu A na rozmaitości $\C^3$  jest także dana przez tę funkcję, jeśli strunową stałą sprzężenia utożsami się jako $q=e^{-g_s}$. Otrzymujemy zatem zaskakujący związek 
$$
Z^{\C^3}_{top} = M(q) = \sum_{\pi} q^{|\pi|}.
$$
W ostatnich latach znaleziono wiele innych równoważności pomiędzy bardziej skomplikowanymi rozmaitościami Calabi-Yau oraz bardziej skomplikowanymi modelami kryształów. W szczególności, amplituda wierzchołka topologicznego jest również związana ze zliczaniem pewnego szczególnego zespołu trójwymiarowych partycji --- takich, które asymptotycznie dążą do ustalonych dwuwymiarowych partycji $R_1, R_2, R_3$ położonych w nieskończoności wzdłuż trzech osi dodatniego oktantu $\mathbb{Z}^3$
$$
C_{R_1 R_2 R_3} \sim M(q)^{-1} \sum_{\pi\to\{R_1,R_2,R_3 \} } q^{|\pi|}.
$$
Powyższe związki zostały po raz pierwszy zaobserwowane w pracy \cite{ok-re-va}, która zapoczątkowała rozwój całej tej dziedziny. %Badanie tego typu związków oraz znalezienie nowych modeli kryształów Calabi-Yau jest głównym tematem niniejszej rozprawy.

\section*{Treść rozprawy}
\addcontentsline{toc}{section}{Treść rozprawy}
 
Głównym tematem niniejszej rozprawy jest realizacja modelu A teorii strun topologicznych jako modeli kryształów Calabi-Yau. W rozprawie wprowadzamy nowe modele kryształu, udowadniamy ich związek z teorią strun topologicznych na odpowiednich rozmaitościach, oraz analizujemy je pod wieloma względami, w szczególności rozważając różne konfiguracje bran i związek ich amplitud z niezmiennikami węzłów. Ponadto, rozwijamy także formalizm wierzchołka topologicznego i wykorzystujemy go przy analizie przestrzeni Calabi-Yau związanych z jednym z wprowadzonych przez nas modeli kryształu. 

Kryształy Calabi-Yau można wprowadzić także dla czterowymiarowych teorii topologicznych z twistem. W tym kontekście modele kryształu są dwuwymiarowe i związane są ze zliczaniem różnego rodzaju dwuwymiarowych partycji. W niniejszej pracy wprowadzamy także takie modele dla teorii supersymetrycznej $\mathcal{N}=4$ z twistem na przestrzeniach ALE typu $\C^2/\Z_k$ i wyprowadzamy związek ich funkcji podziału z charakterami afinicznych algebr Liego wykazany w pracach \cite{naka1,V-W}. W istocie modele dwuwymiarowe wprowadzamy w rozprawie w pierwszej kolejności, w pewnym sensie jako przygotowanie do analizy bardziej skomplikowanych modeli trójwymiarowych.

\bigskip

Plan rozprawy jest następujący. Rozdziały \ref{sec-cy}-\ref{chap-top-strings} mają wstępny charakter, i prezentujemy w nich wszystkie składniki konieczne do zrozumienia dalszych rozważań. W rozdziale \ref{sec-cy}, po ogólnym przedstawieniu rozmaitości Calabi-Yau, wprowadzamy przestrzenie ALE o dwu wymiarach zespolonych, oraz toryczne przestrzenie w trzech wymiarach zespolonych wraz z ich opisem przy pomocy diagramów torycznych. Szczególną uwagę poświęcamy konifoldom i ich przemianom geometrycznym. W rozdziale \ref{chap-top-field} wprowadzamy dwa rodzaje topologicznych teorii pól: teorie typu Schwarza i Wittena (zwane też teoriami kohomologicznymi). Przykładem teorii typu Schwarza jest teoria Cherna-Simonsa, dla której prezentujemy zarówno ścisłe rozwiązanie jak też związek z niezmiennikami węzłów. Następnie wprowadzamy teorie kohomologiczne, konstruowane przez wprowadzenie twistu. W rozdziale \ref{chap-top-strings} dyskutujemy różne aspekty topologicznej teorii strun, począwszy od ich konstrukcji jako modeli sigma z twistem. Następnie skupiamy się na modelu A strun topologicznych: wyjaśniamy jak amplitudy strun zamkniętych mogą być zakodowane w niezmiennikach Gromova-Wittena i Gopakumara-Vafy, oraz w jaki sposób teorie strun otwartych redukują się do teorii Cherna-Simonsa. Ponadto analizujemy przemiany geometryczne: przemianę \emph{flop} oraz przemianę Gopakumara-Vafy. Wreszcie wprowadzamy formalizm wierzchołka topologicznego, którego w dużym stopniu używamy w dalszych rozdziałach pracy.

Pozostałe cztery rozdziały poświęcone są analizie kryształów Calabi-Yau. W szczególności, rozdziały \ref{chap-ale}, \ref{chap-cy-results} i \ref{sec-A-model} zawierają oryginalne rezultaty niniejszej rozprawy. W rozdziale  \ref{chap-ale} wprowadzamy dwuwymiarowe modele kryształu dla teorii supersymetrycznych typu $\mathcal{N}=4$ z twistem na przestrzeniach ALE i dowodzimy ich związku z charakterami afinicznych algebr Liego. W rozdziale \ref{chap-cy-crystals} wprowadzamy trójwymiarowe kryształy Calabi-Yau zdefiniowane jako różne zespoły statystyczne trójwymiarowych partycji. Wyjaśniamy ich związek z topologicznymi teoriami strun, szczególną uwagę poświęcając interpretacji amplitudy wierzchołka topologicznego jako sumy statystycznej kryształu oraz bran jako defektów w kryształach. W rozdziale \ref{chap-cy-results} definiujemy modele kryształu z jedną, dwoma i trzema ścianami, również w obecności defektów reprezentujących brany, i znajdujemy amplitudy dla takich konfiguracji. W przypadkach gdy jest to zasadne, interpretujemy znalezione amplitudy jako niezmienniki węzłów. W rozdziale  \ref{sec-A-model} analizujemy przestrzenie Calabi-Yau odpowiadające modelom kryształu wprowadzonym w rozdziale \ref{chap-cy-results}, i używając formalizmu wierzchołka topologicznego udowadniamy, że amplitudy dla tych przestrzeni są równe sumom statystycznym dla odpowiednich kryształów. Uogólniamy także formalizm wierzchołka topologicznego na przypadek pewnej klasy geometrii oraz używamy tego uogólnienia do analizy tzw. \emph{geometrii zamkniętego wierzchołka topologicznego} i jej przemiany geometrycznej typu flop.

W dodatkach podsumowujemy różne zagadnienia związane przede wszystkim z technikami rachunkowymi używanymi w niniejszej rozprawie. W dodatku \ref{app-star} wprowadzamy następujące obiekty, które wielokrotnie występują w pracy: dwuwymiarowe partycje, związaną z nimi teorię swobodnych dwuwymiarowych fermionów, afiniczne algebry Liego i ich charaktery, oraz funkcje symetryczne. Dodatek \ref{app-topver} zawiera podsumowanie formalizmu rachunkowego wierzchołka topologicznego oraz uproszczone reguły obliczania topologicznych amplitud strunowych dla pewnej szczególnej klasy rozmaitości.

\bigskip

Oto zwięzłe omówienie bibliografii. Pozycje \cite{gsw}-\cite{macdonald} to podręczniki do teorii strun, geometrii algebraicznej, konforemnej teorii pola, teorii węzłów, oraz zagadnień kombinatorycznych. Artykuły przeglądowe \cite{dijkgraaf-topfield}-\cite{pioline} dotyczą różnych aspektów teorii  topologicznych. Pozycje \cite{top-field}-\cite{rhd-ps} traktują o topologicznych teoriach pola oraz ich związkach z rachunkiem instantonowym. Pozycje \cite{top-sigma}-\cite{OSV} to referencje dotyczące topologicznej teorii strun, z uwzględnieniem kilku szczególnych wyników których używamy. Dualność teorii z cechowaniem i grawitacji, a w szczególności jej realizacja w topologicznej teorii strun poprzez związek z teorią Cherna-Simonsa i teorią węzłów, dyskutowane są w pracach  \cite{thooft}-\cite{largeN}. Teoria wierzchołka topologicznego jest rozwinięta w pracach \cite{vertex}-\cite{gen-ver}. Rozwój tematyki kryształów Calabi-Yau oraz innych pokrewnych modeli statystycznych odzwierciedlają publikacje \cite{ok-re-va}-\cite{pearcey}. Pozycje \cite{har-law}-\cite{F-theory} dotyczą kilku szczególnych matematycznych i fizycznych zagadnień związanych z naszymi rozważaniami.
\selectlanguage{english}
%***********************************************************
%***********************************************************

\chapter{Introduction and summary}

%\section{Introduction}

%String Theory is one of the most profound constructions of human mind. It 

What are the fundamental laws of Nature is one of the oldest and still actual questions asked by man. During the twentieth century it has been established that these laws include General Relativity and Quantum Mechanics. General Relativity is responsible for the behaviour of matter on large scales, presumably extending to the size of the Universe. The domain of Quantum Mechanics, or more generally Quantum Field Theory, is restricted to microscales. In particular, at the present state of knowledge, almost all processes in the quantum world are described by the Standard Model, a particular realisation of Quantum Field Theory rules in terms of a gauge theory with $SU(3)\times SU(2)\times U(1)$ gauge group. 

However, it is not known how to reconcile Quantum Mechanics with General Relativity, and despite their great successes there are still many inexplicable phenomena in quantum and gravitational theories which describe Nature. Examples of such phenomena, which undoubtedly belong to the greatest unsolved riddles of Nature, are the breaking of the electroweak symmetry, the low energy description of QCD and high energy description of Einstein's gravity. The electroweak symmetry is related to the  $SU(2)\times U(1)$ gauge group of the Standard Model, and it is known that it is spontaneously broken via the Higgs mechanism, which requires the existence of a scalar Higgs particle. This is the only particle in the Standard Model which has not been discovered in experiment; in fact it is not known whether this is indeed an elementary particle or a composite field, and its small mass as compared to the Planck scale is the origin of the so-called hierarchy problem. QCD is a part of the Standard Model related to the $SU(3)$ gauge group, and one of its main features is the fact that in low energies its coupling constant gets large, so that relying on the perturbative expansion --- the main approach of obtaining quantitative predictions in the Standard Model --- is no longer possible. On the other hand Einstein's gravity is a non-renormalisable theory and it cannot be quantised in any standard way, even though it is strongly believed that in extremely high energy conditions --- which we expect to arise in the Early Universe or inside black-holes --- quantum gravitational effects must play a role. %Therefore these problems belong to the regimes of Quantum Field Theory and General Relativity in which their predictable and computational power is lost. 

Nonetheless, there are some very attractive ideas how to address the above problems. Unfortunately, at present they are not confirmed by any experiment, but there are various arguments suggesting they are serious candidates to describe Nature on the most fundamental level. Quantum Mechanics and General Relativity can be reconciled within String Theory, which unifies the known constituents of matter and principles governing their interactions into states and behaviour of a single entity, the \emph{string}. The most attractive phenomenological features appear to arise in \emph{superstring} theories, which posses an additional symmetry, the so-called \emph{supersymmetry}. The supersymmetry is a very appealing theoretical idea which has also very interesting consequences when applied to field theories, in particular to the Standard Model. Among the others, it predicts that all particles should have their \emph{superpartners} with different spins. Even though these superpartners so far have not been observed, the existence of supersymmetry would have many advantages from the  phenomenological point of view. In particular, it could provide a solution of the hierarchy problem, as well as impose a unification of the Standard Model gauge couplings, so that it is even speculated it might be discovered at the LHC in the near future. 

Another interesting theoretical idea is that the physics in the regimes inaccessible for a description in one theory could be governed by a \emph{dual} theory which do have a predictable power, although its fundamental degrees of freedom are completely different than in the original theory. One proposed realisation of this idea is the \emph{gauge-gravity duality}, which states that in complementary regimes gauge theories should have a dual description as gravitational theories and \emph{vice versa}. In particular, this duality can be realised in String Theory as the so-called \emph{open-closed duality}.

Supersymmetry or gauge-gravity duality are very interesting for yet another reason: it turns out that certain simplified theories with these features are much more tractable or even exactly solvable. It is believed that understanding such theories could provide an essential insight how to tackle the real fundamental theories in the next step, \emph{excluding} these simplifications. This is also the approach we assume in this thesis: it is devoted to studying exact solutions of the so-called \emph{topological} theories, in which many non-trivial phenomena of the quantum world can be explicitly analysed in a simplified setting. 

There are both topological field and string theories. Their simplicity is related to the fact that they usually posses much less physical degrees of freedom than theories which we refer to as \emph{physical} ones, which describe (as gauge theories) or are hoped to describe (as for example ordinary superstring theories) some aspects of Nature. However, there are many motivations for studying topological theories. Firstly, in many cases they are exactly solvable. Secondly, it often happens the amplitudes computed in a topological theory constitute a subsector of amplitudes of a certain physical theory, so the solution of a topological theory provides us automatically with a partial solution of this related physical theory. Apparently, there is also a growing evidence there are some phenomena in Nature described by topological field theories, as for example some exotic condensed matter systems in very low temperatures. Last but not least, topological theories provide new and insightful views and relations between various distinct mathematical theories, and often lead to their new elegant solutions.

Of particular interest in this thesis are topological string theories. There are additional motivations apart from those mentioned above to study them. On one hand, they provide an exactly solvable model of the open-closed duality, upon which many results in this thesis is based. On the other hand, they are deeply related to various four-dimensional theories, and compute such quantities as prepotentials of $\mathcal{N}=2$ supersymmetric gauge theories, superpotentials of $\mathcal{N}=1$ supersymmetric theories, or entropy of extremal black holes. On the mathematical side, topological string theories unify many seemingly remote ideas, for example Kodaira-Spencer theory and theories of Gromov-Witten, Gopakumar-Vafa and Donaldson-Thomas invariants. 

In this thesis we focus on one remarkable aspect of topological theories: in certain favourable situations they reduce to simple statistical models of crystals. These crystal models for field and string theories are related respectively to the counting of two- and three-dimensional partitions, which provides a clear interpretation of topological amplitudes. Moreover, the generating functions of such partitions can be found exactly and they reproduce certain amplitudes of topological theories. The field and string theories for which crystal interpretation is known are defined on the so-called \emph{Calabi-Yau} manifolds, and for this reason the corresponding crystal models are called \emph{Calabi-Yau crystals}. 

To sum up, this thesis is concerned with a realisation of topological theories in terms of Calabi-Yau crystals, with a particular emphasis on topological string theories. As there are many different ideas involved in this programme, we find it appropriate to provide their concise overview in this introductory chapter, starting with a brief summary of String Theory. Apart from general remarks on string theory, all issues mentioned below are explained in much more detail in the following chapters of the thesis. 

%****************************************************

\section{String theory} \label{sec-intro-strings}

\begin{figure}[htb]
\begin{center}
\includegraphics[width=0.3\textwidth]{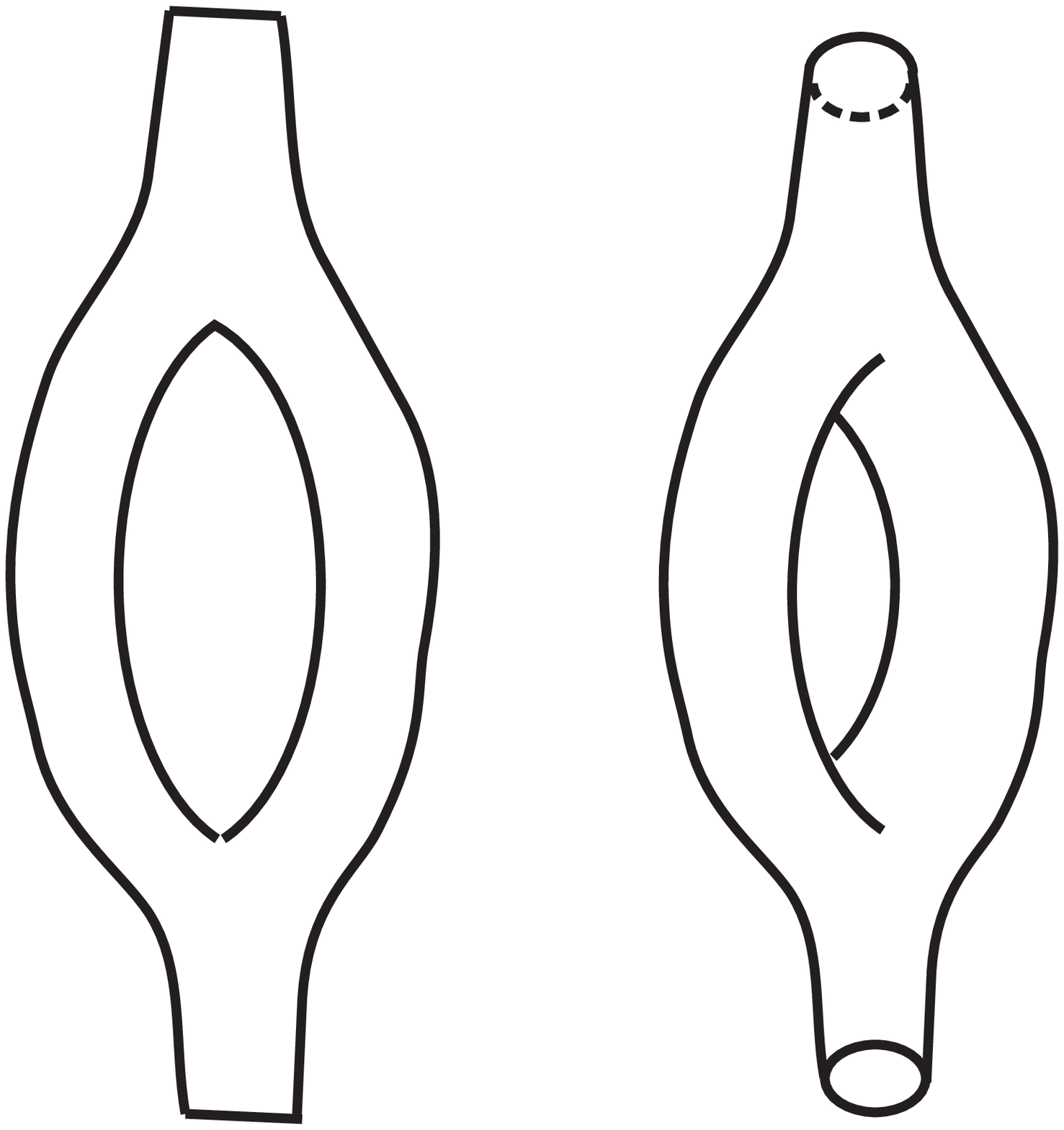}
\caption{Worldsheets of open strings (left) and closed strings (right).} \label{fig-worlds}
\end{center}
\end{figure}

Let us recall basic ingredients of String Theory \cite{gsw,polchinski}. Its most crucial assumption is that all elementary particles arise from quantisation of vibrational modes of one-dimensional strings and all interactions of particles are uniquely determined by the geometry of a two-dimensional \emph{worldsheet} $\Sigma$ swapped when these strings move in spacetime $M$, which is also referred to as a \emph{target space}. There are two types of strings, \emph{closed} and \emph{open}, as shown in figure \ref{fig-worlds}. Worldsheets of closed strings are simply Riemann surfaces of arbitrary genus $g$, possibly with asymptotic ends representing strings incoming from infinity. On the other hand,  worldsheets of open strings are surfaces of genus $g$ and some number of holes $h$. Let us denote coordinates on the worldsheet as $\sigma^a$ and the worldsheet metric as $g_{ab}$ for $a,b=1,2$. The spacetime coordinates $X^{\mu}$ define a map
\be
X^{\mu}:\quad \Sigma \longrightarrow M,   \label{sigma-model}
\ee
and the spacetime metric is $G_{\mu\nu}(X)$. In general a theory describing a behaviour of maps into a nontrivial curved manifold is called a \emph{nonlinear sigma model}, or \emph{sigma model} for short. The classical bosonic closed string action is an example of a sigma model and it can be written as
\be
S_{string} = \frac{1}{4\pi\alpha'} \int_{\Sigma} d^2\sigma\,g^{1/2} g^{ab}G_{\mu\nu}\partial_a X^{\mu} \partial_b X^{\nu},   \label{S-Polyakov}
\ee
which in the case of open strings has to be augmented by appropriate boundary conditions related to the string ends. The only arbitrary parameter of String Theory is $\alpha'$ of dimension $(spacetime\ length)^2$.  This is related to the string tension $T$ as $T=(2\pi\alpha')^{-1}$ and string length is assumed to be of order $\sim\alpha'^{1/2}$. 

One way of quantising strings is to consider the Polyakov path integral in Euclidean signature
$$
Z = \int \mathcal{D}X\,\mathcal{D}g\,e^{-S_{string}}.
$$
The classical action $S_{string}$ is conformally invariant and only if this invariance is preserved upon quantisation will the resulting quantum theory be consistent. This condition imposes a very strong constraints on the target space $M$: its dimension must be 26, and it should satisfy vacuum Einstein equations
$$
R_{\mu\nu}+\mathcal{O}(\alpha') = 0.
$$

However, such a theory is not yet consistent. The lowest string state of such a
theory turns out to have negative mass squared --- it is a tachyon, which
should be absent in a satisfying theory. A way to avoid tachyons in string
spectrum is it consider \emph{supersymmetric} string theories. In the so-called
Ramond-Neveu-Schwarz formulation it amounts to adding to $X^{\mu}$ their
superpartners $\psi^{\mu}$ and considering an extension of the Polyakov action
with $\mathcal{N}=(2,2)$ supersymmetry. A careful analysis shows this not only
eliminates tachyons, but also leads to a supersymmetric spectrum of particles
in spacetime --- a fact which is of great phenomenological interest itself. In
superstring theory vanishing of conformal anomaly implies Einstein equations
still have to be satisfied in spacetime, however the condition on its
dimensionality is modified to $\textrm{dim}\,M = 10$. One approach to obtain a phenomenologically viable model with a four-dimensional spacetime representing a real world is to compactify a theory on a tiny compact manifold. To preserve supersymmetry this manifold must be a \emph{Calabi-Yau three-fold}, i.e. a complex three-dimensional manifold which is Ricci flat. 

After eliminating tachyons, the string spectrum consists of some massless states and an infinite tower of heavy states of masses of order $\sim\alpha'^{-1/2}$. In the point-like limit $\alpha'\to 0$ all heavy states decouple from interactions, and string theory reduces to the effective theory for massless states, described by the so-called \emph{effective action}. For closed strings the massless spectrum consists of spin-2 states which can be identified with gravitons which manifest as $G_{\mu\nu}$ in (\ref{S-Polyakov}), antisymmetric Kalb-Ramond fields $B_{\mu\nu}$, and dilaton $\Phi$ which can be identified with the string coupling as $g_s=\exp\,\Phi$. In superstring theory there are in addition the so-called R-R fields of integer spin and R-NS fields of half-integer spin. The spectrum of open strings is different due to their end-points, which allow to introduce additional degrees of freedom, the so-called \emph{Chan-Paton factors} associated to some Lie group symmetry. The massless state of the bosonic open string theory has spin 1 and its interactions are reproduced by the effective Yang-Mills action with a gauge symmetry associated to this Lie group. Thus this massless state can be identified with the gauge field $A^{\mu}$, and in superstring theory this is augmented with fermionic states. 

There is one more crucial ingredient of String Theory, the so-called \emph{D-branes}. They can be described as submanifolds on which open strings can end. When they extend in time and in $p$ spatial dimensions they are called $Dp$-branes. In fact without $D$-branes String Theory would not be consistent: on one hand they turn out to be sources for R-R fields, and on the other there are some symmetries in String Theory which exchange them with fundamental strings. 

A careful analysis of all conditions mentioned above reveals there are five consistent superstring theories. Of most interest to us will be theories of closed strings known as IIA and IIB. These theories contain R-R fields of a specific spin content, which implies there must exist $D$-branes as their sources of a specific dimensionality as well. In particular, IIA and IIB theories admit $Dp$-branes with respectively even and odd $p$. Other consistent theories are open type I theory with $SO(32)$ symmetry, as well as two heterotic theories with $SO(32)$ and $E_8\times E_8$ symmetries. Moreover, it turns out all these theories are related to each other by various dualities, and it is conjectured they are just a few particular limits of some mysterious eleven-dimensional theory, the so-called $M$-theory.

%****************************************************

\subsection*{Calabi-Yau compactifications and type II theories}

As we already explained, in order to obtain an effective, supersymmetric, four-dimensional theory, superstrings must be compactified on a Calabi-Yau three-fold. Properties of the effective theory, such as a particle content and a form of their interactions, are determined by the geometry of this three-fold, which is therefore associated with one possible vacuum we might live in. It would be appealing if all data about field theories describing elementary particles were indeed encoded in one geometrical object. In fact, the number of all possible Calabi-Yau vacua is very large, so the identification of the one representing our world would be a great challenge. Nonetheless, this field has been actively studied in recent years \cite{susskind,M-landscape}.
%However, it is discauraging to realise a number of such Calabi-Yau vacua is estimated to be of order $10^{500}$, so an identification of the one representing our world would be a great challenge. Nonetheless, the space of those vacua has been actively studied in recent years within the so-called \emph{Landscape} programme \cite{susskind,M-landscape}

To be more specific, let us recall basics of a compactification of type II superstrings on a Calabi-Yau three-fold $M$. In IIA theory this leads to $\mathcal{N}=2$ supergravity in four non-compact dimensions with $h^{1,1}(M)$ vector multiplets and $(h^{2,1}(M)+1)$ hypermultiplets. In type IIB theory numbers of vector and hypermultiplets are exchanged. These multiplets contain complex scalar fields which correspond to the so-called \emph{moduli} of $M$, which specify its geometrical properties, such as shape or size. A non-trivial configuration of such fields in a four-dimensional space corresponds to non-trivial fibrations of the Calabi-Yau space over the spacetime. The behaviour of scalar fields in general is governed by various potentials. As an example, we consider a single modulus $\phi$ corresponding to the overall size of the Calabi-Yau. String Theory predicts a corresponding multiplet in four-dimensional theory is governed by the so-called \emph{K\"ahler potential}
$$
K = -3\,\textrm{ln}(\phi + \phi^*).
$$
Even though scalar fields do not appear often in Nature, there are some notable examples, such as the Higgs field or the \emph{inflaton} responsible for the cosmic inflation, which might originate in Calabi-Yau moduli from String Theory point of view. In phenomenologically viable models, potentials for such moduli should necessarily be more complicated than the one given above. Mechanisms of generating such potentials are actively studied nowadays; some particular example is given in \cite{F-theory}. 

%There are two major problems related to the above picture. On one hand, Nature seems to avoid scalar fields --- the only known serious possibilities seem to be the Higgs field and the \emph{inflaton} responsible for the cosmic inflation. However, for a complicated Calabi-Yau compactification, presumabely related to the Standard Model, a large number of such fields would arise which would have to be interpreted somehow. Such an interpretation usually requires specifying a particular expectation value of a scalar field --- or \emph{fixing a moduli} --- which is the source of the second problem. Its solution might be possible if we knew how appropriate potentials are generated.

There are also attempts to reproduce the whole spectrum of Standard Model from string compactifications. A particularly promising class of models is called \emph{brane world scenarios}, which have a few characteristic ingredients. To include gauge bosons, our four-dimensional spacetime is assumed to lie along a $D3$-brane. Matter fields arise from branes wrapping various cycles and intersecting each other in Calabi-Yau manifold. Even though type II theories give rise to $\mathcal{N}=2$ effective theories, it is possible to break supersymmetry to much more phenomenologically interesting case of $\mathcal{N}=1$, by considering compactifications involving orbifolds. An interesting example of such a model is given in \cite{SM-brane}.

%****************************************************

\subsection*{Gauge-gravity duality}

As mentioned above, in string theory gauge bosons are associated with massless open string states, whereas gravitons with massless closed string states. Moreover, certain dualities are known between open and closed strings, which therefore allow to realise the idea of gauge-gravity correspondence in string theory. This kind of duality was considered for the first time already in 1974 by 't Hooft \cite{thooft}, who argued on general grounds that every gauge theory should be related to some closed string theory. His original motivation was to find a gravitational theory dual to QCD, with a hope it would lead to a solution of QCD for small energies. Even though problem is still unsolved, the gauge-gravity duality enjoyed a lot of revival in recent years due to the famous discovery by Maldacena that $\mathcal{N}=4$ supersymmetric gauge theory has a dual gravitational theory defined on AdS (anti-de-Sitter) background   \cite{adscft,holography}. Subsequently many similar relations were found, which are commonly referred to  as the  AdS/CFT correspondence. $\mathcal{N}=4$ theory is a supersymmetric theory and its bosonic part is the ordinary Yang-Mills action, the same as in QCD, so this discovery gave a real hope the problem of strong interactions could be solved in such an approach. However, the large number of supersymmetries in this case implies $\mathcal{N}=4$ theory is conformal and in particular its coupling does not run, so the very problem of QCD coupling growing large is in fact avoided. Nonetheless, there are still some regimes of QCD in which even supersymmetric theory is believed to give verifiable results; one such example is the behaviour of quark-gluon plasma. Thus we may indeed hope the solution of gauge and gravity theories inspired by the idea of open-closed duality could be relevant for the theories describing Nature.

Apart from the above mentioned theories, there are also other more peculiar string models. A notable example are topological string theories discussed below, which are the main subject of this thesis. A particular case of open-closed correspondence, known  as the Gopakumar-Vafa duality, can be realised and analysed in an exact and quantitative way in these theories. This example is expected to provide insight not only into more complicated models, such as the AdS/CFT duality, but also the nature of strong interactions and quantum gravity.

%****************************************************

\subsection*{Resum{\'e}}

In this section we have summarised a theoretical framework emerging from String Theory. Unfortunately, at present it is not possible to perform an explicit experimental verification whether it indeed describes Nature. Nonetheless, many strong constraints have been mentioned that any theory claiming to be fundamental should be consistent with. Firstly, in appropriate limits such a theory should reproduce the laws already known, in particular those of General Relativity and Quantum Field Theory. Secondly, its mathematical structure must be self-consistent, which imposes quite strong constraints due to sophisticated mathematical structures arising in theories mentioned above. String Theory appears to satisfy these constraints with a surprisingly small and elegant set of assumptions. Moreover, it admits many phenomenologically appealing models. These advantages underlie the belief this is a good candidate for a fundamental theory. 

%****************************************************

\section{Topological field theories}

A quantum theory defined on some manifold $M$ is called \emph{topological} if its observables do not depend on certain degrees of freedom related to continuous deformations of the metric of $M$. In consequence, observables of a topological theory are topological invariants of $M$ or some geometrical objects associated to $M$. There are topological theories of both fields and strings. Two constructions of topological field theories are known, which we now briefly discuss. 

One can consider a theory whose action explicitly does not depend on the metric, what happens for example if it can be written entirely in terms of differential forms. Such a theory is topological if there is no anomaly that would introduce metric dependence in the path integral measure. Chern-Simons theory is an example of such a theory, and its remarkable solution found by Witten \cite{witten-cs} had an immense impact on the mathematical physics. Chern-Simons action for a gauge connection $A$ reads 
\be
S_{CS} = \frac{k}{4\pi} \int_M \textrm{Tr}\,\big(A\wedge dA + \frac{2}{3}A\wedge A\wedge A \big),   \label{CS}
\ee
and the condition of gauge invariance in the quantum theory implies $k$ must be an integer number. 

\begin{figure}[htb]
\begin{center}
\includegraphics[width=0.4\textwidth]{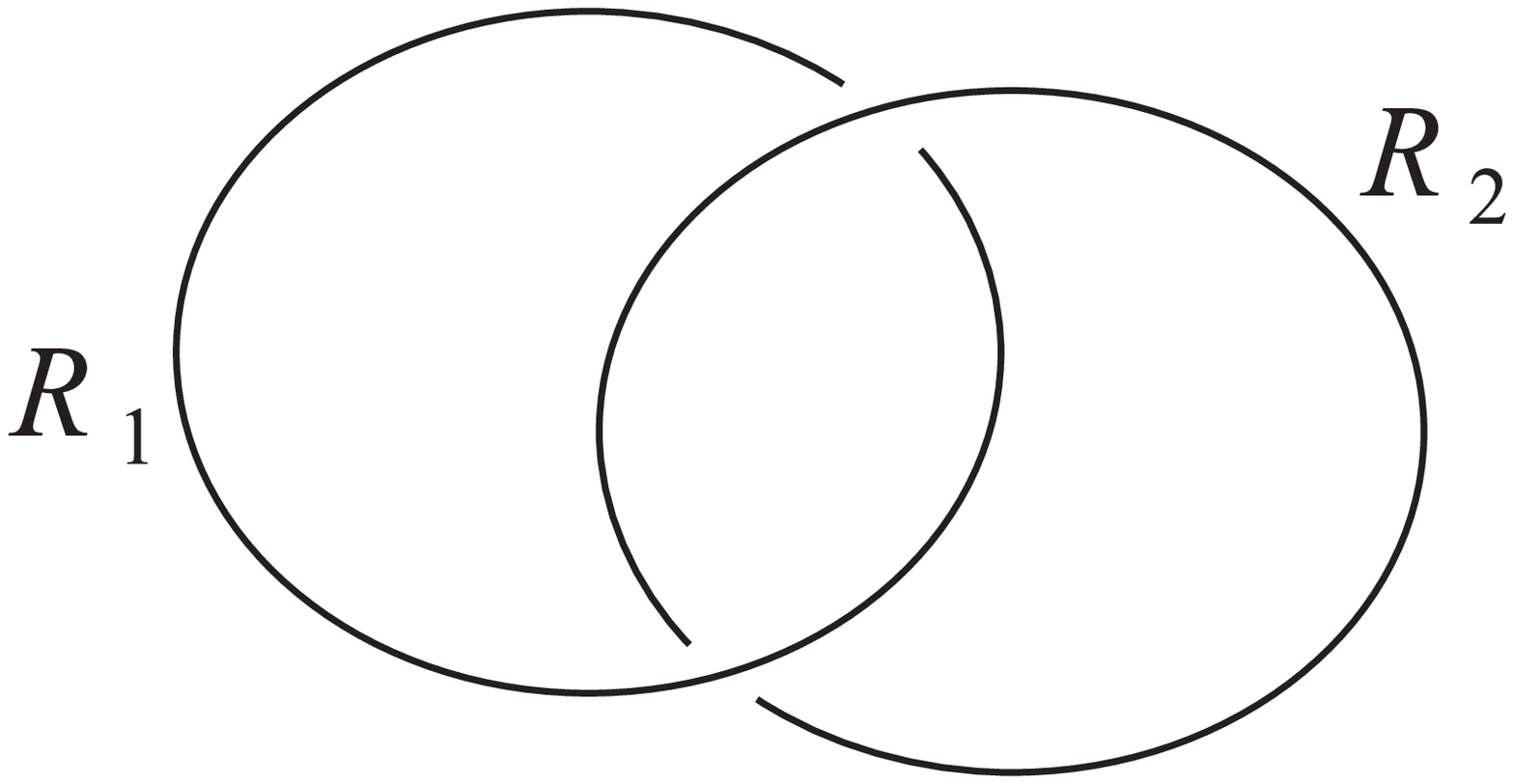}
\caption{Hopf-link.} \label{fig-hopf-intro}
\end{center}
\end{figure}

A very important class of observables of Chern-Simons theory are expectation values of Wilson loops
$$
\mathcal{W}^K_R = \langle \textrm{Tr}_R P e^{\oint_K A} \rangle = \int \mathcal{D}A \, e^{i S_{CS}}\,\textrm{Tr}_R P e^{\oint_K A},
$$
computed along a closed loop $K$ in a representation $R$ of the gauge group. A single loop $K$ in mathematical sense represents a knot, and the independence on continuous deformations of the metric implies this amplitude is a knot invariant. The simplest example of a knot is the so-called \emph{unknot}, a single unknotted closed loop homeomorphic to a circle. Similarly, an expectation value of a product of several Wilson loops, each one associated to a different loop and representation, computes the so-called link invariant. For example, the \emph{Hopf-link} consists of two interlacing unknots shown in figure \ref{fig-hopf-intro}, and its link invariant reads
\be
\mathcal{W}_{R_1  R_2} = \langle \textrm{Tr}_{R_1} P e^{\oint_{K_1} A}\, \textrm{Tr}_{R_2} P e^{\oint_{K_2} A} \rangle, \label{CS-hopflink}
\ee
which reduces to the unknot invariant $\mathcal{W}_{R \bullet}$ if one of its representations is trivial.

%This connection between quantum Chern-Simons theory and knot invariants was made precise by E. Witten in 1989, and it had enormous impact both on physics and mathematics. 

The other method of constructing topological field theories amounts to \emph{twisting} supersymmetric theories \cite{top-field}. The operation of twisting mixes spacetime symmetries with internal $R$-symmetries in such a way that observables of the theory can be identified with cohomology classes of a certain BRST-like operator $Q$, so that all observables involving the metric turn out to be cohomologically trivial. Some results presented in this thesis are related to the $\mathcal{N}=4$ twisted supersymmetric gauge theory introduced in \cite{V-W}.

%****************************************************

\section{Topological string theories}

The procedure of twisting can be applied also to supersymmetric sigma models such as $\mathcal{N}=(2,2)$ extension of (\ref{sigma-model}) \cite{top-sigma,witten-mirror}. This provides a construction of topological string theories. More precisely, topological strings are defined in a genus expansion, such that the genus $g$ amplitude $F_g(t^I)$ arises from a path integral analogous to (\ref{S-Polyakov}), but involving a twisted sigma model. The total closed topological string free energy is defined in terms of  $F_g(t^I)$ as 
\be
F_{top}(t^I) = \sum_{g=0} F_g(t^I) \, g_s^{2g-2},  \label{F-top-g}
\ee
where a generating parameter $g_s$ weights contributions for different genera and can be interpreted as the string coupling. Then the topological string  partition function is simply
\be
Z_{top} = e^{F_{top}(t^I)}.     \label{Z-top-g}
\ee
For open topological strings the construction is analogous, but involves Riemann surfaces with boundaries. As it turns out, various conditions which arise in the construction of topological strings imply this target space must be a Calabi-Yau three-fold. In the above expressions $t^I$ are the same Calabi-Yau moduli as those which appeared in the discussion of string compactifications in section \ref{sec-intro-strings}. 

A moduli space of metrics on a Calabi-Yau manifold can be divided into a space of K\"ahler deformations and complex deformations. The former determine sizes of homology cycles of a Calabi-Yau, whereas the latter their shapes. It turns out all amplitudes in topological string theories depend only on one of these types of deformations. Thus there are two types of topological strings, the so-called \emph{A-model} and \emph{B-model}, such that
\begin{itemize}
\item In the A-model theories, moduli $t^I$ correspond only to K\"ahler deformations of the Calabi-Yau target space, and the amplitudes computed in these theories do not depend on its complex deformations.
\item In the B-model theories, moduli $t^I$ correspond only to complex deformations of the Calabi-Yau target space, and the amplitudes computed in these theories do not depend on its K\"ahler deformations.
\end{itemize}
In open topological string theories the ends of open strings must be attached to branes, similarly as in ordinary string theories. Branes in the A-model are called A-branes, and it turns out they must wrap three-dimensional lagrangian submanifolds of the Calabi-Yau. Branes in the B-model are called B-branes, and they must wrap complex submanifolds of the Calabi-Yau, so in particular they have even real dimension. Moreover, it turns out that A-model and B-model theories are related to each other by a very powerful Mirror Symmetry, according to which Calabi-Yau manifolds arise in pairs, such that K\"ahler and complex moduli of two manifolds in one pair are exchanged under this symmetry. In topological string theories the manifestation of Mirror Symmetry is the equality of amplitudes computed for one Calabi-Yau in A-model with those of the mirror Calabi-Yau computed in B-model. Also A-branes and B-branes are exchanged under Mirror Symmetry.

%****************************************************

\subsection*{Relation to superstring theories}

 The fact the target space of the topological string theory must be a Calabi-Yau three-fold is a reminiscent of a situation in superstring theories, where the consequence of vanishing of a conformal anomaly was the existence of 10 dimensions of a target space. In fact 6 real dimensions compactified on a Calabi-Yau manifold in superstring theory in order to get a four-dimensional spacetime can be identified in a precise sense with 3 complex dimensions of topological strings. This provides an intimate relation between superstring and topological string theories, and in consequence the latter provide a lot of interesting information about effective field theories arising in 4 dimensions orthogonal to the Calabi-Yau space in superstring compactifications. The relation to four-dimensional field theories is another strong motivation to study topological strings, so we briefly review some of its manifestations:
\begin{itemize}
\item The K\"ahler and complex moduli that A and B model topological strings on a Calabi-Yau three-fold depend on, correspond respectively to vector and hypermultiplets in type IIA compactification on the same three-fold. In type IIB compactification the role of K\"ahler and complex moduli is exchanged.
\item The amplitudes $F_g(t^I)$ which arise in (\ref{F-top-g}) turn out to be a subset of superstring amplitudes related to the scattering of gravitons and their superpartners (so called gravi-photons) in the effective $\mathcal{N}=2$ gauge theory \cite{agnt,bcov}. 
\item The free energy $F_{top}$ on a particularly chosen non-compact Calabi-Yau background computes the so-called prepotential which contains the complete non-perturbative information about instanton solutions in certain $\mathcal{N}=2$ gauge theory \cite{Nek,Nek-Ok}. 
\item The topological string free energy $F_{top}$ provides also a contribution to the superpotential for multiplets associated with moduli $t^I$ of the Calabi-Yau background $M$ in  of some $\mathcal{N}=1$ gauge theory
%\be
%W(X^I) = N^I \frac{\partial \mathcal{F}_{top}(X^I)}{\partial X^I} + \tau_I
%X^I,  \label{superpotential}
%\ee
in a presence of certain fluxes %  $N^I$ and $\tau_I$ 
through cycles of $M$ \cite{gvw}. Moreover, by a large $N$ duality \cite{largeN} which exchanges fluxes with branes, %  (\ref{superpotential}) 
this superpotential can be reinterpreted as a superpotential for a glueball field in a related $\mathcal{N}=1$ theory. 
\item In supergravity theories it is possible to provide a microscopic description of the so-called extremal black holes in terms of states of $D$-branes. It has been conjectured that the quantum partition function of certain black holes $Z_{black\ hole}$  is related to the topological string partition function as \cite{OSV}
\be
Z_{black\ hole} = |Z_{top}|^2.     \label{Z-bh}
\ee
\end{itemize}

Even though these relations are not the main subject of this thesis, the results which we derive include various topological string amplitudes and might be of interest in the context of four-dimensional theories as well.

%****************************************************

\subsection*{Gopakumar-Vafa duality}

Similarly as for ordinary string theories, there are effective theories of open and closed topological strings related to their massless modes. In fact, in topological string theories all massive string excitations decouple, which implies that their effective theories are \emph{exact} at the same time. These effective theories also have a topological character and very often are much easier to solve than original topological string theories. This fact underlies in particular many results in this thesis.

As found by Witten in another remarkable paper \cite{cs-string}, the effective exact theory of open topological strings is Chern-Simons theory discussed above. In this thesis we are mainly interested in the A-model theories, for which a more precise statement is the following equivalence
\be
\begin{array}{|ccc|} \hline
\begin{array}{c}
\textrm{open topological strings } \\ 
\textrm{on}\  T^* M
\end{array}
& \iff  &  
\begin{array}{c}
\textrm{Chern-Simons theory} \\ 
\textrm{on}\ M  
\end{array}   
 \\ \hline            
\end{array}               \label{cs-effective}
\ee
where $M$ is a three real-dimensional lagrangian submanifold of a Calabi-Yau three-fold $T^*M$, and there are $N$ A-branes wrapping $M$ in the open string theory. In Chern-Simons theory $N$ becomes the rank of the gauge group. 

%\begin{figure}[htb]
%\begin{center}
%\includegraphics[width=0.5\textwidth]{fat-rules.eps}
%\caption{Gauge propagator and cubic coupling for the Chern-Simons theory (\ref{CS}) in a double-line notation.}  \label{fig-fat-rules-intro}
%\end{center}
%\end{figure}

\begin{figure}[htb]
\begin{center}
\includegraphics[width=0.5\textwidth]{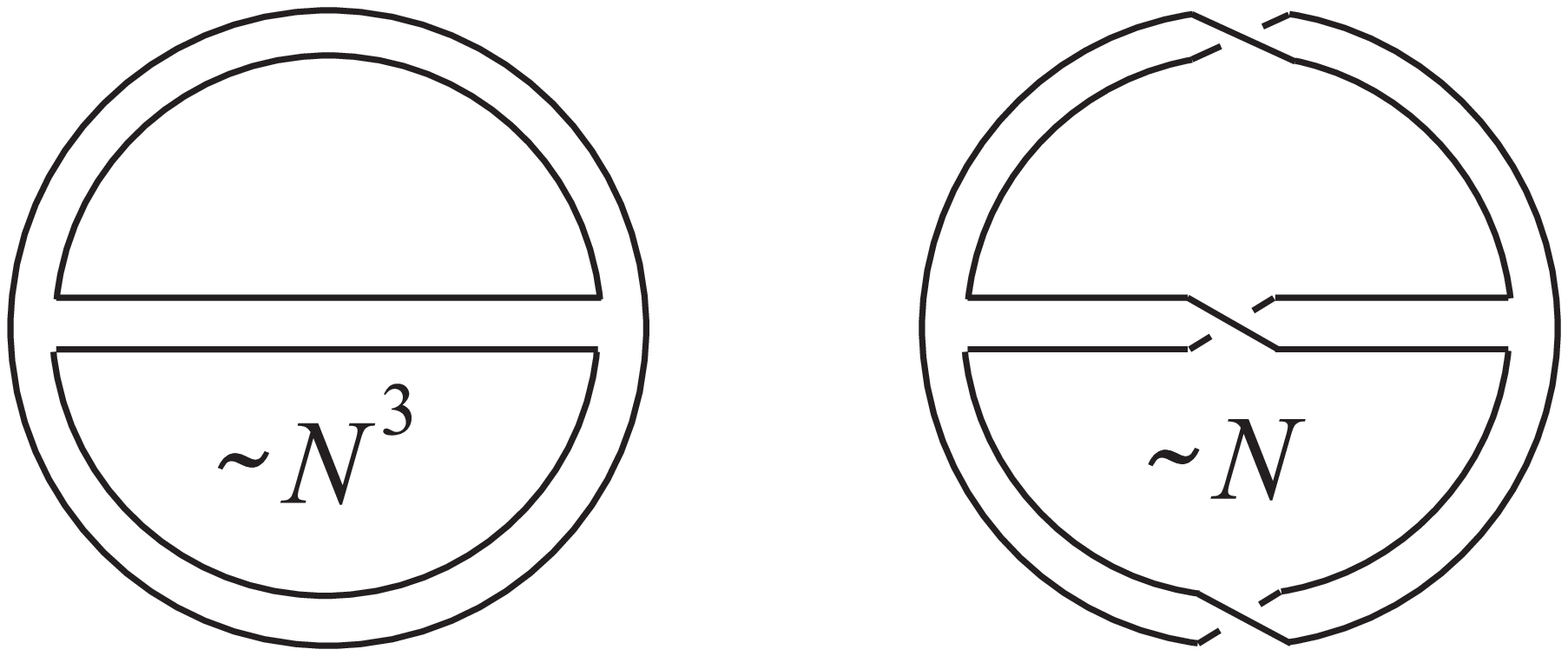}
\caption{A planar fatgraph in the left can be drawn on a sphere, a surface of genus $g=0$. Non-planar fatgraphs can be drawn on a higher genus surfaces; in the example in the right this surface is a torus with $g=1$. A fatgraph amplitude is proportional to $\lambda^{E-V}N^h$, where $E,V$ and $h$ are respectively its number of propagators, vertices and holes.} \label{fig-diagrams-intro}
\end{center}
\end{figure}

Let us recall now 't Hooft large $N$ duality between gauge theory and a certain closed string theory \cite{thooft}. This relation becomes explicit if gauge theory quantities are expanded in powers of the rank of the gauge group $N$. It turns out the so-called \emph{double-line notation} is perfectly suited to perform such an expansion. The diagrams in this notation are often called \emph{fatgraphs}, and they are related to ordinary Feynman diagrams by replacing lines representing gauge fields by ribbons of definite width whose edges represent gauge indices. Examples of fatgraphs are shown in figure \ref{fig-diagrams-intro}. The crucial observation is that each fatgraph encodes a contribution with a definite power of $N$ and a gauge coupling constant $\lambda$ and it can be drawn on a Riemann surface of genus $g$ with $h$ holes. Each such vacuum amplitude $F_{g,h}$ contributes to the perturbative free energy of the theory
$$
F^{pert} = \sum_{g,h}  F_{g,h} \,\lambda^{2g-2} t^h,
$$
where the so-called \emph{'t Hooft coupling} is $t= N\lambda$. 't Hooft argued that in large $N$ limit performing a sum over $h$ in this expression can be interpreted as filling the holes in the corresponding Riemann surface, which leads to a genus expansion related to Riemann surfaces without boundaries. For Chern-Simons theory it can be written as
%\be
%F_{closed}(t) = \sum_{g=0}^{\infty} F_g(t) \lambda^{2g-2},\qquad \textrm{for}
\be
\quad F^{CS}_g(t)=F^{non-pert}_g + \sum_{h=0}^{\infty} F_{g,h}t^h,  \label{F-closed-intro}
\ee
where $F^{non-pert}_g$ are non-perturbative terms which are not captured by the 't Hooft expansion.
Such a genus expansion coincides with a general form of the expansion in a closed string theory, if $F_g(t)$ is identified with a genus $g$ closed string amplitude and parameter $t$ with some modulus of the string theory target space. In particular, if one identifies states of closed strings with gravitational excitations, this leads to a duality between gauge and gravitational theories.

I turns out a very interesting realisation of 't Hooft duality arises in topological theories, as was discovered by Gopakumar and Vafa in \cite{G-V-transition}. They identified the dual theory to Chern-Simons theory on $S^3$ as the A-model closed topological strings on the so-called \emph{resolved conifold}. The resolved conifold is a simple albeit non-trivial Calabi-Yau space which contains a single non-trivial two-cycle $S^2$. Under this correspondence gauge and string theory couplings are mapped to each other and the 't Hooft coupling $t$ is identified with the size of $S^2$ in the resolved conifold. In particular, Gopakumar and Vafa showed explicitly that in this case the topological string expansion (\ref{F-top-g}) is equal to the resummed gauge theory expansion (\ref{F-closed-intro}) 
%$$
%F_{top}(t)=F_{closed}(t) 
%$$
$$
F^{CS}_g(t) = F_{g}^{conifold}(t).
$$
To sum up, for A-model topological strings
\be
\begin{array}{|ccc|} \hline
\begin{array}{c}
\textrm{Chern-Simons theory} \\ 
\textrm{on}\  S^3
\end{array}
& \iff  &  
\begin{array}{c}
\textrm{closed topological strings} \\ 
\textrm{on the resolved conifold} 
\end{array}   
 \\ \hline            
\end{array}               \label{cs-closed}
\ee

\begin{figure}[htb]
\begin{center}
\includegraphics[width=0.8\textwidth]{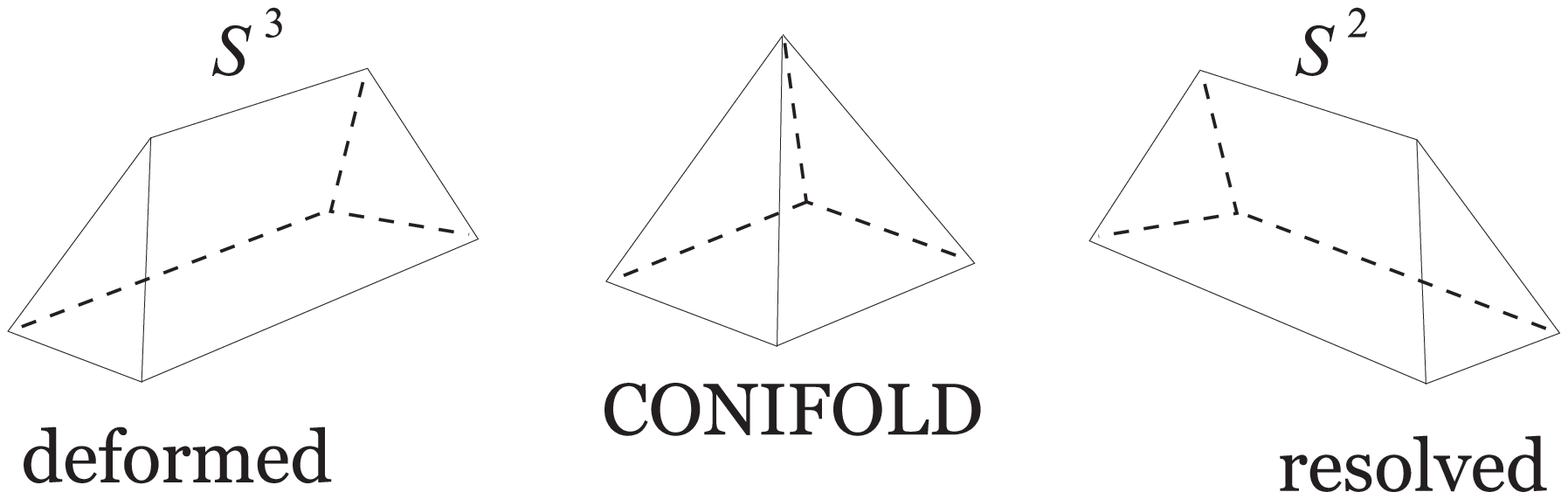}
\caption{The Gopakumar-Vafa geometric transition between A-model open topological strings on the deformed conifold and closed topological strings on the resolved conifold.} \label{fig-transition}
\end{center}
\end{figure}

There is even more remarkable consequence of the above observations. From (\ref{cs-effective}) with $M=S^3$ and (\ref{cs-closed}) we conclude that in the large $N$ limit the A-model \emph{open} topological strings on $T^*S^3$ are equivalent to the A-model \emph{closed} topological strings on the resolved conifold. This is an example of the \emph{open-closed duality} between string theories. In fact, $T^*S^3$ is the so-called \emph{deformed conifold}, which is a Calabi-Yau manifold with a single non-trivial three-cycle $S^3$. Moreover, both resolved and deformed conifolds are two different resolutions of the singular conifold, which is described by the following equation in $\C^4$ with coordinates $(z_1,\ldots,z_4)$
$$
z_1^2 + z_2^2 + z_3^2 + z_4^2 = 0.
$$
This duality therefore can be understood as a smooth process in which a three-sphere $S^3$ inside the deformed conifold shrinks to zero size and subsequently $S^2$ of the resolved conifold blows up, which is accompanied by the vanishing of A-branes wrapping $S^3$ and appearance of the closed string degrees of freedom. This process, often called the Gopakumar-Vafa geometric transition, is shown schematically in figure \ref{fig-transition}.

%*********************************************************

\subsection*{The topological vertex}

\begin{figure}[htb]
\begin{center}
\includegraphics[width=0.3\textwidth]{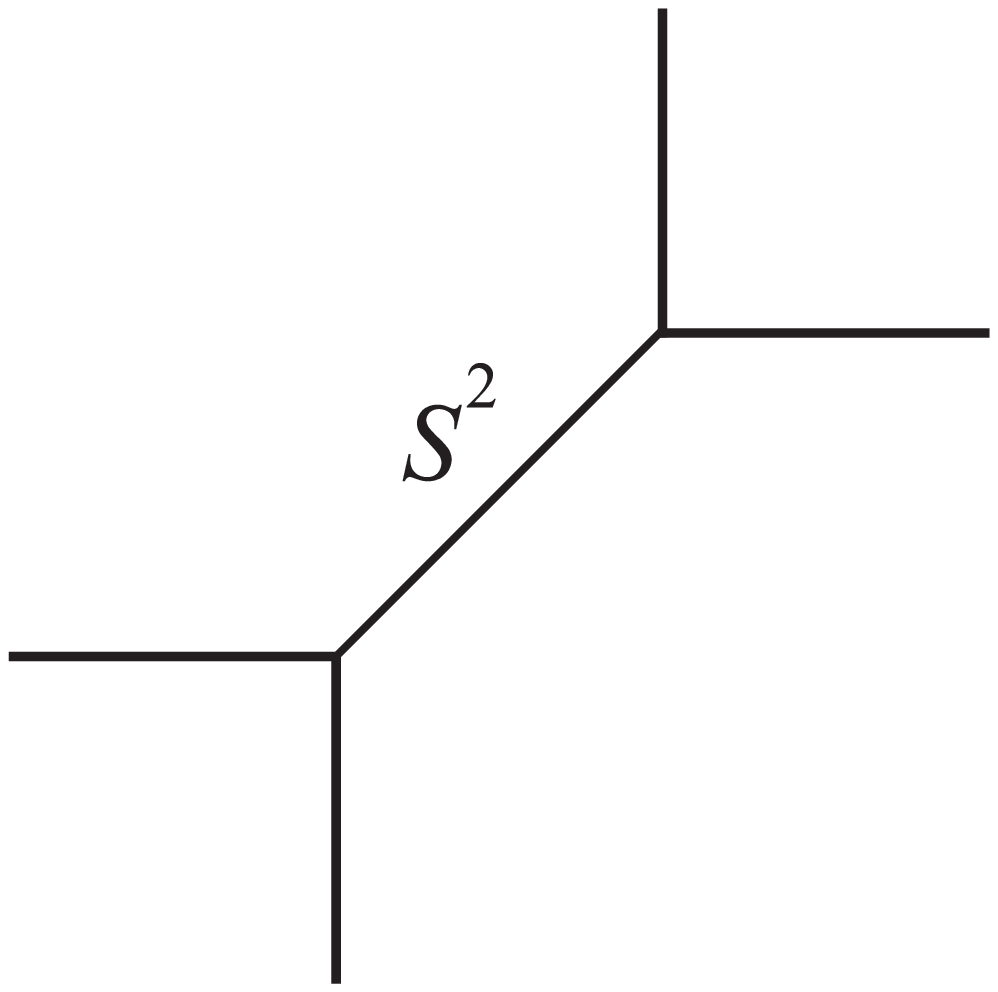}
\caption{The resolved conifold arises from gluing two trivalent vertices (each corresponding to a single $\mathbb{C}^3$ patch) along an interval which represents a non-trivial $S^2$.} \label{fig-coni}
\end{center}
\end{figure}

Gopakumar-Vafa duality relates the closed topological string partition function on the resolved conifold to the partition function of Chern-Simons theory on $S^3$. It turns out this connection can be enormously generalised to allow computations of the A-model topological string amplitudes on a very broad class of non-compact \emph{toric} Calabi-Yau manifolds, essentially in an algorithmic way. Toric manifolds have a structure of a torus bundle over some base, and all data of a toric Calabi-Yau three-fold can be encoded in a two-dimensional graph, whose edges represent loci in the base over which one circle action degenerates. A basic element of such a graph is a trivalent vertex which represents a single $\mathbb{C}^3$ patch, and two patches are glued along a common interval which represents a non-trivial two-sphere $S^2$ inside the manifold. The whole graph encodes information how various $\mathbb{C}^3$ patches are glued together into entire manifold. An example of such a graph for the resolved conifold is shown in figure \ref{fig-coni}.

This prescription for constructing a toric Calabi-Yau manifold can be lifted to a computation of its topological string partition function and even more general amplitudes, by assigning a topological amplitude to each trivalent graph. Such an amplitude is a function of $q=e^{-g_s}$ where $g_s$ is the string coupling constant. It is denoted $C_{R_1 R_2 R_3}$ and it represents open strings ending on three stacks of branes in a single $\mathbb{C}^3$ patch, whose configurations are encoded by Young diagrams $R_i,\  i=1,2,3$. This amplitude is called the \emph{topological vertex} and it was found in \cite{vertex}. Its derivation relies completely on the open-closed Gopakumar-Vafa duality, and it is expressed in terms on large $N$ limit of the Hopf-link invariants (\ref{CS-hopflink}), which makes its relation to Chern-Simons theory transparent. In fact, calculations in terms of the topological vertex resemble usual Feynman rules, with a single diagram representing the entire Calabi-Yau target space, and its vertices and propagators corresponding respectively to strings propagating in $\C^3$ patches and wrapping non-trivial two-spheres inside the manifold. The topological vertex allows also to find topological string amplitudes in a presence of branes. 

The results for closed topological string amplitudes computed from the topological vertex has been compared in many cases with explicit closed string calculations, or checked by other highly non-trivial methods. The agreement found in all such cases is a very strong confirmation of the open-closed duality. Apart from a unique possibility to check quantitatively predictions of such a duality, topological vertex techniques provide enormous simplifications in calculations of topological string amplitudes, which will be used to large extent in this thesis.

%*********************************************************

\section{Calabi-Yau crystals}

We summarised above the construction of the topological string theories and argued that finding their amplitudes is of utmost importance due to many relations with various physical and mathematical theories. We also described how the open-closed duality is realised in topological string theories, and how it leads to the formulation of the topological vertex which allows to find A-model topological string amplitudes on a wide class of non-compact toric Calabi-Yau manifolds. However, at first sight it appears the structure of the topological string amplitudes is very complicated --- already a single topological vertex $C_{R_1 R_2 R_3}$, which is a building block of a total amplitude on a non-trivial Calabi-Yau space, is a very non-trivial function of $q=e^{-g_s}$. Nonetheless, there is a remarkable simplicity encoded in the A-model topological string theories on toric Calabi-Yau manifolds --- it turns out they are closely related to simple statistical models of crystal melting. In particular, topological string amplitudes are reproduced by generating functions of all crystal configurations. This observation allows to interpret toric Calabi-Yau manifolds as \emph{quantum manifolds} in terms of appropriately understood \emph{quantum geometry}. Each crystal configuration can be interpreted as a topology changing quantum fluctuation of the target space, related to some effective target space gravitational theory \cite{foam}. The sum over fluctuating topologies is also called a \emph{quantum foam}, and this gravitational theory, as an effective theory of closed topological strings from figure \ref{fig-transition}, is dual to Chern-Simons theory. This gives a very interesting quantitative example of a gauge-gravity correspondence. To stress the correspondence between statistical models and topological string theories, the former are called \emph{Calabi-Yau crystals} --- and they are the main theme of this thesis.

\begin{figure}[htb]
\begin{center}
\includegraphics[width=0.4\textwidth]{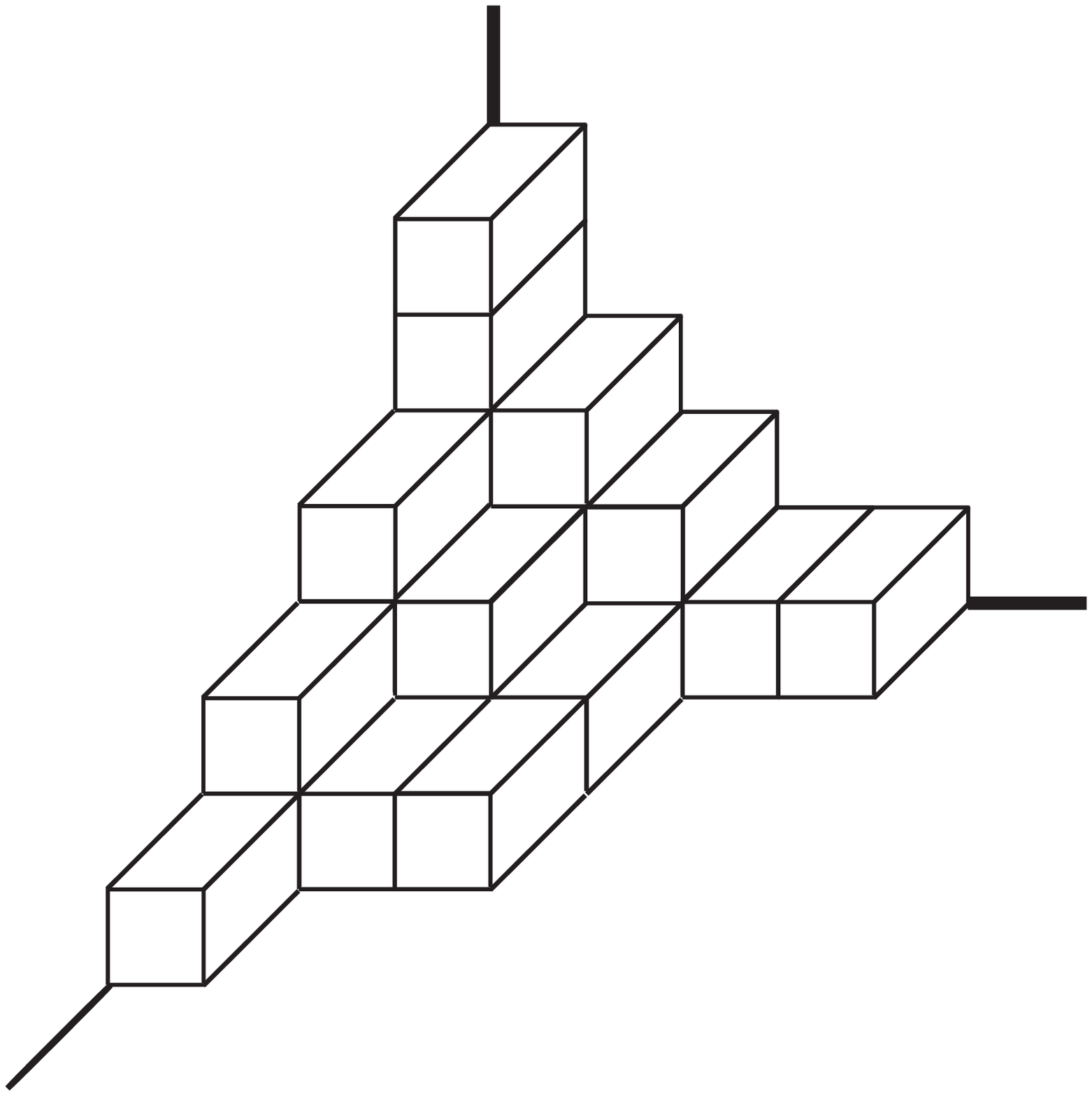}
\caption{A plane partition as an example of a Calabi-Yau crystal configuration.} \label{fig-crystal-3d}
\end{center}
\end{figure}

Calabi-Yau crystal models for topological string theories arise in three real dimensions, which is related to three complex dimensions of Calabi-Yau target spaces. The simplest example of such a crystal is given by the so-called \emph{plane partitions} which fill a positive octant of $\mathbb{Z}^3$. A plane partition can be thought of as a maximally packed configuration of unit boxes localised around a corner of this crystal, as shown in figure \ref{fig-crystal-3d}. Each such configuration $\pi$ represents a melted configuration of a crystal, and its contribution to the crystal generating function is given by its number of boxes $|\pi|$, with each boxes weighted by a parameter $q$. Such a generating function is known as the McMahon function
$$
M(q) = \sum_{\pi} q^{|\pi|} = \prod_{n=1}^{\infty} \frac{1}{(1-q^n)^n},
$$
and remarkably this is equal to the partition function (\ref{Z-top-g}) of the A-model closed topological strings on $\C^3$ if the string coupling $g_s$ is identified as $q=e^{-g_s}$ 
$$
Z^{\C^3}_{top} = M(q) =  \sum_{\pi} q^{|\pi|}.
$$
Many more similar equivalences have been found, which relate more complicated Calabi-Yau manifolds and more complicated ensembles of plane partitions. In particular, the topological vertex amplitude $C_{R_1 R_2 R_3}$ is also related to the counting of a particular set of plane partitions  --- those which asymptote to two-dimensional partitions $R_1, R_2, R_3$ along three axes of the positive octant of $\mathbb{Z}^3$
$$
C_{R_1 R_2 R_3} \sim M(q)^{-1} \sum_{\pi\to\{R_1,R_2,R_3 \} } q^{|\pi|}.
$$
The above relations were observed first in \cite{ok-re-va} and underlie all further developments in this field. Exploring the consequences of the above relations and finding new Calabi-Yau crystal models is the main subject of this thesis.

\section{Contents of the thesis}
 
The main theme of this thesis is the realisation of the A-model topological string theories in terms of Calabi-Yau crystals. We introduce some new crystal models, prove their relation to the topological string theories on appropriate geometries, and analyse them from various points of view, in particular considering various configurations of branes in Calabi-Yau geometries and exhibiting their relation to knot invariants. Moreover, we advance some techniques of the topological vertex and apply them to the analysis of Calabi-Yau geometries related to one of those new crystal models.

In fact, Calabi-Yau crystals arise also for four-dimensional twisted field theories. In this case crystal models are two-dimensional, and are related to the counting of two-dimensional partitions. In this thesis we introduce such models for $\mathcal{N}=4$ twisted gauge theories on ALE spaces of the form $\C^2/\Z_k$, and show their relation to affine Lie algebra characters as originally predicted by Nakajima \cite{naka1,V-W}. We introduce two-dimensional crystal models first, in a sense as a prerequisite for the analysis of three-dimensional models.

\bigskip
%\subsection*{Plan}

The plan of the thesis is as follows. The following three chapters have an introductory character and we present there all ingredients necessary to understand further considerations. In chapter \ref{sec-cy}, after a general presentation of Calabi-Yau manifolds, we introduce ALE spaces in two complex dimensions, as well as toric spaces in three complex dimensions together with their presentation in terms of toric diagrams. We focus a particular attention on conifolds and their geometric transitions. In chapter \ref{chap-top-field} we introduce topological field theories of Schwarz and Witten type. An example of the Schwarz type theory is Chern-Simons theory. We present its exact solution and discuss its relation to knot invariants. Then we present Witten type or cohomological theories based on the procedure of twisting. In chapter \ref{chap-top-strings} we introduce various aspects of topological string theories, explaining first their construction via the twisting of sigma models. Subsequently we focus our attention on A-model theories and present how closed topological string amplitudes are encoded in Gromov-Witten and Gopakumar-Vafa invariants, and how open topological strings reduce to Chern-Simons theory. Further we analyse geometric transitions, in particular the \emph{flop} transition and the Gopakumar-Vafa transition. Finally we introduce the topological vertex, which we use to large extent in calculations in the following chapters.

The remaining four chapters are devoted to the analysis of Calabi-Yau crystals. In particular, chapters \ref{chap-ale}, \ref{chap-cy-results} and \ref{sec-A-model} contain original results of the thesis. In chapter \ref{chap-ale} we introduce two-dimensional crystal models for $\mathcal{N}=4$ twisted gauge theories on ALE spaces and prove their relation to affine Lie algebra characters. In chapter \ref{chap-cy-crystals} we introduce three-dimensional Calabi-Yau crystals as various ensembles of plane partitions. We explain their relation to topological string theories, the connection with the topological vertex, and an interpretation of branes as defects in crystals. In chapter \ref{chap-cy-results} we introduce crystal models with one, two or three walls, also with defects representing branes, and find corresponding amplitudes. When it is possible we relate them explicitly to the knot invariants. In chapter \ref{sec-A-model} we analyse Calabi-Yau geometries corresponding to crystal models introduced in chapter \ref{chap-cy-results}, and using topological vertex techniques we prove their amplitudes are indeed reproduced by crystal generating functions. We also extend topological vertex techniques to the so-called \emph{off-strip} geometries, and use them to analyse the so-called \emph{closed topological vertex geometry} and its flop transition.

In appendices we collect various issues related mainly to calculational methods used in the thesis. In appendix \ref{app-star} we introduce various objects which appear in the thesis in multiple contexts: two-dimensional partitions and closely related free fermions, affine Lie algebras and their characters, and symmetric functions. Appendix \ref{app-topver} contains details of topological vertex calculational techniques and the so-called \emph{rules on the strip}, which vastly simplify topological vertex calculations on a particular class of geometries.

%\subsection*{Guide to bibliography}
\bigskip

Finally, let us briefly summarise the bibliography. Textbooks on String Theory, algebraic geometry, Conformal Field Theory, knot theory, and combinatorial issues are given in \cite{gsw}-\cite{macdonald}. Review articles \cite{dijkgraaf-topfield}-\cite{pioline} concern various aspects of topological theories. Topological field theories and their relation to instanton counting are discussed in \cite{top-field}-\cite{rhd-ps}. General references on topological string theory, together with a few specific results we use, are given in \cite{top-sigma}-\cite{OSV}. Gauge-gravity correspondence in general, as well as its realisation in topological strings via the connection to Chern-Simons theory and knot invariants, is introduced in \cite{thooft}-\cite{largeN}. The theory of the topological vertex is developed in \cite{vertex}-\cite{gen-ver}. The development of Calabi-Yau crystals and related statistical models is commemorated in \cite{ok-re-va}-\cite{pearcey}. Some mathematical and physical issues relevant to our considerations are given in \cite{har-law}-\cite{F-theory}.

%1. strings, idea...
%2. main guidance: 
%   a) known fundamental laws 
%      - gravity, qm -> relation to strings
%   b) understanding and consistency
%      - dualities - what they are, examples:open-closed, S, T etc.
%3. crucial idea: open-closed duality
%   - general idea
%   - meaning: QCD, quantum gravity, etc.
%   - e.g. Gross, AdS/CFT
%4. exactly solvable models: e.g. topological fields and strings; relation to some observables in physical models!!
%5. open-closed for top. strings
%   - reduction to field theories (CS and Kahler)
%   - observables: partition functions and branes
%   - topological vertex - CY crystals and statistical models, top. sectors, quantum foam
%6. this thesis:
%   - new crystal models and top. vertex techniques
%   - resume of contents 

                % shorter version

% \input{intro-old3.tex}                    % this is longer version...

%***********************************************************
%***********************************************************
%***********************************************************

\chapter{Calabi-Yau manifolds and toric geometry}  \label{sec-cy}

% The main subject of this thesis are \emph{Calabi-Yau crystals}. Let us explain what we understand under this name. We consider various quantum topological field and string theories and would like to determine their amplitudes, such as partition functions or expectation values of some operators. On one hand, the theories we consider are defined on Calabi-Yau manifolds. On the other hand, in consequence of a topological character of these theories, computations of their amplitudes reduce to performing a sum over some countable set of elements with prescribed weights. For topological field theories such a set consists of two-dimensional partitions, or Young diagrams. For topological string theories this set consists of plane partitions, which are three-dimensional analogs of Young diagrams. Such ensembles of two- or three-dimensional partitions are considered in statistical mechanics as models of crystals. 

The main subject of this thesis are \emph{Calabi-Yau crystals}, which appear in the context of topological field and string theories defined on Calabi-Yau spaces. In this chapter we introduce Calabi-Yau spaces and toric geometry to the extent necessary to analyse topological theories and their crystal interpretation in further chapters of the thesis. After a concise review of complex and K\"ahler geometry in section \ref{sec-complex-kahler} we introduce Calabi-Yau manifolds and comment on Mirror Symmetry in section \ref{sec-calabi-yau}. There are two particular classes of geometries which will be of particular interest to us: ALE spaces of complex dimension two and toric manifolds of complex dimension three. They are described respectively in sections \ref{ale-spaces} and \ref{ssec-toric}. A family of \emph{conifolds} is described in section \ref{sec-conifolds}. Geometric transitions of conifolds, which are crucial for our considerations concerning topological strings, are analysed in section \ref{sec-transition}. More details concerning complex and toric geometry and various other mathematical structures which appear in the context of topological theories, as well as Mirror Symmetry, can be found in \cite{mirror} and \cite{greene}. More formal mathematical presentation is given in \cite{grif-har}.

\section{Complex and K\"ahler manifolds}    \label{sec-complex-kahler}

In this section we recall basic notions related to complex geometry which underlie all further considerations. 

A \emph{complex $n$-dimensional manifold} $M$ is a topological space which can be covered by charts isomorphic to open sets in $\C^n$ with holomorphic transition functions.
%charts $(U_{\alpha},z_{(\alpha)})$ such that  
%\begin{itemize} 
%\item $U_{\alpha}$ are open subsets of $M$ such that $M=\cup_{\alpha} U_{\alpha}$, and \emph{local coordinates} $z_{(\alpha)}$ are one-to-one maps from $U_{\alpha}$ to $\C^n$.
%\item in every non-empty intersection $U_{\alpha}\cap U_{\beta}$ the map $z_{(\alpha)} z_{(\beta)}^{-1}$ is holomorphic; equivalently, in every such intersection the \emph{transition functions} $f_{\alpha\beta}$ relating local coordinates, $z_{(\alpha)}=f_{\alpha\beta}(z_{(\beta)})$, are holomorphic.
%\end{itemize}
Any complex manifold of dimension $n$ is of course a real manifold of dimension $2n$, but the converse statement in general is not true. An \emph{almost complex structure} $J$ on a real $2n$-dimensional manifold is defined as a global map $J:T_pM\to T_pM$ which squares to identity $J^2 =-1$. For a manifold with an almost complex we can define a Nijenhuis tensor $N$ acting on two vector fields $X,Y$ by $N(X,Y)=[JX,JY]-J[X,JY]-J[JX,Y]-[X,Y]$. A theorem by Nijenhuis states that a manifold with an almost complex structure is in fact a complex manifold if and only if $N=0$, and in such a case $J$ is called a \emph{complex structure}.

A metric on a manifold $M$ is a symmetric map $G:T_pM\times T_pM \to \R$. On a complex manifold it can be linearly extended to a symmetric map from complexified tangent spaces into complex numbers, which we denote by the same symbol $G: T_pM^{\C}\times T_pM^{\C} \to \C$. Such a metric is called \emph{hermitean} if it satisfies the condition $G(X,Y)=G(JX,JY)$ for any vector fields $X,Y$. 

For a hermitean metric we define a corresponding antisymmetric form $k$ by $k(X,Y)=g(X,JY)$. This form is called \emph{K\"ahler} if its exterior derivative vanishes, $dk=0$. In such a case the corresponding metric and the entire manifold are also called \emph{K\"ahler}.

We usually denote local complex coordinates in a fixed chart by $z^j=x^j+iy^j, j=1,\ldots, n$ and their conjugations by $\bar{z}^{\bar{j}}$. Local real coordinates are denoted by $x^{\mu}, \mu=1,\ldots 2n$. The tensors and conditions introduced above can be rewritten using these local coordinates. In real coordinates an almost complex structure can be written as a tensor with mixed indices $J^{\mu}_{\nu}$. If a metric $g_{\mu\nu}$ is introduced indices can be lowered and raised, and the condition that a metric is hermitean reads $J_{\mu\nu}=-J_{\nu\mu}$, where $J_{\mu\nu}=J^{\rho}_{\mu}g_{\rho\nu}$. A hermitean metric written locally in complex coordinates takes the form $G=G_{j\bar{k}}dz^j d\bar{z}^{\bar{k}}$, with all other components vanishing $G_{jk}=G_{\bar{j}\bar{k}}=0$. The corresponding antisymmetric form is $k=\hf J_{\mu\nu} dx^{\mu}\wedge dx^{\nu} =\frac{i}{2}G_{j\bar{k}} dz^j\wedge d\bar{z}^{\bar{k}}$. 

The K\"ahler condition $dk=0$ reads in components as $\partial_i G_{j\bar{k}}=\partial_j G_{i\bar{k}}$, which implies that the metric is determined locally by some function $K$ called \emph{K\"ahler potential}, $G_{i\bar{j}}=\partial_i \partial_{\bar{j}} K$. The K\"ahler potential is defined up to a holomorphic function $f=f(z)$, so that $K+f+\bar{f}$ defines the same metric as $K$. On the K\"ahler manifold, all components of the Levi-Civita connection with mixed complex indices vanish, while the remaining ones are $\G^l_{jk} = G^{l\bar{n}} \partial_{j}G_{k\bar{n}}$ and $\G^{\bar{l}}_{\bar{j}\bar{k}} = G^{\bar{l}n} \partial_{\bar{j}}G_{\bar{k}n}$. It follows the non-trivial components of the Riemann tensor are only $R_{i\bar{j}k\bar{l}} = G_{i\bar{n}} \partial_k \G^{\bar{n}}_{\bar{j}\bar{l}}$ up to permutations of indices consistent with symmetries of the Riemann tensor. Then the Ricci tensor reads $R_{\bar{i}j}=R^{\bar{k}}_{\bar{i}\bar{k}j} = -\partial_j \G^{\bar{k}}_{\bar{i}\bar{k}}$.

In local coordinates the exterior derivative is $d=dx^{\mu}\partial_{\mu}$, which satisfies $d^2=0$. When a differential form $\omega$ of degree $s$ is written in complex coordinates it decomposes into holomorphic and antiholomorphic pieces $\omega=\omega_{\mu_1\ldots \mu_s} dx^{\mu_1}\cdots dx^{\mu_s} = \omega_{i_1\ldots i_p j_1\ldots j_r} dz^{i_1}\cdots dz^{i_p} d\bar{z}^{\bar{j}_1}\cdots d\bar{z}^{\bar{j}_r}$, and it is called a $(p,r)$-form. Consequently, we have a decomposition of the space of forms of degree $s$
$$
\Omega^s(M) = \bigoplus_{s=p+r} \Omega^{p,r}(M).
$$
We also introduce the holomorphic and anti-holomorphic exterior derivatives $\partial = dz^i\partial_i$ and $\bar{\partial} = d\bar{z}^{\bar{i}} \partial_{\bar{i}}$, which act as
$$
\partial : \Omega^{p,r}(M) \to \Omega^{p+1,r}(M),\qquad   \bar{\partial} : \Omega^{p,r}(M) \to \Omega^{p,r+1}(M).
$$
If $M$ is a complex manifold  we have
$$
d=\partial+\bar{\partial},\quad \partial^2=0,\quad \bar{\partial}^2=0,\quad \partial\bar{\partial}+\bar{\partial}\partial=0,
$$
and in addition to de Rham cohomology groups $H^s(M)$ there are the so-called \emph{Dolbeault} cohomology groups $H^{p,r}(M)$ defined using $\bar{\partial}$ (equivalently $\partial$ could be used). De Rham and Dolbeault cohomology groups are defined respectively as
$$
H^s(M) = \frac{\{ \omega\in \Omega^s(M)\, | \, d\omega =0  \} }{ \{\gamma\in\Omega^s(M)\,| \gamma=d\rho  \}}, \qquad  H^{p,r}(M) = \frac{ \{\omega\in \Omega^{p,r}(M)\, | \, \bar{\partial}\omega =0  \} }{ \{\gamma\in\Omega^{p,r}(M)\,| \gamma=\bar{\partial}\rho  \} }.
$$
Their dimensions are denoted as
\be
b_s=\textrm{dim}\,H^s(M),\qquad h^{p,r}=\textrm{dim}\, H^{p,r}(M),   \label{def-hpr}
\ee
and the former are called Betti numbers. 

If $M$ is a manifold with metric, then it is possible to introduce consistently a scalar product of forms $\langle \omega|\gamma\rangle$. Then an adjoint operator $d^{\dag}$ is defined by imposing $\langle \omega |d\gamma \rangle = \langle d^{\dag} \omega | \gamma \rangle $, and it can be shown the de Rham cohomology classes are in one-to-one correspondence with the so-called \emph{harmonic forms}, which are annihilated by the Laplacian $\Delta_d=d^{\dag}d + dd^{\dag}$. 

On a complex manifold $M$ with hermitean metric one defines in an analogous way (e.g. $\langle \omega |\bar{\partial}\gamma \rangle = \langle \bar{\partial}^{\dag} \omega | \gamma \rangle $) adjoint operators 
$$
\partial^{\dag} : \Omega^{p,r}(M) \to \omega^{p-1,r}(M),\qquad   \bar{\partial}^{\dag} : \Omega^{p,r}(M) \to \omega^{p,r-1}(M).
$$
and corresponding Laplacians $\Delta_{\partial}=\partial^{\dag}\partial + \partial\partial^{\dag}$ and $\Delta_{\bar{\partial}}=\bar{\partial}^{\dag}\bar{\partial} + \bar{\partial}\bar{\partial}^{\dag}$. It can be shown the Dolbeault cohomology classes (defined with respect to $\bar{\partial}$) are in one-to-one correspondence with $\bar{\partial}$-harmonic forms, annihilated by $\Delta_{\bar{\partial}}$. If in addition $M$ is K\"ahler one can prove
$$
\Delta_d = 2\Delta_{\bar{\partial}} = 2\Delta_{\partial},
$$
therefore harmonic forms with respect to any of those Laplacians are the same, and in consequence
\be
H^s(M)=\bigoplus_{p+r=s} H^{p,r}(M),\qquad \textrm{and}\quad b_s = \sum_{p+r=s} h^{p,r}.  \label{bs-hpr}
\ee
Moreover, in this case 
\be
h^{p,r}=h^{r,p}, \qquad  h^{p,r}=h^{\textrm{dim}_{\C}M - p,\textrm{dim}_{\C}M - r}.   \label{h-pr}
\ee
Finally, the Euler characteristic of $M$ reads $\chi(M)=\sum_s (-1)^s b_s = \sum_{p,r} (-1)^{p+r} h^{p,r}$.

The existence of a K\"ahler metric will be crucial for our considerations for the following reason. The induced volume form of any holomorphic $m$-dimensional submanifold of $M$ is given by $k^m/m!$. Because $k$ is closed, it determines some cohomology class which can be specified by the so-called real \emph{K\"ahler parameters} 
\be
t_i=\int_{S_i}k,   \qquad  i=1,\ldots,b_2(M),  \label{kahler-param}
\ee
where $S_i$ are a basis of $H_2(M,\Z)$. Partition functions of topological string theories which are main objects of our considerations depend on such K\"ahler parameters.

%*********************************************************
%*********************************************************

\section{Calabi-Yau manifolds}  \label{sec-calabi-yau}

A complex manifold $M$ is called a \emph{Calabi-Yau manifold} if it is K\"ahler and its metric is Ricci-flat, $R_{i\bar{j}}=0$.

In this section we denote $n=\textrm{dim}_{\C}M$. $n$-dimensional Calabi-Yau manifolds are often called $n$-\emph{folds}, as for example \emph{two-folds} or \emph{three-folds}. The Ricci-flatness condition in the above definition can be replaced by one of the following equivalent conditions:
\begin{itemize}
\item the holonomy group of $M$ is $SU(n)$,
\item the first Chern class of the manifold vanishes, $c_1(TM)=0$,
\item the canonical bundle $K_M$, i.e. a bundle of $(n,0)$-forms, is trivial; this means it can be identified with $M\times \C$, and there must exist a global nowhere vanishing holomorphic $(n,0)$-form corresponding to a constant section, denoted as
\be
\Omega \in \Omega^{n,0}(M).  \label{cy-omega}
\ee 
\end{itemize}

In mathematics literature Calabi-Yau manifolds are usually defined to be compact, but we do not impose this condition. In fact in this thesis we are interested mainly in non-compact manifolds. 

A lot of information about a Calabi-Yau manifold is encoded in dimensions of Dolbeault cohomology groups $h^{p,r}$ introduced in (\ref{def-hpr}). For a connected manifold, which is always the case in our considerations, $h^{0,0}=1$. The existence of trivial canonical bundle means $h^{n,0}=1$. The condition of $SU(n)$ holonomy can be shown to imply $h^{p,0}=0$ for $p\neq 0,n$. Together with relations (\ref{bs-hpr}) and (\ref{h-pr}) we are left with a quite restricted set of parameters encoding cohomology structure of the manifold. These parameters are usually presented in the form of the so-called \emph{Hodge diamond}
%$$
%\begin{array}{|c|c|c|c|c|}  \hline
%h^{0,n} &         &         & \cdots   & h^{n,n} \\ \hline
%\vdots  &         &         &    &  \vdots\\ \hline
%h^{0,2} & \vdots  &         &    &   \\ \hline
%h^{0,1} & h^{1,1} & \cdots   &    &  \\ \hline
%h^{0,0} & h^{1,0} & h^{2,0} & \cdots & h^{n,0} \\ \hline
%\end{array}
%$$
$$
\begin{array}{ccccccc}  
  &   &   & h^{0,0}  &   &   &  \\ 
  &   & h^{1,0}  &    & h^{0,1}  &   & \\ 
  & \textrm{\begin{rotate}{66}$\ddots$\end{rotate}} &   & \vdots &  & \ddots & \\
h^{n,0}  &     & \cdots   &  & \cdots & & h^{0,n} \\ 
  & \ddots &   & \vdots &  & \textrm{\begin{rotate}{66}$\ddots$\end{rotate}} & \\
  &   & \ddots &    & \textrm{\begin{rotate}{66}$\ddots$\end{rotate}} &   & \\
  &   &   & h^{n,n}  &   &   &  
\end{array}
$$
In particular, for $n=1,2,3$, using the above relations the Hodge diamonds are respectively 
$$
\begin{array}{ccc} 
  & 1 &  \\ 
1 &   & 1 \\
  & 1 &
\end{array} \qquad \qquad
\begin{array}{ccccc}  
  &   & 1  &   & \\ 
  & 0 &    & 0 & \\ 
1 &   & 20 &   & 1 \\ 
  & 0 &    & 0  & \\
  &   & 1  &    &
\end{array} \qquad \qquad
\begin{array}{ccccccc}  
  &   &   & 1  &   &   &  \\ 
  &   & 0 &    & 0 &   & \\ 
  & 0 &   & h^{2,1} &  & 0 & \\
1 &   & h^{1,1} &   & h^{1,1} & & 1 \\ 
  & 0 &   & h^{2,1} &  & 0 & \\
  &   & 0 &    & 0 &   & \\
  &   &   & 1  &   &   &  
\end{array}
$$
%$$
%\begin{array}{|c|c|}  \hline
%1 & 1 \\ \hline
%1 & 1 \\ \hline
%\end{array}, \qquad \qquad
%\begin{array}{|c|c|c|}  \hline
%1 &  0 & 1 \\ \hline
%0 & 20 & 0 \\ \hline
%1 &  0 & 1 \\ \hline
%\end{array}, \qquad \qquad
%\begin{array}{|c|c|c|c|}  \hline
%1 & 0 & 0 & 1 \\ \hline
%0 & h^{2,1} & h^{1,1} & 0 \\ \hline
%0 & h^{1,1} & h^{2,1} & 0 \\ \hline
%1 & 0 & 0 & 1 \\ \hline
%\end{array}
%$$
A one-dimensional complex manifold is a Riemann surface, and it is K\"ahler,
because its K\"ahler two-form --- as every two-form in two real dimensions --- is
necessarily closed. However, the condition of special holonomy implies the only
compact one-dimensional Calabi-Yau manifold is a torus. Apart from $T^4$, the only non-trivial compact Calabi-Yau two-fold is the so-called $K3$ surface, and the value $h^{1,1}=20$ can be deduced from the knowledge of its Euler characteristic $\chi(K3)=24$. Three-folds are already more complicated, with cohomology structure specified by two parameters $h^{1,1}$ and $h^{2,1}$. 

A very important issue concerning Calabi-Yau manifolds are their \emph{moduli spaces}. A \emph{moduli space} is a space of parameters of a family of Calabi-Yau manifolds, such that any two manifolds in this family can be transformed into each other by smooth deformations. More precisely, we consider smooth deformations of their metrics, which for general manifold $M$ are of the form
$$
\delta G = \delta G_{ij} dz^i dz^j + \delta G_{i\bar{j}} dz^i dz^{\bar{j}} + c.c.
$$
and become very restricted if $M$ is a Calabi-Yau manifold, because in this case the Ricci-flatness condition must hold $R_{i\bar{j}}=0$. As seen in the above formula, there are two types of metric deformations: either with mixed indices $\delta G_{i\bar{j}}$, or with both holomorphic indices $\delta G_{ij}$. The Ricci-flatness implies $G_{i\bar{j}} dz^i \wedge dz^{\bar{j}}$ must be a harmonic form, so it corresponds to a unique element of $H^{1,1}_{\bar{\partial}}(M)$. Therefore deformations with indices of mixed type are interpreted as a change of the K\"ahler class $k$ to a new element of $H^{1,1}_{\bar{\partial}}(M)$, and these deformations constitute some $h^{1,1}$-parameter space. For an $n$-fold with $n>2$ the coordinates on this space can be identified with K\"ahler parameters (\ref{kahler-param}), as in such case there are indeed $b_2=h^{1,1}$ of them.

On the other hand, deformations with mixed indices spoil a hermiticity property of a manifold $M$. However, hermiticity condition is imposed for some fixed complex structure $J$, and it turns out that for allowed $\delta G_{ij}$ one can always modify the complex structure to a new one $J+\delta J$, in which hermiticity holds, although it requires non-holomorphic change of coordinates. These deformations of complex structure are of the form  $\delta J = (\epsilon^i_{\bar{j}}\partial_i)d\bar{z}^{\bar{j}}$. For a three-fold, one can use its holomorphic three-form $\Omega$ to map $\delta J$ uniquely into a holomorphic $(2,1)$-form $\epsilon^i_{\bar{j}} \Omega_{ikl} d\bar{z}^{\bar{j}} \wedge dz^k \wedge dz^l$, therefore complex deformations correspond to elements in $H^{2,1}_{\bar{\partial}}(M)$.

To sum up, a moduli space of Calabi-Yau manifolds splits up into a space of K\"ahler deformations and a space of complex deformations. A family of K\"ahler deformations is always $h^{1,1}$-dimensional. For a three-fold, a family of complex deformations is $h^{2,1}$-dimensional. This provides a complete interpretation of the numbers in a Hodge diamond for a three-fold. 

A very important notion concerning Calabi-Yau manifolds is \emph{Mirror Symmetry}. It is believed Calabi-Yau manifolds arise in pairs, such that Hodge diamonds of two manifolds in such a pair differ only by a reflection along a diagonal (hence the name ``mirror''). In particular, under this symmetry K\"ahler and complex parameters of both manifolds are exchanged with each other. There are various formulations of Mirror Symmetry which differ in strength of their assumptions, and its most general mathematical proof is not known. However, many examples which are already known, as well as very strong physical indications arising from string theory, including proofs at the physical level of rigour, make Mirror Symmetry a very plausible statement. A modern, in-depth overview of Mirror Symmetry is given in \cite{mirror}.

%*********************************************************
%*********************************************************

\section{Two-folds and ALE spaces}      \label{ale-spaces}

As explained above, $K3$ is the only non-trivial compact Calabi-Yau space in two complex dimensions. It can be constructed as the so-called \emph{resolution} of a quotient $T^4/\Z_2$. The quotiening by $\Z_2$ is understood as an identification $\vec{x} \sim -\vec{x}$, where $\vec{x}$ are periodic coordinates on a torus $T^4$. Under this identification 16 singular points arise. A \emph{resolution} of this singular space is a new smooth space, called $K3$, in which singularities are replaced by spheres $S^2$. 

\begin{figure}[htb]
\begin{center}
\includegraphics[width=0.7\textwidth]{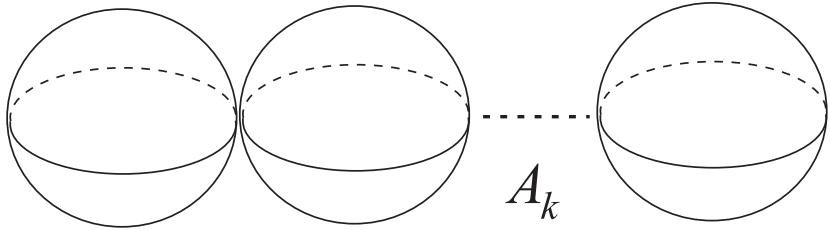}
\caption{ALE space of $A_k$ type contains $k$ spheres whose intersection numbers are given by the Cartan matrix of $A_k$ Lie algebra (\ref{cartan}), and the Dynkin diagram of this algebra (figure \ref{fig-dynkin}) determines relative locations of these spheres. If all spheres blow down to zero size we obtain a singular orbifold space $\C^2/\Z_{k+1}$.} \label{fig-ale}
\end{center}
\end{figure}

There is the whole family of non-compact Calabi-Yau spaces which are related to $K3$ surface and can be thought of as local neighbourhoods of some regions in $K3$. They are called \emph{Asymptotically Locally Euclidean} spaces, or ALE spaces for short. They can be constructed as resolutions of orbifolds of the form
\be
\mathbb{C}^2/\Gamma,         \label{ale-space}
\ee
where $\Gamma$ is a finite subgroup of $SU(2)$. Indeed, at infinity they asymptote locally to the Euclidean space $\C^2\sim\R^4$, whereas the fact we divide by a subgroup of $SU(2)$ ensures the resulting space has $SU(2)$ holonomy, which is the Calabi-Yau condition. Globally, the boundary of ALE space is the so-called Lens space $S^3/\Gamma$. Similarly as in $K3$ case, in the process of resolution the singular orbifold points of (\ref{ale-space}) are replaced by spheres $S^2$. The geometry of ALE space is determined by a relative locations of these spheres, which is in fact very peculiar: their intersection numbers are equal to the entries of a Cartan matrix of a certain Lie algebra $\mathfrak g$. This provides a mapping between spheres $S^2$ and a nodes in a Dynkin diagram of $\mathfrak g$. Moreover, it turns out that irreducible representations of $\Gamma$ are in one-to-one correspondence with nodes of the extended Dynkin diagram, and their dimensions $d_i$ are equal to the corresponding dual Dynkin indices. This provides a one-to-one mapping between finite subgroups of $SU(2)$ and Lie algebras
\be
SU(2) \supset \Gamma \leftrightarrow \mathfrak g,            \label{mckay}
\ee
known as McKay correspondence. In particular, under this correspondence abelian groups $\Gamma=\Z_{k+1}$ are mapped to $A_{k}$ Lie algebras, dihedral groups are mapped to Lie groups of $D$ type, and symmetry groups of regular solids are mapped to exceptional Lie algebras of $E$ type. ALE space of $A_k$ type is shown schematically in figure \ref{fig-ale}.

ALE spaces and McKay correspondence will be the main subject of chapter \ref{chap-ale}. More detailed introduction to ALE spaces and McKay correspondence can be found in \cite{mckay}.

%*********************************************************
%*********************************************************

\section{Three-folds and toric geometry}  \label{ssec-toric}

A \emph{toric} manifold is a manifold which has a fibration structure with torus fibres. Now we wish to restrict our attention to non-compact toric Calabi-Yau three-folds, as they will of great interest to us when we consider topological string theories. More precisely, we consider three-folds with a structure of $T^2\times \R$ fibration over $\R^3$ following \cite{branes-toric,vertex}. 

\subsection*{$\C^3$}

It turns out all the information how $\C^3$ patches are glued into a toric three-fold can be encoded in a simple two-dimensional graph. The edges of this graph represent loci of the base space $\R^3$ over which certain cycles of $T^2$ fibre degenerate. A basic building block of these graphs is a planar trivalent vertex, which represents a single $\C^3$ patch. Graphs corresponding to various $\C^3$ patches of the entire manifold are glued together along common edges, and Calabi-Yau conditions ensure all vertices we glue lie in a common plane.

Let us start with a single $\C^3$ patch, which we interpret as a phase space with canonical coordinates $z_i, i=1,2,3$, momenta $\bar{z}_i$,  and a symplectic form 
\be
\omega=\sum_{i=1}^3 dz_i\wedge d\bar{z}_i.  \label{c3-sympl}
\ee
We define three Hamiltonians
\bea
r_{\alpha} & = & |z_1|^2 - |z_3|^2, \nonumber \\
r_{\beta}  & = & |z_2|^2 - |z_3|^2, \\               \label{c3-r-toric}
r_{\gamma} & = & \textrm{Im}(z_1 z_2 z_3), \nonumber
\eea
associated to three ``times'' $\epsilon=\alpha,\beta,\gamma$. These Hamiltonians define $\R^3$ basis of the fibration we are looking for. The Poisson bracket $\{\cdot,\cdot\}$ associated to $\omega$ generates three flows on $\C^3$
$$
\partial_{\epsilon} z_i = \{ r_{\epsilon}, z_i\},
$$
which determine fibres of the fibration. The $T^2$ fibre is generated by
\be
e^{i\alpha r_{\alpha}+ i\beta r_{\beta}}:\ (z_1,z_2,z_3) \to (e^{i\alpha} z_1, e^{i\beta}z_2, e^{-i(\alpha+\beta)}z_3)  \label{toric-fiber}
\ee
with $\alpha,\beta \in [0,2\pi [$, and $\R$ fibre is generated by $r_{\gamma}$. We denote cycles generated by $r_{\alpha}$ and $r_{\beta}$ respectively as $(0,1)$ and $(1,0)$. 

\begin{figure}[htb]
\begin{center}
\includegraphics[width=0.4\textwidth]{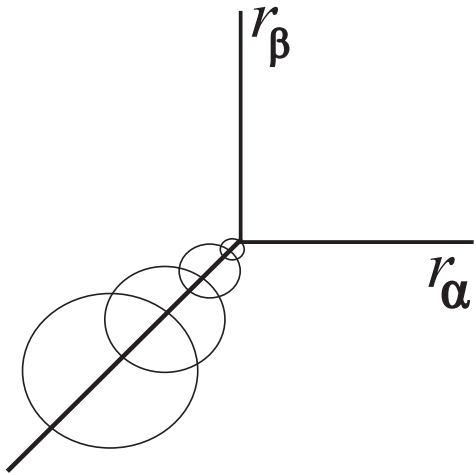}
\caption{A \emph{two-dimensional} graph representing $\C^3$ as a toric manifold. Over each edge one circle in the fibration degenerates.} \label{fig-c3-toric}
\end{center}
\end{figure}

The cycle parametrised by $\alpha$ degenerates over the subspace $r_{\alpha}=r_{\gamma}=0, r_{\beta}>0$, or equivalently $z_1=z_3=0$. The cycle parametrised by $\beta$ degenerates over the subspace $r_{\beta}=r_{\gamma}=0, r_{\alpha}>0$, or equivalently $z_2=z_3=0$. The cycle parametrised by $\alpha+\beta$ degenerates over $r_{\alpha}-r_{\beta}=r_{\gamma}=0, r_{\alpha}\leq 0$, or equivalently $z_1=z_3=0$. This degeneration structure can be encoded in a trivalent vertex drawn in a plane $r_{\gamma}=0$, such that a cycle degenerating over a locus $p r_{\alpha} + q r_{\beta}=0$ is represented by a line in a direction $(-q,p)$. In the present example the degenerating loci correspond to cycles of $T^2$ which we denote $v_1=(-1,-1)$, $v_2=(0,1)$ and $v_3=(1,0)$, and the corresponding vertex is shown in figure \ref{fig-c3-toric}. More generally, we could use $SL(2,\Z)$ symmetry of $T^2$ to transform $(0,1)$ and $(1,0)$ cycles into $(a,b)$ and $(c,d)$ with $a,b,c,d\in\Z$ and $ad-bc=1$. Such a transformation would rotate edges into new ones $v_1,v_2,v_3$. After any such $SL(2,\Z)$ transformation these vectors have to satisfy
$$
\sum_{i=1}^3 v_i =0.
$$

%**********************************************************************

\subsection*{Gluing $\C^3$}

General toric three-fold $M$ can be constructed from gluing $\C^3$ patches in the following way. We introduce complex coordinates $z_1,\ldots,z_{N+3}$, so that triples of them describe various $\C^3$ patches $U_a=(z_{i_a},z_{j_a},z_{k_a})$. Gluing of patches is achieved by imposing $N$ linear identifications of the form
\be
\sum_j Q_j^A |z_j|^2 = t^A,\qquad \textrm{for}\quad A=1,\ldots, N     \label{toric-glue}
\ee
and then dividing a space of solutions of these equations by $U(1)^N$ action acting on coordinates as
\be
z_j \mapsto \exp(i\, Q^A_j \alpha_A) \, z_j.   \label{toric-glue-divide}
\ee
The Ricci-flatness condition for a Calabi-Yau manifold arises in the present context as 
$$
\sum_{i=1}^{N+3} Q_i^A = 0.
$$

\begin{figure}[htb]
\begin{center}
\includegraphics[width=0.4\textwidth]{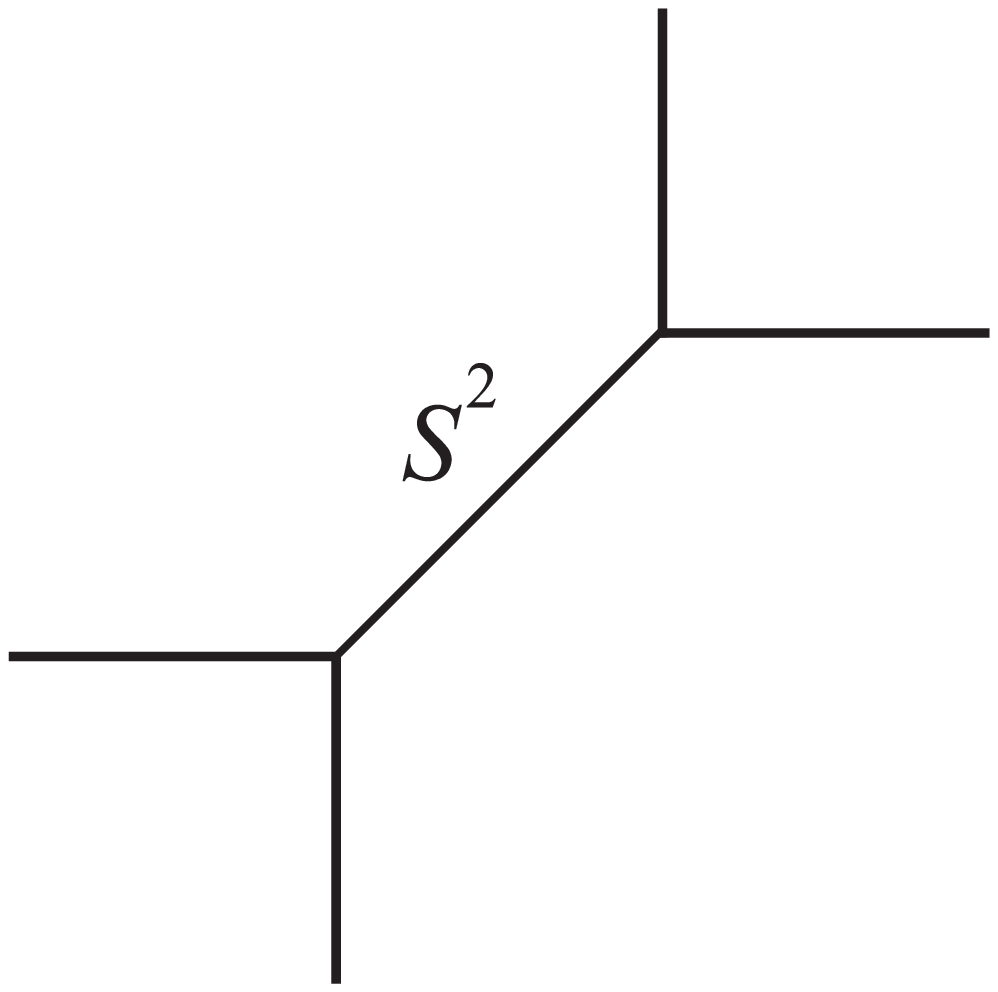}
\caption{Resolved conifold constructed from two $\C^3$ patches. The finite interval represents non-trivial $S^2$ in the total geometry, which arises from $S^1$ in the fibre shrinking to zero size in two vertices.} \label{fig-conifold}
\end{center}
\end{figure}

In this setting the entire three-fold $M$ can be presented as a fibration over $\R^3$ base with a \emph{global} $T^2$ fibre defined as above. In particular in $U_1$ patch it takes the form
\bea
r_{\alpha} & = & |z_{1_1}|^2 - |z_{3_1}|^2, \nonumber \\
r_{\beta}  & = & |z_{2_1}|^2 - |z_{3_1}|^2, \nonumber
\eea
and in other patches actions of $r_{\alpha}$ and $r_{\beta}$ can be found solving (\ref{toric-glue}). Furthermore, $\R$ fibre is defined patch by patch as
$$
r_{\gamma} = \textrm{Im}\, \prod_{i=1}^{N+3} z_i.
$$
This allows to draw a graph representing entire three-fold $M$. Each $\C^3$ patch is represented by a single vertex, and the Calabi-Yau conditions ensure all vertices lie in one plane. They are glued along common edges, which have to point in opposite directions, and after gluing become a finite interval. Such an interval represents $S^2$ embedded in $M$, arising from $S^1$ fibre shrinking to zero size at two vertices which are ends of the interval. The sizes of such $S^2$'s are given by $t^A$'s in (\ref{toric-glue}), which are therefore precisely the K\"ahler parameters introduced in  (\ref{kahler-param}). An example of a toric manifold glued from two $\C^3$ patches, the so-called \emph{resolved conifold}, is given in figure \ref{fig-conifold}. We will discuss this example in more detail in the next section.

%**********************************************************************

\subsection*{Lagrangian submanifolds} 

Let $M$ be $2n$-dimensional symplectic manifold with symplectic form $\omega$. Its submanifold $L$ is called a \emph{lagrangian submanifold} if it is $n$-dimensional and $\omega$ vanishes on a tangent space to $L$
\be
\omega|_{L} = 0 \qquad \iff \qquad \forall_{p\in L}\forall_{X,Y\in T_pM}\  \omega(X,Y) = 0.   \label{lagr-subm}
\ee
This is a general definition of a lagrangian submanifold. For K\"ahler manifolds, we can identify symplectic two-form with a K\"ahler form $k$.  

There are some interesting examples of lagrangian submanifolds in toric Calabi-Yau manifolds considered above. They were introduced in \cite{har-law} and will be of particular interest to us. Let us describe the geometry of these submanifolds in $\C^3$ first. Using the notation introduced in (\ref{c3-r-toric}), they are given by half-lines parametrised by $r_{\gamma}\geq 0$ in the base $\R^3$ together with $T^2$ in the fibre. Such manifolds might have a boundary at $r_{\gamma}=0$. We wish to consider submanifolds without a boundary, which is possible if the half-line $r_{\gamma}\geq 0$ in the base ends on some edge of the toric diagram, where one circle of $T^2$ fibre degenerates. Therefore a topology of these lagrangian submanifolds is $S^1\times \C$. For $\C^3$ there are therefore three continuous families of them
\bea
L_1: & & r_{\alpha}=r_{\beta}=u, \quad r_{\gamma}\geq 0, \nonumber \\
L_2: & & r_{\alpha}=u, \quad r_{\beta}=0, \quad r_{\gamma}\geq 0, \\   \label{c3-lag-brane}
L_3: & & r_{\alpha}=0, \quad r_{\beta}=u, \quad r_{\gamma}\geq 0, \nonumber
\eea
with $u=const$ being a modulus which corresponds to a position along an edge where the half-line in the base ends. 

For toric manifolds with arbitrary toric diagram there are analogous classes of lagrangian submanifolds, given by half-lines in the base and $T^2$ in the fibre. The half-line must end on the edge a toric diagram, and this position becomes a modulus of this submanifold.

\begin{figure}[htb]
\begin{center}
\includegraphics[width=0.4\textwidth]{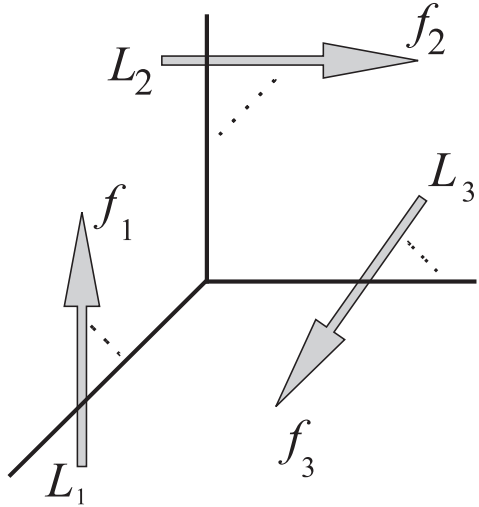}
\caption{$\C^3$ with lagrangian submanifolds in the canonical framing.} \label{fig-c3-branes}
\end{center}
\end{figure}

In the context of topological string amplitudes, one often considers slightly modified submanifolds, with additional $(p,q)$ cycle in the $T^2$ fibre degenerating at infinity. This cycle is specified by an additional \emph{framing vector} $w=(-q,p)$ in $\R^3$ base. With this modification, a lagrangian brane is $T^2$ fibration over a half-line in the base with two cycles degenerating at the ends of this half-line. Therefore it would have a topology of $S^3$, if these two cycles of the torus were not homologous. This happens if 
\be
w\times v = w_1v_2 - v_1w_2 = 1,   \label{wxv}
\ee
where $v$ is a vector specifying an axis the submanifold is attached to, $w$ is its framing vector, and $\times$ denotes a usual vector product projected on the axis perpendicular to the toric diagram. In fact all vectors $w-fv$ with $f\in\Z$ satisfy (\ref{wxv}) as long as $w$ does. Therefore a framing of a lagrangian submanifold can be specified by a single number $f$ with respect to some reference framing vector. For $\C^3$ with axes $v_1,v_2,v_3$ we define such reference framing for each axis 
\be
w_1=v_2,\qquad w_2=v_3,\qquad w_3=v_1.   \label{canon-frame}
\ee
This is called the \emph{canonical framing} and it is shown in figure \ref{fig-c3-branes}.

%**********************************************************************

\subsection*{Dual graphs}

\begin{figure}[htb]
\begin{center}
\includegraphics[width=0.4\textwidth]{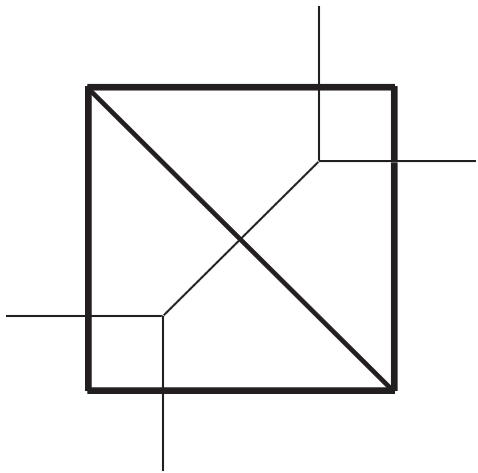}
\caption{A dual diagram for the resolved conifold. The internal interval corresponds to $S^2$.} \label{fig-conifold-dual}
\end{center}
\end{figure}

A structure of a toric manifold can be equivalently encoded in terms a graph which is dual to a toric diagram. For a three-fold, the dual diagram is obtained by replacing faces of the toric diagram by points, and connecting these points by intervals which intersect intervals in the original diagram. A dual diagram also encodes a lot of information about the manifold. Its vertices correspond to divisors (submanifolds of complex codimension 1), whereas internal intervals to compact curves, each one arising as an intersection of two divisors represented by the ends of this interval. An example of the dual diagram for the resolved conifold from figure \ref{fig-conifold} is shown in figure \ref{fig-conifold-dual}.

%**********************************************************************
%**********************************************************************

\section{Conifolds}   \label{sec-conifolds}

We consider now two interesting non-compact Calabi-Yau three-folds: \emph{the deformed conifold} and \emph{the resolved conifold} \cite{mirror,grif-har,greene,marcos,conifolds}. They are nice examples of ideas introduced above and their properties will be essential in the following considerations. These manifolds are resolutions of the so-called \emph{conifold singularity}. We describe them below as subsets of $\C^4$ with coordinates $z=(z_1,\ldots,z_4)$.

The \emph{conifold singularity} is a singular set described by the equation
$$
z_1^2 + z_2^2 + z_3^2 + z_4^2 = 0.
$$
If we write $z = x + i y$, the above equation is equivalent to
$$
x^2 = y^2 = \frac{r^2}{2},\qquad \sum_j x_jy_j=0,\qquad \textrm{for}\quad r^2 = x^2 + y^2\geq 0.
$$ 
For fixed $r$ we can interpret $x$ as living on a sphere $S^3$. For a fixed $x$ the allowed $y$, being perpendicular to $x$, must lie on $S^2$. As any fibration of $S^2$ over $S^3$ is necessarily trivial, the singular conifold is a cone over $S^2\times S^3$, and the very singularity corresponds to a point $x=y=0$, the tip of this cone.

The \emph{deformed conifold} is a subset of $\C^4$ described by the equation
\be
z_1^2 + z_2^2 + z_3^2 + z_4^2 = R^2 \qquad \iff \qquad x^2-y^2 = R^2,\quad \sum_j x_jy_j=0.   \label{def-coni}
\ee
Asymptotically this is $S^2\times S^3$ as in the singular case. Again $x$ lie on a sphere $S^3$, however the radius of this sphere cannot be smaller than $R$, which happens for $y=0$. Therefore this is a smooth manifold, with a singular point replaced by $S^3$ of radius $R$. This can also be interpreted as the total bundle $T^*S^3$. This is in fact the whole family of manifolds, parametrised by $R$. As a change of the parameter $R$ changes a polynomial defining a manifold, this is associated to a deformation of the complex structure. 

\begin{figure}[htb]
\begin{center}
\includegraphics[width=0.4\textwidth]{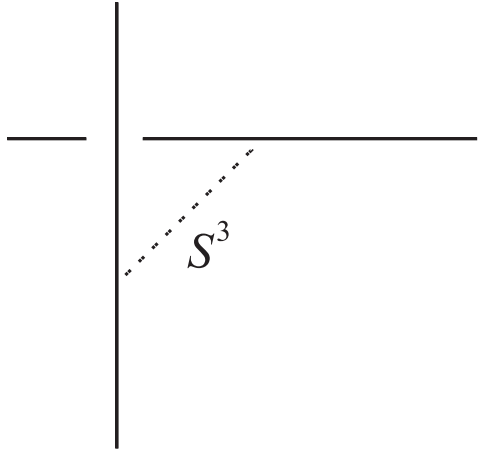}
\caption{Deformed conifold as $T^2\times\R$ fibration over $\R^3$. The solid lines represent degeneration loci in the $\R^3$ base where one of the two homology cycles of $T^2$ degenerates. A torus $T^2$ is fibred over the dashed interval, whose two opposite cycles degenerate at the ends of this interval, revealing minimal $S^3$ in the geometry.} \label{fig-def-coni-toric}
\end{center}
\end{figure}

The deformed conifold cannot be presented in terms of a planar toric diagram, although it also has a structure of $T^2\times \R$ fibration over $\R^3$. By a linear change of coordinates from $z_i$ into $\zeta_i$ the equation (\ref{def-coni}) can be transformed into 
$$
\zeta_1\zeta_2 - \zeta_3\zeta_4 = R^2,
$$ 
invariant under the action 
$$
(\zeta_1,\zeta_2,\zeta_3,\zeta_4) \mapsto (e^{-i\alpha} \zeta_1,e^{i\alpha} \zeta_2,e^{-i\beta} \zeta_3,e^{i\beta} \zeta_4)
$$
where $\alpha$ and $\beta$ parametrise respectively $(1,0)$ and $(0,1)$ cycles of $T^2$. Each of these cycles degenerates respectively along $\zeta_1=\zeta_2=0$ or $\eta_3=\zeta_4=0$ locus. There are two cylinders associated to these loci, $\zeta_3\zeta_4=-R^2$ and $\zeta_1\zeta_2=R^2$. The fibration structure is given by parametrising $\R^3$ base by axes of these cylinders and $\zeta=\textrm{Re}(\zeta_1\zeta_2)$, and $T^2\times\R$ fibre by $\alpha,\beta$ and $\textrm{Im}(\zeta_1\zeta_2)$. Then the degeneration loci are located respectively at $\zeta=0$ and $\zeta=R^2$, and $T^2$ fibred over an interval connecting the two loci has opposite circles shrinking at its ends, which reveals non-trivial $S^3$ in the geometry. This degeneration structure is shown in figure \ref{fig-def-coni-toric}.

It is also possible to invoke a K\"ahler deformation to replace the singular conifold by another smooth manifold, the so-called \emph{resolved conifold}. This is a toric manifold, and we describe it using the toric language introduced above. We consider one constraint (\ref{toric-glue}) in $\C^4$
\be
- |z_1|^2 - |z_2|^2 + |z_3|^2 + |z_4|^2 = t,   \label{res-coni}
\ee
with the corresponding $U(1)$ action (\ref{toric-glue-divide})
$$
(z_1,z_2,z_3,z_4) \mapsto (e^{-i\alpha} z_1,e^{-i\alpha} z_2,e^{i\alpha} z_3,e^{i\alpha} z_4).
$$
This geometry is asymptotically again of the form $S^2\times S^3$, but now there is a minimal sphere $S^2$ corresponding to $z_1=z_2=0$. Its area is proportional to $t$, which is a K\"ahler parameter (\ref{kahler-param}). The above equations mean the resolved conifold is in fact the $\mathcal{O}(-1)\times\mathcal{O}(-1)\to\mathbb{P}^1$ bundle, with $\mathbb{P}^1$ identified with $S^2$. The two $\mathcal{O}(-1)$ patches are $U_{123}=(z_1,z_2,z_3)$ with $z_4\neq 0 $ and $U_{124}$ with $z_3\neq 0$. Using the toric language, in the $U_{123}$ patch the global Hamiltonians can be chosen as in (\ref{c3-r-toric})
$$
r_{\alpha} = |z_{1}|^2 - |z_{3}|^2, \qquad   r_{\beta} = |z_{2}|^2 - |z_{3}|^2.
$$
The $T^2$ fibre they generate is the same as in (\ref{toric-fiber}), and this patch is the same as the one in figure \ref{fig-c3-toric}. In $U_{124}$ patch the global Hamiltonians can be rewritten as
$$
r_{\alpha} = -t + |z_{4}|^2 - |z_{2}|^2, \qquad  r_{\beta} = -t + |z_{4}|^2 - |z_{1}|^2,
$$
and they generate the action
$$
e^{i\alpha r_{\alpha}+ i\beta r_{\beta}}:\ (z_1,z_2,z_4) \to (e^{-i\alpha} z_1, e^{-i\beta}z_2, e^{i(\alpha+\beta)}z_4), 
$$
with degeneration loci $z_1=z_4=0$ with $r_{\beta}=-t$ where $(-1,0)$ cycle degenerates, $z_2=z_4=0$ with $r_{\alpha}=-t$ where $(0,-1)$ cycle degenerates, and $z_1=z_2=0$ with $r_{\alpha}=r_{\beta}$ and $(1,1)$ cycle degenerating. This gives the second vertex, rotated by an angle $\pi$ with respect to the first one. The two vertices are connected by a finite interval of size $t$ representing the minimal $S^2$, as shown in figure (\ref{fig-conifold}).

%**********************************************************************
%**********************************************************************

\section{Geometric transitions}   \label{sec-transition}

Calabi-Yau manifolds often arise in families parametrised by certain complex or K\"ahler moduli. Generically, changing slightly these moduli affects the size and the shape of a given manifold, but it does not change its topological properties. However, sometimes it happens various families with distinct topological properties have common points, so it is possible to move smoothly from one family to another. Such phenomena are called \emph{geometric transitions}. They may affect the topology of a manifold in various ways --- either more drastically, changing the Hodge diamond of the manifold, or rather mildly, leaving the Hodge diamond intact, albeit changing other invariants such as intersection numbers \cite{greene}. 

Two different geometric transitions may arise in the above example of conifolds, commonly known as the \emph{flop transition} and the \emph{conifold transition}. They will be very important for us when we discuss topological string theory. With this perspective in mind, we describe them now just from the geometric point of view. 

\begin{figure}[htb]
\begin{center}
\includegraphics[width=0.7\textwidth]{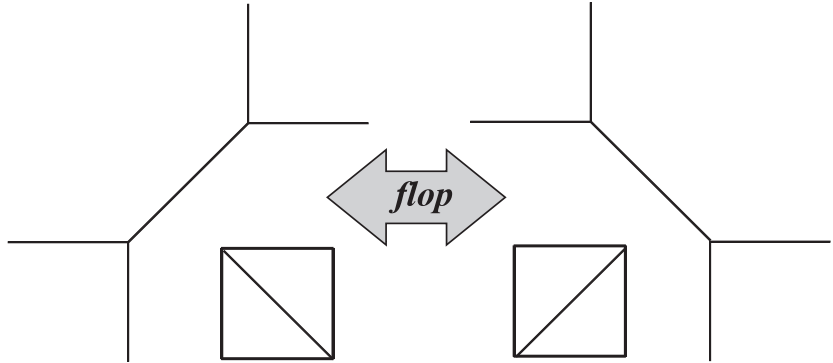}
\caption{Flop transition. Spheres $S^2$ in resolved conifolds on two sides of the transition extend in orthogonal directions, which is represented as a tilt of a diagonal in the dual toric diagram.} \label{fig-flop-conifold}
\end{center}
\end{figure}

The flop transition is a process in which the size $t$ of $\mathbb{P}^1$ inside the resolved conifold is formally continued to negative values. We start from a manifold given by (\ref{res-coni}) with positive $t$, which corresponds to $\mathbb{P}^1$ extended in directions $z_3-z_4$. Next we shrink or \emph{blow down} this $\mathbb{P}^1$ to zero size. Finally, the continuation to negative $t$ means that $\mathbb{P}^1$ is \emph{blown up} in orthogonal directions, in $z_1-z_2$ plane. The geometries on both sides of the transition are resolved conifolds with one non-trivial two-cycle, although these two-cycles are different, as they lie in orthogonal directions. The flop is shown in figure \ref{fig-flop-conifold}. In terms of a dual diagram it corresponds to a tilt of its diagonal.

\begin{figure}[htb]
\begin{center}
\includegraphics[width=0.7\textwidth]{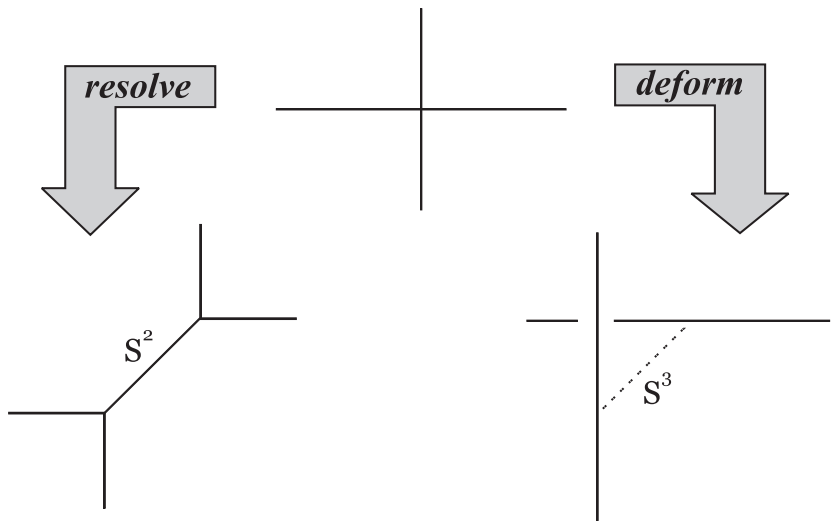}
\caption{Geometric transition between resolved (left) and deformed (right) conifolds, with a singular conifold in an intermediate state.} \label{fig-transition-toric}
\end{center}
\end{figure}

The conifold or Gopakumar-Vafa transition is a process in which $\mathbb{P}^1$ inside the resolved conifold shrinks to zero size, and subsequently the singular conifold is replaced by the deformed one by enlarging $R$ in (\ref{def-coni}). Therefore the homology structure changes drastically during this transition: the initial resolved conifold contains non-trivial two-cycle and no three-cycles, whereas the deformed conifold after the transition contains a non-trivial three-cycle and no two-cycles. This transition is shown in figure \ref{fig-transition-toric}.

%***********************************************************
%***********************************************************
%***********************************************************

\chapter{Topological field theories}  \label{chap-top-field}

The main subject of this thesis are topological \emph{string} theories. However, it is advantageous to discuss first topological \emph{field} theories. On one hand, the construction of topological string theories involves the so-called twisting, which is easier to introduce and understand on the example of cohomological field theories. Moreover, instanton counting in cohomological theories is explicitly related to topological string amplitudes. On the other hand, the knowledge of Chern-Simons topological field theory is a prerequisite for all our further considerations, as by a chain of powerful dualities it allows to solve topological string theories on a very wide class of non-trivial backgrounds. Moreover, Calabi-Yau crystal models arise for both topological field and string theories, and it is useful to introduce them first in the former case. Apparently, in the latter case crystal models are also intimately related to knot invariants arising in Chern-Simons theory. 

Let us consider a quantum field theory defined on a manifold $M$. Such a theory is called \emph{topological} if its observables are topological invariants, which means they do not change under continuous variations of the metric of $M$. In particular the partition function, a fundamental quantity in every quantum theory, should provide some topological invariant of the manifold $M$. Another interesting class of observables computed by topological theories are expectation values of Wilson loops, and invariance under continuous changes of their shapes implies they should be related to knot and link invariants. This is indeed so, as we will see when we consider particular examples of topological theories. 

There are two known classes of topological field theories. The so-called theories of Schwarz type are explicitly independent of the metric of the underlying manifold $M$. The main example of such a theory is Chern-Simons theory discussed in section \ref{sec-cs}. Theories in the second class, the so-called Witten or cohomological type theories, are constructed from originally metric-dependent supersymmetric theories via the operation of \emph{twisting}. In section \ref{sec-cohom} we consider twisted $\mathcal{N}=4$ supersymmetric gauge theory as an example of such topological theory. More detailed introductions to topological field theories can be found in \cite{dijkgraaf-topfield,dijkgraaf-hierarchies}.

%***********************************************************
%***********************************************************

\section{Chern-Simons theory}  \label{sec-cs}

Chern-Simons theory as a quantum gauge theory was analysed in-depth by Witten in a seminal paper \cite{witten-cs}. Its concise review can be found for example in \cite{labastida-cs,marcos}. Chern-Simons theory is a gauge theory in three dimensions with an action which can be written entirely in terms of differential forms, and therefore it does not depend explicitly on the metric of the underlying manifold $M$
\begin{equation}
S^{CS} = \frac{k}{4\pi} \int_M \textrm{Tr}\,\big(A\wedge dA + \frac{2}{3}A\wedge A\wedge A \big).  \label{CS-action}
\end{equation}
This is a necessary condition for a theory to be of Schwarz type, but to obtain topological observables one must ensure the path integral measure does not introduce metric dependence in the process of quantisation.  This is indeed so up to the subtlety of \emph{framing} which in an integer number $f \in \mathbb{Z}$. This means for a given manifold $M$ there is a family of Euclidean partition functions
$$
Z^f_M = \int \mathcal{D}A \, e^{i S^{CS}}
$$
all of which are topological invariants of $M$, and a choice of $f$ uniquely specifies one of them. There is a distinguished framing for $f=0$ called the \emph{canonical framing}, which we usually refer to in what follows. We also have to specify the gauge group $G$. In principle this could be arbitrary, but usually we assume $G=U(N)$ or $G=SU(N)$.

In fact, the action (\ref{CS-action}) is not completely invariant under gauge transformations, which can introduce an additional term $2\pi i k$. Thus consistency of the quantum theory requires $k\in \mathbb{N}$, which has very important consequences. Moreover, the so-called \emph{quantum shift} $k\to k+N$ arises upon renormalisation in the relation between $k$ and the gauge coupling constant $\lambda$ used in perturbative expansion 
\be
\lambda = \frac{2\pi i}{k+N}. \label{cs-lambda}
\ee

An interesting class of observables in Chern-Simons theory are vacuum expectation values of Wilson loops. If Wilson loop is computed in a representation $R$ of the gauge group along a \emph{knot} $K$, its vacuum expectation value gives a \emph{knot invariant} of $K$ 
\be
\mathcal{W}^K_R  =  \langle \textrm{Tr}_R P \exp \oint_K A \rangle = \int \mathcal{D}A \, e^{i S^{CS}}\,\textrm{Tr}_R P e^{\oint_K A}.  \label{wilson-loop}
\ee

\begin{figure}[htb]
\begin{center}
\includegraphics[width=0.5\textwidth]{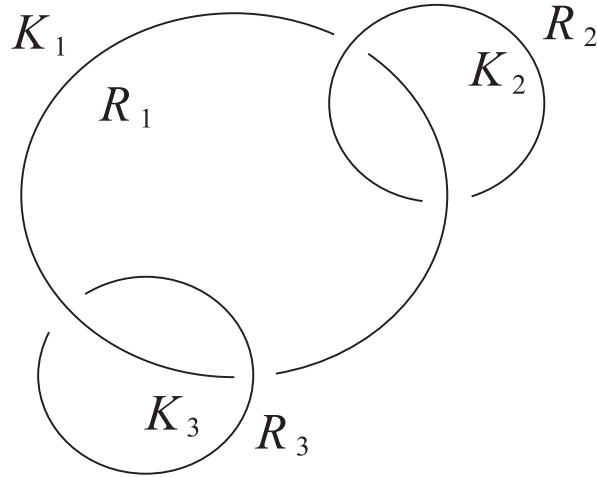}\label{fig-dbl-hopf}
\caption{A link consisting of 3 unknots $K_i$ associated to representations $R_i$.}
\end{center}
\end{figure}

A set of knots $K_1,\ldots,K_L$ constitutes a \emph{link}, and similarly Chern-Simons theory computes link invariants as expectation values of such sets of Wilson loops with some choice of representations $R_1,\ldots, R_L$
$$
\mathcal{W}^{K_1\ldots K_L}_{R_1\ldots R_L} = \langle \textrm{Tr}_{R_1} P e^{ \oint_{K_1} A}\cdots \textrm{Tr}_{R_L} P e^{ \oint_{K_L} A} \rangle.
$$
The simplest knot is the so-called \emph{unknot}, a closed loop homeomorphic to a circle. An example of a link consisting of 3 unknots is shown in figure \ref{fig-dbl-hopf}.

There are many excellent introductions to knot theory, for example \cite{kaufman,lickorish}.

%***************************************************************

\subsection{Quantisation and knot invariants} \label{ssec-cs-quant-knots}

Following \cite{witten-cs}, let us consider Chern-Simons theory on a manifold $M$ with boundary $\partial M = \Sigma$, possibly in a presence of some operator $\mathcal{O}$. In general $\Sigma$ is a two-dimensional Riemann surface. Wave functional of the values of the fields on the boundary $A|_{\Sigma}=A_0$
$$
\Psi_{M}(A_0,\mathcal{O}) = \int_{A|_{\Sigma}=A_0} \mathcal{D} A \,\mathcal{O}\, e^{i S^{CS}}
$$
corresponds to a state $|\Psi_M(\mathcal{O}) \rangle$ in the Hilbert space $\mathcal{H}_{\Sigma}$ which arises in canonical quantisation of Chern-Simons theory on $\Sigma \times \mathbb{R}$
$$
|\Psi_M(\mathcal{O}) \rangle \in \mathcal{H}_{\Sigma}.
$$

\begin{figure}[htb]
\begin{center}
\includegraphics[width=0.5\textwidth]{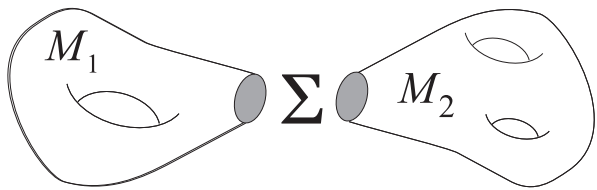}
\caption{The surgery of two manifolds $M_1$ and $M_2$ along their homeomorphic boundaries $\Sigma$.} \label{fig-surgery}
\end{center}
\end{figure}

The amplitude on a closed manifold $M$ can be found by performing the so-called \emph{surgery}, which is the following process. Firstly, $M$ is divided into two parts $M=M_1 \cup_f M_2$ with a homeomorphic boundaries $\Sigma$, as schematically shown in figure \ref{fig-surgery}. $f$ is the homeomorphism which identifies the boundaries
\be
f: \partial M_1\to \partial M_2.        \label{surgery}
\ee
This homomorphism can be lifted to an operator acting in the Hilbert space
\be
U_f: \mathcal{H}_{\Sigma} \to \mathcal{H}_{\Sigma},   \label{surgery-U}
\ee
which allows to identify the amplitude on $M$ as 
$$
Z_M = \langle \Psi_{M_1} | U_f | \Psi_{M_2} \rangle.
$$

The crucial fact presented in \cite{witten-cs} is that the above Hilbert space can be identified with a space of conformal blocks in Wess-Zumino-Witten model at level $k$ on $\Sigma$
\be
\mathcal{H}_{\Sigma} = \textrm{\{conformal blocks in WZW$_{level=k}$ model on $\Sigma$\}}.   \label{wzw}
\ee
%\be
%\begin{array}{|c|} \hline
% \mathcal{H}_{\Sigma} = \{ \textrm{conformal blocks in  WZW$_{level=k}$  model on}\ \Sigma\}.
% \\ \hline            
%\end{array}               \label{wzw}
%\ee
We do not introduce Wess-Zumino-Witten models (or WZW models for short) in this thesis. However, all information required to provide the solution of Chern-Simons theory in cases we are interested in is encoded in properties of affine Lie algebras related to WZW models. Relevant affine Lie algebras and their properties are discussed in appendix \ref{app-characters}, and below we implicitly refer to notation and facts introduced in this appendix. A detailed overview of WZW models and affine Lie algebras can be found in \cite{cft}.

It follows from the identification (\ref{wzw}) that $\mathcal{H}_{\Sigma}$ is of finite dimension. In particular, for $\Sigma=S^2$ this space is one-dimensional. Let us consider in greater detail the case $\Sigma=T^2$, for which conformal blocks of WZW model are in one-to-one correspondence with integrable highest weights of affine Lie algebra associated to $G$ at level $k$. We focus now on $G=SU(N)$, when --- as discussed in appendix \ref{app-characters} --- $\widehat{su}(N)_k$ integrable weights are parametrised by Young diagrams $R$ with less than $N$ rows and at most $k$ columns, $R\in\cY_{N-1,k}$, and the dimension of $\mathcal{H}_{\Sigma}$ is equal to the number of those weights (\ref{nr-weights}). By  $|R\rangle$ we denote a state corresponding to an integrable weight associated with a diagram $R$, and by $\lambda_R$ the corresponding affine integrable weight. In particular the vacuum state of WZW model $|\bullet\rangle$ corresponds to the $k$'th multiplicity of the basic integrable weight $k\widehat{\omega}_0$ associated with the trivial diagram $R=\bullet$. The states $|R\rangle$ can be chosen to be orthonormal
\be
\langle R | R' \rangle = \delta_{RR'}.   \label{orthonormal}
\ee
There is the $SL(2,\mathbb{Z})$ group of homeomorphisms of the torus, with the usual generators
\be
T = \left[\begin{array}{cc}
1 & 1 \\
0 & 1 
\end{array} \right], \qquad \qquad S = \left[\begin{array}{cc}
0 & -1 \\
1 & 0 
\end{array} \right]       \label{torus-ST}
\ee
acting on the homology cycles $(0,1)$ and $(1,0)$, representing respectively contractible and non-contractible circle of the torus. Denoting a modulus of $T^2$ as $\tau$, these actions can be identified with modular transformations (\ref{modular-TS}) which extend to the full Hilbert space as modular matrices $\mathcal{T}_{\lambda_R \lambda_{R'}}$ and $\mathcal{S}_{\lambda_{R} \lambda_{R'}}$ given in (\ref{S-modular}).

Now we consider a solid torus $\mathbb{T}=\mathbb{D}\times S^1$, which is a product of a disk $\mathbb{D}$ and a circle $S^1$. This is a three-dimensional manifold with boundary $T^2$. Path integral over $\mathbb{T}$ without any operator corresponds to the vacuum state $|\bullet\rangle$. When two such solid tori are glued along their boundaries (\ref{surgery}) identified under the identity map $f=id$, two disks combine into a sphere $S^2$ and we get
$$ 
\mathbb{T} \cup_{id} \mathbb{T} = S^2 \times S^1.
$$
This identity map is lifted to the identity operator in Hilbert space (\ref{surgery-U}), and in consequence of the orthonormality (\ref{orthonormal}) the partition function is just
$$
Z_{S^2 \times S^1} = \langle \bullet | \bullet \rangle = 1,
$$
which can be regarded as a normalisation for partition functions of all manifolds which can be engineered in a similar way. 

Let us analyse a more involved case of $S^3$, which can also be obtained by gluing two solid tori, but this time using the homeomorphism $f=S$ given in (\ref{torus-ST}), which exchanges two one-cycles of $T^2$
$$
\mathbb{T} \cup_S \mathbb{T} = S^3.
$$
Using the modular transformations (\ref{S-modular}) together with Weyl denominator formula (\ref{weyl})
we get
\be
Z_{SU(N)}^{S^3} =  \langle \bullet | \mathcal{S} | \bullet \rangle = \mathcal{S}_{\bullet \bullet} = \frac{N^{-1/2}}{(k+N)^{\frac{N-1}{2}}} \prod_{\alpha\in \Delta_+} 2\sin \frac{\pi\,(\alpha|\rho)}{k+N}. \label{Z-SUN-S3}
\ee
This is a very important relation, which underlies many quantitative results in this thesis.

%\frametitle{Unknot}

\begin{figure}[htb]
\begin{center}
\includegraphics[width=0.4\textwidth]{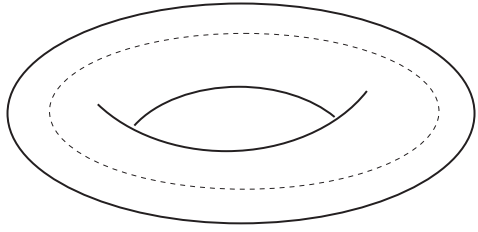}
\caption{Wilson loop along the unknot in the non-contractible cycle of the solid torus.} \label{fig-torus-unknot}
\end{center}
\end{figure}

In the present example it is also quite simple to construct explicitly some knot invariants. Let us consider a Wilson loop $U= P \exp \oint_K A$ along the unknot lying in the non-contractible cycle of one solid torus $\mathbb{T}$, in a representation given by a partition $R$, as shown in figure \ref{fig-torus-unknot}. It can be shown the amplitude for such a configuration determines WZW state associated with a representation $R$
\be
|\Psi_{S^3}(\textrm{Tr}_R U) \rangle =  |R \rangle \in \mathcal{H}_{T^2}. \label{wilson-state-R}
\ee
Performing now the surgery with the homeomorphism $S$ with another empty solid torus the Wilson loop becomes the unknot in $S^3$ and the corresponding amplitude gives a knot invariant associated to the unknot
$$
\mathcal{W}^{unknot}_R =  \langle \bullet|\mathcal{S}|R\rangle = \mathcal{S}_{\bullet R} = Z_{SU(N)}^{S^3} \cdot Q\, s_R(q^{-1/2},\ldots,q^{-N+1/2}) \nonumber
$$
where
$$
q=e^{-\frac{2\pi i}{k+N}},\qquad Q=q^N,
$$
and $s_R$ are Schur symmetric polynomials introduced in appendix \ref{app-schur}. In particular, for $R=\Box$ we get HOMFLY polynomial
\be
Z_{SU(N)}^{S^3}\cdot \mathcal{W}^{unknot}_{\Box} =  Q\,s_{\Box}(q^{-1/2},\ldots,q^{-N+1/2}) = \frac{1-Q}{1-q}. \label{unknot-homfly}
\ee

\begin{figure}[htb]
\begin{center}
\includegraphics[width=0.5\textwidth]{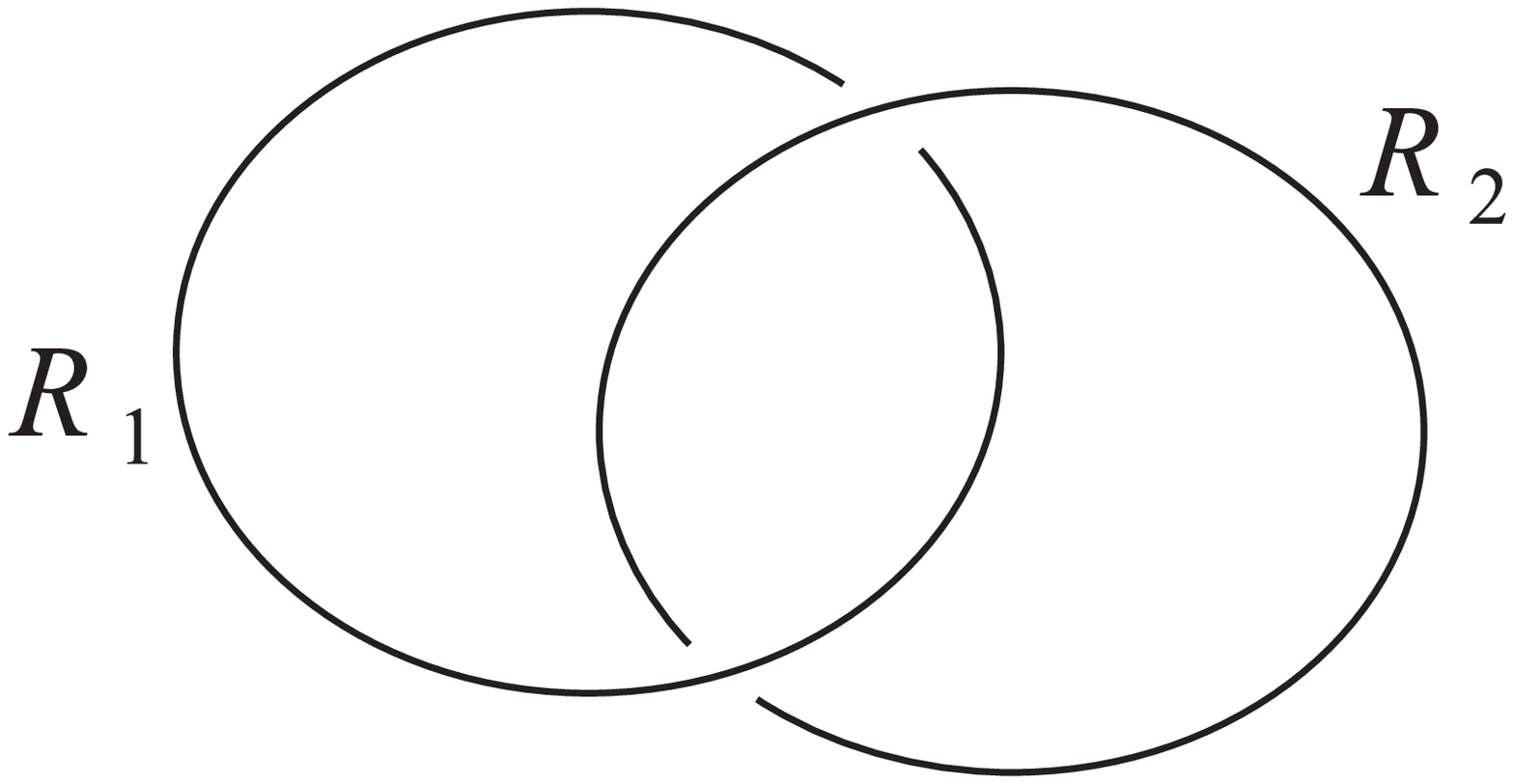}
\caption{Hopf-link.}  \label{fig-hopf}
\end{center}
\end{figure}

The simplest non-trivial example of a link is the so-called \emph{Hopf-link} shown in figure \ref{fig-hopf}, which consists of two interlacing unknots. It can be obtained by performing surgery with homeomorphism $S$ of two solid tori, with Wilson loops along non-contractible cycles of both of them. Each torus produces then a state of the form (\ref{wilson-state-R}), and the Hopf-link invariant $\mathcal{W}_{R_1 R_2}$ is determined by the modular transformation
\be
\langle R_1 | \mathcal{S} | R_2 \rangle = \mathcal{S}_{\lambda_{R_1} \lambda_{R_2}} = Z_{SU(N)}^{S^3}\cdot \mathcal{W}_{R_1 R_2},  \label{hopf-link-modular}
\ee
and it is also possible to write it in terms of Schur functions. We do not need this expression in full generality, and as it is slightly complicated we refer the interested reader to references \cite{marcos,ml-hopf}. 

%\frametitle{Hopf link}

Expectation values for more complicated knots can also be computed from the knowledge of simpler ones. One such operation is called a \emph{direct sum}, and it allows to merge two links along a common component. For example the link in figure \ref{fig-dbl-hopf} can be obtained from merging two Hopf-links with components $(K_1,K_2)$ and $(K_1,K_3)$ respectively, along a common unknot $K_1$. It can be shown the amplitude for such a link indeed can be expressed by Hopf-link invariants and it is equal to
\be
\mathcal{W}_{R_1 R_2 R_3} =  \frac{\mathcal{W}_{R_2^t R_1}\mathcal{W}_{R_3 R_1}}{\mathcal{W}_{R_1 \bullet}}.  \label{dbl-hopf}
\ee

On the other hand, to find the amplitude for a product of two Wilson loops $U$ corresponding to the same knot but in different representations we can apply the \emph{fusion rule}
\be
\langle \textrm{Tr}_{R_1} U \, \textrm{Tr}_{R_2} U \rangle = \sum_R N^R_{R_1 R_2} \langle \textrm{Tr}_R U \rangle,  \label{lit-rich-quant}
\ee
where $N^R_{R_1 R_2}$ are \emph{fusion coefficients} from the underlying conformal field theory. In our applications $k$ and $N$ are always larger than sizes of Young diagrams corresponding to representations $R,R_1,R_2$. In such a case $N^R_{R_1 R_2}$ reduce to usual Littlewood-Richardson coefficients $c^R_{R_1 R_2}$ given in (\ref{lit-rich}).
%which satisfy a classical relation
%\be
%\textrm{Tr}_{R_1\otimes R_2} U = \textrm{Tr}_{R_1} U  \, \textrm{Tr}_{R_2} U = \sum_R N^R_{R_1 R_2} \textrm{Tr}_R U.  \label{lit-rich-clas}
%\ee

\begin{figure}[htb]
\begin{center}
\includegraphics[width=0.8\textwidth]{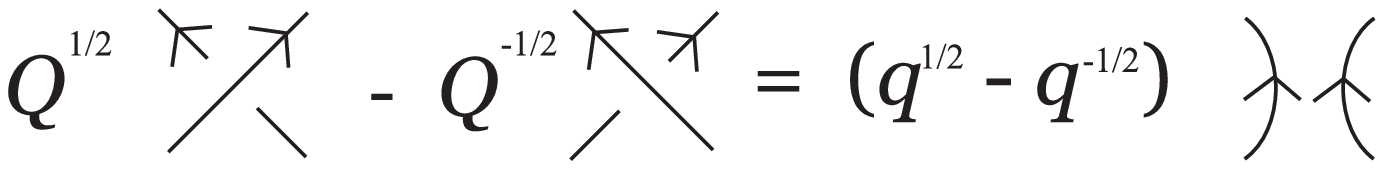}
\caption{Skein relation. For knots or links which differ only in a neighbourhood of one crossing as shown in the figure, it relates their HOMFLY polynomials.}  \label{fig-skein}
\end{center}
\end{figure}

The fact that for $R=\Box$ in example (\ref{unknot-homfly}) we obtained a HOMFLY polynomial discovered by mathematicians is not accidental. The defining property of a HOMFLY polynomial is the so-called \emph{Skein relation} shown in figure \ref{fig-skein}, which is a recurrence relation which allows to determine invariants of more complicated knots in terms of simpler ones. It can be shown the Skein relation arises from the above WZW formalism as long as representations of all Wilson loops correspond to the same diagram $\Box$. Therefore Chern-Simons theory provides a huge generalisation of knot invariants which had not been known to mathematicians before: we automatically get various sets of knot invariants just by changing the gauge group of Chern-Simons theory, its rank, or a particular representation. Moreover, this formalism is valid in principle for any underlying manifold $M$. Let us note some simplest cases of knot invariants easily obtained from Chern-Simons theory:
\begin{itemize}
\item for $SU(N)$ gauge group and $R=\Box$ we get HOMFLY polynomial...
\item ... which for $N=2$ reduces to Jones polynomial,
\item for $SU(2)$, and representation $R$ corresponding to arbitrary spin $j$ we get Akatsu-Wadati polynomials,
\item for $SO(N)$ we get Kauffman polynomial. 
\end{itemize}

Wilson loop observables in Chern-Simons are in fact not uniquely defined. They involve a \emph{framing ambiguity}, which requires specifying an integer number for each component of a link. This ambiguity originates in the fact that in the quantum theory one has to replace widthless components of a link by ribbons of a finite width to regularise the theory properly. A ribbon my be twisted around its axis before it is glued into a closed loop, and the number of such twists is measured by $f_i\in\Z$ for $i$'th component of a link. The so-called \emph{canonical framing} corresponds to taking all ribbons in a link untwisted. If $\mathcal{W}_{R_1\ldots R_L}$ is a link invariant in some particular framing, it is modified by twisting its $i$'th component $f_i$ times into 
\be
\mathcal{W}_{R_1\ldots R_L} \to q^{\hf \sum_{i=1}^L f_i\kappa_{R_i}} Q^{\hf \sum_{i=1}^L f_i|R_i|} \mathcal{W}_{R_1\ldots R_L}, \label{frame-knot}
\ee
where a size $|R|$ and $\kappa_{R}$ are defined respectively in (\ref{R-size}) and (\ref{kappa}).

Quantities we will often come across are leading terms of knot invariants in $N\to\infty$ limit, which will be denoted as $W_R$ instead of $\mathcal{W}_R$. In particular, for the Hopf-link (\ref{hopf-link-modular}) such a leading term is given by
\be
W_{R_1 R_2} = \lim_{N\to\infty} Q^{\frac{|R_1|+|R_2|}{2}}\mathcal{W}_{R_1 R_2} = s_{R_2}(q^{\rho}) s_{R_1}(q^{R_2+\rho}),  \label{hopf-limit}
\ee
and it has a relatively simple expression in terms of Schur functions introduced in appendix \ref{app-schur}.

%***************************************************************

\subsection{'t Hooft expansion in Chern-Simons theory}

Our main motivation for studying Chern-Simons theory is to reveal its relation with certain closed string theory. 
% In this duality worldsheets of closed strings are related to open string worldsheets. The letter are given by Riemman surfaces with holes, which can also be interpreted as diagrams of some gauge theory in the so-called 't Hooft or $1/N$ expansion. 
This relation can be understood if the so-called 't Hooft or $1/N$ expansion of gauge theory is performed, where $N$ is the rank of the gauge group. In the usual perturbative approach to gauge theory the amplitudes are expanded in powers of a small coupling constant $\lambda$. However, for large $N$ these amplitudes can as well be expanded in powers of $1/N$. The general form of this expansion and its relation to certain closed string theory was analysed by 't Hooft in a famous work \cite{thooft}, and its particular form in the case of Chern-Simons theory was derived in \cite{G-V}.

\begin{figure}[htb]
\begin{center}
\includegraphics[width=0.5\textwidth]{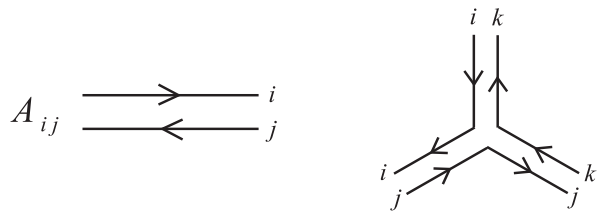}
\caption{A propagator and a cubic gauge coupling in a double-line notation.}  \label{fig-fat-rules}
\end{center}
\end{figure}

Let us consider first the general features of 't Hooft expansion in $1/N$ for an arbitrary gauge theory. The usual Feynman rules are not well-suited to perform such an expansion, as in general a single Feynman diagram captures a nontrivial dependence on $N$. However, it is possible to expand gauge theory amplitudes in terms of a special set of diagrams which contribute a definite power of $N$. These diagrams are called \emph{fatgraphs} and they represent gauge propagators and vertices in terms of ribbons of definite widths (as opposed to lines in usual Feynman expansion). The boundaries of these ribbons represent gauge indices and they combine into closed loops in the fatgraphs, so that each such loop produces a single power of $N$. This expansion is also called a \emph{double line notation} and an example of a propagator and a cubic gauge coupling is shown in figure \ref{fig-fat-rules}.

\begin{figure}[htb]
\begin{center}
\includegraphics[width=0.5\textwidth]{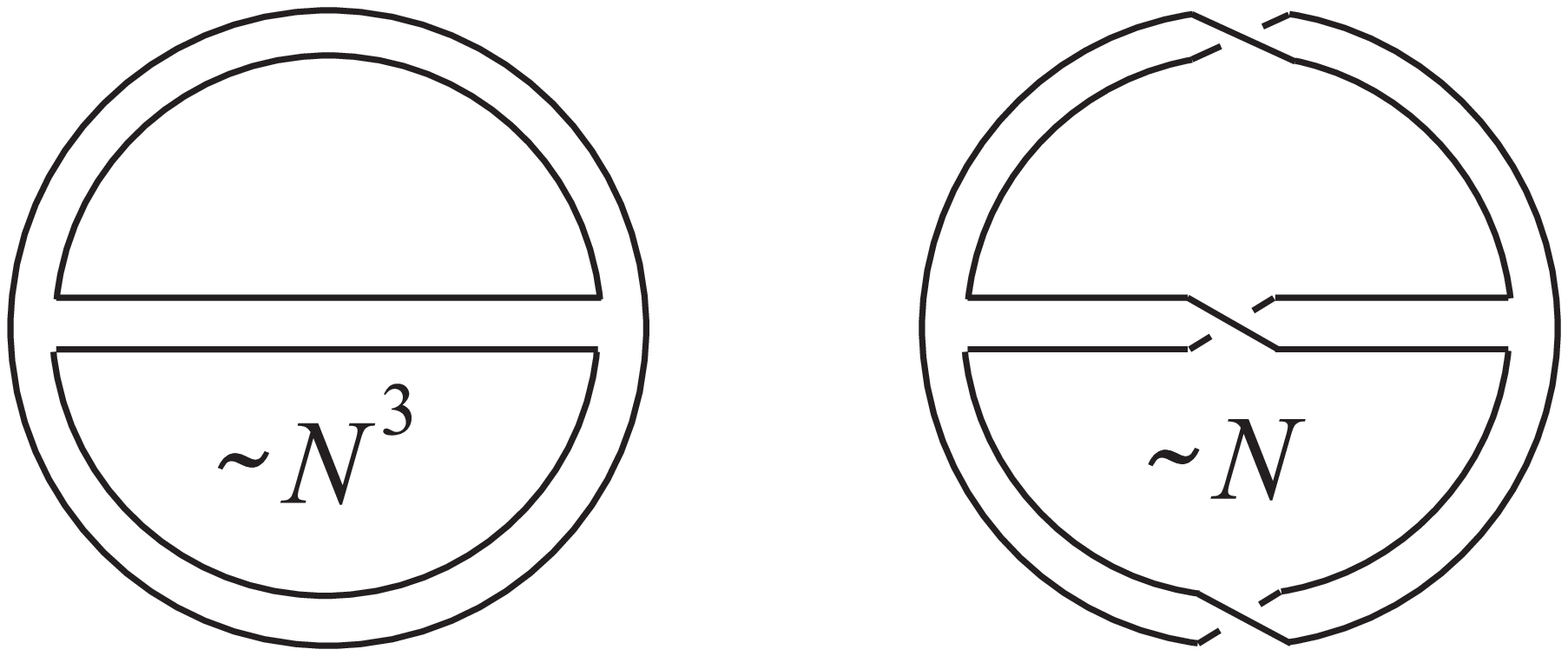}
\caption{Planar (left) and non-planar (right) fatgraphs.} \label{fig-diagrams}
\end{center}
\end{figure}

Let us consider $U(N)$ gauge theory and a fatgraph which consists of $V$ vertices, $E$ edges and $h$ closed loops. By Euler formula this fatgraph can be drawn on a Riemann surface of genus $g$ such that
$$
2g-2 = E-V-h.
$$
To the quantum amplitude represented by such a graph each vertex and each edge contribute respectively one power of gauge coupling and its inverse, whereas each closed loop gives a power of $N$. These contributions can be rewritten using Euler formula as
\be
\lambda^{E-V}N^h = t^{2g-2+h} N^{2-2g} = \lambda^{2g-2} t^h,  \label{fat-cntrb}
\ee
and the middle form makes the dependence on $1/N$ explicit. The so-called 't Hooft coupling $t$ is
$$
t = N\lambda,
$$
and 't Hooft limit which leads to the closed string interpretation is defined as
$$
N\to \infty,\qquad t=const.
$$
This means in particular that $\lambda$ gets small as $N$ grows large, so the leading contribution to the amplitude (\ref{fat-cntrb}) arises from fatgraphs with $g=0$. These fatgraphs can be drawn on a surface of a sphere and are thereby called \emph{planar}. All other fatgraphs for $g>0$ are called \emph{non-planar}. An example of a planar and non-planar (with $g=1$) fatgraphs are given in figure \ref{fig-diagrams}.

Let us consider the partition function $Z$ of a gauge theory and a related free energy $F$ which consists of a perturbative $F^{pert}$ and non-perturbative $F^{non-pert}$ parts
\be
Z = e^F,\qquad F=F^{pert}+F^{non-pert}.   \label{Z-F}
\ee 
The perturbative part admits 't Hooft expansion
\be
F^{pert} = \sum_{g=0}^{\infty} \sum_{h=0}^{\infty} F_{g,h} \lambda^{2g-2} t^h,  \label{F-pert}
\ee
where $F_{g,h}$ represents a contribution from a Riemann surface with $g$ handles and $h$ holes. The non-perturbative piece can also be presented in a genus expansion
\be
F^{non-pert} = \sum_{g=0}^{\infty} F^{non-pert}_g(t) \lambda^{2g-2}.   \label{F-npert}
\ee
According to 't Hooft arguments, performing now the sum over holes in $F^{pert}$ is supposed to have an interpretation of filling them as it results in the sum over surfaces without holes
\be
F = \sum_{g=0}^{\infty} F_g(t) \lambda^{2g-2},\qquad \textrm{for}\quad F_g(t)=F^{non-pert}_g(t) + \sum_{h=0}^{\infty} F_{g,h}t^h.  \label{F-closed}
\ee
This result coincides with a general form of the expansion in a closed string theory, with parameter $t$ corresponding to some modulus of a target space the closed strings propagate on, $F_g(t)$ a genus $g$ closed string amplitude, and $\lambda$ identified with string coupling constant. %The identification what closed string theory is related to Chern-Simons gauge theory is our first aim in what follows.

We focus now on 't Hooft expansion in Chern-Simons theory with gauge group $U(N)$. In principle one could compute its partition function in perturbative expansion using fatgraphs, although it turns out to be a difficult task \cite{labastida-cs}. However, we already know the exact answer (\ref{Z-SUN-S3}), which can be explicitly expanded into the 't Hooft form. That result was derived for $SU(N)$ theory, but it can be shown that for $U(N)$ it should just be multiplied by the partition function of $U(1)$ factor which is equal to $(1+k/N)^{-1/2}$. Then the partition function can be written explicitly as
\bea
Z^{S^3}_{U(N)} & = & \frac{1}{(k+N)^{\frac{N}{2}}} \prod_{\alpha\in \Delta_+} 2\sin \frac{\pi\,(\alpha|\rho)}{k+N} = \nonumber \\
& = & \exp\Big(-\frac{N}{2}\textrm{ln}(k+N) + \sum_{j=1}^{N-1}(N-j)\,\textrm{ln}\big(2\sin\frac{\pi j}{k+N}\big) \Big), \label{Z-S3-cs}
\eea
and the perturbative terms in (\ref{F-pert}) can be read off as \cite{marcos,G-V}
\begin{eqnarray}
F_{0,h\leq 3} & = & 0,\nonumber \\
F_{0,h>3} & = & -\frac{|B_{h-2}|}{(h-2)h!}, \nonumber \\
F_{1,h} & = & \frac{1}{12} \frac{|B_h|}{h\,h!}, \nonumber \\
F_{g>1,h} & = & \frac{\zeta(2g-2+h)}{(2\pi)^{2g-3+h}} {2g-3+h \choose h}  \frac{B_{2g}}{2g(2g-2)},  \label{CS-Fgh}
\end{eqnarray}
where $B_k$ are Bernoulli numbers. In addition the non-perturbative piece can be written as
\be
F^{non-pert} =  -\frac{N^2}{2}\textrm{ln}(k+N)  + \frac{N(N-1)}{2}\textrm{ln}\,2\pi + \sum_{j=1}^{N-1} (N-j)\,\textrm{ln}\, j,             % = \textrm{ln}\frac{(2\pi\lambda)^{N^2 /2}}{vol(U(N))},
\ee
and as a function of $N=t/\lambda$ it can also be asymptotically expanded as in (\ref{F-npert}), which gives
$$
F^{non-pert}_{g>1} = \frac{B_{2g} t^{2-2g}}{2g(2g-2)}.
$$.

With the above data the summation in (\ref{F-closed}) can be performed \cite{marcos,G-V} with the result
\be
F^{CS}_g(t) =   \frac{(-1)^g |B_{2g}B_{2g-2}|}{2g(2g-2)(2g-2)!} - \sum_{k=1}^{\infty}\frac{|B_{2g}|}{2g(2g-2)!} \textrm{Li}_{3-2g}(e^{-t}), \label{hooft-cs-g}
\ee
which we would like to identify with amplitudes of some closed string theory. As we discuss in chapter \ref{chap-top-strings}, a string theory with such amplitudes is known, and its relation to Chern-Simons theory underlines all considerations in this thesis.

%Remarkably such a string theory is explcitly known: this is the topological A-model closed string theory on resolved conifold, whose details will be explained in the next section.

%***********************************************************
%***********************************************************

% Cohomological field theories

\section{Cohomological field theories}         \label{sec-cohom}

In this section we present the second class of topological theories, namely \emph{cohomological} field theories. They can be constructed from theories with extended supersymmetry via the so-called \emph{twisting}, and for this reason are often called \emph{twisted}. Initially introduced by Witten in \cite{top-field} in the case of $\mathcal{N}=2$ supersymmetry, these theories evolved into an immense field of research. Recent analysis of twisted $\mathcal{N}=2$ theories \cite{Nek,Nek-Ok} led to an explicit solution of Seiberg-Witten theory \cite{S-W-1} and revealed deep connections with topological string theory. In fact, instanton counting in these theories is also related to topologial string theories \cite{iqbal-amir-1,iqbal-amir-2}. Remarkable analysis of twisted $\mathcal{N}=4$ theory revealed consequences of $S$-duality in field theory \cite{V-W}, and recently it proved to be related to the esoteric Langlands programme in mathematics \cite{langlands}. Anticipating interpretation in terms of crystal models, after a brief general discussion of cohomological theories below we introduce in more detail twisted $\mathcal{N}=4$ gauge theory on ALE spaces following \cite{V-W}.

Let us consider a supersymmetric theory for a set of fields denoted collectively $\phi$ with action $S$ and supersymmetry transformations generated by a supercharge $Q$, defined on a manifold $M$ with metric $g_{\mu\nu}$. Such a theory is called called \emph{cohomological} if $Q$ is nilpotent $Q^2 = 0$, it transforms as a scalar under Lorentz rotations, and the energy-momentum tensor $T_{\mu\nu}$, generally defined as a variation of the action with respect to the underlying metric on $M$, can be written as a commutator
$$
T_{\mu\nu} = \frac{\delta S}{\delta g^{\mu\nu}} = \{ Q,\Lambda_{\mu\nu} \},
$$
where $\Lambda_{\mu\nu}$ is an arbitrary field. In cohomological theories $Q$ is identified as a BRST operator, and any physical operator $\mathcal{O}$ must be $Q$-invariant, $\{ Q,\mathcal{O}\}=0$. Moreover,  $\langle \{ Q,\mathcal{O}\} \rangle=0$ for any  $\mathcal{O}$ as we consider a supersymmetric theory, and the fact that $Q$ is a Lorentz scalar guarantees such a theory is supersymmetric even on an arbitrary manifold $M$. The topological properties of twisted theories follow from the above assumptions. In particular, the partition function of such a theory is defined as the Euclidean path integral
$$
Z = \int \mathcal{D}\phi\,e^{-S},
$$
and its topological invariance is easily proved
\be
\frac{\delta}{\delta g^{\mu\nu}} Z = \int \mathcal{D}\phi\, \frac{\delta S}{\delta g^{\mu\nu}}\,e^{-S} = \langle T_{\mu\nu} \rangle = \langle\{Q,\Lambda_{\mu\nu}\}\rangle = 0,   \label{cohom-prove}
\ee
where the path integral measure $\mathcal{D}\phi$ is assumed to behave well. A partition function is a non-trivial quantity due to the existence of topologically non-trivial classical configurations of gauge fields, the so-called \emph{instantons}. It can be proved all these contributions are captured in the limit of a small coupling constant where indeed only classical configurations matter, and in consequence the partition function is equal to the Euler characteristic of instanton moduli spaces. It is also simple to prove topological invariance of physical $Q$-trivial operators. 

To construct cohomological theories satisfying the above assumptions one can \emph{twist} supersymmetric theories, which in a standard formulation have a supercharge which is a spinor. \emph{Twisting} amounts to redefining spins of all fields in the theory in such a way that this supercharge becomes a scalar, and it can be achieved by declaring a new rotation group which is embedded in a product of the original rotation group and $R$-symmetry group. For a flat space the \emph{twisted} theory is equivalent to the original, \emph{physical} one. 

For example, symmetries of $\mathcal{N}=2$ Yang-Mills theory on a Euclidean space include rotation group $E=SO(4)\sim SU(2)_L\times SU(2)_R$ and $R$-symmetry $SU(2)_I$ group. Twisting is achieved by replacing $SU(2)_R$ by $SU(2)^{tw}_R$ which is a diagonal combination of $SU(2)_R\times SU(2)_I$, and then defining $E^{tw}=SU(2)_L\times SU(2)^{tw}_R$ as a new rotation group. 

We focus now on $\mathcal{N}=4$ theory, for which we will construct the so-called Calabi-Yau crystal models in chapter \ref{chap-ale}.

%*******************************************************

\subsection{$\mathcal{N}=4$ Yang-Mills theory}  \label{ssec-N4}

$\mathcal{N}=4$ supersymmetric gauge theory can be thought of as $\mathcal{N}=2$ theory with a chiral multiplet and a massless hypermultiplet. Its beta-function vanishes, which implies the theory should be invariant under $S$-duality group $SL(2,\mathbb{Z})$ which acts on the gauge coupling $\tau$ in exactly the same way as underlying modular transformation of affine characters (\ref{SL-2Z}).
%$$
%\tau_0 \to \frac{a\tau_0 + b}{c\tau_0 + d},\qquad \textrm{for} \
%\left[ \begin{array}{cc}
%a & b \\
%c & d
%\end{array}\right] \in SL(2,\mathbb{Z})
%$$
In particular the partition function of the theory should be invariant under this transformation, which puts very strong constraints on its form. These consequences were established in \cite{V-W}. 

In order to consider topological properties of the theory, we have to consider its twisted version. Twisting in this case requires specifying how the rotation $E=SO(4)$ group is embedded in the $SU(4)$ $R$-symmetry group. There are three twists which lead to the scalar supercharge, and we will consider the one corresponding to $({\bf 1},{\bf 2})\oplus({\bf 1},{\bf 2})$ representation of $E$, following considerations in \cite{V-W}. After the twisting the theory becomes topological and the partition function is expected to depend just on the gauge group $G$, the topology of the four-dimensional manifold and the gauge coupling $\tau$. On a K\"ahler manifold $M$ it takes the form
\be
Z(\tau,G) = q^{-c/24} \sum_{k=0}^{\infty} c_k q^k,         \label{Z-gauge}
\ee
where $c_k = \chi(\mathcal{M}_k)$ is the Euler number of instanton moduli space of topological charge $k$, $c$ is some constant and $q=e^{2\pi i \tau}$. Under $S:\tau \to -1/\tau$ element of the $S$-duality group (\ref{modular-TS}) the partition function should transform as a modular form
$$
Z(-1/\tau,G) \sim \tau^{w/2} Z(\tau,\widehat{G}),
$$
where $\widehat{G}$ is the dual group  (whose weight lattice is dual to the weight lattice of $G$) and $w$ is a certain weight. Such a behaviour is a reminiscent of a transformation property of affine characters (\ref{affine-TS}). It does not necessarily mean the partition functions have to be equal to those characters, but in fact they do in the known cases. 

It is also important to stress that the coefficients $c_k$ are expected to be integers. If $Z(\tau,G)$ coincides with some affine character this is automatically ensured, as these coefficients are multiplicities of weights in certain representations. 

For example, the partition function of $U(N)$ theory on $\mathbb{R}^4$ is equal simply to
\be
Z^{\R^4}(\tau,U(N)) = \eta(q)^{-N} = q^{-N/24} \Big(\sum_{n=0} p(n)q^n  \Big)^N.    \label{Z-N4}
\ee
The coefficients $p(n)$ are given in (\ref{euler-fn}) and they are indeed integers, which count the number Young diagrams with $n$ boxes. For $N>1$ the above partition function is a generating function of coloured partitions defined in appendix \ref{app-coloured}. 

%*******************************************************

\subsection{$\mathcal{N}=4$ theory on ALE spaces}  \label{ssec-N4-ale}

Aspects of $\mathcal{N}=4$ $U(N)$ theories on ALE spaces were elucidated in \cite{V-W}, following earlier analysis in \cite{naka1}. In this thesis we will be particularly interested in theories on ALE spaces of $A_{k-1}$ type. ALE spaces, as introduced in more detail in section \ref{ale-spaces}, in general arise as resolutions of orbifolds $\C^2/\Gamma$ where $\Gamma$ is a finite subgroup of $SU(2)$. The boundary of ALE space is a Lens space $S^3/\Gamma$, so the $U(N)$ gauge field can approach some non-trivial flat connection at infinity. Such flat connection is labelled by the $N$-dimensional representation of $\Gamma$
$$
\lambda \in \textrm{Hom} (\Gamma,U(N)),
$$
which can be decomposed into irreducible representations $\rho_i$ of $\Gamma$ as
$$
\lambda = \sum_i N_i \rho_i,
$$
with $N_i$ non-negative integers satisfying the relation
$$
\sum_i N_i\, \textrm{dim}\, \rho_i = N.
$$

Introducing ALE spaces in section \ref{ale-spaces}, we discussed McKay correspondence which relates finite subgroups $\G$ to corresponding simple Lie algebras $\mathfrak g$ of ADE type, such that dimensions $d_i$ of irreducible representations of $\G$ can be identified with the dual Dynkin indices of the extended Dynkin diagram. This suggests we can think of each $\lambda$ as determining an integrable highest-weight representation of the the affine extension $\widehat{\mathfrak g}$  at level $N$. We will denote this representation as $V_{\widehat{\lambda}}$, where $\widehat{\lambda}$ is the corresponding integrable weight. In particular for $\Gamma=\Z_k$ and ALE space of $A_{k-1}$ type we get a representation of $\widehat{su}(k)_N$. The relevant extended Dynkin diagram is given in figure \ref{fig-aff-dynkin}, and in this case all dual Dynkin indices $d_i=1$ for $i=0,\ldots,k-1$. With the boundary condition $\lambda$ at infinity we get a vector-valued partition function whose components are of the form
$$
Z_{N,\lambda} = \sum_{n} d(N,n;\lambda) q^{n-c/24}.
$$
In a remarkable work \cite{naka1} Nakajima proved that on the middle dimensional cohomology of 
the moduli space of gauge theory one can actually realise the affine algebra
$\widehat{\mathfrak g}$, and in consequence the partition function can be identified with the affine character of $\widehat{\lambda}$
\be
Z_{N,\lambda} = \Tr_{V_{\widehat{\lambda}}}\left(q^{L_0-c/24} \right) = \chi^{\widehat{\mathfrak g}}_{\widehat{\lambda}}(q).   \label{Z-VW}
\ee

Affine characters can be explicitly calculated in a variety of ways, for example using a relation to $\Theta$-functions (\ref{aff-char}) which we discuss in appendix \ref{app-characters}. This opens up a wide perspective of testing various gauge theory results. In chapter \ref{chap-ale} we will construct the so-called Calabi-Yau crystal models, which compute partition functions of $\mathcal{N}=4$ theory in terms of affine characters.

%***********************************************************
%***********************************************************
%***********************************************************

\chapter{Topological string theories}    \label{chap-top-strings}

%Topological string theories are constructed similarly to cohomological field theories, by a procedure of twisting which redefines a rotation group by mixing it with $R$-symmetry group. In topological strings, the twisting is performed on the string worldsheet with $\mathcal{N}=(2,2)$ supersymmetry. Again the crucial object in this construction is a scalar nilpotent supercharge $Q$, such that energy-momentum tensor can be written as an anti-commutator with $Q$. This scalar supercharge ensures supersymmetric sigma model is properly defined on an \emph{arbitrary} curved worldsheet, which is necessary to \emph{couple the theory to gravity} by performing path integral over all worldsheet metrics. 

In this chapter we introduce \emph{topological string theories}. There are various string models with topological properties which are conventionally referred to as different \emph{theories}. They differ either on a more fundamental level, as for example in the case of open and closed or A-model and B-model theories, or in more specific details, for example in the choice of the target space manifold. The main and common ingredient of those various theories are underlying topological sigma models introduced in \cite{top-sigma,witten-mirror}, whose construction involves an application of twisting (described for cohomological field theories in chapter \ref{chap-top-field}). 

In past twenty years topological strings evolved into a vast area of research related to many other problems in theoretical physics, so it would not be possible even to mention in this thesis all their aspects and applications. In this chapter we review the construction of topological strings to the extent required to understand their relation to Calabi-Yau crystals, focusing mainly on A-model theories on toric manifolds. More details about the construction of topological strings can be found in many reviews \cite{mirror,dvv-review,marcos, marcos-matrix,tops-review,vonk}. 

In sections \ref{sec-closed-tops}, \ref{sec-A-closed} and \ref{sec-open-top} we introduce various aspects of closed and open topological string theories. A far reaching consequences has a behaviour of  A-model topological strings under a conifold geometric transition, which is described in section \ref{sec-open-closed}. In particular, this leads to the construction of the topological vertex formalism, which allows to compute topological string amplitudes on an arbitrary toric manifold, as discussed in section \ref{chap-top-vertex}. These amplitudes will be reformulated in chapters \ref{chap-cy-crystals}-\ref{sec-A-model} in terms of Calabi-Yau crystal models.

\section{Construction of closed topological strings}       \label{sec-closed-tops}

To construct closed topological string theory one considers a sigma model with $\mathcal{N}=(2,2)$ supersymmetry on a fixed Riemann surface $\Sigma$ of genus $g$ playing a role of a string worldsheet and performs a procedure of twisting, which redefines a rotation group by mixing it with $R$-symmetry group. Then this twisted theory is \emph{coupled to gravity} by performing a path integral over all worldsheet metrics, or more precisely an integral over the moduli space of genus $g$ Riemann surfaces. This produces a genus $g$ amplitude $F_g$. The crucial object in this construction is again a scalar nilpotent supercharge $Q$, such that the worldsheet energy-momentum tensor can be written as an anti-commutator with $Q$. This scalar supercharge ensures supersymmetric sigma model is properly defined on an \emph{arbitrary} curved worldsheet integrated in the path integral. Finally, the closed topological string partition function is defined as a generating function of $F_g$'s, with a generating parameter identified with a string coupling constant $g_s$. In this and the next section we describe these steps in more detail, with a particular emphasis on A-model theories.

%****************************************************************

\subsection{$\mathcal{N}=(2,2)$ supersymmetry}

Let us introduce general aspects of $\mathcal{N}=(2,2)$ supersymmetric field theory. In our considerations such theories arise on a two-dimensional worldsheet which describes propagation of a string. In what follows indices $a,b=1,2$ refer to Euclidean coordinates on this worldsheet $x_1,x_2$. 

On the classical level $\mathcal{N}=(2,2)$ supersymmetric field theory is defined in terms of a supersymmetric action involving certain fields, with the following symmetries, conserved quantities and currents:
\begin{itemize}
\item two-dimensional Poincare invariance implying conservation of energy, momentum and angular momentum, which are Noether charges for time translations, space translations and rotations, denoted respectively as 
\be
H,P,L,         \label{N22-HPL}
\ee
and their corresponding currents being encoded in the energy-momentum tensor $T_{ab}$; in Euclidean signature a group o rotations is just $U(1)_E$, with index $E$ for ``Euclidean'',

\item $\mathcal{N}=(2,2)$ supersymmetry giving rise to two pairs of fermionic supercharges
\be
Q_{\pm}, \overline{Q}_{\pm},            \label{N22-susy}
\ee
with corresponding supercurrents $G^{a}_{\pm}, \overline{G}^a_{\pm}$,

\item the so-called \emph{vector} and \emph{axial} R-symmetries related to the above supercharges, giving rise to charges
\be
F_V, F_A,              \label{N22-Fav}
\ee
with corresponding $U(1)$ currents $J^a_V, J^a_A$.

\end{itemize}

In a quantum theory the above conserved quantities are promoted to operators generating symmetry transformations, which satisfy the following $\mathcal{N}=(2,2)$ supersymmetry algebra (adapted for the Euclidean worldsheet)
\bea
Q_{\pm}^2 = \overline{Q}_{\pm}^2 = 0, & & \{Q_{\pm}, \overline{Q}_{\pm} \} =  H \pm P, \nonumber \\
\left[L , Q_{\pm} \right] = \mp Q_{\pm}, & & \left[ L , \overline{Q}_{\pm} \right] = \mp \overline{Q}_{\pm}, \nonumber \\
\left[F_V,Q_{\pm}\right] = - Q_{\pm}, & & \left[F_V,\overline{Q}_{\pm}\right] = \overline{Q}_{\pm}, \nonumber \\
\left[F_A,Q_{\pm}\right] = \mp Q_{\pm}, & & \left[F_A,\overline{Q}_{\pm}\right] = \pm \overline{Q}_{\pm}, \nonumber
\eea
with other (anti-)commutators vanishing.

% In particular infinitesimal a supersymmetry transformation of any operator $\mathcal{O}$ is generated as
%$$
%\delta\mathcal{O} = [\epsilon_+ Q_+ ???,\mathcal{O}],
%$$

%****************************************************************

\subsection{Twisting of topological sigma models}

Let us consider a \emph{sigma model} on a K\"ahler manifold $M$, which is a theory of maps from a genus $g$ Riemann surface $\Sigma$ into $M$
\be
\phi^{i}:\quad \Sigma \longrightarrow M, \qquad i=1,\ldots,\textrm{dim}_{\C}\,M.   \label{sigma-model-twist}
\ee
A theory of such maps with the Polyakov action (\ref{S-Polyakov}) is a starting point of a construction of the bosonic string. Now we consider its extension with $\mathcal{N}=(2,2)$ supersymmetry. It is convenient to introduce it using a superspace language, with coordinates $x_1,x_2$ on $\Sigma$ augmented with a pair of Grassman coordinates $\theta^{\pm}$ and their complex conjugates $\overline{\theta}^{\pm}$. The maps (\ref{sigma-model-twist}) become then the lowest components of chiral multiplets $\Phi^i$, the other components being fermionic spinor fields $\psi^i_{\pm}$ and an auxiliary fields $F^i$
$$
\Phi^i(x^a,\theta^{\pm},\overline{\theta}^{\pm}) = \phi^i(y^{\pm}) + \theta^+\psi^i_+(y^{\pm}) + \theta^-\psi^i_-(y^{\pm}) + \theta^+\theta^-F^i(y^{\pm}),
$$
where $y^{\pm} = (x^1\pm i x^2) - i\theta^{\pm}\overline{\theta}^{\pm}$. We define \emph{vector} and \emph{axial R-rotations} of a superfield respectively as
\bea
U(1)_V:\quad \Phi^i(x^a,\theta^{\pm},\overline{\theta}^{\pm}) \mapsto e^{i\alpha q_V} \Phi^i(x^a,e^{-i\alpha}\theta^{\pm},e^{i\alpha}\overline{\theta}^{\pm}) ,\nonumber \\
U(1)_A:\quad \Phi^i(x^a,\theta^{\pm},\overline{\theta}^{\pm}) \mapsto e^{i\alpha q_A} \Phi^i(x^a,e^{\mp i\alpha}\theta^{\pm},e^{\pm i\alpha}\overline{\theta}^{\pm}),\nonumber
\eea
where $q_V, q_A$ are called vector and axial R-charges.

The supersymmetric sigma model action can be written compactly as
\be
S = 2 \lambda \int_{\Sigma} d^2x\,d^4\theta\,\Phi^i\overline{\Phi}^i,    \label{sigma-action}
\ee
where $\lambda$ is a coupling constant.

Now we twist the theory, which amounts to a redefinition of the Lorentz symmetry. In Euclidean signature the Lorentz group on $\Sigma$ is just $U(1)_E$ rotation group with generator $L$ given in (\ref{N22-HPL}). One can define a modified Lorentz symmetry $U(1)_E^{tw}$ as a diagonal subgroup of a product of $U(1)_E$ with one of the above R-rotations (\ref{N22-Fav}), which leads to two possible twists with the following rotation generators $L_{tw}$
\bea
\textrm{A-twist}: & &  U(1)_{E} \longrightarrow U(1)_E^{tw} = U(1)_{E}  \times U(1)_V,\qquad L_{tw} = L+F_V \nonumber \\ % \label{A-twist} \\
\textrm{B-twist}: & & U(1)_{E} \longrightarrow U(1)_E^{tw} = U(1)_{E}  \times U(1)_A,\qquad L_{tw} = L+F_A   \nonumber
\eea
and the modified theories are called respectively \emph{A-model} and \emph{B-model}. The twisting modifies spins of the component fields.  In fact $\phi$ transforms trivially under both Lorentz and R-symmetry groups, so it is still a scalar in the twisted theory. However fermions turn into scalars and one-forms, so we rename them accordingly as $\chi$ and $\rho_z,\rho_{\bar{z}}$ to make their spins explicit:
\bea
\textrm{A-model}: & & \left\{ \begin{array}{cc}
\chi^i:= \psi^i_-,&  \chi^{\bar{i}}:= \overline{\psi}^{\bar{i}}_+ \\
\rho^{\bar{i}}_z:= \overline{\psi}^{\bar{i}}_-, & \rho^i_{\bar{z}}:= \psi^i_+ \\
\end{array} \right. \label{A-fields} \\
\textrm{B-model}: & & \left\{ \begin{array}{cc}
\chi^{\bar{i}}:= \overline{\psi}^{\bar{i}}_-,&  \overline{\chi}^{\bar{i}}:= \overline{\psi}^{\bar{i}}_+ \\
\rho^{i}_z:= \psi^i_-, & \rho^i_{\bar{z}}:= \psi^i_+ \\
\end{array} \right. \nonumber 
\eea

The twisting also affects other quantities in the theory. In particular the original $\mathcal{N}=(2,2)$ supersymmetry includes two pairs of fermionic supercharges (\ref{N22-susy}). They are
spinors under $U(1)_E$, but become scalars or vectors under $U(1)_E^{tw}$
after twisting. The following combinations of scalar supercharges
respectively in A- and B-model are nilpotent (on-shell, i.e. modulo equations of motion), and can be regarded as BRST operators, analogously as was the case in cohomological field theories
\bea
\textrm{A-model}: & & Q_A = \overline{Q}_+ + Q_-,\qquad Q_A^2 = 0,   \nonumber \\ %\label{A-charge} \\
\textrm{B-model}: & & Q_B = \overline{Q}_+ + \overline{Q}_-,\qquad Q_B^2 = 0. \nonumber
\eea

The topological character of twisted sigma models arises from the fact their energy-momentum tensor can be written as an anti-commutator with this charge
\be
T_{ab}=\{Q,\ldots\},     \label{T-topological}
\ee
with $Q=Q_A$ or $Q=Q_B$ respectively in A- and B-model. This topological character can be proved similarly as in cohomological field theories (\ref{cohom-prove}).

Sigma models described above have quantum anomalies related to the presence of fermions. These anomalies impose severe constraints on a theory to be consistent. In particular, consistent A- or B-model theories arise if $M$ is a Calabi-Yau threefold, which is the main reason we introduced Calabi-Yau manifolds in chapter \ref{sec-cy}. In fact non-trivial amplitudes in twisted sigma models arise only for genus $g=0,1$. To obtain non-trivial amplitudes for higher genus surfaces these models must be coupled to gravity,  which gives rise to topological string theories.

%****************************************************************

\subsection{Coupling to gravity and topological strings}

The topological strings are defined by coupling topological sigma models to two-dimensional gravity. Let us consider a Riemann surface of fixed genus $g$. By the \emph{coupling to gravity} we understand considering all possible metrics on this Riemann surface, which can be done by integrating certain observables over a moduli space of genus $g$ Riemann surfaces $\mathcal{M}_g$. To identify which observables could lead to nontrivial amplitudes, it is very helpful to follow an analogy to the bosonic string. 

In the bosonic string the genus $g$ amplitude is defined as
\be
\int_{\mathcal{M}_g} \langle |\prod_{i=1}^{3g-3} b(\mu_i) |^2 \rangle. \label{bosonic-Fg}
\ee
There are $3g-3$ anti-holomorphic one-forms $\mu_i$ with values in the holomorphic tangent bundle, called \emph{Beltrami differentials}, which give rise to
$$
b(\mu_i) = \int_{\Sigma} d^z\,b_{zz} (\mu_i)^z_{\bar{z}},
$$
where $b$ is a bosonic string antighost. It can be shown the algebra appearing in the bosonic string is isomorphic to the superconformal algebra of the topological sigma model, with antighost $b$ corresponding to the current $G_-$ related to the charge $Q_-$. Thus the topological genus $g$ free energy can be defined as in (\ref{bosonic-Fg}), with antighost replaced by the current $G_-$
\be
F_g = \int_{\mathcal{M}_g} \langle |\prod_{i=1}^{3g-3} G_-(\mu_i) |^2 \rangle. \label{Fg-int}
\ee
The total free energy is defined as a generating series
\be
 F_{top} = \sum_{g=0} F_g \, g_s^{2g-2}, \label{F-top-sum}
\ee
where we also include genus $0$ contribution. The generating parameter $g_s$ is interpreted as the string coupling. The topological string partition function is then defined as
$$
Z_{top} = e^{F_{top}}.
$$

%****************************************************************
%****************************************************************

\section{A-model closed topological strings}        \label{sec-A-closed}

We have explained above in general how closed topological strings are constructed. In this section we reveal special properties of the A-model topological strings which will be of main interest for us. In particular A-model amplitudes can be encoded either in \emph{Gromov-Witten invariants} or \emph{Gopakumar-Vafa invariants}, which have very interesting properties.

%****************************************************************

\subsection{A-model}

We focus on A-model quantities, firstly writing the action (\ref{sigma-action}) explicitly in terms of A-model fields (\ref{A-fields}). Performing integrations over $\theta$'s we get
\be
S = 2 \lambda \int_{\Sigma} d^2 z\big( \hf g_{ij}\partial_z \phi^i \partial_{\bar{z}}
\phi^j + i G_{i\bar{j}} \rho^i_{\bar{z}} D_z \chi^{\bar{j}} + i G_{\bar{i} j}
\rho^{\bar{i}}_z D_{\bar{z}} \chi^j - R_{i\bar{k}j\bar{l}} \rho^i_{\bar{z}}
\rho^{\bar{k}}_z \chi^j \chi^{\bar{l}}  \big) \label{A-action}
\ee
with spacetime metric $G_{ij}$ and curvature $R_{ikjl}$ and the covariant derivative 
$$
D_{\alpha} \chi^i = \partial_{\alpha}\chi^i +
\partial_{\alpha}\phi^i \G^i_{jk} \chi^k.
$$
This action can be written as
\be
S = -i\{Q_A,V\} - \lambda \int_{\Sigma} \phi^*(K), \label{sigma-action-V}
\ee
where the K\"ahler class of $M$ is $K=iG_{i\bar{j}} d\phi^i\wedge
d\phi^{\bar{j}}$ and
$$
V = \lambda \int_{\Sigma} d^2z \, G_{i\bar{j}} (\rho^i_{\bar{z}}\partial_z
\phi^{\bar{j}} + \partial_{\bar{z}}\phi^i \rho^{\bar{j}}_z).
$$

The second term in (\ref{sigma-action-V}) is topological and measures a homotopy type of the map $\phi:\Sigma\to M$. This implies that after covariantising (\ref{A-action}) by considering arbitrary metric $g_{ab}$ on $\Sigma$, the energy-momentum tensor can be written as
$$
T_{ab} = \Big\{ Q_A, \frac{\delta V}{\delta g^{ab}}\Big\},
$$
as anticipated in (\ref{T-topological}). In consequence the A-model is a topological theory and does not depend on the
metric $g_{ab}$. Moreover, its partition function does not depend on $\lambda$, so
it can be computed for $\lambda\to\infty$ which gives a semi-classical result, which
is nonetheless exact. In the semi-classical limit the allowed field
configurations are \emph{holomorphic} maps $\phi:\Sigma\to M$, such that 
$$
\partial_{\bar{z}} \phi^i = \partial_z\phi^{\bar{i}} = 0.
$$
These maps minimise the bosonic action. There are two classes of
them. Firstly, for constant --- or \emph{classical} --- maps the surface $\Sigma$ is mapped just to a
point in $M$. Secondly, there are nontrivial maps called \emph{worldsheet instantons} and
they include various sectors classified by homology classes $\beta$ of $M$, which we will often characterise by integers $n_i$
\be
\beta = \phi_*[\Sigma] = \sum_{i=1}^{b_2(M)} n_i [S_i] \in H_2(M,\Z),     \label{class-beta}
\ee
where $[S_i]$ constitute a basis of $H_2(M,\Z)$.%, and it what follows we will denote 
%$$
%t_i = \int_{S_i} K.
%$$ 

It is interesting to consider A-model observables. Similarly as in cohomological field theories they
must be $Q_A$-closed but not $Q_A$-exact combinations of the form 
$$
\mathcal{O}_{\phi} = \phi_{i_1\ldots i_p} \chi^{i_1}\cdots \chi^{i_p},
$$
and due to anticommuting character of $\chi^i$'s they can be identified with differential forms which are
representatives of de Rham cohomology classes $H^p(M)$. In consequence of the anomalies mentioned above 
in a consistent theory on a Calabi-Yau threefold $M$
non-trivial observables $\langle \mathcal{O}_{\phi_1} \ldots \mathcal{O}_{\phi_s} \rangle$ arise for fields satisfying 
\be
\sum_{a=1}^s \textrm{deg}\, \mathcal{O}_{\phi_a} = 6(1-g), \label{A-select}
\ee
where $\textrm{deg}\, \mathcal{O}_{\phi_a}$ denotes a degree of the
corresponding differential form. For $g=0$ one can thus consider
$\mathcal{O}_{\phi}$ corresponding to two-forms with nontrivial expectation
values arising for $s=3$ in the above sum. These two-forms can be represented
by four-cycles $D_i$ by the Hodge duality. It implies that contributions from constant maps
give just intersection numbers of these four-cycles $a_{ijk}=D_i\cap D_j\cap D_k$ in the form
$$
\int_M k\wedge k\wedge k = \sum_{i,j,k} \frac{1}{6}a_{ijk}t_i t_j t_k + P_2(t_i),
$$ 
where $k$ is a K\"ahler form on $M$, $t_i$ are K\"ahler parameters (\ref{kahler-param}) and 
$P_2(t_i)$ is an ambiguous polynomial in $t_i$'s of degree two. Worldsheet instantons contribute terms
proportional to $\exp (- n_i t_i)$, so that entire information about genus zero
amplitudes can be encoded in the so-called \emph{prepotential}
\be
F_{0,total} = \Big( \sum_{i,j,k} \frac{1}{6}a_{ijk}t_i t_j t_k + P_2(t_i)\Big) + \sum_{\beta=\sum_i n_i[S_i] \in H_2(M)}  N^{0}_{\beta}  e^{-n_i t_i},  \label{F0}
\ee
where a subscript $total$ is to emphasise it contains contributions from both classical maps and worldsheet instantons. The former are characterised by intersection numbers $a_{ijk}$ and the latter by the so-called \emph{Gromov-Witten invariants} $N^0_{\beta}$. 

For genus $g>0$ A-model amplitudes are less interesting, as can be easily deduced from (\ref{A-select}). For $g=1$ the only nontrivial quantity is the partition function itself, whereas for $g>1$ all amplitudes vanish. However, if the A-model is coupled to gravity (\ref{Fg-int}) non-trivial amplitudes arise for arbitrary genus, specified by more general Gromov-Witten invariants.

%****************************************************************

\subsection{Gromov-Witten invariants}

Let us summarise a general form of the A-model topological string partition function on a Calabi-Yau manifold $M$. It can be written as
\begin{eqnarray}
Z_{top} = e^{F_{top}} & = & e^{\sum_{g=0} F_g \, g_s^{2g-2}} = M(q)^{\chi(M)/2} e^{F_{class}+F}, \label{Z-top} \\
F_{class} & = & \sum\frac{1}{6g_s^2} a_{ijk}\,t_i t_j t_k + \sum \frac{1}{24} b_i t_i, \label{F-class} \\
F & = & \sum_{g\geq 0} \sum_{\beta} g_s^{2g-2}\, N^{g}_{\beta}\, Q^{\beta}, \label{grom-wit}
\end{eqnarray}
with the following notation. The entire amplitude depends on a string coupling $g_s$ and  K\"ahler parameters $t_i$ (\ref{kahler-param}) of $M$. $F_{class}$ with polynomial dependence on $t_i$ are contributions from constant maps, which for genus $g=0$ encode classical intersection numbers $a_{ijk}$ and for genus $g=1$ terms are related to $c_2(M)$ as $\sum b_i t_i = \int_M k \wedge c_2(M)$. Constant maps for $g>1$ can be shown to give rise to powers of McMahon function $M(q)= \prod_{n=1}\, (1 - q^n)^{-n}$, where $q=e^{-g_s}$ and $\chi(M)$ is Euler characteristic of $M$. 

Contributions from worldsheet instantons are given by $F$, and are determined by the Gromov-Witten invariants $N^{g}_{\beta}$, which roughly  count maps from a genus $g$ Riemann surface into a curve in a Calabi-Yau manifold of a certain class $\beta\in H_2(M,\Z)$ specified as in (\ref{class-beta}). These invariants are integer numbers for isolated curves, or fractional numbers for smooth families of curves. We also denote $Q^{\beta}=\prod Q_i^{n_i}$, where $Q_i=e^{-t_i}$.

Even though the notion of Euler characteristic of $M$ may be ambiguous for non-compact manifolds, the Gromov-Witten invariants can be rigorously computed in some cases. The invariants for the simplest case of $M=\C^3$, predicted first by physical reasoning in \cite{G-V} have been derived mathematically in \cite{localP1}
\be
N^g_0 = \frac{(-1)^g |B_{2g}B_{2g-2}|}{2g(2g-2)(2g-2)!},\qquad \textrm{for}\ g\ge 2.
\ee
This leads to the result
\be
Z(q)^{\C^3}_{top} = M(q) = \prod_{n=1}^{\infty}\, (1 - q^n)^{-n}, \label{Z-c3}
\ee
which implies we should assume $\chi(\C^3)=2$ in the topological string amplitude (\ref{Z-top})

As another example let us consider topological strings on the resolved conifold $M=\mathcal{O}(-1)\oplus
\mathcal{O}(-1) \to \mathbb{P}^1$ introduced in section \ref{sec-conifolds}. 
Its second homology group is generated by a
single $\mathbb{P}^1$, so that $\beta=n\,[\mathbb{P}^1]$ can be identified with a
number $n\in\Z$. The contributions from constant maps can be identified with
$N^g_{n=0}$, and for $g\ge 2$ they arise from the expansion of $M(q)^{\chi/2}$
in (\ref{Z-top}). The Euler characteristic for the resolved conifold,
even though it is a non-compact manifold, should be taken as $\chi=2$ to agree
with computations in Gromov-Witten theory. The proper Gromov-Witten invariants
for $n\neq 0$ have also been computed in \cite{localP1}
\be
N^g_n = - n^{2g-3} \frac{|B_{2g}|}{2g\,(2g-2)!}. \label{GW-P1}
\ee
Thus the free energy for the resolved conifold can be written as
\be
F_{top}^{conifold} = F_{class} +\sum_{g\geq 0} g_s^{2g-2}\Big(\frac{(-1)^g
  |B_{2g}B_{2g-2}|}{2g(2g-2)(2g-2)!} - \frac{|B_{2g}|}{2g\,(2g-2)!} \textrm{Li}_{3-2g}(Q)\Big), \label{F-conifold-GW}
\ee
where a polylogarithm is defined as $\textrm{Li}_k(Q)=\sum_{n\geq 1}
\frac{Q^n}{n^k}$, and $Q=e^{-t}$ where $t$ is the size of $\mathbb{P}^1$. The
contributions from worldsheet instantons are captured by terms in the second sum in the bracket.

%****************************************************************

\subsection{Gopakumar-Vafa invariants}
%\subsection{Relation to superstring theories}

There is an intimate relation between topological strings, superstrings and supergravity \cite{agnt,bcov} which is also essential in further considerations. For definiteness let us restrict to the A-model and type IIA superstrings. We recall a compactification of Type IIA
superstring theory on a Calabi-Yau threefold $M$ leads to four-dimensional
$\mathcal{N}=2$ supergravity in non-compactified dimensions with $h^{1,1}(M)$
vector multiplets \cite{gsw,polchinski}. Each such multiplet contains a complex scalar field $\phi^i$ which corresponds to one K\"ahler modulus of $M$. This supergravity
contains of course a gravity multiplet, which includes a graviton $R$ and
gravi-photon $F_{ph}$. Now all supergravity data is specified by string
theory and details of a manifold $M$. String theory effective action reproduces
supergravity action, which in particular contains terms of the form
\be
\int d^4 x\, F_g(\phi^i)\,R^2_+ F^{2g-2}_{ph,+}   \label{sugra-intro}
\ee
for $g\geq 1$, where the subscript $_+$ denotes a self-dual part of a
corresponding field. This term describes a scattering of gravitons and
gravi-photons, whose details are encoded in a function $F_g$. This function can be shown 
to be equal precisely to genus $g$ A-model topological string amplitude (\ref{Z-top}) on $M$, if the scalar fields are identified with moduli $t_i$ of the manifold $M$. Under this identification genus zero amplitude $F_0(\phi^I)$ turns out to compute $\mathcal{N}=2$ gauge theory prepotential which determines the gauge coupling constant $\tau_{ij}$ as
$$
\tau_{ij} = \partial_i \partial_j F_0.
$$

%****************************************************************

%\subsection{Gopakumar-Vafa invariants}

%The above relation of topological strings to supergravity theories has
These relations have
very far-reaching consequences for the structure of topological string amplitudes, discussed first in \cite{G-V}. One can consider a background of a constant graviphoton field identified as $F_{ph,+} = g_s$, so that (\ref{sugra-intro}) can be written as
\be
\int d^4 x\, \,R^2_+ F_{top}(t_i).  \label{sugra-4d}
\ee
One can ask if such terms in the effective supergravity action can be derived from the spacetime point of view. It is indeed possible, and they must arise from integrating out some hidden degrees of freedom. From the string theory point of view these degrees of freedom can be identified with solitons in a form of $D0$ and $D2$ branes. Their contribution is analogous to a contribution obtained from integrating out a charged scalar in a constant electromagnetic field considered by Schwinger. Therefore a famous Schwinger computation can be performed in the present context as well, and its result can be reinterpreted from the point of view of topological strings. A careful analysis in \cite{G-V} revealed the instanton part $F$ of the topological string free energy (\ref{Z-top}) must be necessarily of the form
\be
F(t_i) = \sum_{\beta\in H_2(M)} \sum_{g=0}^{\infty} \sum_{d=1}^{\infty} n^g_{\beta} \frac{Q^{d\beta}}{d[d]^{2-2g}}, \label{gop-vafa}
\ee
where $[d] = e^{d/2}-e^{-d/2}$ and other notation is as in (\ref{Z-top}). The
numbers $n^g_{\beta}$ characterise the amplitude in a novel way and are called
Gopakumar-Vafa invariants. The crucial point is they count states which were
integrated out from the four-dimensional point of view and thereby should be
integer. From the point of view of topological string theory this is a very
non-trivial statement. These integer numbers must of course encode fractional Gromov-Witten invariants and carry equivalent information. Comparison between (\ref{grom-wit}) and (\ref{gop-vafa}) shows that instanton contributions to $F_g$, encoded in Gromov-Witten invariants $N^g_{\beta}$, are determined by Gopakumar-Vafa invariants $n^k_{\beta}$ with $k\leq g$. In particular, expanding (\ref{gop-vafa}) we find that genus $g=0$ instanton contributions to the free energy (\ref{F0}) are of the form
\be
F_{0} = \sum_{\beta \in H_2(M)} \sum_{d=1}^{\infty}  \frac{n^{0}_{\beta}}{d^3}  Q^{d \beta}.
\ee

Even though there is a rigorous mathematical theory of Gromov-Witten invariants, there is no mathematical proof that it has a reformulation in terms of integer invariants. In fact, an outstanding problem is even to formulate a general mathematical definition of Gopakumar-Vafa invariants. However, in all cases where Gromov-Witten invariants are known, it has been checked their redefinition according to (\ref{gop-vafa}) indeed leads to integer $n^g_{\beta}$.

Let us consider the simplest example of a Gopakumar-Vafa expansion for some manifold $M$, which arises in the case of a single invariant $n^0_1=-1$. Here the subscript $\beta=1$ is identified with a two-sphere which must be a generator of a one-dimensional second homology group. The instanton part of the free energy and its genus expansion take respectively the form
\be
F^{conifold} = \sum_{n\geq 1}\frac{-Q^{n}}{n [n]^{2}} = -\sum_{g\geq 0} g_s^{2g-2} \frac{|B_{2g}|}{2g\,(2g-2)!} \textrm{Li}_{3-2g}(Q). \label{F-conifold}
\ee
In this genus expansion we recognise explicitly the instanton part of the free energy of the resolved conifold (\ref{F-conifold-GW}), so we can identify $M=\mathcal{O}(-1)\oplus \mathcal{O}(-1) \to \mathbb{P}^1$. Thus all the information encoded in an infinite number of Gromov-Witten invariants (\ref{GW-P1}) for the resolved conifold is now hidden in a single Gopakumar-Vafa invariant $n^0_1$.

%****************************************************************
%****************************************************************

\section{Open topological strings}   \label{sec-open-top}

Open topological strings were analysed in \cite{branes-cy,branes-toric,branes-mirror,cs-string}. In a paragraph below we review their construction and in the next one --- their target space interpretation as Chern-Simons theory.

\subsection{Construction}

To define open topological strings we parallel the construction of closed strings described above, now with the ends of open strings taken into account. These ends of strings give rise to holes in Riemann surfaces representing worldsheets of propagating strings. Thus now our starting point is $\mathcal{N}=(2,2)$ sigma model for maps from a genus $g$ Riemann surface $\Sigma$ with $h$ holes into a K\"ahler manifold $\phi^{i}:\ \Sigma \to M, \qquad i=1,\ldots,\textrm{dim}_{\C}\,M$, such that a boundary of $\Sigma$ is mapped to a certain submanifold $J\subset M$
\be
\phi^i: \quad \partial \Sigma \longrightarrow J\subset M. \label{sigma-model-open}
\ee
Ends of open strings, apart from being restricted to $J$, can also be coupled in the usual way to Chan-Paton degrees of freedom leading to $U(N)$ gauge symmetry. We interpret these as strings in a presence of $N$ D-branes wrapping the cycle $J$. On a worldsheet with boundary a translational symmetry in a direction perpendicular to the boundary is broken, which breaks half of supersymmetries. The other half of supersymmetries must be preserved on the boundary, which imposes strong conditions both on fields $\phi^i$  and their superpartners $\psi^i_{\pm}$, as well as on D-branes themselves. It turns out in A- and B-models we have respectively the so-called 
\begin{itemize}
\item A-branes, for which $J$ must be middle dimensional lagrangian submanifold,
\item B-branes, for which $J$ must be a complex submanifold of $M$, so in particular it must have even real dimension.
\end{itemize} 

Let us assume $H_1(J)$ is generated by a single non-trivial cycle $\gamma$ and consider a worldsheet with $h$ holes $C_i,\ i=1,\ldots h$. Under the map (\ref{sigma-model-open}) each $\phi_*(C_i)$ can be assigned a winding number $w_i\in\mathbb{Z}$ and we can define a generating functional
\be
F_{w,g}(t_i) = \sum_{\beta\in H_2(M,J)} N_{w,g,\beta} Q^{\beta}  \label{grom-wit-open}
\ee
where $w=(w_1,\ldots,w_h)$, $H_2(M,J)$ is a relative homology class, and $Q^\beta$ is defined as in (\ref{grom-wit}). Such an amplitude is characterised by open Gromov-Witten invariants $N_{w,g,\beta}$ which roughly count number of maps from a Riemann surface $\Sigma$ into $M$ with lagrangian boundary conditions on $J$. To define the total free energy which assembles all sets of winding numbers $w$, a matrix $V$ must is introduced, so that
\be
F(V) = \sum_{g=0}^{\infty} \sum_{h=1}^{\infty} \sum_{w=(w_1,\ldots,w_h)} \frac{g_s^{2g-2+h}}{h!} F_{w,g}(t_i) \textrm{Tr}\,V^{w_1}\cdots \textrm{Tr}\,V^{w_h}.  \label{open-inst}
\ee
In fact contributions from sectors which differ by a permutation of $w_i$'s are the same, so for each $w$ we specify $k=(k_1,k_2,\ldots)$, such $k_j$ is the number of $w_i$'s equal to $j$, i.e. $k_j=\#(w_i=j)$. Then $h=\sum_j k_j$, $l=\sum_i w_i=\sum_j jk_j$, and $k$ defines a conjugacy class $C(k)$ of the symmetric group $S_{l}$. For a given $k=(k_j)$ there are $h!/\prod_j k_j!$ equivalent $w$'s (differing only by a permutation of their $h$ components). Finally, the free energy can be written as
$$
F(V) = \sum_{g=0}^{\infty} \sum_{k=(k_j)} \frac{g_s^{2g-2+h}}{\prod_j k_j!} F_{k,g} \, \prod_{j=1} (\textrm{Tr}\,V^j)^{k_j}.
$$
Now the Frobenius formula can be used to write a product of $\textrm{Tr}\,V^j$ as a linear combination of $\textrm{Tr}_R V$
\be
\prod_{j=1}^{\infty} (\textrm{Tr}\,V^j)^{k_j} = \sum_R \chi_R(C(k)) \textrm{Tr}_R V,   \label{frobenius-sums}
\ee
where $\chi_R(C(k))$ is the character of $S_l$ in the representation $R$ for the conjugacy class $C(k)$, so that $F(V)$ can be written as a sum over representations
$$
F(V) = \sum_{R} F_R(g_s,t_i) \textrm{Tr}_R V.
$$
As usual the partition function is defined by an exponential of the free energy, which also can be expanded as a sum over representations
\be
Z(V) = e^F = \sum_R Z_R \textrm{Tr}_R V.  \label{Z-brane}
\ee
If there are $L$ independent one-cycles in $M$ along which boundaries $\partial \Sigma$ can wrap, the partition function takes the form
\be
Z(V_1,\ldots,V_L) = \sum_{R_1,\ldots,R_L} Z_{R_1,\ldots,R_L} \prod_{i=1}^L \textrm{Tr}_{R_i}V_i. \label{Z-branes}
\ee

In what follows we will consider Calabi-Yau threefolds $M$ with a number of A-branes topologically equivalent to $\C\times S^1$. The partition function of such a system with one brane will be exactly of the form (\ref{Z-brane}), whereas with $L$ branes of the form (\ref{Z-branes}). In fact, these partition functions are specified only up to \emph{framing ambiguity}, which is related to the framing ambiguity of knot invariants (\ref{frame-knot}) and requires specifying an additional integer number $f_i$ for $i$'th brane in the system. Let us recall a particular class of branes discussed in section \ref{ssec-toric}, which are attached to an axis specified by a vector $v$ in $\R^3$ base of a toric manifold, and a homology cycle of a fibred $T^2$ determined by a framing vector $w$ degenerates at infinity. If $Z_{R_1,\ldots,R_L}$ is a partition function for a configuration with $L$ such branes in framings $w_i$, then a partition function in framings $w_i-f_iv_i$ with $f_i\in\Z$ is
\be
Z_{R_1\ldots R_L} \to q^{\hf \sum_{i=1}^L f_i\kappa_{R_i}} (-1)^{\sum_{i=1}^L f_i|R_i|} Z_{R_1\ldots R_L}. \label{frame-brane}
\ee

Let us note the open Gromov-Witten invariants can be rewritten in the representation basis as $N_{R,g,\beta}$. Moreover, one can introduce open Gopakumar-Vafa invariants $n_{R,g,\beta}$ in analogy with (\ref{gop-vafa}) which also have integrality properties \cite{marcoscs1}-\cite{marcoscs4}. Nonetheless, they do not arise in our considerations so we do not analyse them in more detail.

%****************************************************************

\subsection{Open topological strings and Chern-Simons theory} \label{ssec-stringfield}

It is believed the most fundamental description of any open string theory should be given in terms of \emph{string field theory}, which describes all open string configurations in terms of a functional $\Psi$. An abstract formulation of a string field theory has been found by Witten in \cite{string-field}. In general the dynamics of the string field is governed by the string field action
\be
S^{string\ field} = \frac{1}{2g_s} \int \big(\Psi \star Q_{BRST}\Psi + \frac{2}{3}\Psi\star\Psi\star\Psi \big),  \label{string-field}
\ee
where $Q_{BRST}$ is BRST charge, whereas the noncommutative and associative \emph{star product} $\star$ and the \emph{integration} $\int$ are defined in some precise way. $\Psi$ describes all tower of string excitations in a very nontrivial way, which in fact is a mixed blessing, because it requires solving the entire system at once which is seldom possible. 

However, string field theory approach leads to a very elegant solution of the A-model topological strings \cite{cs-string}. We focus first on a situation when a target space is the total space of a bundle $T^*M$, where $M$ is a manifold of real dimension three. In this case $M$ is a lagrangian submanifold of $T^*M$, so one can wrap A-branes around it upon which ends of open strings can end. In topological string theory all higher excitations decouple and the string field $\Psi$ describes only a single degree of freedom, the ground state. This ground state is a gauge connection $A$, which is valued in $u(N)$ Lie algebra, where $N$ is the number of A-branes wrapped around $M$. This reduction of $\Psi$ is accompanied by a reduction of the string field integration and the star product --- they become respectively the usual integration on $M$ and the exterior product. Moreover, in topological theory $Q_{BRST}$ is naturally identified with a topological charge, and together with a decoupling of higher excitations it reduces to the exterior differentiation. The reduction of these various objects in string field theory is given in a table below.
$$
\begin{array}{|c|c|}  \hline
\textrm{String field theory} & \textrm{Chern-Simons theory}  \\ \hline\hline
\Psi & A  \\ \hline
Q_A & d    \\ \hline
\star & \wedge  \\ \hline
\int & \int_M  \\ \hline
\end{array}
$$
Eventually, after these substitutions the string field action (\ref{string-field}) for open topological strings reduces to the usual Chern-Simons action (\ref{CS-action}) on $M$
\be
S^{CS}(A) = \frac{1}{2g_s} \int_M \textrm{Tr}\,\big(A\wedge dA + \frac{2}{3}A\wedge A\wedge A \big), \label{cs-stringfield}
\ee
and moreover the relation between string coupling and gauge coupling is found 
$$
g_s = \frac{2\pi i}{k+N},
$$
where $k$ gets shifted $k\to k+N$ upon renormalisation. 

One can also analyse a more general situation, when A-branes wrap a lagrangian cycle $M$ inside an arbitrary Calabi-Yau manifold $X$, not necessarily of the form $T^*M$. Then the effective action for the gauge field also contains a Chern-Simons term on $M$, together with additional contributions arising from worldsheet instantons which can end on $M$. %These worldsheet instantons are absent in the case $X=T^*M$, but in more general situation they give rise to additional contributions. 
In particular, if $H_1(J)$ is generated by a single non-trivial cycle $\gamma$, these contributions can be shown to be equal to $F(V)$ given in (\ref{open-inst}), with $V$ identified with a Wilson loop along the cycle $\gamma$
$$
V = P\,\exp \oint A.
$$
Therefore the total effective action on the worldvolume of A-branes wrapping a lagrangian submanifold $M$ takes the form
\be
S(A) = S^{CS}(A) + F(V).  \label{CS-F-brane}
\ee

We have already analysed Chern-Simons theory in section \ref{sec-cs}, therefore we can use results obtained there in the context of topological string theories. As we will see this has very powerful consequences.

%****************************************************************
%****************************************************************

\section{Topological strings and open-closed duality}       \label{sec-open-closed}

In this chapter we have already introduced open and closed topological strings with the emphasis on the A-model constructions. We also discussed how open topological strings on $T^*M$ reduce to Chern-Simons theory on $M$. In fact we have assembled all ingredients necessary to exhibit the open-closed duality for in the context of A-model topological strings, often referred to as the Gopakumar-Vafa duality. This duality was analysed first in \cite{G-V-transition} which we follow below, and was generalised to the case of Wilson loop observables in \cite{ov}. An interesting complementary derivation of the Gopakumar-Vafa duality was given in \cite{G-V-worldsheet}. A discovery of this duality triggered a lot of activity which eventually led to the discovery of the topological vertex which we introduce in the next section.

In the simplest setting the Gopakumar-Vafa is a statement that the A-model theory of open topological strings on the deformed conifold $T^*S^3$ is equivalent to the theory of closed topological strings on the resolved conifold. This duality can be realised in two steps. The first step is the equivalence of open topological strings on $T^*S^3$ and Chern-Simons theory on $S^3$. The second step is a realisation of 't Hooft duality (explained in section \ref{sec-cs}) in a particular case of Chern-Simons theory on $S^3$ and closed topological string on the resolved conifold. In fact we have already discussed the first step in the previous section, so to derive the duality we have to argue for the second step, which we do in the paragraph below. Then we discuss Wilson loop observables.

\begin{figure}[htb]
\begin{center}
\includegraphics[width=0.8\textwidth]{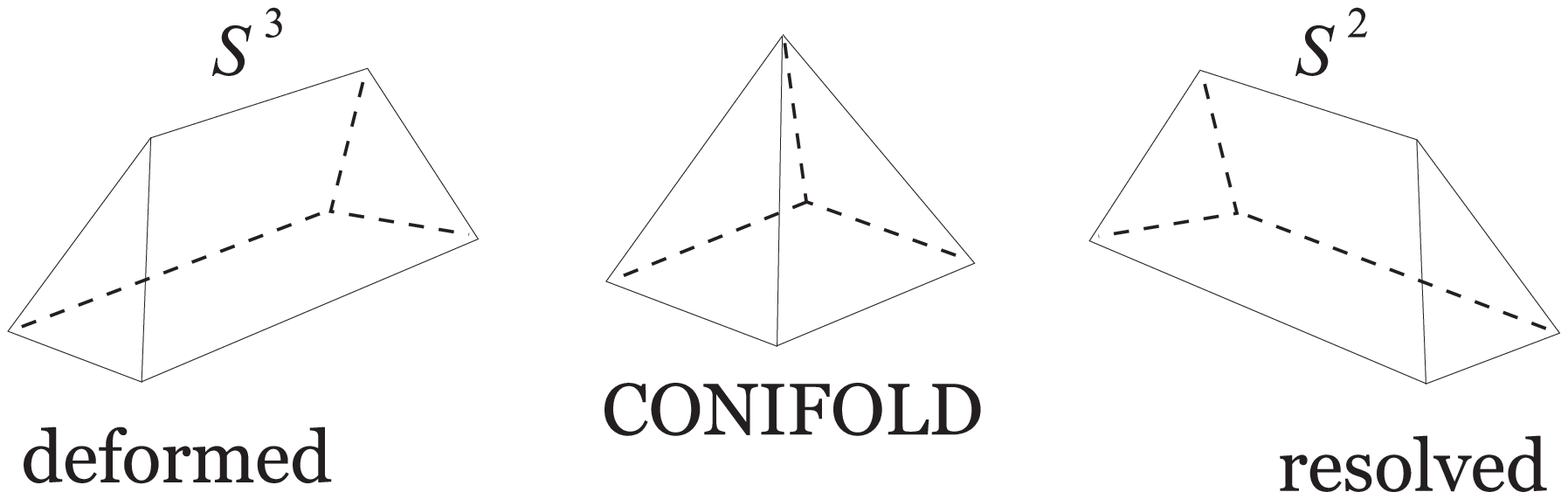}
\caption{The Gopakumar-Vafa duality as a geometric transition between open topological strings the deformed conifold (left) and closed topological strings on the resolved conifold (right), with a singular conifold in an intermediate state. All conifolds are asymptotically isomorphic to $S^2\times S^3$, with $S^2$ and $S^3$ represented by two sides of a rectangle in the base. The singularity is represented by a tip of the cone. The deformed and the resolved conifolds have respectively minimal $S^3$ and $S^2$, represented by orthogonal intervals replacing the tip of the cone.} \label{fig-transition-top}
\end{center}
\end{figure}

\subsection{Equality of partition functions}

We present now the open-closed duality on the level of partition functions, focusing on the equivalence between Chern-Simons theory on $S^3$ and closed topological string on the resolved conifold.
We recall the amplitudes of Chern-Simons theory on $S^3$ can be computed in
terms of fatgraphs, and each fatgraph can be drawn on a Riemann surface of
genus $g$ and with $h$ holes. We denote now a contribution of such a fatgraph
to the free energy as $F_{g,h}$. We already presented this fatgraph
expansion in (\ref{CS-Fgh}). According to 't Hooft idea, a resummation of
this series (\ref{F-closed}) over a number of holes $h$
should lead to some closed string expansion, with $F^{CS}_g(t)$ being a
contribution to the amplitude from worldsheets of genus $g$, and 't Hooft
coupling is $t= N\lambda = \frac{2\pi i N}{k+N}$. We presented the resummed series in (\ref{hooft-cs-g}). 

On the other hand, the genus $g$ terms in the expansion of the A-model closed topological strings on the resolved conifold with a sphere of size $t$ are given in (\ref{F-conifold-GW}), and rejecting classical contributions
$$
F_{g}^{conifold} = \frac{(-1)^g   |B_{2g}B_{2g-2}|}{2g(2g-2)(2g-2)!} - \frac{|B_{2g}|}{2g\,(2g-2)!} \textrm{Li}_{3-2g}(e^{-t}).
$$

Comparing the expansion in (\ref{hooft-cs-g}) and the above conifold expansion
we find a perfect agreement 
$$
F^{CS}_g(t) = F_{g}^{conifold}(t).
$$
 This equality of partition functions is a very strong
confirmation of the Gopakumar-Vafa duality. 

We have to stress that we have not only realised  't Hooft idea of interpreting
fatgraphs of Chern-Simons theory on $S^3$ as worldsheets of a closed string theory. 
This Chern-Simons theory is equivalent to open topological strings on $T^*S^3$ as explained 
in the previous section, so we found explicitly an
open string theory which corresponds to the closed string theory. In the
present case it is possible to give an a very beautiful explanation and interpretation of the equivalence
between these open and closed string theories in terms of the geometric conifold transition 
introduced in section \ref{sec-transition}. We consider A-model
open topological strings on the deformed conifold $T^*S^3$, with string
coupling $g_s$ and with $N$ A-branes wrapped on $S^3$ which is a lagrangian
submanifold. The geometric process amounts to shrinking $S^3$ to zero size
until the singular conifold is reached. Subsequently the singularity is
resolved, and the sphere $\mathbb{P}^1$ of size $t=Ng_s$ blows up. The
A-branes vanish together with vanishing of $S^3$, but their degrees of freedom
are transformed into degrees of freedom of closed strings on the resolved
conifold. The transition in geometry is local --- both deformed and resolved
conifolds asymptote to $S^2\times S^3$ at infinity, and they differ only in the
neighbourhood of the origin. This transition is shown picturesquely in figure \ref{fig-transition-top}.

%****************************************************************

\subsection{Wilson loops and A-branes}   \label{ssec-O-V}

A wide class of observables in Chern-Simons theory are Wilson loops. Their expectation values are knot invariants, as we explained in paragraph \ref{ssec-cs-quant-knots}. It turns out the Gopakumar-Vafa duality manifests itself also in this case, and its quantitative check can be performed for the case of Wilson loops in $S^3$ \cite{ov}.
%We suppose $S^3$ with embedded Wilson loop should undergo a geometric transition in an analogous way to the one described above, so from the closed strings point of view we would expect to obtain the resolved conifold again, although with a modification related to the presence of the Wilson loop in the original configuration. It turns out this modification is the presence of additional A-branes in the resolved conifold geometry, and partition functions corresponding to these configurations on both sides of the duality are again equal.

\begin{figure}[htb]
\begin{center}
\includegraphics[width=0.3\textwidth]{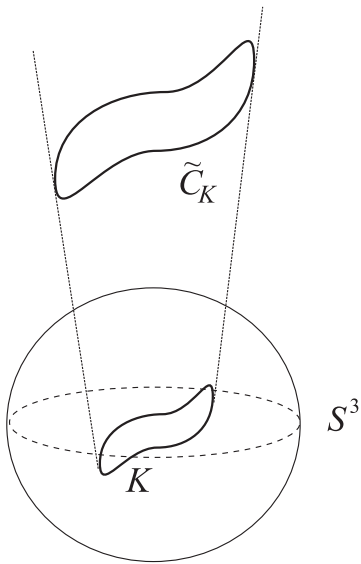}
\caption{A lagrangian submanifold $\tilde{C}_K$ intersects $S^3$ inside $T^*S^3$ along a knot $K$. We wrap $N$ branes on $S^3$ and $M$ branes on $\tilde{C}_K$. Under the geometric transition $S^3$ vanishes together with $N$ branes wrapped on it, and instead $S^2$ blows up. $\tilde{C}_K$ is transformed into a lagrangian submanifold $C$ in the resolved conifold geometry, whereas $M$ branes survive the transition and wrap $C$ afterwards. } \label{fig-lag-knot}
\end{center}
\end{figure}

Let us consider Chern-Simons theory on $S^3$ with a Wilson loop along a knot $K$. Firstly we have to understand what is the corresponding configuration in the open topological string theory on $T^*S^3$. Let us parametrise a knot $K$ by $q(s)\in S^3$ for $s\in [0,2\pi[$. There is a canonical construction of the so-called \emph{conormal bundle} $\tilde{C}_{K}$
$$
\tilde{C}_{K} = \{(q(s),p) \,|\ p_i \frac{dq^i}{ds} = 0, \ p_i\in T^*_qS^3,\ s\in [0,2\pi[,\ i=1,2,3  \},
$$
which associates to every point $q\in K$ a two-dimensional submanifold of a cotanget space $T^*_q S^3$ orthogonal to $dq/ds$. $\tilde{C}_K$ is a three-dimensional submanifold of $T^*S^3$, it has a topology of $\C\times S^1$, and intersects $S^3$ along $K$, as shown in figure \ref{fig-lag-knot}. The crucial point is it satisfies a lagrangian condition, so we can wrap $M$ A-branes around it, which we consider together with $N$ A-branes wrapped on $S^3$.

Now we would like to take a large $N$ limit and follow a geometric transition. We expect $N$ branes should vanish and $S^3$ they wrapped should be replaced by $S^2$. On the other hand, we would like $M$ branes wrapping $\tilde{C}_K$ to survive the transition and still be present in the system. This means they should be wrapping some lagrangian manifold $C$ in the resolved conifold geometry, which arises from $\tilde{C}_K$ in the process of the transition. The branes such as those wrapping $C$, which arise in the background of the closed string geometry --- in this case the resolved conifold --- are called \emph{probe branes}.

To sum up, we consider the following configurations on two sides of the transition:
\begin{itemize}
\item Before the transition: $T^*S^3$ with a knot $K$ and a lagrangian submanifold $C_K$ such that $S^3\cap \tilde{C}_K = K$; there are $N$ A-branes wrapped on $S^3$ and $M$ branes wrapped on $C_K$
\item After the transition: a closed string geometry of the resolved conifold arises, in the presence of $M$ probe branes wrapping $C$
\end{itemize}
%Both configurations contain a system of $M$ A-branes wrapped on a lagrangian submanifolds which are mapped to each other during the transition. 
In both cases effective degrees of freedom should be $U(M)$ gauge fields localised on $M$ A-branes. To check the duality we should identify the effective actions for these gauge fields and show they are equal.

We consider first the configuration before the transition. The massless modes of this system are Chern-Simons $U(N)$ gauge field $A$ on $S^3$, $U(M)$ gauge field $\tilde{A}$ on $\tilde{C}_K$, as well as a scalar $\phi$ in a bifundamental representation of $U(N)\times U(M)$ living on the intersection $S^3\cap \tilde{C}_K = K$. Let us denote the Wilson loops of gauge fields $A$ and $\tilde{A}$ along $K$ respectively as
$$
U=P\,\exp \oint_K A,\qquad V=P\,\exp \oint_K \tilde{A}.
$$
We would like to derive the effective action for $\tilde{A}$ on $\tilde{C}_K$. The string field action for the entire system is a sum of Chern-Simons actions (\ref{cs-stringfield}) for both gauge fields $S^{CS}(A), S^{CS}(\tilde{A})$ and the action for the scalar --- therefore we have to integrate out $A$ and $\phi$. The action for the scalar is just a covariantised kinetic term $\overline{\phi} d\phi$ 
\be
\oint_K \textrm{Tr}\,\overline{\phi}(d+A-\tilde{A})\phi.    \label{str-field-scalar}
\ee
and integrating this out introduces the so-called Ooguri-Vafa operator
\be
\exp \sum_{n=1}^{\infty}\frac{1}{n}\textrm{Tr}U^n \textrm{Tr}V^n  = \sum_R \textrm{Tr}_R U \, \textrm{Tr}_R V,  \label{ov-operator}
\ee
where the equality is a variant of the Frobenius relation (\ref{frobenius-sums}). Keeping $V$ fixed while integrating out $A$ leads to
\be
\int \mathcal{D}A e^{-S_{CS}(A)} \sum_R \textrm{Tr}_RU \textrm{Tr}_R V = \sum_R \mathcal{W}^K_R \textrm{Tr}_R V, \qquad \textrm{where} \quad  \mathcal{W}^K_R = \langle \textrm{Tr}_R P\,\exp \oint_K A \rangle  \label{int-out-A}
\ee
where the expectation value refers to the Chern-Simons path integral. Therefore $\mathcal{W}^K_R$ is precisely a knot invariant (\ref{wilson-loop}) in representation $R$, and the partition function for this open topological string configuration is a generating function of knot invariants for all possible representations. On the other hand, the effective action for the gauge field $\tilde{A}$ on $\tilde{C}_K$ reads,
\be
S^{open}(\tilde{A}) = S^{CS}(\tilde{A}) + F^{open}(\tilde{A}),   \label{Seff-open}
\ee
where $ F^{open}(\tilde{A})$ are corrections which arise from integrating out $\phi$ and $A$ and they can read off from (\ref{int-out-A}) before using Frobenius formula. 

We argued how to determine the effective action for $\tilde{A}$ on one side of the duality. On the resolved conifold side we already know the effective action is (\ref{CS-F-brane})
$$
S(\tilde{A}) = S^{CS}(\tilde{A}) + F(V),
$$
which is also Chern-Simons actions with some corrections $F(V)$. However, the corrections on the closed string side have very different origin: they arise from worldsheet instantons which wrap $S^2$ and end on $\tilde{C}$. In \cite{ov} these two types of corrections on both sides of the duality were analysed in detail for $K$ being the unknot, and it has been shown they are equal
\be
F^{open}(V)|_{K=unknot} = F(V)|_{K=unknot}.          \label{ov-F-unknot}
\ee
This gives another strong check of the duality, and it is believed this relation should extend to all knots. This brane configuration for the unknot will also be of particular interest to us --- we will analyse it quantitatively in section \ref{oogurivafa} from the point of view of Calabi-Yau crystals.

We should note the above observations lead to the following important conclusion. Suppose we would like to find a partition function $Z$ for a system of probe branes in a closed string background, which contains $S^2$. We assume this background is open-closed dual to a configuration with several Wilson loops along components $K_1,\ldots,K_L$ of some link in $S^3$. Then the partition function (\ref{Z-branes}) is a generating function of link invariants with $Z_{R_1,\ldots,R_L}=\mathcal{W}^{K_1,\ldots,K_L}_{R_1,\ldots,R_L}$
\be
Z(V_1,\ldots,V_L) = \sum_{R_1,\ldots,R_L} \mathcal{W}^{K_1,\ldots,K_L}_{R_1,\ldots,R_L} \prod_{i=1}^L \textrm{Tr}_{R_i}V_i.   \label{Z-branes-knot}
\ee
We will see many examples of such relations in what follows. On the other hand, if a closed string configuration cannot be mapped to some link, the partition function coefficients are still some topological invariants $Z_{R_1,\ldots,R_L}$, which are not related to any link invariant.

%***********************************************************

\section{The topological vertex}   \label{chap-top-vertex}

%So far in this chapter we have reviewed a general construction of topological strings, emphasizing the theory of the A-model, its relation to superstring and supergravity theories, and the idea of the Gopakumar-Vafa transition. 

In general a computation of topological string amplitudes is a difficult task. As they are defined in a perturbative expansion, at first sight it is not obvious how to determine them beyond a few lowest genera. However, it turns out there is a remarkable method which allows to find exact, all-genus A-model topological string amplitudes on a wide class of non-compact toric geometries discussed in section \ref{sec-cy}. A crucial ingredient of this method is the so-called \emph{topological vertex}
%A crucial ingredient in this method is the Gopakumar-Vafa transition and the relation of topological strings to Chern-Simons theory. 

The topological vertex for A-model topological strings was introduced  in \cite{vertex}, and subsequently in \cite{B-vertex} for mirror B-model theories. It was actively studied for example in \cite{HIV,strip,flop,gen-ver,ps,ps-cube}. Its rigorous mathematical formulation was given in \cite{math-vertex}. In this section we introduce the topological vertex formalism. It will be used to large extent in chapter \ref{sec-A-model} in checking Calabi-Yau crystal results. 

The method of the topological vertex is based on the following idea.
A-model closed string amplitudes compute a number of holomorphic maps from a
string worldsheet into some Calabi-Yau three-fold $M$. There are two kinds of
such maps: constant ones for which the image of the worldsheet is a point in
$M$, and worldsheet instantons which can wrap holomorphic two-cycles of
$M$. We focus only on the latter case now. If $M$ is toric its only
non-trivial two-cycles, which can be wrapped by worldsheet instantons, are
spheres $S^2$, represented by finite intervals in a toric diagram. It is possible to split a given closed
string worldsheet into a configuration of open strings by introducing  into $M$ branes, which wrap
lagrangian cycles of the type discussed in section
\ref{sec-cy}. These branes are attached to intervals in a toric diagram and
open string worldsheets can end on them from both sides in such a way, that
when glued together they reproduce a given closed string configuration. If
such branes are introduced for each finite interval of the toric diagram, this
diagram is divided into separate trivalent vertices which represent distinct $\C^3$ patches of $M$. 
The topological vertex is an open string amplitude for a single $\C^3$ patch with branes attached to its three axes. Together with appropriate rules of gluing such vertices, it provides a basic building block for topological string amplitudes on arbitrary toric geometries. In fact, the topological vertex method of computing topological string amplitudes on toric geometries resembles usual Feynman rules: the toric diagram is analogous to a single Feynman diagram, its trivalent vertices represent interactions, and gluing rules correspond to propagators.

As an open string amplitude, the topological vertex has a general form given in (\ref{Z-branes}). It is conventionally written in the representation basis as 
\be
Z(V_1, V_2, V_3) =  \sum_{R_1, R_2, R_3}  C_{R_1 R_2 R_3}\, {\rm Tr}_{R_1} V_1\, {\rm Tr}_{R_2} V_2\, {\rm Tr}_{R_3} V_3,   \label{C-top-vertex}
\ee
where each representation $R_i$ corresponds to a stack of branes wrapping a lagrangian cycle $L_i$ fixed to $i$'th axis of $\C^3$ patch, as shown in figure \ref{fig-c3-branes}, and $V_i$ are the corresponding sources (holonomy matrices), which in general can be given by infinite matrices. The vertex amplitude $C_{R_1R_2R_3}$, as explained in the next paragraph, can be derived using the Gopakumar-Vafa transition, by mapping this configuration of branes into the corresponding observable in Chern-Simons theory. It is derived in the next paragraph and it reads 
\be
C_{R_1 R_2 R_3} = q^{\frac{\kappa_{R_2} + \kappa_{R_3}}{2}} \sum_{P_1,P_2,P_3} c^{R_1}_{P_3 P_1} c^{R_3^t}_{P_3P_2} \frac{W_{R_2^t P_1}W_{R_2 P_2}}{W_{R_2 \bullet}},       \label{vertex-hopf}
\ee
where $c^{R}_{P Q}$ are tensor product coefficients (\ref{lit-rich}). At this point let us just note that it is expressed by the expectation value (\ref{dbl-hopf}) of a link shown in figure \ref{fig-dbl-hopf}, for the reasons which will become clear below. 

The topological vertex amplitude can be presented in a variety of ways. In computations it is very convenient to write the vertex in terms of Schur functions, which can be easily done using the relation (\ref{hopf-limit}) and properties of symmetric functions given in appendix \ref{app-schur}. This leads to the formula  \cite{ok-re-va}
\begin{equation}
C_{R_1 R_2 R_3} = q^{\frac{1}{2}(\kappa_{R_2}+\kappa_{R_3})} s_{R_{2}^{t}}(q^{\rho})
\, \sum_{P} s_{R_{1}/P}(q^{ R_{2}^{t}+\rho}) s_{R_{3}^{t}/ P}(q^{ R_{2}+\rho}).   \label{vertex-chap}
\end{equation}

In paragraph \ref{ssec-vertex-derive} we discuss a derivation of the topological vertex amplitude in more detail. Further details concerning the topological vertex, in particular the calculational framework in terms of Schur functions, are discussed in detail in appendix \ref{app-topver}. Even though in this thesis we consider mainly A-model topological vertex, in section \ref{sec-B-model} B-model version will also be briefly introduced for completeness.

We also note that many computational techniques have been advanced to simplify topological vertex calculations. In particular, in \cite{strip} a set of rules have been derived for geometries whose dual diagrams are represented as a triangulation of a rectangle or a \emph{strip}. This simplification is quite convenient for a part of the calculations we perform, so we briefly recall the \emph{rules on the strip} in appendix \ref{app-strip}. In fact, we also need to perform some calculations which can be understood as \emph{moving off the strip}. In order to do that we generalise the method of \cite{strip} to some more general geometries, which is discussed in section \ref{move-off}.

%***********************************************************************
%***********************************************************************

\subsection{Derivation}  \label{ssec-vertex-derive}

Following \cite{vertex} we determine the topological vertex amplitude for a configuration of three stacks of branes wrapping lagrangian submanifolds in $\C^3$ shown in figure \ref{fig-c3-branes}. In paragraph \ref{ssec-O-V} we explained that topological string amplitudes reduce to generating functions of knot or link invariants (\ref{Z-branes-knot}) if a brane configuration can be mapped to a configuration of knots in $S^3$ via the geometric transition. Even though the configuration of branes in figure \ref{fig-c3-branes} cannot be mapped explicitly to any knot or link, there is an indirect way to find such a relation and to express $C_{R_1 R_2 R_3}$ in terms of knot invariants. This can be done in four steps, which we discuss below.

\begin{figure}[htb]
\begin{center}
\includegraphics[width=0.9\textwidth]{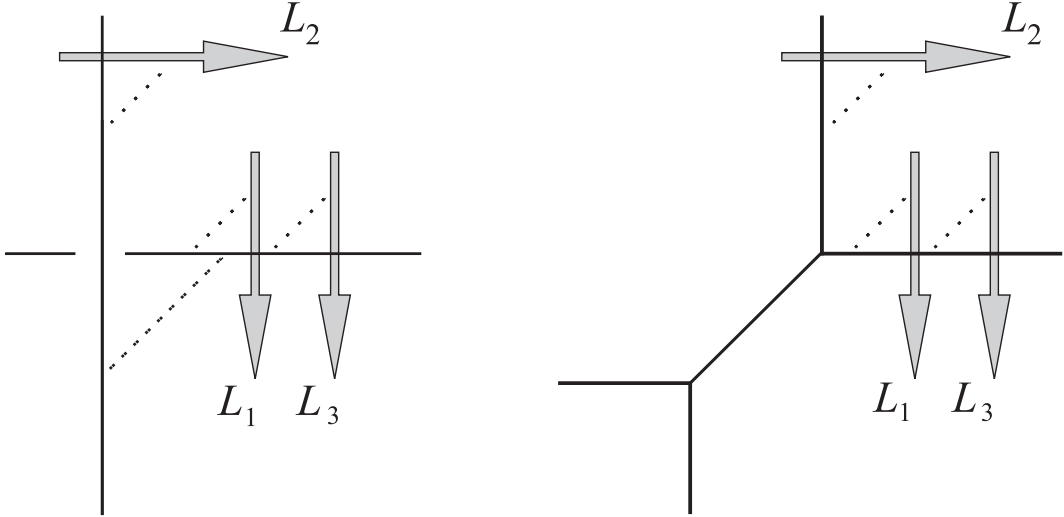}
\caption{Gopakumar-Vafa transition in the presence of branes.} \label{fig-transition-branes}
\end{center}
\end{figure}

In the first step we consider the deformed conifold $T^*S^3$ shown in figure \ref{fig-def-coni-toric} with a stack of $N$ branes wrapped over $S^3$ represented by a dashed line. Additionally, we wrap three stacks of branes along three lagrangian submanifolds $L_1,L_2,L_3$, as shown in the left picture in figure \ref{fig-transition-branes}. Arrows determine the framing of $L_i$ submanifolds at infinity, as explained in section \ref{ssec-toric}. We would like to find an open topological string amplitude for such a configuration. It is possible by a reasoning similar as in section \ref{ssec-O-V}, taking the string field theory point of view and using the relation to Chern-Simons theory. It can be shown the string field degrees of freedom reduce to gauge fields on each stack of branes, and a scalar field analogous to (\ref{str-field-scalar}) arising from open strings stretched between two stacks of branes. In general, for gauge fields $A$ and $\tilde{A}$ on such two branes, the action for this scalar field takes the form
\be
\oint \textrm{Tr}\,\overline{\phi}(d+A-\tilde{A} - r)\phi,  
\ee
and now a distance $r$ between branes has been taken into account. It can also be shown that for strings stretched between parallel branes this scalar field has a fermionic statistics, and integrating this out leads to an operator
\be
\exp \Big[-\sum_{n=1}^{\infty}\frac{e^{-nr}}{n}\textrm{Tr}U^n \textrm{Tr}V^n \Big] = \sum_R e^{-|R|\,r}(-1)^{|R|} \, \textrm{Tr}_R U \, \textrm{Tr}_{R^t} V, . \label{ov-operator-r}
\ee
where $V$ and $U$ and holonomies for gauge fields $A$ and $\tilde{A}$ respectively. Moreover, such scalar fields can arise only for strings stretched in parallel to degeneration loci of torus actions in $\R^3$ base, i.e. only between $L_1-S^3, L_2-S^3, L_3-S^3$ and $L_1-L_3$ branes. 

Let us introduce holonomies for gauge fields on $L_i$ branes denoted respectively as $V_i$ and computed in representations $R_i$, which we will treat as fixed sources. We have to take into account a holonomy $V^{-1}_1$ of opposite orientation than $V_1$, which will arise from integrating out strings stretched between $L_1-L_3$ branes, computed in representation $R$.  We also introduce holonomies $U_1, U_2$ for gauge fields on $S^3$ which we will integrate out. Taking all these factors into account, and in particular including operators (\ref{ov-operator-r}) from each scalar field mentioned above, the amplitude for the configuration in the left picture in figure \ref{fig-transition-branes} reads
\bea
Z(V_1, V_2, V_3) & = & \frac{1}{Z^{S^3}}\sum_{R,R_1, R_2, R_3}  (-1)^{|R_1|+|R_2|+|R|} \langle {\rm Tr}_{R_1^t} U_1\,{\rm Tr}_{R_2} U_2\,{\rm Tr}_{R_3^t} U_1\,  \rangle\, \cdot \nonumber \\
& & \cdot {\rm Tr}_{R_1^t} V_1\, {\rm Tr}_{R_2} V_2\, {\rm Tr}_{R\otimes R_3} V_3\, {\rm Tr}_{R_1} V^{-1}_1,  \label{C-step1}
\eea
where we explicitly factor out $Z^{S^3}$ , a partition function of $S^3$ given in (\ref{Z-S3-cs}). From the effective string field theory point of view the expectation value $\langle {\rm Tr}_{R_1^t} U_1\,{\rm Tr}_{R_2} U_2\,{\rm Tr}_{R_3^t} U_1\,  \rangle$ can be computed as an observable in Chern-Simons theory, related to the link consisting of circles which are boundaries of worldsheets of open strings ending in $S^3$. These boundaries constitute a link shown in figure \ref{fig-dbl-hopf}, whose expectation value is given in (\ref{dbl-hopf}), and therefore 
\be
\langle {\rm Tr}_{R_1^t} U_1\,{\rm Tr}_{R_2} U_2\,{\rm Tr}_{R_3^t} U_1\,  \rangle = Z^{S^3}\,\frac{\mathcal{W}_{R_2^t R_1}\mathcal{W}_{R_3 R_1}}{\mathcal{W}_{R_1 \bullet}}   \label{2hl-vertex}
\ee

In the second step we argue the amplitude computed in (\ref{C-step1}) should not change under the geometric transition, in which $S^3$ blows down and instead $S^2$ blows up. This is analogous to the reasoning in paragraph\ref{ssec-O-V}, and leads to a configurations of branes $L_1,L_2,L_3$ in the background of the resolved conifold, as shown in the right picture in figure \ref{fig-transition-branes}.

\begin{figure}[htb]
\begin{center}
\includegraphics[width=0.9\textwidth]{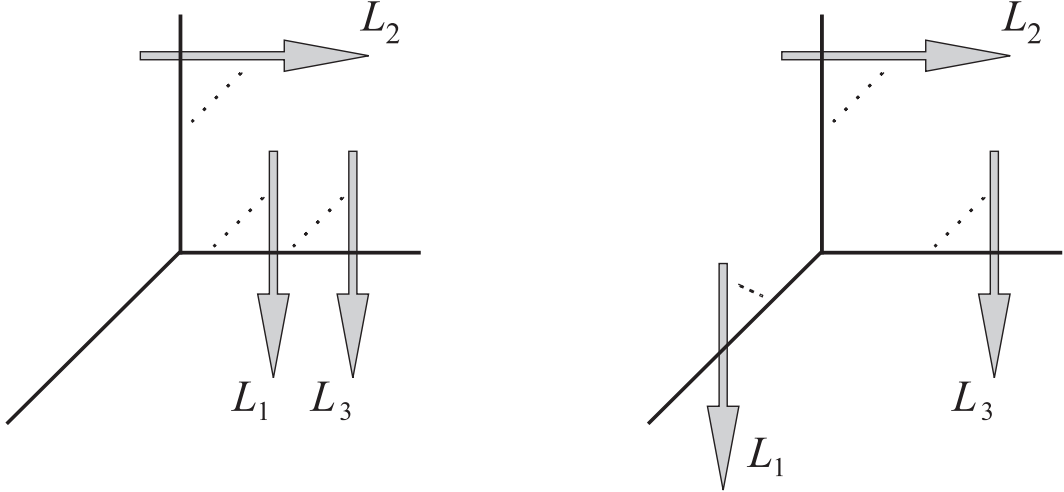}
\caption{Moving a brane from one axis to the other is accompanied by an appropriate analytic continuation of the topological string amplitude for this configuration.} \label{fig-move-brane}
\end{center}
\end{figure}

In the third step we relate the resolved conifold background to $\C^3$ we are really interested in. This amounts to taking a limit of infinite K\"ahler parameter associated to $S^2$ in the resolved conifold. In Chern-Simons theory this limit extracts the leading term of Hopf-link invariants (\ref{hopf-limit}). In this limit the interval representing $S^2$ becomes infinitely long and the toric diagram reduces to the one representing $\C^3$ with three stacks of branes $L_i$, as in the left picture in figure \ref{fig-move-brane}. This is almost the configuration we are interested in.

In the fourth step we analytically continue the lagrangian brane $L_1$ from the edge $v_3=(1,0)$ to the edge $v_1=(-1,-1)$. The configuration we obtain is shown in the right picture in figure \ref{fig-move-brane}. It can be shown this continuation amounts to replacing
$$
(-1)^{|R_1|} W_{R_2 R_1^t}\, {\rm Tr}_{R_1} V^{-1}_1\,  {\rm Tr}_{R_2} V_2 \ \to\ W_{R_2^t R_1}\, {\rm Tr}_{R_1} V_1\,  {\rm Tr}_{R_2} V_2
$$
in the large K\"ahler limit of (\ref{C-step1}). However, this is not yet the final result --- to obtain an amplitude for a configuration in figure \ref{fig-c3-branes} we need to change framings of branes according to (\ref{frame-brane}), in order to rotate the arrows representing framing of branes as required. Performing the steps described above and tracing the framing dependence carefully, formulae (\ref{C-step1}) and (\ref{2hl-vertex}) lead to the result given in (\ref{vertex-hopf}).

%***********************************************************************
%***********************************************************************

%***********************************************************
%***********************************************************
%***********************************************************

\chapter{Two-dimensional Calabi-Yau crystals}   \label{chap-ale}

In this chapter we introduce in a novel way Calabi-Yau crystal models for topological field theories in four dimensions \cite{rhd-ps}. The idea is that certain amplitudes in gauge theories are reproduced by simple models of crystal melting known from statistical mechanics. The name \emph{Calabi-Yau crystals} arises because the field theories we consider are defined on Calabi-Yau manifolds. Crystal models for four-dimensional field theories we consider in this chapter are defined in two dimensions. Even more interesting class of three-dimensional crystal models arise for topological string theories, which we discuss in further chapters.

The entire crystal in two-dimensional models is identified with a lattice given by a positive quarter of $\Z^2$ or its certain subset. Individual points in this lattice are identified as atoms in the crystal which can melt away in a non-zero temperature. A given melted configuration of the crystal, with some places vacant due to melted atoms, can be assigned a statistical weight defined by the number of melted atoms. Denoting such a configuration by $R$ and a number of vacant boxes it contains by $|R|$, the partition or generating function of this system reads
\be
Z=\sum_{R}q^{|R|},           \label{generating-2d}
\ee
where $q=e^{-\mu/T}$ is related to the chemical potential $\mu$ and the temperature $T$ from the statistical physics point of view. One can consider of course various rules which allow an individual atom to melt away. In particular, one can impose a constraint that the only allowed melted crystal configurations are represented by two-dimensional partitions introduced in appendix \ref{sec-partitions}. Then $|R|$ becomes an ordinary weight of a partition $R$, defined by its number of boxes (\ref{R-size}). In fact, a remarkable work of Nekrasov \cite{Nek} and \cite{Nek-Ok} revealed many partition functions of supersymmetric gauge theories are related to the counting of various ensembles of two-dimensional partitions, with weights depending on the details of a given theory. These partitions can be mapped to states of free fermions as explained in appendix \ref{app-fermion}, so that their generating functions can also be rewritten as various amplitudes in a free fermion theory. The two-dimensional partitions which arise in this correspondence can be shown to encode information about moduli spaces of instantons in a gauge theory, and the relation (\ref{generating-2d}) arises from an identification of the gauge theory partition function with the Euler characteristic of instanton moduli spaces (\ref{Z-gauge}). On the other hand, the physical origin of the associated free fermions is not completely clear. 

As the simplest example of a two-dimensional Calabi-Yau crystal model, let us consider $\mathcal{N}=4$ twisted $U(1)$ gauge theory on $\mathbb{R}^4$. Its partition function is given by the inverse $\eta(q)$ function (\ref{Z-N4}), which apart from a factor of $q^{-1/24}$ is a generating function of all two-dimensional partitions weighted by the number of boxes, in accordance with (\ref{generating-2d}).

In this chapter we wish to generalise this counting to the case of $\mathcal{N}=4$ theories on ALE spaces, whose partition functions are given by affine characters corresponding to different affine weights  (\ref{Z-VW}). It turns out such a generalisation may be rephrased in terms of a special class of partitions which we introduce. We call them \emph{orbifold crystals} or \emph{generalised partitions}. At least for $U(1)$ theory on $A_{k-1}$ singularity, the counting of these orbifold crystals $R$ indeed reproduces affine characters computed at some special values, and the partition function (\ref{generating-2d}) reads
$$
Z=\sum_{R} q^{|R|} = \chi^{\widehat{su}(k)_N}_{\widehat{\lambda}}(q).
$$
We also interpret orbifold crystals explicitly in terms of a two-dimensional fermionic system. Let us remark some other relation between gauge theories on ALE spaces and partitions was discussed in \cite{ALEspaces}.

%***********************************************************
%***********************************************************

\section{Orbifold crystals} \label{sec-crystals}

An ordinary two-dimensional partition $\lambda$ can be determined by an ideal of functions $\mathcal{I}=\{f(x,y)\}$ generated by a set of monomials $x^i y^j$ for $i,j \geq 0$, in such a way that a box $(m,n)\in \lambda$ if and only if $x^m y^n \notin \mathcal{I}$.

We introduce the following $\mathbb{Z}_k$ orbifold action on $\mathbb{C}^2$
\begin{equation}
(x,y) \to (\omega x,\overline{\omega}y),\qquad \textrm{for}\ \omega=e^{2\pi i /k}.  \label{Zk-action}
\end{equation}
Let us consider ideals of functions having definite transformation properties under this action. A given monomial $x^i y^j$ transforms as $\omega^{i-j}$, and all monomials with the same transformation property can be represented as a periodic sub-lattice of $\mathbb{Z}^2$. In particular, there is a set of invariant monomials which we refer to as \emph{invariant} (or \emph{non-twisted}) sector (these necessarily include constant functions represented by $(0,0)$ point on the lattice). All the other classes of monomials will be called as \emph{twisted} sectors. 

We define two types of partitions, which we call \emph{orbifold crystals} or \emph{generalised  partitions}:
\begin{itemize}
\item firstly, we consider ordinary two-dimensional partitions, however with all boxes corresponding to points in $\mathbb{Z}^2$ lattice with a definite transformation properties distinguished; we define a weight of such a partition as the number of these distinguished boxes (and \emph{not all boxes} in this partition),
\item secondly, we consider diagrams which consists \emph{only} of these distinguished boxes, with a weight given by their number.
\end{itemize}
We draw these distinguished boxes in black in the figures below. In the former case, there are generally many partitions with the same set of distinguished boxes, but with different positions of remaining (``weightless'') boxes. In the latter case, a given set of distinguished boxes defines one and only one partition. One can also think of the partitions of the latter type as equivalence classes of partitions of the former type, such that all elements in one class have the same set of distinguished boxes. %The generalised partitions will be sometimes called $\Gamma$-partitions (in the present case $\Gamma=\mathbb{Z}_k$), orbifold crystals, or orbi-crystals for short.

%To specify a generalised partition of the former type, we have to specify an ordinary partition in one of the standard ways together with a particular sector we are interested in. A generalised partition of the latter type is specified 

\begin{figure}[htb]
\begin{center}
\includegraphics[width=0.8\textwidth]{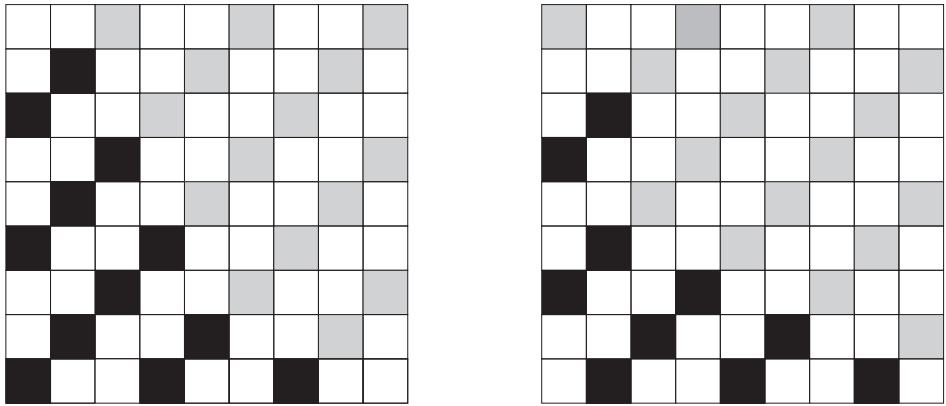}
\caption{Sample generalised partitions of the latter type (consisting only of black boxes) in invariant (left) and twisted (right) sectors of $\mathbb{C}^2/\mathbb{Z}_{3}$ orbifold.} \label{fig-orbi-Z3}
\end{center}
\end{figure}

Examples of generalised $\mathbb{Z}_3$-partitions of the latter type are given in figure \ref{fig-orbi-Z3}. The left picture corresponds to the invariant sector, and the right one to the twisted sector corresponding to monomials transforming with a factor of $\omega=\exp(2\pi i/3)$. The corresponding lattices are denoted by grey boxes; the box in the left-bottom corner has coordinates $(0,0)$ and corresponds to constant functions. Sample partitions in both sectors are denoted in black, and white boxes are immaterial (examples of the former type partitions arise if we take white boxes into account as well). A proper way to define a generalised partition (of both types and in any sector) is as follows: if a box $(m,n)$ belongs to a partition, then all the boxes $(i,j)$ from the (grey) sub-lattice such that $i\leq m$ and $j \leq n$ must also belong to this partition; this property is easily seen in the figure \ref{fig-orbi-Z3}.

To work with generalised partitions of the second type it is crucial to denote them in a convenient way, which takes into account only distinguished boxes. The usual convention to write number of distinguished boxes in each row is not the best choice, as these numbers are not decreasing; for example partitions in figure \ref{fig-orbi-Z3} correspond respectively to sequences $(3,2,1,2,1,1,1,1)$ and $(3,2,2,1,0,1,1)$. An additional condition which states when a number of boxes in the next row may increase must be introduced in this case; this condition is simple but awkward.

\begin{figure}[htb]
\begin{center}
\includegraphics[width=0.5\textwidth]{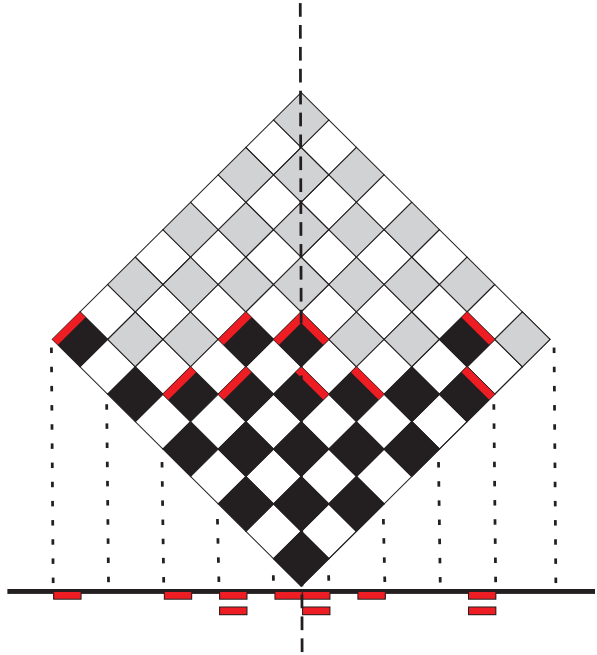}
\caption{A generalised partition of the latter type in an invariant sector of $\mathbb{C}^2/\mathbb{Z}_{2}$ orbifold as a state in a parafermi sea.} \label{fig-orbi-Z2-fermion}
\end{center}
\end{figure}

It turns out a better idea is to use the so-called Frobenius notation introduced in appendix \ref{sec-partitions}. For definiteness, let us focus on an invariant sector. We slice a given partition $\lambda$ diagonally and introduce two sequences of numbers $(a_i)$ and $(b_j)$, which denote number of boxes in rows to the right and to the left of the diagonal
$$
\lambda = \left( \begin{array}{lllll} 
a_1 & a_2 & \ldots & a_{d(R)} \\
b_1 & b_2 & \ldots & b_{d(R)} \\
\end{array} \right)
$$
where $d(R)$ is a number of boxes on the diagonal. For ordinary partitions, the sequences $(a_i)$ and $(b_i)$ must be strictly decreasing. For generalised $\mathbb{Z}_k$-partitions, our crucial observation is that this condition is relaxed: each number $a_i$ and $b_i$ can occur at most $k$ times --- and apparently this coincides with a definition of \emph{generalised Frobenius partitions}, which were introduced by Andrews in \cite{Andrews}. Let us note that such a partition can be presented as a state of a Fermi sea with a generalised statistics: there may be at most $k$ fermions at a given position (such objects are called \emph{parafermions}, and their unusual form of the Pauli principle a \emph{parastatistics}). In figure \ref{fig-orbi-Z2-fermion} the partition
$$
\left( \begin{array}{lllll} 
3 & 3 & 1 & 0 & 0 \\
4 & 2 & 1 & 1 & 0 \\
\end{array} \right)
$$
is presented for $\mathbb{C}^2/\mathbb{Z}_2$ orbifold, together with the corresponding state in the parafermi sea. Indeed, in this case at most two parafermions can sit in the same place. Generalisation of this setup to other sectors is straightforward. 

%***********************************************************
%***********************************************************

\section{The former type partition counting} \label{sec-crystals-resum-0}

In this section we wish to compute generating functions of the generalised partitions of the former type for ALE spaces of $A_{k-1}$ type
$$
\tilde{Z}^k_r = \sum_{former\ gen.\ partitions} q^{\# (black\ boxes)},
$$
in all sectors $r=0,\ldots,k-1$, where $r$ specifies the power of $\omega$ in (\ref{Zk-action}).

As we show below, these generalised partitions are related to states in the Fock space of $k$ free fermions, and the number of black boxes gives weights of such states. As is well known, a system of $k$ free fermions provides a representation of $\widehat{u}(k)_1$ affine Kac-Moody algebra. For this reason $\tilde{Z}^k_r$ can be expressed in terms of affine characters. On the other hand, the black boxes transform in a definite way under $\mathbb{Z}_k$ orbifold group of $A_k$ ALE space, which provides a relation to Nakajima results.

We start with the observation that generalised partitions of the former type can be identified with \emph{blended} partitions, which describe a state of a Fermi sea of several fermions. Let us consider $k$ complex fermions with charges $p_1,\ldots,p_k$, and fixed total charge $p$
$$
p=\sum_{i=1}^{k} p_i.
$$
For $i$'th of these fermions there is a corresponding Fock space $\mathcal{F}_i$, and elements of its basis can be represented in a standard way as states of a Fermi sea, or usual Young diagrams with a specified charge $p_i$. The total Fock space $\mathcal{F}$ is a tensor product of $k$ Fock spaces corresponding to individual fermions
$$
\mathcal{F} = \bigotimes_{i=1}^{k} \mathcal{F}_i.
$$
Basis elements of $\mathcal{F}$ are obtained by tensoring basis elements of $\mathcal{F}_i$. A tensor product of states corresponds to a coloured partition (introduced in appendix \ref{sec-partitions}) $\vec{{\bf R}}=\{ R_{(i)} \}$ with charges $p_i$, which also determines the so-called blended partition. A \emph{blended partition} ${\bf R}$, associated to a coloured partition $\vec{{\bf R}}$, is defined by a set of integers
\be
\{ k(p_i+R_{(i),m} - m) +i-1\ |\ m\in\mathbb{N}  \} = \{ p+{\bf R}_K -K \ | \ K\in\mathbb{N}  \}, \label{blend}
\ee
with a finite number of non-zero ${\bf R}_K$ and ordered such that ${\bf R}_1 \geq {\bf R}_2 \geq \ldots$. The total number of boxes of such a partition is equal to
\be
| {\bf R} | = \sum_i \Big( k|R_{(i)}| + \frac{k}{2} p_i^2 + ip_i \Big) - \frac{(k+1)p}{2} - \frac{p^2}{2}. \label{blend-nr}
\ee

\begin{figure}[htb]
\begin{center}
\includegraphics[width=0.9\textwidth]{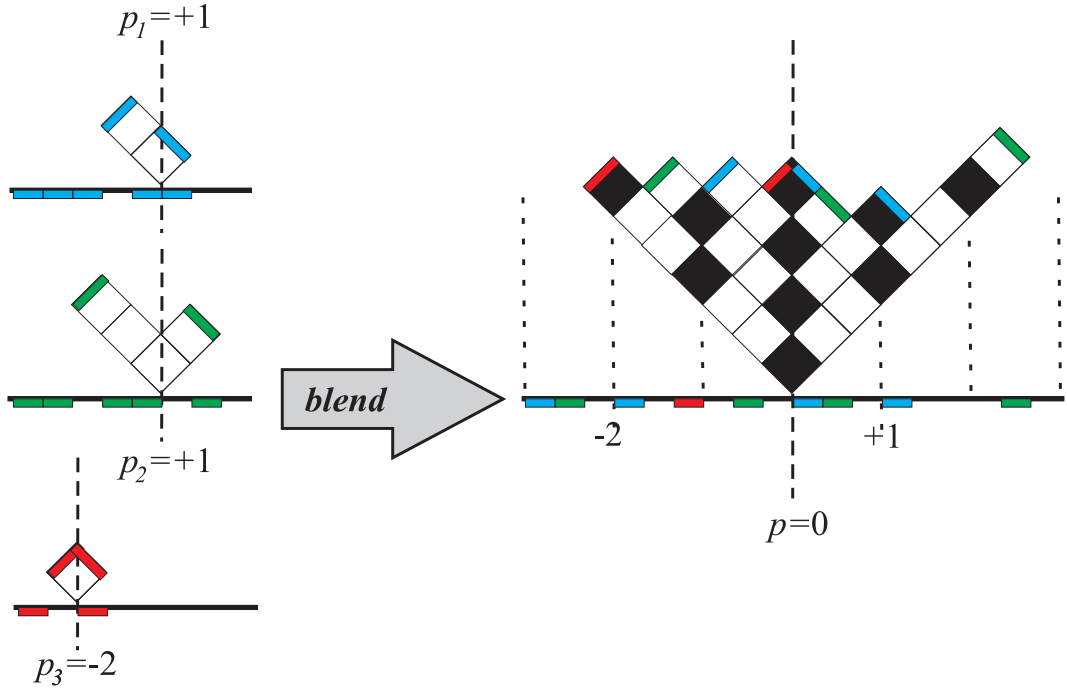}
\caption{Blended partition obtained from $k=3$ fermions is in fact equivalent to generalised partition. The total fermion weight $q^{\sum(|R_{(i)}| + p_i^2/2)} = q^{10}$ is determined by the number of black boxes.} \label{fig-orbi-Z3-blend}
\end{center}
\end{figure}

Now we claim the generalised $\mathbb{Z}_k$-partitions of the former type are in one-to-one correspondence with blended partitions ${\bf R}$ obtained from $k$-coloured partitions $\vec{{\bf R}}$, such that
\begin{itemize}
\item a generalised partition has the same shape as the corresponding blended partition,
\item a weight of a generalised partition (as given by the number of distinguished boxes it contains) is equal to the total weight of a state of $k$ fermions related to $\vec{{\bf R}}$. 
\end{itemize}
A generalised partition corresponding to a certain blended one is shown in figure \ref{fig-orbi-Z3-blend}. 

\begin{figure}[htb]
\begin{center}
\includegraphics[width=0.6\textwidth]{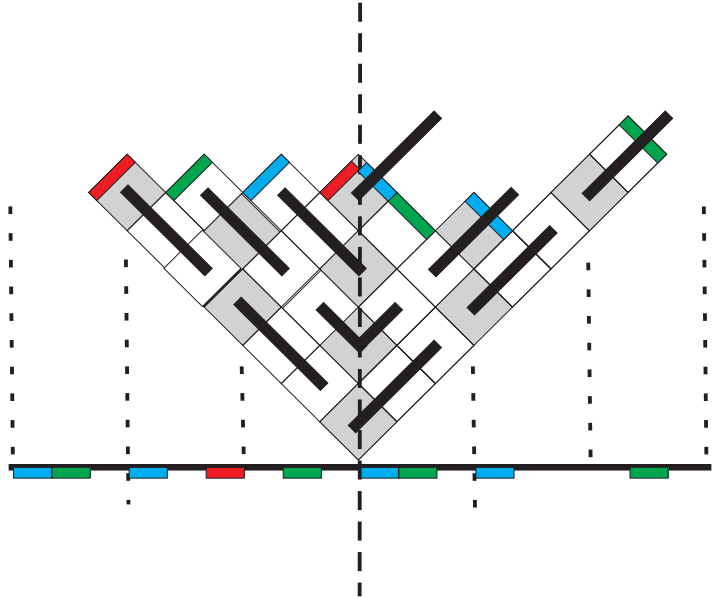}
\caption{Counting of distinguished boxes of a generalised partition from figure \ref{fig-orbi-Z3-blend}. This partition consists of 4 hooks attached to the diagonal. The number of ``excessive'' boxes is $-1-1+1-2=-3$, which can be computed (\ref{excessive}) as $\sum_i ip_i = 1+2*1+3*(-2)$ (in this case $n=r=0$).} \label{fig-blend-inv}
\end{center}
\end{figure}

The total weight of a state of  $k$ fermions is equal to their contribution to the character (\ref{u-k-1})
\be
\sum_i \Big( |R_{(i)}| + \frac{p_i^2}{2} \Big).  \label{black-boxes}
\ee
To find the number of distinguished boxes in a generalised partition ${\bf R}$ (of the same shape as a blended
partition) it is helpful to represent it as a set of hook
partitions (having precisely a single row and a single column), fixed to
diagonal elements of ${\bf R}$, and divided into intervals of at most $k$
boxes. An example of such a construction for a situation in figure
\ref{fig-orbi-Z3-blend} is given in figure \ref{fig-blend-inv}, with these
intervals drawn in black. Each interval of $k$ boxes contains precisely one
distinguished box. To count distinguished boxes, we will count all boxes
contained in those black intervals and divide their number by $k$. However,
there are some ``excessive'' boxes: in fact some of those black intervals
contain less than $k$ boxes (when an interval sticks out of the partition); or some white boxes don't belong to any black interval (when all
distinguished boxes in a given hook have already been matched to some black
interval). Let us rewrite charges $p_i$ in terms of $n,r,n_1,\ldots,n_{k-1}$ as in (\ref{p-decompose}), which is
always possible in a unique way. In particular
$$
n = [p/k],\qquad r=p-kn,
$$
where $[\cdot]$ denotes the integer part of a real number. Let us also
introduce
$$
p'_i = p_i - n -\delta,\qquad \textrm{where}\quad \delta=\left\{
\begin{array}{cl}
1 & \textrm{for}\ i=1,\ldots,r \\
0 & \textrm{for}\ i=r+1,\ldots,k
\end{array}\right.
$$
Now it is straightforward to show the number of these excessive boxes is
\be
\sum_i ip'_i = \sum_i ip_i -\frac{(N+1)N n}{2} - \frac{(r+1)r}{2}.  \label{excessive}
\ee
Subtracting this from the total number of boxes $| {\bf R}|$ and dividing by
$k$ we conclude the number of distinguished boxes is equal to
$$
\frac{|{\bf R}| - \sum_i ip'_i}{k} = \sum_i \big( |R_{(i)}| + n_i^2 - n_i
n_{i+1}  \big) + \frac{r^2}{2} + n_1 r - \frac{r}{2},
$$
and summing over all diagrams $R_{(i)}$ with the total fixed charge $p=kn+r$ we get
\bea
\tilde{Z}^k_r & = & \sum_{former\ gen.\ partitions} q^{\# (black\ boxes)} = \frac{q^{k/24}}{\eta(q)^k}
\sum_{n_1,\ldots,n_{k-1}} q^{\sum_i (n_i^2 - n_i n_{i+1}) + \frac{r^2}{2} +
  n_1 r - \frac{r}{2}} = \nonumber \\
& = & \frac{q^{\frac{k}{24} + \frac{r^2}{2k} - \frac{r}{2}}}{\eta(q)} \chi^{\widehat{su}(k)_1}_r(z_i=0)  \label{Z1-k-r}
\eea
which reproduces the $\widehat{su}(k)_1$ characters (\ref{su-char})
computed for $z_i=0$. Now we recall that $\widehat{u}(k)_1$ character decomposes (\ref{char-u-su}) into $k$ level 1 affine characters (\ref{su-char}) indexed by $r=0,\ldots,k-1$, weighted by $\widehat{u}(1)_k$ characters (\ref{u1-char})
\begin{equation}
\chi^{\widehat{u}(k)_1}(x_i) = \sum_{r=0}^{k-1} \chi^{\widehat{u}(1)_k}_r(\tilde{x})\, \chi^{\widehat{su}(k)_1}_r(\tilde{x}_i),
\end{equation}
where $x_i$, $i=1,\ldots,k$ are specialisation points (\ref{char-u-vars}), which in particular determine variables $y_j$, $j=1\ldots k-1$ in which $\widehat{su}(k)_1$ characters are naturally expressed (\ref{char-su-vars}). We see that generating functions for generalised partitions indeed combine into $\widehat{u}(k)_1$ character computed at values of $x_i=\exp(2\pi i z_i) = 1$. Summing (\ref{Z1-k-r}) over allowed $r$ we reproduce the $n=0$ sector of $\widehat{u}(k)_1$ character (\ref{uNchar})
\be
\chi^{\widehat{u}(k)_1}(z_i=0)|_{n=0} = q^{-k/24} \sum _{r=0}^{k-1}
q^{r/2}\,\tilde{Z}^k_r.  \label{Z1-uk-chi}
\ee
Thus the counting of states of $k$ fermions with fixed total charge is
equivalent to the counting of generalised partitions of the former type. 

Let us finally note that we always can think of the distinguished boxes as belonging to the \emph{invariant} sector, i.e. such that their
corresponding monomials $x^i y^j$ are invariant under $\mathbb{Z}_k$
action. If $p$ is a multiplicity of $k$, the box at position $(0,0)$ of a blended partition always transforms invariantly. But if the total charge $p$ is not a multiplicity of $k$, the corner of
the blended partition is fixed at the position $p$ of the total Fermi sea, and the box $(0,0)$ does not transform invariantly; thus those boxes which do transform invariantly may be thought of as belonging to twisted sectors of the generalised partition.

%***********************************************************
%***********************************************************

\section{The latter type partition counting} \label{sec-crystals-resum}

Now we consider generalised partitions of the latter type. We prove their
generating functions are also given by affine characters, this time computed at
some special values of parameters $y_i$. Thus they can be identified with equivalence classes of blended partitions mentioned above.

To start with, for a given $k$ we consider the following function of two parameters introduced by Andrews in \cite{Andrews}
\begin{eqnarray}
G_k(z,q) & = & \prod_{n=0}^{\infty} (1+Q^{n+1} +\ldots z^k q^{k(n+1)})(1+z^{-1}q^n +\ldots z^{-k} q^{k n} ) \nonumber \\
& = & \sum_{r\in\mathbb{Z}} Z^{k}_r(q) z^{r},  \label{G-N}
\end{eqnarray}
where the second line defines implicitly functions $Z^{k}_r(q)$.

The crucial observation is that this expression encodes generating functions for all non-twisted and twisted sectors of  $\mathbb{C}^2/\mathbb{Z}_{k}$ orbifold: a partition function for a sector twisted by $\omega^r$ (for $r=0,\ldots,k-1$) is given by  $Z^{k}_{-r}(q)$, i.e. a term proportional to $z^{-r}$. It is understood that the invariant sector corresponds to $i=0$, i.e. $z$-independent term. Thus we claim
$$
Z^k_{-r} = \sum_{latter\ gen.\ partitions} q^{\# (black\ boxes)}.
$$ 
We recall the generalised partitions can be interpreted as blended partitions, and then there is always only one type of distinguished boxes, related to invariant monomials from the total Fock space $\mathcal{F}$ point of view. But for $r\neq 0$ the corner of this blended partition is fixed at such a position that these distinguished invariant boxes belong to the twisted sector of the partition.

In particular, for $k=1$ the expression (\ref{G-N}) reduces to the standard Jacobi triple product identity (\ref{Jacobi}), with a single sector with $Z^{k=1}_{0}(q) \sim \prod (1-q^n)^{-1} = q^{-1/24} \eta(q)$. This reproduces of course the partition function for $\mathcal{N}=4$ theory on $\mathbb{R}^4$ given in (\ref{Z-N4}). We now present how it generalises to ALE spaces of $A_{k-1}$ type for arbitrary $k$.

As mentioned above, our generalised partitions from invariant sector of $\mathbb{C}^2/\mathbb{Z}_k$ lattice are in one-to-one correspondence with generalised Frobenius partitions with $k$ repetitions allowed. Their generating function was shown in \cite{Andrews} to be given by $Z^k_0(q)$. We have to extend this observation to all other sectors in order to include instanton contributions for all possible flat connections at infinity.

\begin{figure}[htb]
\begin{center}
\includegraphics[width=\textwidth]{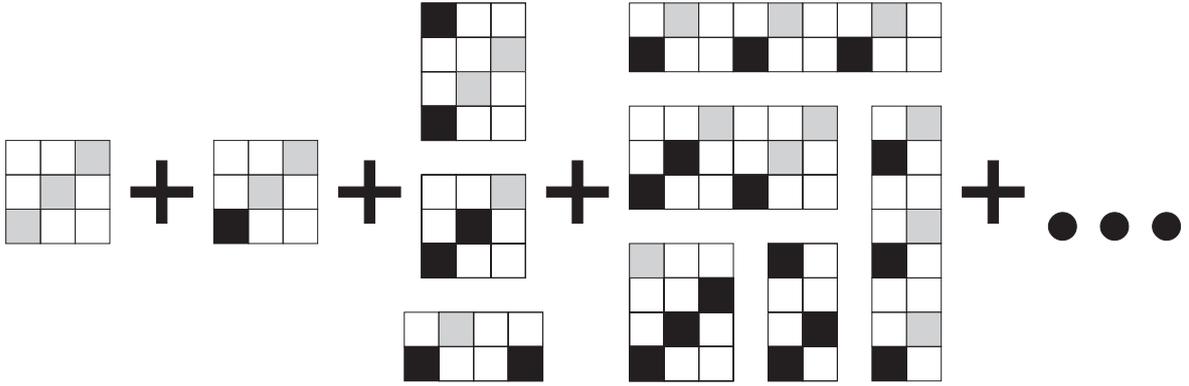}
\caption{First terms of a generating function for an invariant sector of $\mathbb{C}^2/\mathbb{Z}_{3}$ orbifold $Z^3_0=1+q+3q^2+5q^3+\ldots$} \label{fig-orbi-Z3-sum}
\end{center}
\end{figure}

\begin{figure}[htb]
\begin{center}
\includegraphics[width=\textwidth]{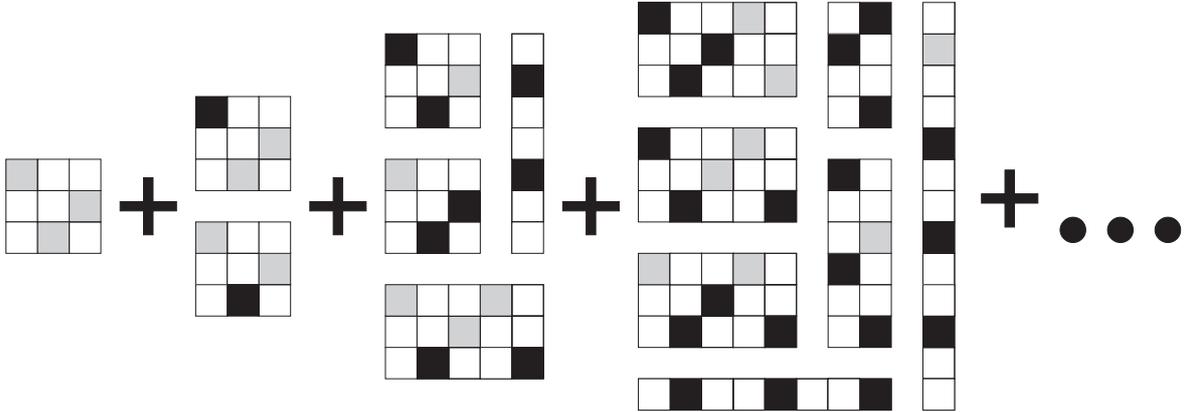}
\caption{First terms of a generating function for twisted sectors of $\mathbb{C}^2/\mathbb{Z}_{3}$ orbifold $Z^3_{-1}=Z^3_{-2}=1+2q+4q^2+7q^3+\ldots$} \label{fig-orbi-Z3-sum-twist}
\end{center}
\end{figure}

Thus, our present aim is to compute generating functions $Z^{k}_{-r}(q)$ for each $r=0,\ldots,k-1$. A symmetry under reflection along the diagonal $x=y$ implies the relation
$$
Z^k_{-r}=Z^k_{-(k-r)}.
$$
Anticipating the result, this is also an important property of $\widehat{su}(k)$ affine characters at level 1. Several lowest terms in generating functions for invariant and twisted sectors of $\mathbb{C}^2/\mathbb{Z}_{3}$ orbifold, together with all relevant partitions, are shown in figures \ref{fig-orbi-Z3-sum} and \ref{fig-orbi-Z3-sum-twist}.

The function (\ref{G-N}) can be rewritten as follows
\begin{eqnarray}
G_k(z,q) & = & \prod_{j=1}^{k} \Big(\prod_{m=1}^{\infty}(1-\zeta^j z q^m)\,\prod_{n=0}^{\infty}(1-\zeta^{-j} z^{-1} q^n) \Big) = \label{G-fermions} \\
& = & \prod_{i=1} \frac{1}{(1-q^i)^k} \prod_{j=1}^{k} \sum_{m_j\in \mathbb{Z}} (-1)^{m_j} q^{m_j(m_j+1)/2} z^{m_j} \zeta^{j m_j} \nonumber
\end{eqnarray}
where $\zeta = e^{2\pi i / (k+1)}$ (note it's different from $\omega$!). The first line above reveals a relation to a system of $k$ fermions with phases $\zeta^{\pm j}$. The second line allows to extract a term $Z^k_{-r}(q)$ we are looking for by imposing the condition
\begin{equation}
-r = m_1+m_2+\ldots + m_k.
\end{equation}
After a little  algebra this leads to the main result:
\begin{equation}
Z^k_{-r} = (-1)^r q^{r(r-1)/2} \zeta^{- k r} \frac{q^{k/24}}{\eta(q)^k} \times \label{Z-N-k}
\end{equation}
$$ 
\times \sum_{m_1,\ldots,m_{k-1}\in\mathbb{Z}} q^{\sum_{i=1}^{k-1} m_i^2  +  \sum_{i<j} m_i m_j + r\sum_{i=1}^{k-1} m_i}  \zeta^{-(k-1)m_1 - (k-2)m_2 -\ldots -m_{k-1}},
$$
where $r=0,\ldots,k-1$.
% and Euler function is
%$$
%\varphi(q) = \prod_{i=1}^{\infty} (1-q^i).
%$$

Let us prove now the generating functions we obtained (\ref{Z-N-k}) indeed can be written as affine $\widehat{su}(k)_1$ characters, and moreover they encode entire $\widehat{u}(k)_1$ character,  similarly as in the former case (\ref{Z1-uk-chi}).
%The notion and notation for characters are given in appendix \ref{app-characters}. 
%$\widehat{u}(k)_1$ character decomposes (\ref{char-u-su}) into $k$ level 1 affine characters (\ref{su-char}) indexed by $r=0,\ldots,k-1$, weighted by $\widehat{u}(1)_k$ characters (\ref{u1-char})
%\begin{equation}
%\chi^{\widehat{u}(k)_1}(x_i) = \sum_{r=0}^{k-1} \chi^{\widehat{u}(1)_k}_r(\tilde{x})\, \chi^{\widehat{su}(k)_1}_r(\tilde{x}_i),
%\end{equation}
%where $x_i$, $i=1,\ldots,k$ are specialisation points (\ref{char-u-vars}), which in particular determine variables $y_j$, $j=1\ldots k-1$ in which $\widehat{su}(k)_1$ characters are naturally expressed (\ref{char-su-vars}). 
%We now show the generating functions for generalised partitions are given by these characters computed at particular values of $x_i$ (or $y_j$).

Firstly, let us redefine summations in (\ref{su-char}) by introducing $n_i = \sum_j M_{i j}m_j$, for $M$ having $1$ on and over the diagonal (and zeros otherwise). We get
$$
\chi^{\widehat{su}(k)_1}_r(y_i) = \frac{1}{\eta(q)^{k-1}} q^{\frac{r^2}{2}\frac{k-1}{k}}\zeta^{-r(k-1)/2} \times
$$
$$    
\times \sum_{m_1,\ldots,m_{k-1}} q^{\sum_i m_i^2 + \sum_{i<j} m_i m_j + r\sum m_i} \zeta^{-m_1(k-1)\ldots -m_{k-1}}.
$$
To match to $Z^k_{-r}$ in (\ref{Z-N-k}) we have to choose 
$$
y_i = \zeta^{a_i},\quad \textrm{for}\ a_i = -\sum_j A^{-1}_{ij} = -\frac{(k-i)i}{2},
$$
which immediately determine ratios of $x_i$'s in (\ref{char-u-vars})
$$
y_i = \zeta^{a_i} \iff \tilde{y}_i = \frac{x_i}{x_{i+1}} = \zeta^{-1}.
$$
Moreover, prefactors in (\ref{Z-N-k}) match $n=0$ factor of $\widehat{u}(1)_k$ character if we identify
$$
\tilde{x} = x_1\cdots x_k = \zeta^{-k(k+1)/2},
$$
which altogether determines
$$
x_i = \zeta^{-k-1+i} = \zeta^{i},
$$
where we used $\zeta^{k+1}=1$.

Now the precise relation reads between characters and orbifold crystals reads
\be
Z^k_{-r} = \sum_{latter\ gen.\ partitions} q^{\# (black\ boxes)} = f_{k,r}(q,\zeta)\,\chi^{\widehat{su}(k)_1}_r(y_i=\zeta^{a_i}), \label{Z-chi}
\ee
where
$$
f_{k,r} =  (-1)^r\,\zeta^{-\frac{r}{2}-\frac{rk}{2}}\, \frac{q^{\frac{k}{24}+\frac{r^2}{2k}-\frac{r}{2}}}{\eta(q)} = (-1)^{r} q^{\frac{k}{24}-\frac{r}{2}}\,\chi^{\widehat{u}(1)_k}_r \big(\tilde{x}=\zeta^{-\frac{k(k+1)}{2}}  \big)|_{n=0},
$$
where $|_{n=0}$ denotes the corresponding term in (\ref{u1-char}). 

Finally, the full partition function arises from different sectors and should be given by the overall $\widehat{u}(k)_1$ character as in (\ref{char-u-su}). From the above identifications we get
\be
\chi^{\widehat{u}(k)_1}(x_i=\zeta^i)|_{n=0} = q^{-k/24} \sum_{r=0}^{k-1} (-1)^r q^{r/2} \, Z^k_{-r}.  \label{Z-uk-chi}
\ee

%***********************************************************
%***********************************************************

\section{The latter type examples}  \label{sec-examples}

The generating functions we obtained can also be presented in terms of infinite products by applying the Jacobi identity
\begin{equation}
\prod_{n=1}^{\infty} (1-q^n)(1+z q^n)(1+z^{-1}q^{n-1}) = \sum_{m \in \mathbb{Z}}z^m q^{m(m+1)/2}.  \label{Jacobi}
\end{equation}
Some examples of particular computations corresponding to the formula (\ref{Z-N-k}) for the latter type partitions for orbifolds $\mathbb{C}^2/\mathbb{Z}_k$ for $k=2,3,4$ are given below.

\subsection*{$\mathbb{C}^2/\mathbb{Z}_2$}

The invariant sector:
\begin{eqnarray}
Z^{2}_0(q) & = & \prod_{n=1} \frac{1}{(1-q^n)(1-q^{12n-10})(1-q^{12n-9})(1-q^{12n-3})(1-q^{12n-2})} = \nonumber \\
& = & \prod_{n=1} \frac{1}{1-q^n} \exp \sum_{n>0} \frac{q^{-4n}+q^{-3n}+q^{3n}+q^{4n}}{(-1)\cdot n[12n]} = \\
& = & 1+q+3q^2+5q^3+9q^4+14q^5+24q^6+35q^7+55q^8+81q^9+\dots \nonumber
\end{eqnarray}
The twisted sector (terms proportional to $\omega =e^{2\pi i /2} =-1$):
\begin{eqnarray}
Z^{2}_{-1}(q) & = & \prod_{n=1} \frac{1+q^{2n-1}}{(1-q^n)(1-q^{12n-6})} = \\
& = & \Big( \prod_{n=1} \frac{1}{1-q^n} \Big)\  \exp \sum_{n>0} \frac{q^{-5n}+q^{-3n}+q^{-n}-(-1)^n+q^n+q^{3n}+q^{5n}}{(-1)^n\cdot n[12n]} = \nonumber \\
& = & 1+2q+3q^2+6q^3+10q^4+16q^5+26q^6+40q^7+60q^8+90q^9+\dots \nonumber
\end{eqnarray}

\subsection*{$\mathbb{C}^2/\mathbb{Z}_3$}

The invariant sector (compare with figure \ref{fig-orbi-Z3-sum}):
\begin{eqnarray}
Z^{3}_0(q) & = & \prod_{n=1} \frac{(1-q^{6n})(1-q^{12n-6})}{(1-q^n)(1-q^{2n})(1+q^{6n})(1-q^{6n-3})^2} = \nonumber \\
& = & \prod_{n=1} \frac{1}{(1-q^n)} \\
& &  \exp \sum_{n>0} \frac{q^{-4n}+2q^{-3n}+q^{-2n}-1+(-1)^n+q^{2n}+2q^{3n}+q^{4n}+(-1)^nq^{6n}}{(-1)\cdot n[12n]} = \nonumber \\
& = & 1+q+3q^2+6q^3+11q^4+18q^5+31q^6+49q^7+78q^8+119q^9+\dots \nonumber
\end{eqnarray}

Both twisted sectors (terms proportional to $\omega=e^{2\pi i /3}$ and $\overline{\omega}$) have the same generating functions, as is geometrically obvious; this function in fact is a sum of two infinite products (compare with figure \ref{fig-orbi-Z3-sum-twist}):
\begin{eqnarray}
Z^{3}_{-1;-2}(q) & = & -i \prod_{n=1} \frac{(1-q^{6n})(1-q^{2n})}{(1-q^n)^3}
\Big(\prod_{n=1} (1+iq^{6n-1})(1-iq^{6n-5})(1+q^{4n-2}) + \nonumber \\
& & +(i-1)\prod_{n=1} (1+iq^{6n-4})(1-iq^{6n-2})(1+q^{4n})   \Big) \\
& = & 1+2q+4q^2+7q^3+13q^4+22q^5+36q^6+57q^7+90q^8+137q^9+\dots \nonumber
\end{eqnarray}

\subsection*{$\mathbb{C}^2/\mathbb{Z}_4$} \label{sss-C2Z4}

We present just several terms in the expansion. For the invariant sector:
$$
Z^{4}_0(q) = 1+q+3q^2+6q^3+12q^4+20q^5+35q^6+56q^7+92q^8+142q^9+\dots
$$

Twisted sectors proportional to $\omega=e^{2\pi i /4}=i$ and $\overline{\omega}=-i$ have the same generating functions:
$$
Z^{4}_{-1;-3}(q) =  1+2q+4q^2+8q^3+14q^4+25q^5+42q^6+68q^7+108q^8+168q^9+\dots
$$

The twisted sector proportional to $\omega^2=-1$:
$$
Z^{4}_{-2}(q) = 1+2q+5q^2+8q^3+16q^4+26q^5+45q^6+72q^7+115q^8+176q^9+\dots
$$

\subsection*{$\mathbb{C}^2/\mathbb{Z}_4$ and affine characters}

Let us consider an example of a relation between orbi-crystals and characters for the case of $\mathbb{C}^2/\mathbb{Z}_4$. Using
$$
\zeta=e^{\frac{2\pi i}{5}},\qquad \tilde{y}_1=\tilde{y}_2=\tilde{y}_3 = \zeta^{-1},\qquad x_i = \zeta^i,
$$
in formulae (\ref{Z-chi}) and (\ref{su-char}), one can immediately rederive the above expansions for $\mathbb{C}^2/\mathbb{Z}_4$. Moreover, using these expansions, the decomposition of $\widehat{u}(4)_1$ character (\ref{u-k-1}) indeed matches the result (\ref{Z-uk-chi})
$$
\chi^{\widehat{u}(4)_1}(x_i=\zeta^i)|_{n=0} = q^{-1/6}\Big( Z^4_{0} - q^{1/2}Z^4_{-1}+qZ^4_{-2}-q^{3/2}Z^4_{-3} \Big) =
$$
$$
 = q^{-1/6} \big( 1 -q^{1/2}+2q-3q^{3/2}+5q^2-6q^{5/2}+11q^3-12q^{7/2}+20q^4-22q^{9/2}+\ldots \big).
$$

%***************************

\section{Summary} \label{sec-crystals-review}

In this chapter we derived explicit relations between affine characters and orbifold crystals (\ref{Z1-uk-chi}) and (\ref{Z-uk-chi}). As $\widehat{u}(k)_1$
characters, these two expressions differ only in their arguments, and both can
be understood as generating functions of generalised partitions. We identified these
generalised partitions with blended partitions representing $k$
fermions with charges $p_i$ and fixed total charge $p=\sum_i p_i$. In the affine
character these fermions are weighted by a power of $q$, which turns out to be
equal to the number of distinguished (black) boxes in corresponding
generalised partition. In the former case each blended partition arises once,
so in general there are several partition with the same set of black boxes,
but differing in positions of white ``weightless'' boxes. In the latter case
the effect of a particular value of $x_i=\zeta^i$ is such that coefficients of
all generalised partitions with the same configuration of black boxes add up
exactly to 1, so their counting reduces to the counting of the generalised Frobenius
partitions introduced by Andrews.

%***********************************************************
%***********************************************************
%***********************************************************

\chapter{Three-dimensional Calabi-Yau crystals --- rudiments}   \label{chap-cy-crystals}

Topological string theories introduced in chapter \ref{chap-top-strings} are a very interesting area of research. Being simple enough to allow computations of exact amplitudes, they exhibit a rich phenomena of the quantum world. An example of such a phenomenon is the open-closed duality, whose detailed understanding is believed to give insight not only into more complicated models, such as AdS/CFT correspondence, but also into quantum gravity and non-perturbative regimes of QCD. Topological string amplitudes constitute also a subset of superstring amplitudes and are related in numerous ways to four-dimensional gauge and supergravity theories. All these connections provide a strong motivation for studying topological string theories. 

Nonetheless, at first sight it appears the structure of the topological string amplitudes is highly involved. For example, on a wide class of toric Calabi-Yau manifolds, A-model amplitudes can be computed using the topological vertex formalism introduced in section \ref{chap-top-vertex}, by gluing vertices $C_{R_1 R_2 R_3}$ which represent various patches of the Calabi-Yau space. Already a single vertex amplitude $C_{R_1 R_2 R_3}$ is a very non-trivial function of $q=e^{-g_s}$. Even though we have also shown how topological amplitudes are encoded in Gromov-Witten and Gopakumar-Vafa invariants, it is usually a difficult task to determine these invariants and the corresponding formulae (\ref{Z-top}) and (\ref{gop-vafa}) still do not provide a very clear interpretation of A-model partition functions.

%The simplicity of topological strings does not actually mean the computation of their amplitudes is simple. A-model amplitudes we focus on in this thesis can be computed in a rigorous way in the Gromov-Witten theory, which is however a highly involved branch of mathematics and it provides answers only for particular Calabi-Yau target spaces. The physical method of the topological vertex introduced in section \ref{chap-top-vertex}, which is based on the open-closed duality and relation to Chern-Simons theory,  allows to solve topological strings on a wide class of toric geometries, although the results are quite complicated --- already a single topological vertex $C_{R_1 R_2 R_3}$ is a very non-trivial function of $q=e^{-g_s}$.

However, there is a remarkable simplicity hidden in topological A-model amplitudes on toric Calabi-Yau manifolds --- it turns out they are reproduced by generating functions of simple models of crystal melting known from statistical mechanics. This relation was found first in \cite{ok-re-va}. This allows to interpret toric Calabi-Yau manifolds as \emph{quantum manifolds} in terms of appropriately understood \emph{quantum geometry}. In this context these models are called \emph{Calabi-Yau crystals}. 

Crystal partition functions can also be given a \emph{quantum foam} interpretation as sums over fluctuating geometries in some effective gravitational theory on the Calabi-Yau space. This gravitational theory was identified with a six-dimensional twisted $U(1)$ gauge theory in \cite{foam}. Such an effective theory of A-model closed strings is generally called \emph{K\"ahler gravity} \cite{kahler-gravity}, and it is analogous to Kodaira-Spencer theory describing B-model closed strings \cite{bcov} or Chern-Simons theory describing A-model open topological strings \cite{cs-string}. In this sense the open-closed duality in A-model topological strings reduces to the gauge-gravity correspondence between Chern-Simons theory and K\"ahler gravity represented by Calabi-Yau crystals.

Crystal models were activly \cite{va-sa}-\cite{crystal-bubble}. Moreover, they turn out to be closely related to \emph{dimers}, which constitute another class of statistical models. Dimers encode in a concise way a lot of information about supersymmetric gauge theories and mirror symmetry \cite{dimer}-\cite{dimer-mirror}. There is certainly many more fascinating relations hidden in this circle of ideas which are still to be found.

Crystal models related to Calabi-Yau manifolds are of interest also from many purely mathematical points of view. Firstly, they were related to Donaldson-Thomas invariants in \cite{mnop}. Secondly, they are defined in terms of the so-called \emph{plane partitions} studied in \cite{mcmahon,wedge-partitions,schur-process,pearcey}. In the next section we explain what plane partitions are and how they relate to two-dimensional fermions. In the following sections in this chapter we introduce other aspects of Calabi-Yau crystals: their interpretation as quantum manifolds and the relation to the topological vertex and branes. In some sense, these all are \emph{rudiments} of Calabi-Yau crystals already known in literature. In chapter \ref{chap-cy-results} we will introduce and analyse various new crystal models, and their relation to topological string theories will be proved in chapter \ref{sec-A-model}.

\begin{figure}[htb]
\begin{center}
\includegraphics[width=0.4\textwidth]{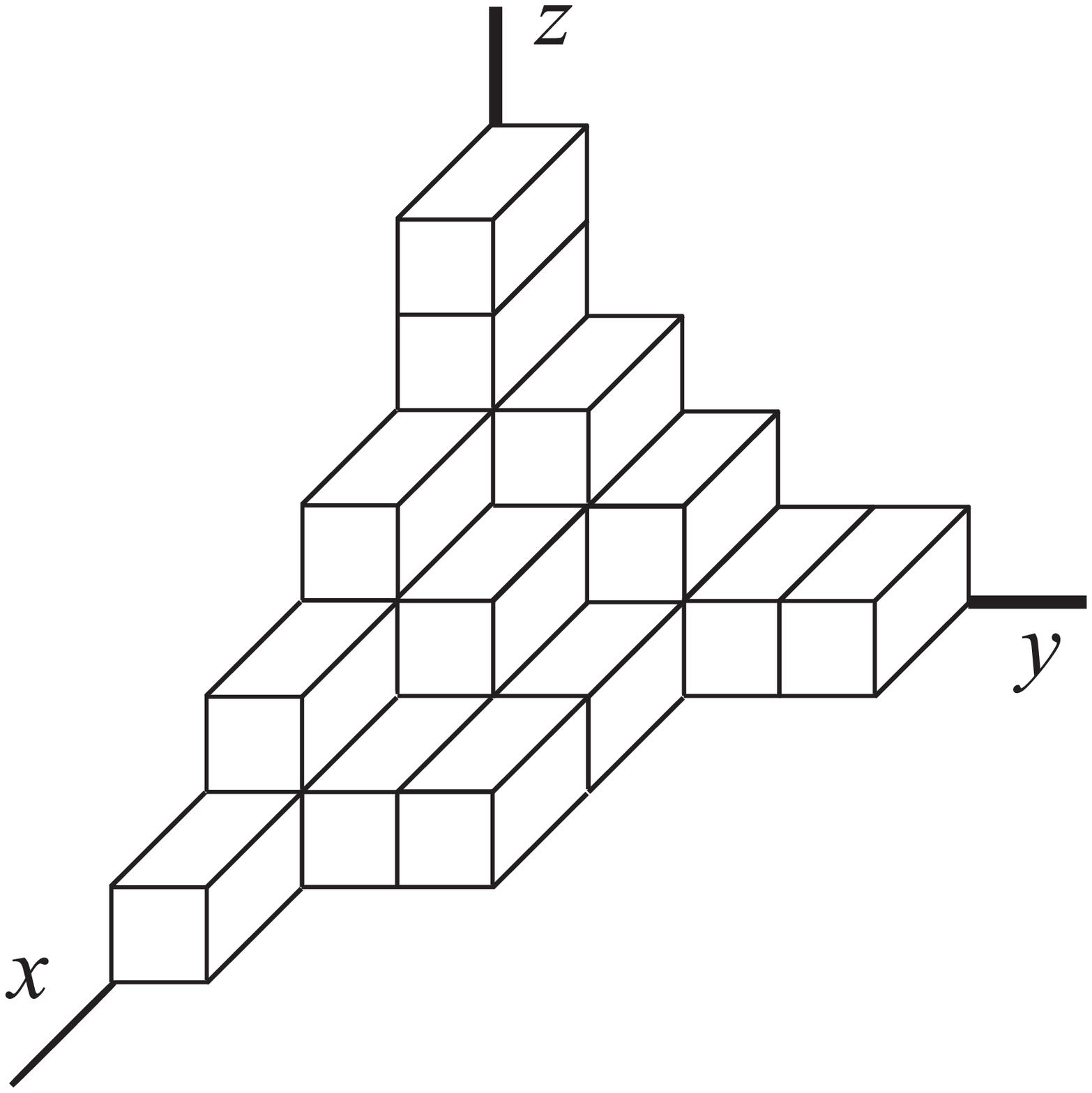}
\includegraphics[width=0.4\textwidth]{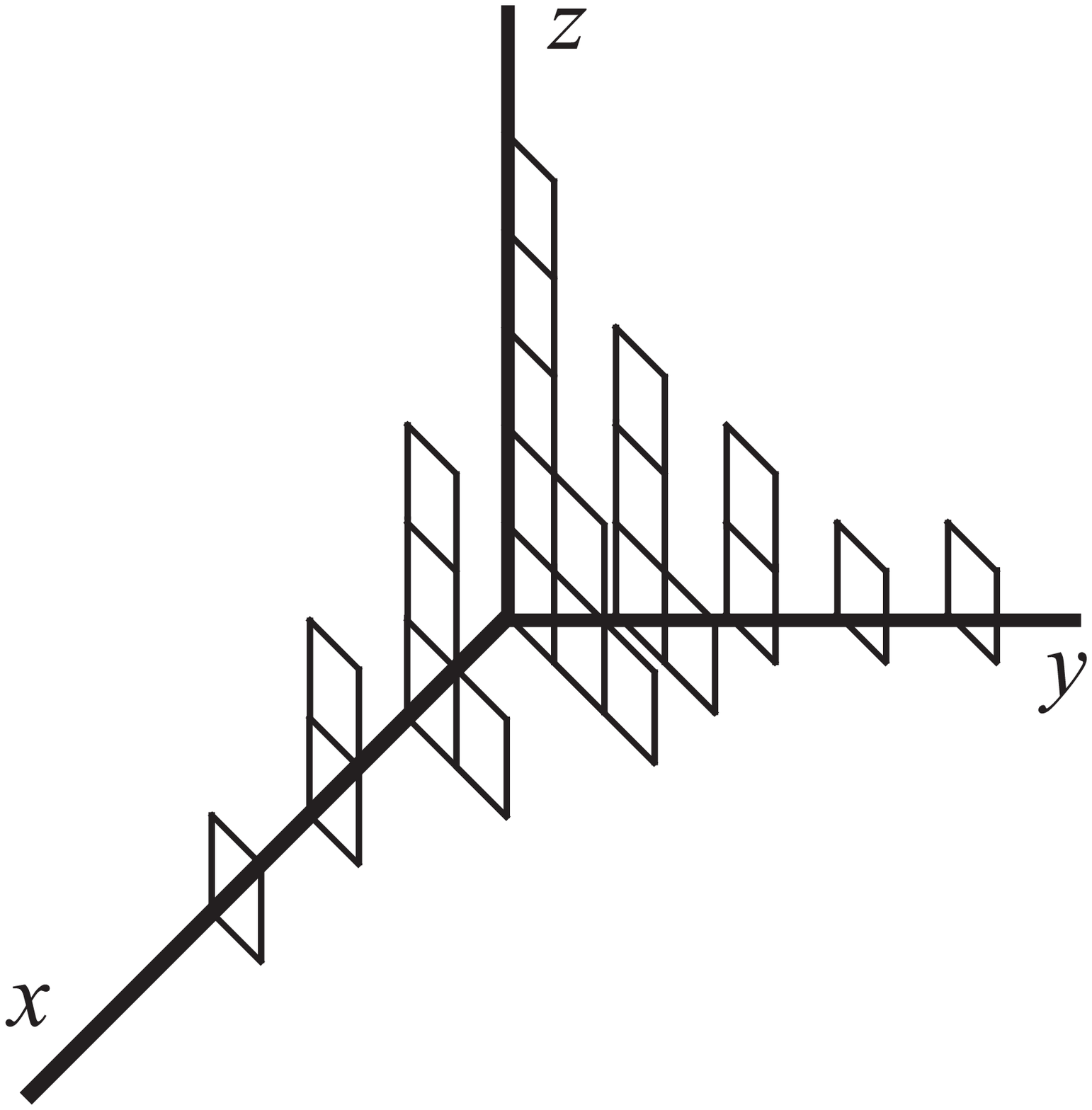}
\caption{Plane partition (left) and its presentation in terms of interlacing Young diagrams (right). This plane partition represents a particular configuration of a Calabi-Yau crystal which initially (before melting) entirely filled a positive octant of $\Z^3$. It is understood all boxes (atoms) this plane partition consists of have melted away, and remaining non-melted atoms (not drawn) remain in their original positions. As discussed in the text, a unit distance along each direction is measured in $g_s$ units from the topological strings point of view.} \label{fig-crystal3d}
\end{center}
\end{figure}

\section{Plane partitions}

We wish to consider a simple statistical models of crystal melting, in which it is assumed that the crystal consists of atoms represented by unit boxes located in integer points of a three-dimensional lattice. Usually this lattice is some subset of $\Z^3$, and melting is localised only in a neighbourhood of crystal corners. This lattice can be thought of as a container initially filled entirely with atoms, which depart from the crystal in the process of melting. A given melted configuration of the crystal can be represented by the so-called \emph{plane partition}, which represents atoms already melted, so that their original positions in the crystal are vacant. An example of a plane partition is shown in figure \ref{fig-crystal3d}. Intuitively, plane partitions can be thought of as maximally packed configurations of unit boxes. Below we introduce their proper definition and discuss most important properties.

Let us remark first that \emph{plane partitions} are also called \emph{three-dimensional partitions} and they are intimately related to two-dimensional partitions or Young diagrams, which on the other hand are in one-to-one correspondence with states of free fermions in two dimensions. Young diagrams and free fermions are introduced in appendix \ref{sec-partitions}, where also their properties, which we use to large extent in what follows, are discussed in detail. Below we often use the notation introduced in that appendix.

A plane partition $\pi$ of shape $R$ is defined as a map from a two-dimensional partition $R$ into positive integers $\pi:\ R \longrightarrow \Z_+ $
such that 
\be
\pi(m_1,n_1)\geq \pi(m_2,n_2) \quad \iff \quad (m_1\leq m_2\ \textrm{ and}\  n_1\leq n_2).  \label{condition-3d}
\ee
One can think of such a plane partition as a set of stacks of boxes of heights $\pi(m,n)$, such that $(m,n)\in R$ represents a box in the two-dimensional partition $R$ lying in $z=0$ plane. These stacks of boxes fill in a positive octant of $\Z^3$. This is an example of a plane partition of shape $R=(5,3,3,1)$ 
\be
\pi = \left( \begin{array}{ccccc} 5 & 3 & 2 & 1 & 1  \\
3 & 2 & 1 & & \\
2 & 1 & 1 & & \\
1 &   &   & & \\
\end{array} \right)    \label{example-3d}
\ee
presented also in the left picture in figure \ref{fig-crystal3d}. 

A plane partition can also be presented as a set of two-dimensional partitions $\pi(t)$ obtained from a diagonal slicing of $\Z^3$ by planes $y=x-t$ with $t\in\Z$. A definition of $x,y,z$ axes together with an example of such a slicing is shown in the right picture in figure \ref{fig-crystal3d}. In such a presentation the condition (\ref{condition-3d}) is equivalent to the fact that two-dimensional partitions $\pi(t)$ \emph{interlace} 
\be
\ldots \prec \pi(-3) \prec \pi(-2) \prec \pi(-1) \prec \pi(0) \succ \pi(1) \succ \pi(2) \succ \pi(3) \succ \ldots.   \label{interlace-3d}
\ee
The interlacing relation between two-dimensional partitions is defined in (\ref{interlace}).

In a statistical model each crystal configuration represented by a plane partition $\pi$ is assigned a statistical weight defined by the number of melted atoms, or the number of unit boxes which fill a positive octant of $\Z^3$. This number is denoted by $|\pi|$ and is given as
$$
|\pi| = \sum_{(m,n)\in R} \pi(m,n).= \sum_{t\in\Z} |\pi(t)|.
$$
Here $|\pi(t)|$ is a size of a two-dimensional partition  (\ref{R-size}) located at slice $t$. For example, the weight of the plane partition given in (\ref{example-3d}) and shown in figure \ref{fig-crystal3d} is equal to $|\pi| = 23$.

The partition function (also called the generating function) of the crystal model is defined as a sum over all possible configurations $\pi$ of the crystal, weighted by a number of melted atoms $|\pi|$
\be
Z=\sum_{\pi}q^{|\pi|},           \label{generating}
\ee
where $q=e^{-\mu/T}$ is related to a chemical potential $\mu$ and temperature $T$ from the statistical physics point of view. In general the crystal begins to melt in large temperature, and in $T\to\infty$ limit it approaches a smooth limit shape, which can be identified after rescaling the octant of $\Z^3$ in all directions by a factor $1/T$. Such limit shapes are related to the geometry of the mirror Calabi-Yau space and they were analysed in \cite{ok-re-va,dimer-mirror,schur-process,pearcey}. 

The partition function $Z$ can be found in a very elegant way in the so-called \emph{transfer matrix formalism}, which is based on the relation to two-dimensional partitions. To apply this formalism it is necessary to represent crystal as a string of two-dimensional partitions satisfying interlacing conditions (\ref{interlace-3d}). Then, by the relation to a free fermion system, all possible plane partitions can be built from the charge $p=0$ vacuum using a string of vertex operators $\Gamma_{\pm}$ introduced in appendix \ref{app-fermion}, as follows from their property (\ref{gamma-1-build}). To weight a contribution of each two-dimensional partition properly we have to insert the operator $q^{L_0}$ at each slice $t$ and use (\ref{L0}). Then the partition function can be computed using the commutation relation (\ref{commute}) and the property (\ref{gamma-z-1}) 
\bea
M(q) & = & q^{\sum_{t\in\Z} |\pi(t)|} = \langle0| \Big(\prod_{t=0}^{\infty} q^{L_0} \G_+(1)\Big) q^{L_0} \Big(\prod_{t=-1}^{-\infty} \G_-(1) q^{L_0}  \Big) |0\rangle = \nonumber \\
& = & \langle 0|\prod_{n>0}\Gamma_{+}(q^{n-1/2})\,\prod_{m>0}\Gamma_{-}(q^{-(m-1/2)})|0\rangle = \nonumber \\
& = & \prod_{n,m>0}\Big(1-\frac{q^{n-1/2}}{q^{-m+1/2}}\Big)^{-1}=\prod_{k>0}(1-q^{k})^{-k} = e^{\sum_{k>0} \frac{1}{k[k]^2}}. \label{McMahon}
\eea
The expectation value in the first line of (\ref{McMahon}) is called a \emph{transfer matrix}, and its evaluation is an example of the \emph{transfer matrix formalism}. We will use this formalism to large extent in further chapters, therefore the above derivation is a nice prerequisite to more difficult calculations. This particular partition function of plane partitions in a positive octant of $\Z^3$ is denoted $M(q)$ and is called the McMahon function to honour a mathematician who found it. In fact, McMahon derived also much more general generating function of plane partitions restricted to lie inside a cube of size $M\times L\times N$. It is given by the formula (\ref{Z1Z2}) and we discuss it in detail in the context of Calabi-Yau crystals in section \ref{sec-closed-ver}.

Above we have discussed the simplest Calabi-Yau crystal model. One can also introduce more complicated models, either by  gluing crystals together or by imposing some additional constraints. In the former case, one can prove a remarkable fact that a certain generating function of plane partitions in a positive octant of $\Z^3$ is closely related to the topological vertex amplitude $C_{R_1 R_2 R_3}$ (\ref{vertex-hopf}). Therefore the topological vertex gluing rules may be understood as a prescription for composing together such crystals, which in principle allows to interpret the whole class of toric manifolds discussed in section \ref{ssec-toric} in a crystal language. However, their gluing does not have any obvious three-dimensional interpretation.
 
In this thesis we mainly focus on the latter class of Calabi-Yau crystals, which arise from imposing some additional constraints in $\Z^3$ lattice. These constraints do have a straightforward three-dimensional interpretation, even though they are known only in a few particular cases and are not related explicitly to topological vertex gluing rules. In these models the entire crystal can be thought of as a container of a non-trivial shape, so that the class of allowed plane partitions is restricted, as they cannot extend beyond this container. 
%A particular shape of the container is related to the Calabi-Yau target space geometry, and 
Parameters which determine its shape correspond to K\"ahler moduli of the Calabi-Yau target space from the topological string point of view. 

In this thesis we show that generating functions of such models reproduce the partition functions of topological string theories on Calabi-Yau manifolds. In fact, Calabi-Yau crystals also encode much more interesting information. On one hand, Gopakumar-Vafa invariants are encoded in a very explicit way in crystal amplitudes. On the other hand, branes in topological theories can be easily interpreted in crystal models, and the connection between branes and Chern-Simons theory leads to a correspondence between crystals and knot invariants. We discuss all these intricacies in what follows.

       % shorter version and much better version

% \input{crystal-intro-old.tex}          % this is longer version...
% \input{partition-3d.tex}           % ...together with this

\section{Quantum Calabi-Yau manifolds}

In this section we explain the basic relation between A-model topological string theory on toric geometries and statistical models. Let us start with a very interesting observation: the generating function of plane partitions in the positive octant of $\Z^3$ given by the McMahon function (\ref{McMahon}) is equal to the topological string partition function $Z_{top}^{\C^3}$ for the target space geometry of $\C^3$ given in (\ref{Z-c3})
\be
Z_{top}^{\C^3} = M(q) = \sum_{\pi} q^{|\pi|}.     \label{quantum-cy}
\ee

How can this equality be explained? Let us first note that A-model topological string amplitudes (\ref{Z-top})-(\ref{grom-wit}) can be interpreted as amplitudes for a \emph{quantum} Calabi-Yau manifold $M$, with the following understanding of the \emph{quantum geometry} term. The \emph{classical} part of the total amplitude can be identified with contributions in $\beta=0\in H_2(M)$ class which are constant maps sending entire string worldsheet into a single point in the Calabi-Yau space. Apart from powers of McMahon function, their genus zero part encodes classical triple intersection numbers of Calabi-Yau in the cubic couplings. Then the \emph{quantum} piece of the total amplitude is represented by worldsheet instantons terms characterised by $\beta\neq 0$. This quantum point of view can be made precise in mathematical terms, where classical intersection numbers define the classical cohomology ring of the manifold, which can be deformed into the so-called \emph{quantum cohomology ring} when worldsheet instantons are taken into account.

Even though we do not analyse quantum cohomology rings in any more detail, we wish to take the idea of quantum geometry seriously. In this respect we propose now how $\C^3$ Calabi-Yau manifold can be quantised explicitly, and how it leads to the crystal picture. We recall $\C^3$ with canonical coordinates $z_i$ is a toric geometry, and in section \ref{ssec-toric} its structure as $T^2\times\C$ fibration over $\R^3$ base has been revealed with respect to the symplectic form $\omega$ (\ref{c3-sympl}). However, this symplectic form can also be written as
\be
\omega=\sum_{j=1}^3 dz_j\wedge d\bar{z}_j = \sum_{j=1}^3 d|z_j|^2 \wedge d\theta_j = \sum_{j=1}^3 d p_j \wedge d\theta_j,
\ee
which provides another torus fibration structure of $\C^3$: now $(x,y,z) = (p_1,p_2,p_3) = (|z_1|^2, |z_2|^2, |z_3|^2)$ parametrise the positive octant of $\R^3$ which becomes a base of this fibration, and $\theta_i$ are coordinates in $T^3$ fibres. We can treat $\theta_i$ as positions and $p_i$ as momenta, and we impose the canonical quantisation condition with $g_s$ playing a role of the Planck constant $\hbar$
\be
[\theta_j, p_k ] = ig_s\, \delta_{jk}.   \label{quantize-crystal}
\ee
A basis for the Hilbert space can be taken as 
$$
\psi_{n_1,n_2,n_3}(\theta_k) = \exp(i\sum_{j=1}^3 n_j \theta_j)
$$
with $0\leq n_j\in \Z$. $n_j$ cannot be negative as the momentum $p_j=|z_j|^2$ is necessarily non-negative. Therefore these states can be identified with boxes in the positive octant of the three-dimensional lattice at integer positions $(n_1,n_2,n_3)\in\Z^3$, where the unit distance is determined by the quantum parameter $g_s$. This quantum Calabi-Yau is frozen in the temperature $T=0$ and correspondingly the positive octant of $\Z^3$ is completely filled, which encodes the classical geometry of $\C^3$. However, in non-zero temperatures the manifold experiences quantum fluctuations, which is manifested by allowing configurations described by plane partitions to melt away from the corner of the positive $\Z^3$ octant. Remarkably, the statistical sum over all melted configurations reproduces the topological string partition function of $\C^3$.

\section{Relation to the topological vertex}    \label{sec-crys-topver}    

We have shown the topological string partition function for $\C^3$ is equal to the  McMahon function, which is a generating function of plane partitions in the positive octant of $\Z^3$. For a nontrivial manifold made of a few $\C^3$ patches, we might assume each patch is represented by an ensemble of plane partitions which altogether give rise to some power of McMahon function. This would correspond to contributions from constant maps for genus $g>1$ in (\ref{Z-top}). However, there are also contributions from worldsheet instantons associated with Gromov-Witten invariants, which give rise to the second part of the topological string partition function (\ref{Z-top}). Are these worldsheet instanton contributions related to some crystal model as well? Remarkably, the answer is \emph{yes}, at least for the class of toric manifolds discussed in section (\ref{ssec-toric}). These worldsheet instanton contributions for toric manifolds arise from gluing topological vertex amplitudes $C_{R_1 R_2 R_3}$ together. Therefore relating the topological vertex to some crystal model would provide a relation between worldsheet instantons and crystals as well.

We now prove the vertex amplitude $C_{R_1 R_2 R_3}$ is closely related to the generating function of a special class of plane partitions --- those which asymptote to Young diagrams $R_1,R_2,R_3$ along three axes of $\Z^3$. In other words, these plane partitions fill a special container, which is a positive octant of $\Z^3$ with three infinite ``cylinders'' with bases $R_1,R_2,R_3$ removed along its three axes. Let us denote the generating function of these partitions as $P_{R_1,R_2,R_3}$. %To compute it, it is convenient to consider a container which is finite in two directions with lengths $M,L\in\Z$, so that the basis of this former container is $([0,M] \times [0,L]) / R$. Denoting a generating function of plane partitions in such a container as $P^{M,L}_{R_1,R_2,R_3}$ we have
%$$
%P_{P,Q,R} = \lim_{M,L\to\infty} P^{M,L}_{P,Q,R},.
%$$
%and $P^{M,L}_{P,Q,R}$ can be computed in the transfer matrix formalism as 
This can be found using the transfer matrix formalism
$$
P^{M,L}_{R_1,R_2,R_3} = q^{-{R_1\choose 2} -{R_2^t \choose 2}} \langle R_1^t| \Big(\prod q^{L_0} \G_{\pm}(1)\Big) q^{L_0} \Big(\prod \G_{\pm}(1) q^{L_0}  \Big) |R_2\rangle,
$$
where infinite number of $\G_{\pm}$ operators is acting on both vacua, and their order is determined by the shape of the partition $R$. The prefactor in this expression arises from the fact that in the transfer matrix formalism we slice $\Z^3$ diagonally, which increases the volume we are interested in by ${R_1 \choose 2}$ at the end of the cylinder along $x$ axis, and analogously by ${R_2 \choose 2}$ for the cylinder along $y$ axis. Using now (\ref{gamma-skew}) and commutation relations (\ref{commute}) the above correlator is evaluated as 
\be
P_{R_1,R_2,R_3} = N(R)\, q^{-{R_1\choose 2} -{R_2^t \choose 2} - |R_1|/2 - |R_2|/2} \sum_{\mu} s_{R_1^t/\mu}(q^{- R_3-\rho}) s_{R_2/\mu}(q^{- R_3^t-\rho}),   \label{P-PQR}
\ee
where $N(R)$ is a factor which depends only on the shape of $R$. It can be derived using an obvious cyclic symmetry
$$
P_{\bullet,\bullet,R_3} = P_{R_3,\bullet,\bullet},
$$
and applying $\ref{P-PQR}$ for both sides of this equality. This implies
$$
N(R_3) = M(q) q^{-{R_3\choose 2} - |R_3|/2} s_{R_3^t}(q^{-\rho}),
$$
and using properties (\ref{young-choose}) and (\ref{kappa}) we finally get
\be
P_{R_1,R_2,R_3}(q^{crystal}) = q^{-\frac{||R_1^t||^2 + ||R_2^t||^2 + ||R_3^t||^2}{2}}M(q) C_{R_1 R_2 R_3}(q),   \label{Pcrys-Cver}
\ee
where we introduced
\be
q^{crystal} = q^{-1}.   \label{q-crystal-vertex}
\ee
This relation between crystal and topological vertex parameters will also appear in other crystal models. 

To summarise, we have shown the generating function of plane partitions which asymptote to Young diagrams $R_1,R_2,R_3$ is equal to the topological vertex amplitude up to a normalisation to $M(q)$ (which counts all plane partitions), inversion of $q$ and a prefactor which is a simple power of $q$. 

\section{Defects and branes}         \label{sec-crys-brane}

$D$-branes are intrinsic ingredients of string theory. We recall in A-model theory they have to wrap lagrangian submanifolds of real dimension 3. In the class of toric geometries we discussed before, an interesting set of such submanifolds is given by $T^2$ fibred over a half-line in the toric base ending on some edge of a toric diagram. We would like to understand whether these branes can somehow be interpreted in the crystal picture. As we already know which crystal model corresponds to topological strings on $\C^3$, we will first try to find a crystal description of branes in $\C^3$. 

Lagrangian submanifolds $\C^3$ which can be wrapped by these branes are given in (\ref{c3-lag-brane}). Let us focus on a brane $L_3$ which ends on the axis denoted by $y$. We can rewrite the equation of this brane in the present notation, and using (\ref{c3-lag-brane}) and Hamiltonians (\ref{c3-r-toric}) it reads
\be
y = x + u = z + u, \qquad  \theta_1+\theta_2+\theta_3=0,  \qquad u= g_s (N+ \hf) > 0.   \label{brane-crystal}
\ee
A position of this brane along $y$ axis is specified by the modulus $u$ which is quantised in $g_s$ units, according to the quantisation condition related to the crystal picture (\ref{quantize-crystal}). Additional $g_s/2$ is inserted for convenience and it ensures compatibility with topological string results. 

\begin{figure}
\begin{center}
\includegraphics[width=0.5\columnwidth]{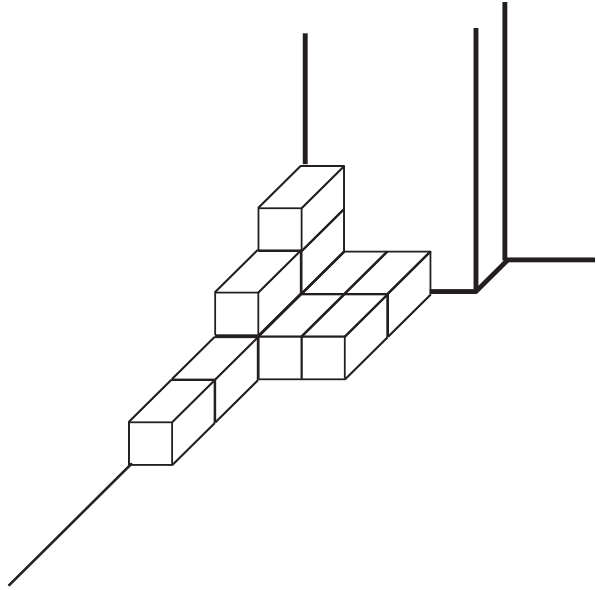}
\end{center}
\caption{Lagrangian brane in $\C^3$ is represented by insertion of $\Psi_{D}(q^{-N-\hf}) = \Gm^{-1}(q^{-N-\hf}) \Gp(q^{-N-\hf})$ operator in the crystal picture, which produces a defect --- an additional corner at position $Ng_s$ --- in the crystal configuration.}  \label{fig-brane} 
\end{figure}

It is known that a presence of an A-brane affects the geometry of the underlying manifold by changing periods of the K\"ahler form $k$ by 
\be
\Delta\int_{\Sigma} k = g_s,      \label{backreaction}
\ee
where $\Sigma$ is a two-cycle linking the worldvolume of the brane. This effect is also known as the \emph{backreaction of geometry on brane insertion}. For a brane defined by (\ref{brane-crystal}) this two-cycle can be chosen as a difference $\Sigma=\hat{\sigma}_1-\hat{\sigma}_2$, where $\hat{\sigma}_1, \hat{\sigma}_2$ are planes given by $\theta_1$ cycle fibred over the following half-lines in the toric $\R^3$ base
$$
\sigma_1=(x,y_1 +x,z_*),\qquad \sigma_2=(x,y_2 +x, z_*)
$$
which are parametrised by $x$, whereas $y_1,y_2,z_* = const$. Then the period of the K\"ahler form along such $\Sigma$ is equal to
$$
\Delta\int_{\Sigma} k = (|\sigma_2|-|\sigma_1|)\, g_s, 
$$
where $|\sigma_i|$ is the number of integer points (crystal atoms) on the line $\sigma_i$. Therefore, in the presence of a single brane at position $a$ on $y$ axis, numbers of atoms on these parallel lines must differ by $1$, which means there is an additional corner (or a jump, or a defect) in the crystal structure at position $u$, as shown in figure \ref{fig-brane}. 

To get convinced more quantitatively these defects in crystals indeed represent branes, we will show now the generating function for $\C^3$ crystal with such a defect is equal to the A-model topological string partition function in the presence of an appropriate brane. It is not difficult to represent such a crystal in the transfer matrix formalism. We note a shape of pure $\C^3$ crystal is specified just by to infinite strings of $\G_+$ and $\G_-$ operators in (\ref{McMahon}) along $x$ and $y$ axes. Each $\G_+$ or $\G_-$ operator corresponds respectively to a unit step of length $g_s$ in $x$ or $y$ direction. To introduce an additional corner along $y$ axis we have to add one $\G_+$ operator at position $N+1$ slice along $y$ axis, and at the same time to cancel $\G_-$ operator which is initially present in this position in (\ref{McMahon}). This can be represented by an operator
\be
\Psi_{D}(a^{-1}) = \Gm^{-1}(a^{-1}) \Gp(a^{-1}), \qquad \qquad \textrm{for} \quad  a=e^{-u}=e^{-(N+\hf)g_s}=q^{N+\hf}.      \label{A-brane}
\ee
Comparing with (\ref{boson}), (\ref{fermion}) and (\ref{gamma}) we recover this operator is just a fermion written in the bosonised language, with zero-mode removed. For this reason we call it $\Psi_{D}$ and subscript $D$ means it corresponds to a $D$-brane. 

At this point we find another interesting analogy. As will be explained in section \ref{sec-B-model}, under Mirror Symmetry A-branes in toric geometry are mapped to B-branes represented as free fermions on a Riemann surface (\ref{B-brane}). Now we have found a realisation of A-model lagrangian brane as a fermion in the transfer matrix formalism (\ref{A-brane}). B-model topological string partition function with brane can be reproduced as the expectation value of the B-model fermion (\ref{B-amplitude}). Let us demonstrate an analogous statement is also true in A-model. Inserting $\Psi_{D}(a^{-1})$ in the $(N+1)$'th slice along $y$ axis we get
\begin{eqnarray}
 \langle \Psi_{D}(q^{-N-1/2}) \rangle &=& \langle 0|\prod_{n=1}^{\infty}\Gamma_{+}(q^{n-1/2})\ \prod_{m=1}^{N+1}\Gamma_{-}(q^{-m+1/2})\ \Psi_{D}(q^{-N-1/2}) \nonumber\\
& & \prod_{m=N+2}^{\infty} \Gamma_{-}(q^{-m+1/2})|0\rangle = \frac{M(q)}{\varphi(q)} L(a,q), \nonumber
\end{eqnarray}
where $\varphi(q)$ is Euler function (\ref{euler-fn}) and $L(a,q)$ denotes the quantum dilogarithm
\be
L(a,q) = \prod_{n=1}^{\infty} ( 1 - q^{n + N}) = \sum_{n=0}^{\infty} a^n h_n(q^{\rho}), \qquad \textrm{for} \quad  a=q^{N+\hf}, \label{def-dilog}
\ee
which we expressed in terms of complete symmetric polynomials $h_n$ defined in appendix \ref{app-schur}. This indeed agrees with topological string amplitudes for $\C^3$ with a single brane inserted \cite{B-vertex}
$$
Z^{\C^3}_{D}(q) = M(q)L(a,q) = \varphi(q)\, \langle \Psi_{D}(q^{-N-1/2}) \rangle 
$$
up to a factor of $\varphi(q)$. It can be shown this factor introduces non-perturbative change in the topological string free energy in $g_s\to 0$ limit, so it cannot be detected in perturbative topological string expansion \cite{va-sa}. Therefore it does not spoil the prediction of crystal computation, but rather indicates what the non-perturbative string amplitude might be. 

We can also compute the expectation value for a brane inserted at position $N+1$ along $x$ axis
$$
\langle \Psi_{D}(q^{N+1/2}) \rangle = \langle 0|\prod_{n=-\infty}^{-N-1}\Gamma_{+}(q^{-n+1/2})\,\Big[\,\Gamma_{-}^{-1}(q^{N+1/2})\Gamma_{+}(q^{N+1/2})\,\Big]
$$
\be
 \cdot \prod_{n=-N}^{0}\Gamma_{+}(q^{-n+1/2})\,\prod_{m=1}^{\infty}\Gamma_{-}(q^{-(m-1/2)})|0\rangle = \frac{\varphi(q)\,M(q)}{L\Big(q^{(N +\frac{1}{2})g_s},q\Big)}, \label{c3-brane-x}
\ee
which also agrees with topological string amplitude for a single B-brane, albeit in framing $-1$. 

To sum up, a brane ending on the $y$ axis at distance $u=g_s(N+\hf)$ is described inserting a fermionic operator $\Psi_{D,y}(a^{-1})$ at the $(N+1)$'th slice for $a=e^{-u}=q^{N+\hf}$. Similarly, a brane at distance $v=g_s(N+\hf)$ on the $x$-axis is described by  $\Psi_{D, x}(b)$ at the negative side of the diagonal at slice $-(N+1)$, where $b=e^{-v}=q^{N+\hf}$.

\begin{figure}
\begin{center}
\includegraphics[width=0.45\columnwidth]{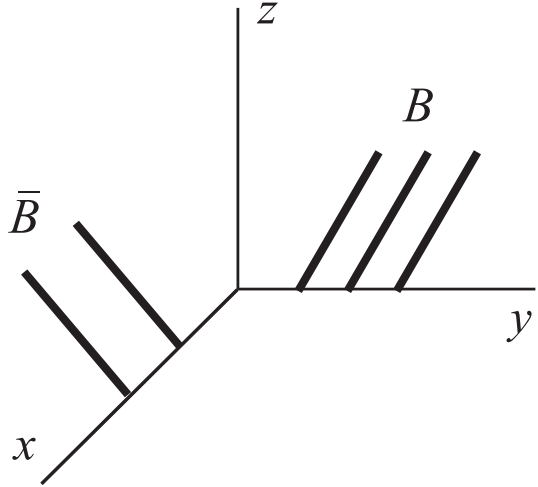}
\end{center}
\caption{\label{branesc3} Three operators representing branes inserted on the positive slice and two representing anti-branes inserted in the negative slice of the ${\C}^3$ crystal.}
\end{figure}

At this point it is simple to realise what operator corresponds to anti-branes --- this is inverse operator to (\ref{A-brane}), which reads
\be
\Psi_{\bar{D}}(a^{-1})=\Gamma_{-}(a^{-1})\Gamma^{-1}_{+}(a^{-1}), \label{anti-brane}
\ee
and when inserted on $y$ axis its gives expectation value equal to (\ref{c3-brane-x}). A sample configuration involving fermionic insertions in $\C^3$ crystal corresponding to branes and anti-branes in $\C^3$ geometry is shown in figure \ref{branesc3}.

Finally we recall topological string amplitudes for branes are defined with respect to the framing factor $p\in\Z$ (\ref{frame-brane}). It is possible to define also crystal framing factor $p_{crystal}$, so that crystal results could be modified to agree with topological string results in arbitrary framing. However, usually the results obtained in the transfer matrix formalism are given in $p_{crystal}=0$ framing. We do not introduce arbitrary crystal framing --- when it is necessary, we equivalently change topological brane framing to compare with crystal results with $p_{crystal}=0$.

In the next chapter we will show in much more general situations that generating functions for plane partitions in containers with various defects agree with corresponding topological string partition functions with lagrangian branes.

%***********************************************************
%***********************************************************

\chapter{Three-dimensional Calabi-Yau crystals --- intricacies}     \label{chap-cy-results}

In chapter \ref{chap-cy-crystals} we introduced basics of Calabi-Yau crystal models. In this chapter we analyse Calabi-Yau crystals in much more detail, following results published in \cite{ps} and \cite{ps-cube}. In section \ref{crystal} we still consider a model for $\C^3$ geometry with arbitrary number of defects representing branes. In sections \ref{sec-conifold}, \ref{sec-CY-2walls} and \ref{sec-closed-ver} we introduce a class of models with restricted containers, with additional bounds imposed respectively by one, two and three walls. It turns out that in the corresponding Calabi-Yau geometry each such wall corresponds to an additional non-trivial $\mathbb{P}^1$, represented by a compact interval in a toric diagram.

Moreover, in sections \ref{crystal} and \ref{sec-conifold} we recover explicit relation between crystal amplitudes and knot invariants. Before we proceed, let us comment on this relation. In (\ref{Pcrys-Cver}) we have shown that a certain crystal partition function is related in a simple way to the topological vertex amplitude $C_{R_1 R_2 R_3}$ and is expressed in terms of Hopf-link invariants (\ref{vertex-hopf}) for a particular set of representations $R_1,R_2,R_3$. Hopf-link invariants are associated to lagrangian branes on the string theory side and arise from a relation to Chern-Simons theory, in which they are expectation values of Wilson loops. However, partition functions in a presence of branes are generating functions of knot invariants (\ref{Z-branes-knot}), but not expectation values for some particular set of representations. Therefore we expect that partition functions for crystals with defects should provide generating functions of knots invariants as well. This is indeed the case, as we show in the next two sections.

We have to stress that no general method is known how to determine a particular shape of a crystal container corresponding to a given Calabi-Yau manifold. For this reason, while computing various crystal amplitudes in this chapter, \emph{a priori} we do not know if they agree with topological string amplitudes for some geometries. In the next chapter we prove, using the topological vertex techniques, that crystal generating functions found in this chapter are indeed equal to the A-model topological string partition functions on appropriate geometries.

\section{Knot invariants from the crystal} \label{crystal}

In this section we continue our analysis of $\C^3$ crystal model for more general brane configurations. Let us insert $m$ Lagrangian branes at distances $g_s (N_i + \hf)$ for $i=1 \ldots m$ on the $y$ axis, and $n$ Lagrangian anti-branes on $x$ axis at distances $g_s(M_j + \hf)$ for $j=1 \ldots n$. Crystal partition function for such a configuration reads
$$
Z(a_1, \ldots a_n; b_1 \dots b_m; q) 
= \left\langle \, \Psi_{{\bar D}, x} (b_1) \ldots \Psi_{{\bar D}, x} (b_n) \,
\Psi_{D,y} (a_1^{-1}) \ldots \Psi_{D,y} (a_m^{-1})\,  \right\rangle
$$
\be
= M(q) \left(  \prod_{i=1}^m \prod_{j=1}^n {L(a_i, q)   L(b_j, q) \over  (1 - a_i b_j)} \right) 
\left(\prod_{i>j} (1- a_i /  a_j) 
 (1- {b_i /  b_j}) \right). \label{nmbranes}
\ee
This result has a simple and nice interpretation. Each brane or anti-brane in the system is represented by the quantum dilogarithm (\ref{def-dilog}), whereas factors of type $(1-a_i /a_j)$
and $(1- a_i b_j)$ correspond to strings stretched between branes. In paragraph \ref{ssec-2axes} we will explicitly see how these factors arise in the topological vertex computation. 

Of course we could as well insert branes on $x$ axis, which would be related by a change of framing to the situation we consider. In this and other results with branes involved, we often omit inessential factors of $\varphi(q)$. In the following we will reinterpret the above expression as a generating function of certain knot invariants in arbitrary representations. %As $\C^3$ can be thought of as a limit of the conifold when its K\"ahler parameter $t =g_s N\rightarrow \infty$, by geometric transition we expect to see the leading part of knot invariants.We are then probing the invariants of $U(\infty)$ Chern-Simons theory.  More precisely, from the geometric picture of the Lagrangian branes with topology $S^1 \times \R^2$ we expect to find unknot and Hopf-link invariants.
%As the crystal result is written entirely
%in terms of dilogarithms and simple prefactors from the stretched strings, it is not immediately
%obvious that these would provide the generating functions for more complicated
%link invariants, for example for Hopf-link invariants in tensor product representations. 
%It will turn out that the simplicity of crystal results is partly due to a particularly
%natural framing choice.

\subsection{Single unknot}

Consider first a single brane on the $y$ axis. In this case, we have
$$
Z(a, q) = M(q) L(a, q).
$$
It is indeed the leading part of the generating function 
for unknot invariants normalised by $M(q)$, as follows from the expansion  
\bea
L(a, q) &=& e^{\sum_{n=1}^{\infty} {a^n \over n [n]}} = 1 + {a \over 
(q^{\hf} - q^{-\hf}) } + a^2 {q^2 \over (q^2-1) (q-1)} + \ldots \nn 
&=& \sum_{R-one\ row} W_{R\bullet} a^{|R|},
\eea
where $[n] = q^{n/2} - q^{-n/2}$ and $W_{R\bullet}$ are leading terms of unknot invariants (\ref{hopf-limit}). The summation runs over \emph{one row representations} only, i.e. such representations whose Young diagrams consist of only one row of boxes, e.g. $\square$, $\square\!\square$, etc. %Therefore the partition function for a single brane inserted in the $\C^3$ crystal is equal to the generating function of the leading part of unknot invariants in one row representations. 

\subsection{Hopf-link}

Inserting an antibrane on $x$-axis and a brane on $y$-axis gives the generating
function of Hopf-link invariants in single row representations.
Expanding the normalised part of the partition function
\be
{\tilde Z}(a, b, q) = {Z(a ,b ,q) \over M(q)} = 
{L(a,q) L(b,q) \over (1- a b)} \label{c3-antiDx-Dy}
\ee
and comparing with (\ref{hopf-limit}) gives
\be
\tz(a, b, q) =  \sum_{R, P\,-\,one\ row}  q^{{\kappa_{R} + \kappa_{P} \over 2}} W_{R^t P^t} a^{|R|} b^{|P|} \label{c3-antiDx-Dy-HL}
\ee
where $\kappa_R$ is given in (\ref{kappa}) and the summation again runs only over one row representations.  In section \ref{sec-c3-amplitudes} we will prove this expansion agrees with the topological vertex calculation, and from that point of-view the $q$-dependent prefactors will be interpreted as vertex framing factors $(-1, 0)$, which represent a brane in framing $-1$ and an antibrane in framing $0$. 
Alternatively, when expressed in terms of $q^{-1}$ this expansion gives Hopf-link coefficients with a knot framing $(-1,-1)$.

%\subsection{Hopf-link in representations with many rows} 

In the general case, for $n$ branes on the $y$-axis and $m$ antibranes on the $x$-axis 
the normalised partition function (\ref{nmbranes}) generates the leading part of Hopf-link 
coefficients with $(n, m)$ rows
\be
\tz(a_1, \ldots a_n; b_1, \ldots b_n;q) =  \sum_{R_1, \ldots R_n} \sum_{P_1, \ldots P_m}  q^{{\kappa_{R} + \kappa_{P} \over 2}} W_{P^t R^t} a_1^{|R_1|} \ldots a_n^{|R_n|} b_1^{|P_1|} \ldots b_m^{|P_m|}  \label{claim-rhs}
 \ee
%\bea
%&& \tz(a_1, \ldots a_n; b_1, \ldots b_n;q) = \nn 
% && \sum_{R_1, \ldots R_n} \sum_{P_1, \ldots P_m}  
%q^{{\kappa_{R} + \kappa_{P} \over 2}} W_{P^t R^t} a_1^{|R_1|} \ldots a_n^{|R_n|} 
%b_1^{|P_1|} \ldots b_m^{|P_m|}  \label{claim-rhs}
% \eea
 where $R=(R_n, \ldots, R_1)$ and $P=(P_m,\ldots P_1)$ are $n$ and $m$ row representations
 respectively. Apart from ordinary Young diagrams, the last summation contains also a finite number of \emph{improper} Young diagrams, which we understand as diagrams $(R_n, \ldots, R_1)$ for which the condition  $R_1 \leq R_2 \ldots  \leq  R_n\leq 0$ is not satisfied. Nonetheless, these improper contributions can still be formally manipulated using definitions of Schur functions and coefficients $\kappa_R$.  In the next section contributions from improper partitions are analysed and a proof of the formula \ref{claim-rhs} is given in one simplified case.

In this section we discussed one sense in which crystal partition functions are generating functions of Hopf-link invariants in arbitrary representations. It turns out the same crystal partition functions for configurations of several branes inserted on each axis can also be interpreted as 
generating functions for more complicated Hopf-link invariants, corresponding to tensor products of one-row representations. This latter interpretation will become clear in section \ref{sec-A-model}, where we discuss brane configurations from the topological vertex point of view. 

%***********************************************************

%************************************************************************
%************************************************************************

\subsection{Proof of the Hopf-link expansion} \label{induct}

In this paragraph we prove the formula (\ref{claim-rhs}) for the Hopf-link expansion in representations with many rows, in the simplified case of $n$ branes only on the positive slice at positions $a_1,\ldots, a_n$.  As follows from (\ref{nmbranes}), the normalised crystal partition function in this case is
\begin{eqnarray}
\tz(a_1,\ldots,a_n) &=&  L(a_1,q)\ldots L(a_n,q) \cdot \label{crys-branes} \\
& & \cdot
\big(1-\frac{a_1}{a_2}\big)\big(1-\frac{a_1}{a_3}\big)\ldots\big(1-\frac{a_1}{a_n}\big)\ldots\big(1-\frac{a_{n-1}}{a_n}\big),
\nonumber
\end{eqnarray}
and in consequence the statement (\ref{claim-rhs}) takes the form
\begin{equation}
\tz(a_1,\ldots,a_n)=\sum_{R_1,\ldots,R_n} a_1^{R_1}\ldots a_n^{R_n} \
s_{(R_n,R_{n-1},\ldots,R_1)}(q^{\rho}). \label{claim-one-leg}
\end{equation}

As mentioned above, this expansion contains Schur functions corresponding also to \emph{improper} partitions, either with a negative number of boxes or not decreasing in length. Nonetheless,  due to the very structure of Schur functions encoded in their definition in terms of determinants, these improper partitions are taken into account in the proof below automatically.

We prove (\ref{claim-one-leg}) by induction on the number of branes $n$. The first step in
this induction is the expression for the dilogarithm 
\begin{equation}
L(a,q)=\sum_{R-one\ row} a^{|R|} h_R(q^{\rho}), \label{L-h}
\end{equation}
where in the case of a single variable $a$ the sum automatically runs only over one-row partitions $R$, and then Schur functions reduce to complete symmetric functions $s_{(R,0,\ldots)}=h_{|R|}$.

In the second induction step, let us assume that
$\tz (a_1,\ldots,a_n)$ is given by (\ref{claim-one-leg}), and we add one more
brane at $a_0$. Then
\be
\tz(a_1,\ldots,a_n,a_{0}) = \tz(a_1,\ldots,a_n)L(a_{0},q)  \cdot \big(1-\frac{a_1}{a_{0}}\big)\ldots \big(1-\frac{a_n}{a_{0}}\big).
\label{C3-leg-recursive}
\ee
If we expand this expression with respect to $a_i$'s and use (\ref{claim-one-leg}) and (\ref{L-h}), the
coefficient at $a_0^{R_0}\cdots a_{n}^{R_{n}}$ is equal to 
\begin{equation}
h_{R_{0}}s_{(R_m,\ldots,R_1)} - h_{R_{0}+1}\sum_{i=1}^{m}s_{(\hat{i})} +
h_{R_{0}+2}\sum_{i\neq j}^{m}s_{(\hat{i},\hat{j})} - \ldots h_{R_{0}+n}s_{(R_n
-1,\ldots, R_1 -1)}, \label{proof-sum-1}
\end{equation}
where we skipped arguments of symmetric functions $(q^{\rho})$ temporarily; $\hat{i}$ means that $i$'th variable $R_i$ is replaced by $(R_i -1)$, for example
\begin{equation}
s_{(\hat{i},\hat{j})} = s_{(R_n,R_{n-1},\ldots,R_{i}-1,\ldots,R_{j}-1,\ldots,R_1)}.
\end{equation}
In the first term in this expression no variable is reduced by 1, and in the last
term all $m$ variables are reduced. In other terms several variables are reduced,
and the sums are over all possible combinations of choosing this number of variables
from the set $(R_1,\ldots,R_n)$.

The final observation is that (\ref{proof-sum-1}) is the Laplace expansion of the
determinant defining $s_{(R_0,R_n,\ldots,R_1)}$ along the first row (\ref{s-det})
\begin{equation}
s_{(R_0,R_n,\ldots,R_1)} = \textrm{det}\,
\left[ \begin{array}{cccc}
h_{R_0} & h_{R_{0}+1} & \ldots & h_{R_{0}+n} \\
h_{R_{n}-1} & h_{R_{n}} & \ldots & h_{R_{n}+n-1} \\
\vdots & \vdots & \ddots & \vdots \\
h_{R_{1}-m} & h_{R_{1}-n+1} & \ldots  & h_{R_{1}}
\end{array} \right]  \label{sKm-K1}
\end{equation}
 This completes the induction and proves (\ref{claim-one-leg}). In the more general case for branes on both axes an analogous proof can be constructed.

\subsubsection{Improper partitions}

It should be stressed that in the above expansion not all $W_{PR}$ correspond
 to Hopf-link invariants. This is so only
in the case when $P$ and $R$ are proper partitions, i.e. if $R_n \geq R_{n-1} \geq\ldots
\geq R_1 \geq 0$, and similarly for the representation $P$. Otherwise $W_{PR}$ are just
coefficients resulting from the expansion, but generally these cannot be thought of as Hopf
link invariants. Nonetheless, functions $s_P$ involved in $W_{PR}$ are still given
by the determinant $(\ref{s-det})$, and thus we will call them \emph{improper Schur
functions}.

Moreover, the summations over $R_i$ and $P_i$ in (\ref{claim-rhs})
 do not start from 0, because in the crystal partition function expansions there 
are also terms with negative powers of $a_i,\ b_i$. These negative powers arise only from
prefactors for strings stretched between branes on the same axis, which are of the
form $(1-a_i/a_j)$, and there is always a finite number of such terms.

In fact, the easiest way to take care of them is to understand the summations in
(\ref{claim-rhs}) as running over all integers, positive and
negative. The very structure of Schur functions, together with the fact that
$h_i=0$ for $i<0$, will assure that only relevant terms will be non-zero and we get
the correct result. In particular, this means that there will be partitions $R$ with
'negative number of boxes' in some rows, $R_i<0$.  So if we expand determinant
(\ref{s-det}) for the corresponding \emph{improper Schur functions} $s_R$, and use
$h_{i<0}=0$, we are left with Schur functions for partitions with
lower number of rows, now only of positive length. These new functions can also be
proper or not, according to whether lengths of their rows are properly decreasing.

Thus, if we put $n$ branes on one leg, the crystal expansion in fact contains
information about all proper knot invariants, and finite number of improper knot invariants 
for partitions with all number of rows $1,\ldots,n$.

%***********************************************************
%***********************************************************

\section{Crystal model for the conifold} \label{sec-conifold}

\begin{figure}
\begin{center} 
\includegraphics[width=0.45\columnwidth]{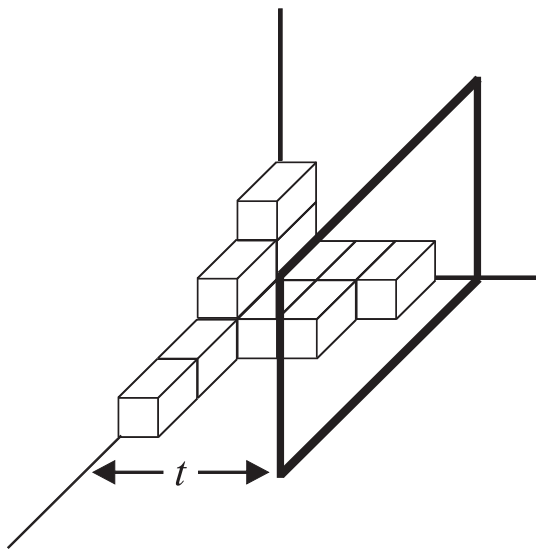}
\end{center}
\caption{\label{wall1} The toric geometry of the conifold crystal ending in a wall at the 
$y$ axes at distance corresponding to the K\"ahler parameter $t$.}
\end{figure}

In this section we introduce a crystal model which encodes A-model topological strings on the resolved conifold geometry. This model is of the latter type, according to the description in the beginning of chapter \ref{chap-cy-crystals}: its defining property is a modification of the container which represents $\C^3$ crystal. This modification has a straightforward geometric interpretation of inserting a wall at the positive slice $N$, beyond which crystal cannot melt. The allowed melting region of length $N$ in one direction represents the conifold geometry with K{\"a}hler parameter $t=g_s N$, which we often refer to as $Q=e^{-t}=q^N$. This crystal model is shown in figure \ref{wall1}. We will usually consider the wall on $y$ axis, and then refer to $x$ as \emph{external} or \emph{non-compact} axis or \emph{leg}, and to $y$ axis as \emph{internal} or \emph{compact} axis. This model was first proposed in \cite{okuda}, however the analysis of brane configurations we present is original.

The effect of the wall at slice $N$ is represented by inserting only first $N$ operators $\G_-$ acting on right vacuum in (\ref{McMahon}), and thus the partition function of this crystal model is
\begin{equation}
Z^{\mathbb{P}^1}(q,N)=\langle 0|\prod_{m=1}^{\infty} 
\Gamma_{+}(q^{m-1/2})\,\prod_{n=1}^{N}
\Gamma_{-}(q^{-(n-1/2)})|0\rangle = 
M(q)\,e^{-\sum_{k}\frac{Q^{k}}{k[k]^{2}}}. \label{P1-formula}
\end{equation}
Apart from a factor of $M(q)$, we indeed rederive the resolved conifold result (\ref{F-conifold}), which is also in agreement with the topological vertex result (\ref{Z-P1}). Taking the K{\"a}hler parameter $t \rightarrow \infty$ gives back the partition function of the $\C^3$ crystal.

This model can be derived by rewriting the expression for Chern-Simons partition function on $S^3$ (\ref{Z-S3-cs}), which was shown in \cite{okuda}.

%The fact it describes the A-model topological strings on the conifold geometry follows from the equivalence of Chern-Simons theory on $S^3$ to open topological strings on the deformed conifold, and then the open-closed geometric transition. 

\subsection{Full unknot invariant}

Lagrangian branes in the crystal with a wall are again defects described by
fermionic operators. Unknot invariants with many row representations can be generated by inserting 
a number of branes on the non-compact axis of the conifold crystal. 
This is analogous to the topological vertex picture as we will see in
section \ref{sec-A-model}. Since now we consider the full conifold geometry, we
will get the full unknot invariants, contrary to the $\C^3$ geometry which
could only detect the leading  $t\rightarrow\infty$ part of knot invariants (\ref{hopf-limit}).

Let us consider $m$ anti-branes on the $x$ axis at positions $a_i=q^{N_i+1/2}$, $i=1 \dots m$, with a wall inserted at position $N$ at $y$ axis (this is equivalent to branes on $x$ axis and a wall on $y$ axis). The partition function for this configuration is obtained by inserting fermionic brane operators at appropriate positions in (\ref{P1-formula}) and leads to the result 
\be
\tz^{\mathbb{P}^1}_{D}(a_1,\ldots a_n) = \Big[
\prod_{i<j}^{m}(1-\frac{a_i}{a_j})\Big] \prod_{i=1}^{m} \frac{L(a_i,q)}{L ( a_i Q,q )},
\label{p1-brane-noncompact}
\ee
which is already normalised to the conifold partition function, and irrelevant $\varphi(q)$ factors have been dropped.
Taking a single brane first at $a=q^{N_1+1/2}$ gives the full 
unknot generating function for single row representations 
by the rearrangement
\be
\tz^{\mathbb{P}^1}_D(a) =  \frac{L(a,q)}{L ( aQ,q )} = \sum_{n=0}^{\infty} a^n \Big(\sum_{i=0}^n h_i(q^{\rho}) h_{n-i}(Qq^{-\rho}) 
\Big) = \sum_{n=0}^{\infty} a^n h_n(q^{\rho},Qq^{-\rho}).\label{conifold-brane}
\ee
In the first equality the expression of dilogarithm in terms of symmetric
polynomials is used (\ref{def-dilog}) and
in the second
equality (\ref{schur-sum1}) is used. The final coefficient
$h_n(q^{\rho},Qq^{-\rho})$ is precisely the full unknot invariant 
(quantum dimension) for one-row representation, $R=(n,0,\ldots)$.
Taking $m$ branes and expanding in their positions $(a_1, \ldots, a_m)$
gives similarly unknot invariants with $m$-row representation. The proof of
this is completely analogous to the induction included in paragraph \ref{induct}.

%We note that the full unknot invariants were extracted before in \cite{okuda},
%from branes inserted in the compact
%axis of $\mathbb{P}^1$. Our procedure is different and is motivated by the topological
%vertex picture as will be discussed in more detail below.

\subsection{Ooguri-Vafa generating function} \label{oogurivafa}

In section \ref{ssec-O-V} we analysed the geometric transition for the conifold geometry with probe branes following \cite{ov}. We argued the free energy for probe branes should be equal to the unknot expectation value in Chern-Simons theory (\ref{ov-F-unknot}). From the toric point of view these probe branes are encoded by half-lines ending on the compact interval which represents $\mathbb{P}^1$ in the resolved conifold. Now we rederive these results in the case of a single brane in the crystal model for the conifold.

Let us recall first the Ooguri-Vafa generating function. For a single unknot it is given as \cite{ov}
$$
Z_{{\rm OV}} = \exp{ \left[- \sum_{n=1}^{\infty} {(e^{{n t \over 2}} - 
e^{{-nt \over 2}})
\over n [n]}  a_{{\rm OV}}^{-n} \right] }
$$
where $[n]=q^{n/2} - q^{-n/2}$ as before, and $a_{OV}$ is the parameter of the
one-dimensional holonomy matrix, corresponding to the representation $R$ given by Young diagram with one row. After analytic continuation, denoted by an upper index $a$, it is rewritten as
\be
Z^{a}_{{\rm OV}} = 
\exp{ \left[ {(a_{{\rm OV}}^n + a_{{\rm OV}}^{-n}) \over n [n]} 
e^{- n t/2} \right] } 
\ee
and this result agrees perfectly with the topological string 
amplitude of a probe brane inserted in the conifold geometry. This
topological amplitude can also be derived considering the relevant open
topological string amplitude from the M-theory point of view
\cite{G-V}. Alternatively, it can be computed in the topological
vertex formulation, which we do in chapter \ref{sec-A-model}.

We wish to show now that inserting a brane in the compact axis of the conifold 
crystal reproduces the Ooguri-Vafa generating function. In fact we insert an anti-brane fermion operator, because when matching the crystal to the topological vertex result, we will choose the convention $q_{vertex} =1/q_{crystal}$, which turns an anti-brane in the crystal to a brane in the vertex. Alternatively, brane and anti-brane differ by a change of framing as discussed in section \ref{sec-crys-brane}, so choosing one of them is only a matter of convention. The expectation value for this anti-brane operator on the compact axis of the crystal at the positive slice at $a=q^{N_0 +1/2}$ reads
\bea
 Z^{\mathbb{P}^1}_{D,y}(q,N_0, N) & = & 
M(q)\varphi(q)^{-1}\,e^{-\sum_{n>0}\frac{q^{n(N+1)}}{n[n]^{2}}}\,
e^{\sum_{n>0}\frac{q^{n(N_{0}+1/2)}+q^{n(N-N_{0}+1/2)}}{n[n]}} \nn
&=& Z^{\mathbb{P}^1}\varphi(q)^{-1}\,L(a,q)L(Q/a,q),
\label{p1-brane}
\eea
where $Z^{\mathbb{P}^1}$ is given in (\ref{P1-formula}). This is indeed the Ooguri-Vafa generating function with identifications
$$
a_{{\rm OV}} = q^{N_{\rm OV}+\hf}, \quad N_{{\rm OV}} = N_0 - {N \over 2}, \quad  Q=q^{N+1}.
$$
The redefinition of $ N_{{\rm OV}}$ does not represent any physical effect, and just means that the position of the brane in the geometry is measured from the middle point of the compact axis, and not from one of its ends. Much more interesting is the last equality which means the K\"ahler parameter of the conifold is shifted by $g_s$ from its original value $g_s N$. This is precisely the expected back-reaction of the geometry on the brane insertion (\ref{backreaction}). 

Inserting more branes on the compact axis would correspond to inserting more
stacks of branes in the geometry. The generating function can be easily 
computed on the crystal side. On the other hand,  in the crystal geometry it is not clear how to incorporate increasing the number of branes in a single stack (thus increasing the holonomy matrix of probe).

It is a natural question whether inserting a number of branes on each 
axis of the conifold crystal would provide complete Hopf-link invariants
with many rows, similarly to the leading part of Hopf-link invariants
obtained from $\C^3$. For example, inserting a brane on the compact axis
at position $a$ and an antibrane
on the non-compact axis at position $b$ in the conifold crystal 
one gets
\be
Z^{\mathbb{P}^1}_{D;\,\bar{D}}(q,N) = Z^{\mathbb{P}^1}{L(a,q) L(b, q) L(Q/a, q) \over L(bQ, q) (1 - ab)},
\label{p1-Dy-antiDx}
\ee
where again the K{\"a}hler parameter gets shifted to $t=g_s(N+1)$ due to brane insertion on compact axis.
Expanding in $a$ and $b$ does not naturally give many row Hopf-link invariants.
The reason is seen better in the language of topological vertex, where Hopf-link invariants are associated to having two branes inserted, each
on a non-compact axis of the conifold geometry \cite{marcos}. In the conifold crystal model,
a Hopf-link would naturally arise from placing branes on the non-compact
$x$-axis and another on the non-compact $z$-axis. The latter branes are not natural in the crystal picture.
Working out the operators for insertion of such branes and generating full 
Hopf-link invariants from the crystal is an interesting open problem. 

%***********************************************************
%***********************************************************

\section{Crystal model with two walls} \label{sec-CY-2walls}

\begin{figure}
\begin{center}
\includegraphics[width=0.45\columnwidth]{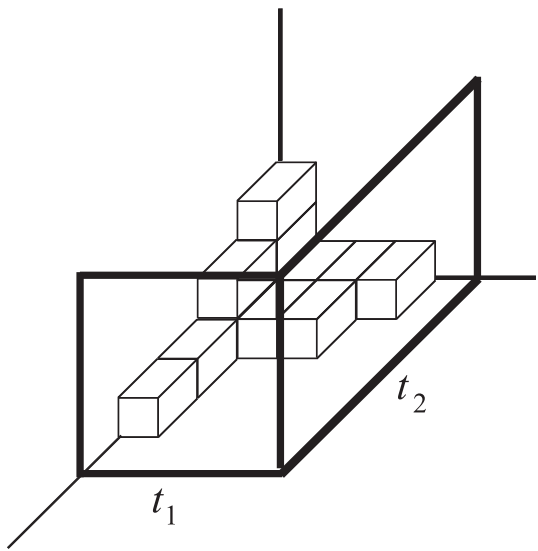}
\end{center}
\caption{\label{doublewall} The crystal model for a geometry of two neighbouring $\mathbb{P}^1$'s (the so-called double-$\mathbb{P}^1$) is given by two walls at distances corresponding to the K\"ahler parameters $t_1$ and $t_2$.}
\end{figure}

The crystal model with one wall introduced in the previous section can be easily 
generalised to represent the geometry with two neighbouring $\mathbb{P}^1$'s shown in figure \ref{fig-double-P1}.
This is naturally described by a crystal with two walls, on both the
positive and negative slice, at distance $t_1 =N_1 g_s$ and 
and $t_2=N_2 g_s$ respectively, which are the two K\"ahler parameters 
of the geometry. This model is shown in figure \ref{doublewall}.

The partition function is computed as
\begin{eqnarray}
Z^{2\, walls}(q,N_1, N_2) & = & \langle 0|\prod_{n=1}^{N_1}\Gamma_{+}(q^{n-1/2})\,\prod_{m=1}^{N_2}\Gamma_{-}(q^{-(m-1/2)})|0\rangle = \nonumber \\
& = & \exp \sum_{k>0} \frac{(1-q^{kN_1})(1-q^{kN_2})}{k[k]^2}, \label{crystal-P1P1}
\end{eqnarray}
The factors in the exponent represent (apart from the unity giving McMahon function (\ref{McMahon})) worldsheets wrapping each of the spheres independently, and then both of them simultaneously.

Let us now insert a brane on the right compact axis at a position given by $a=q^{N_0+1/2}$, which
gives
\be
Z^{2\, walls}_{D,y} = Z^{2\, walls}\,\xi(q)\, L(Q_1/a, q )\,\frac{L(a,q)}{L(a Q_2,q)}, \label{crystal-D-P1P1}
\ee
where again there is a shift of the K\"ahler parameter corresponding to the axis the brane is put on.
It is also interesting to compare this result with a brane in the resolved conifold (\ref{p1-brane}). The essential difference is the dilogarithm 
in the denominator, which represents worldsheet wrapping a part of right 
$\mathbb{P}^1$ (of length $N_0$) and the whole left $\mathbb{P}^1$ 
(of length $N_2$).

%***********************************************************
%***********************************************************

\section{Crystal model for the closed topological vertex geometry} \label{sec-closed-ver}

\subsection{The closed topological vertex geometry}

\begin{figure}[htb]
\begin{center}
\includegraphics[width=0.4\textwidth]{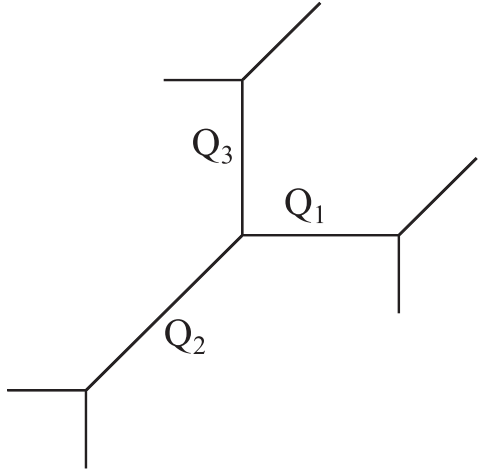}
\caption{The closed topological vertex $\mathcal{C}$.} \label{fig-closed-ver}
\end{center}
\end{figure}

The closed topological vertex $\mathcal{C}$ is a toric Calabi-Yau threefold, consisting of three $\mathbb{P}^1$'s meeting in one point, with local neighbourhood of each sphere being isomorphic to a sum of line bundles $\mathcal{O}(-1)\oplus \mathcal{O}(-1)$. This geometry was discussed in \cite{cremona,GW-curves}, and can be understood as a particular resolution of $\mathbb{C}^3/\mathbb{Z}_2\times\mathbb{Z}_2$ singularity \cite{branes-resolve,F-theory}. It is convenient to introduce quantities $Q_i=e^{-t_i}$ related to K\"ahler parameters $t_i$, $i=1,2,3$ which correspond to sizes of the spheres. The toric diagram of the closed topological vertex is shown in figure \ref{fig-closed-ver}. 

The main object of considerations in this section is the partition function of
$\mathcal{C}$. To start with, we briefly quote some of the results already
known in literature in order to set later computations in a proper context.

 The elements of the second homology group of the closed topological vertex
 $H_2(\mathcal{C})$ can be encoded in a vector $\beta=(d_1,d_2,d_3)$, with
 degree $d_i$ corresponding to the $i$'th sphere, and the corresponding
 Gromov-Witten worldsheet instanton expansion (\ref{grom-wit}) reads
$$
F^{\mathcal{C}} = \sum_{g\geq 0} \sum_{d_1,d_2,d_3} g_s^{2g-2}\, N^{g}_{d_1,d_2,d_3}\, Q_1^{d_1} Q_2^{d_2} Q_3^{d_3}. \label{grom-wit-closed}
$$
These Gromov-Witten invariants $N^{g}_{d_1,d_2,d_3}$ can be derived rigorously
via the so-called Cremona transform, by identification with known invariants of a relevant blow-up of a projective space. This approach was originally introduced in this context in \cite{cremona-K3}, and applied to the closed topological vertex in \cite{cremona} with the result
\begin{eqnarray}
N^{g}_{d,0,0} & = & N^{g}_{0,d,0} = N^{g}_{0,0,d} = N^{g}_{d,d,d} = -N^{g}_{d,d,0}  = -N^{g}_{d,0,d} = -N^{g}_{0,d,d} = \nonumber \\
& = & - d^{2g-3} \frac{|B_{2g}|}{2g\,(2g-2)!}, \qquad \textrm{for}\ d\neq 0, \label{GW-P1-cube}
\end{eqnarray}
and all other $N^{g}_{d_1,d_2,d_3}$ vanishing. Apparently,
$N^{g}_{d,0,0}=N^g_d$ is equal to the invariant $N^g_d$ for the resolved
conifold (\ref{GW-P1}).

The structure of the closed topological vertex partition function is similar
to $F^{conifold}$ given in (\ref{F-conifold}), in a sense that there is also only a finite number of
non-vanishing Gopakumar-Vafa invariants (\ref{gop-vafa})
$$
n^0_{1,0,0}= n^0_{0,1,0} = n^0_{0,0,1} =  - n^0_{1,1,0} = - n^0_{1,0,1} = - n^0_{0,1,1} = n^0_{1,1,1} = -1
$$
and they correspond respectively to single spheres, each possible pair of them and the entire triple. Thus the instanton part of the partition function for $\mathcal{C}$ can be written as
\begin{equation}
Z^{\mathcal{C}} = \exp\sum_{n>0}\frac{-Q_1^n-Q_2^n-Q_3^n + Q_1^n Q_2^n + Q_1^n Q_3^n + Q_2^n Q_3^n - Q_1^n Q_2^n Q_3^n}{n[n]^2}. \label{closed-ver}
\end{equation}
Of course this result is consistent with (\ref{GW-P1-cube}). This can also be written as a product formula, as described in general in \cite{HIV}. In fact, it turns out that the crystal model we introduce in the next paragraph naturally computes the result in the form related both to the above Gopakumar-Vafa expansion and a product formula, which is a fact already stressed in \cite{ps}. 

We note the partition function $Z^{\mathcal{C}}$ has been computed in the
'physical' and 'mathematical' topological vertex formalisms in \cite{GW-curves}, and was shown to agree with (\ref{closed-ver}). In section \ref{s-vertex} we will also compute $Z^{\mathcal{C}}$ using 'physical' topological vertex, but in a way which is simpler and faster than it is done in \cite{GW-curves}. In fact the main motivation behind the calculation we present is it makes an immediate connection with the crystal model which we present next. And then --- last but not least --- the method we introduce suggests how to generalise formalism of \cite{strip} to 'off-strip' geometries.

So far we focused only on the instanton contributions $F^{\mathcal{C}}$ which
have already been derived in literature. The classical part
$F_{class}^{\mathcal{C}}$ (\ref{F-class}) will be discussed in section \ref{subsec-flop} together with the analysis of the flop transition.

%*******************************************************

\subsection{The cube model}

\begin{figure}[htb]
\begin{center}
\includegraphics[width=0.5\textwidth]{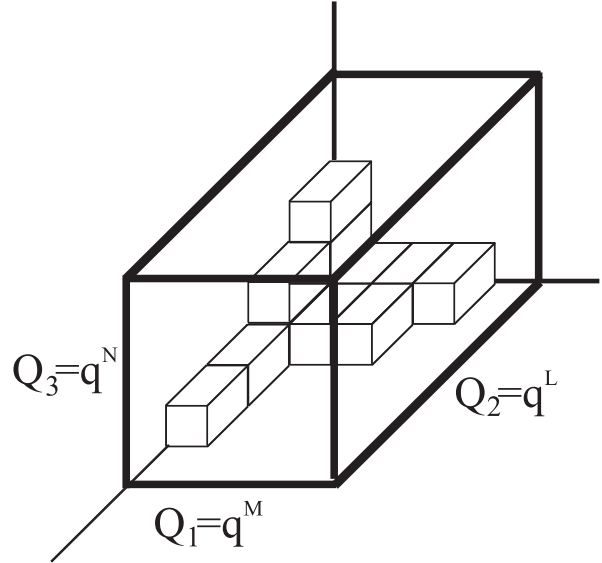}
\caption{The cube crystal model of finite size  $M \times L \times N$.} \label{fig-cube}
\end{center}
\end{figure}

The model we wish to introduce is a natural extension --- or rather a
truncation --- of models with walls presented in sections \ref{sec-conifold} and \ref{sec-CY-2walls}. We consider all 3-dimensional partitions which fit into a finite cube of size $M \times L \times N$, and ask what is the corresponding generating function $Z^{cube}$, given by (\ref{generating}) with the present restriction on $\pi$. Let us remark there is only a finite number of terms in $Z^{cube}=1+\ldots + q^{LMN}$, the last one corresponding to the highest power of $q$. In other words, we introduce three 'walls' at positions $x=M$, $y=L$, $z=N$, as illustrated in figure \ref{fig-cube}.

In this case the partition function can also be written in the transfer matrix
formalism, with two walls represented by finite number of $\Gamma_{\pm}$ and
the third wall by the projection operator $\mathbf{1_{d^t\leq N}}$
\begin{equation}
Z^{cube} = \langle 0|\prod_{n=1}^{L}\Gamma_{+}(q^{n-1/2})\,\mathbf{1_{d^t\leq N}}\, \prod_{m=1}^{M}\Gamma_{-}(q^{-(m-1/2)})|0\rangle. \label{crystal-cube}
\end{equation}

Our claim is this generating function is equal to the closed topological vertex partition function (\ref{closed-ver}) up to the McMahon function
\begin{equation}
Z^{cube} = M(q) Z^{\mathcal{C}}, \label{claim}
\end{equation}
with identification of parameters
\begin{equation}
Q_1 = q^M,\qquad Q_2 = q^L, \qquad Q_3 = q^N.  \label{Q123-MLN}
\end{equation}

The explicit evaluation of the correlator (\ref{crystal-cube}) is nontrivial, due to a complicated form of the projector $\mathbf{1_{d^t\leq N}}$. Fortunately, the generating function for plane partitions in a finite box has been well known since originally derived by combinatorial methods by McMahon \cite{mcmahon}; another combinatorial proof is presented in \cite{macdonald}. Thus we can just use this result, which reads:
\begin{equation}
Z^{cube} =  Z_1 Z_2 = \prod_{(i,j)\in M^L} \frac{1-q^{N+i+j-1}}{1-q^{h(i,j)}}, \label{Z1Z2} 
\end{equation}
where for later convenience we introduce two factors
\begin{eqnarray}
Z_1 & = & \prod_{(i,j)\in M^L} \frac{1}{1-q^{h(i,j)}}, \nonumber \\
Z_2 & = & \prod_{(i,j)\in M^L} (1-q^{N+i+j-1}), \nonumber 
\end{eqnarray}
and $M^L$ denotes 2-dimensional partition with L rows of the same length $M$,
which is a base of the cube, and $h(i,j)$ is the hook-length of its $(i,j)$
element (\ref{hook}). Of course (\ref{Z1Z2}) reduces to (\ref{McMahon}) in the
limit $M,N,L\to\infty$. Using the elementary series
$$
\log(1-a)=-\sum_{k>0}\frac{a^{k}}{k}
$$
to rewrite products as exponentials we arrive at the following form of the above factors
\begin{eqnarray}
Z_1 & = & \exp\sum_{n>0}\frac{1-Q_1^n -Q_2^n + Q_1^n Q_2^n}{n[n]^2}, \label{Z1-exp}\\
Z_2 & = & \exp\sum_{n>0}\frac{-Q_3^n + Q_1^n Q_3^n + Q_2^n Q_3^n - Q_1^n Q_2^n Q_3^n}{n[n]^2}. \label{Z2-exp}
\end{eqnarray}
Remarkably, the product (\ref{Z1Z2}) of the above two factors indeed
reproduces the closed topological vertex partition function (\ref{closed-ver})
up to the McMahon function, as claimed in (\ref{claim}). In section
\ref{s-vertex} we derive this result from the topological vertex point of view, in a way which makes relation to the crystal result (\ref{Z1Z2}) explicit.

The crystal model in a cube we introduced above is interesting for yet another reason. We speculate that gluing such cubes might correspond to building more complicated crystal models related to more general toric Calabi-Yau spaces. In this sense a crystal model for the closed topological vertex geometry would have an analogous status as the ordinary topological vertex amplitude. These more general crystal models obtained from gluing would have two characteristic scales: one of order $g_s$ related to a size of an elementary box, and the other of order of K\"ahler parameters of Calabi-Yau, similarly as $Q_1 = q^M, Q_2 = q^L, Q_3 = q^N$ above. Results in \cite{crystal-bubble} --- even though derived with a different motivation --- seem to confirm such an interpretation, at least for threefolds whose toric diagrams consists of a string of intervals representing $\mathbb{P}^1$'s. Understanding such a general framework of Calabi-Yau crystals is still an open problem.

% chain!!!!!!!!!!!!!!!! (okuda&gomis)

%***********************************************************
%***********************************************************

\chapter{Derivation of crystal results from the topological vertex}   \label{sec-A-model}

In chapter \ref{chap-cy-results} new amplitudes have been computed in
crystal models already known and new Calabi-Yau crystal models have been
introduced. However, apart from general arguments presented in chapter
\ref{chap-cy-crystals}, we have not given any proof that these amplitudes
indeed agree with topological string results. In the
present chapter, following results published in \cite{ps} and \cite{ps-cube}, we provide such a proof by computing topological string
amplitudes for the corresponding topological string configurations. In all
cases, using topological vertex formalism, we reproduce our earlier crystal results.
 Admittedly, in same cases these computations are non-trivial and
lengthy, and all facts from appendices \ref{app-vertex} and \ref{app-strip}
have to be used. Moreover, we also extend topological
vertex methods to deal with more general situations than considered so
far. In particular this concerns the so-called \emph{off-strip geometries}
and is discussed is section \ref{move-off}.

To match crystal and vertex results a few important issues have to be taken
into account. Firstly, the topological vertex is normalised in such a way that
the McMahon function $M(q)$ of $\C^3$ does not arise from
calculations. Secondly, we need to choose some particular framing; usually
this is $(-1)$ framing for branes on one axis and the canonical one for branes
on another axis. Then, we have to take holonomy matrices $V_i$ to be one dimensional 
$$ 
V_i=a_i=q^{N_i+1/2} \label{brane-at-a}
$$ 
so in this sense the crystal can see only a fraction of what the full vertex computes. On the other hand, the crystal calculations are much simpler, so this is quite an advantage of using it.

If we have a single brane on one axis, then the above $a_i$ become simply
moduli seen in the crystal.
For more branes on one axis, we have to introduce parameters which give their positions, which must be combined with holonomy matrices
appropriately. We will see examples of this in what follows.

Finally, we perform  substitution 
\be
q\to \frac{1}{q}=q_{crystal} \label{inverse-q}
\ee
in vertex results to map crystal-branes to vertex-branes. Apparently, in topological strings such an operation exchanges branes to antibranes \cite{vertex}, and what we call branes and antibranes can be regarded just as a convention. Not performing the $q$ inversion would result in mapping crystal branes to vertex antibranes. We choose the former point of view. In fact, the $q$ inversion is important only for configurations with branes; 
the partition functions without any branes is invariant under $q\to 1/q$. We
have already seen one example of the above inversion, when we related
the topological vertex amplitude to generating function of plane partitions in (\ref{Pcrys-Cver}).

Let us note that while some of the topological vertex expressions we consider here were already
written in literature in terms of sums over representations, it is not at all obvious that they can be resummed into compact expressions, involving just dilogarithms and simple polynomials (as we have seen from crystal point of view). Nonetheless, by a clever choice of framing, we manage to rederive all these crystal results.

By construction, the topological vertex includes the correct worldsheet instantons which can appear in any toric construction, with or without lagrangian probe branes. The contributions from specific instantons which stretch between probe branes can be read off from the form of the free energy. Specifically, the ${\rm Li}_1$ function in the factor

\be
(1-a b)={\rm exp} \left( {\rm Li}_1 (ab)\right),
\ee
appearing in all calculations involving more than one probe brane, shows that this is a contribution from annuli instantons and not of any higher genera instantons.

%*********************************

\section{Resolved conifold results}

We shall first test the amplitudes found in the crystal model for the
conifold. The results for $\C^3$ crystal will follow then naturally from
taking the K\"ahler parameter to infinity. The partition function for the
resolved conifold, or equivalently its free energy (\ref{F-conifold}), can be
computed using topological vertex formalism as shown in (\ref{Z-P1}). This
result is in agreement with the crystal result (\ref{P1-formula}) up to
McMahon function. 

Let us then compute brane configurations corresponding to those found in the crystal language. We start with a single brane on the external axis of the conifold in 
canonical framing
\begin{equation}
Z^{\mathbb{P}^1}_{D-ext}(V) = \sum_{P,R} C_{\bullet R^{t} P} (-Q)^{|R|} C_{R\bullet \bullet}\, Tr_{P}V. \label{Z-brane-full}
\end{equation}
Using identities on Schur functions we get (see also \cite{marcos})
\begin{equation}
Z^{\mathbb{P}^1}_{D-ext}(V) = Z^{\mathbb{P}^1} \sum_P s_P(Qq^{-\rho},q^{\rho})\, Tr_{P}V. \label{Z-P1-brane-ext}
\end{equation}
Taking the matrix $V$ to be one 
dimensional $V=a=q^{N_0 + 1/2}$, and using $Tr_R(a)=s_R(a)$ and formula (\ref{schur-sum3}), we obtain
\be
Z^{\mathbb{P}^1}_{D-ext} 
=  Z^{\mathbb{P}^1} \sum_R s_R(Qq^{-\rho},q^{\rho})\, s_R(a)
=Z^{\mathbb{P}^1} \frac{L(aQ,q_{crystal})}{L(a,q_{crystal})}.
\label{Z-D-L-v}
\ee
It is important to note that the sum is in fact performed over representations
corresponding to diagrams with only one row (for a single number
$a$ and for any representation given by a diagram with more than one row
$s_{rep.\ with \ >1\ rows}(a)=0$). Taking into account the mapping (\ref{inverse-q}) we find perfect agreement 
with the normalised crystal result for antibranes (\ref{p1-brane-noncompact}).

A single brane can also be situated on the compact axis of the conifold at position $g_sD$
$$
Z^{\mathbb{P}^1}_{D-int} = \sum_{R,Q^L,Q^R} C_{\bullet\bullet R\otimes Q^L} 
(-1)^s q^f e^{-L} C_{R^t\otimes
Q^R\bullet\bullet}\, Tr_{Q^L }V\, Tr_{Q^R}V^{-1}.  \label{comp-leg-brane-res}
$$
It is possible to perform resummation for $V=a=q^{N_0+1/2}$ and if $(-1)$ 
framing is chosen. 
If we follow the crystal convention and set the size of the compact axis to be $N+1$ (the shift  is responsible for brane insertion), and absorb the brane position into its modulus by defining 
\begin{equation}
N_0'=D+N_0, \qquad a'=q^{N'_0+\frac{1}{2}},
\end{equation}
we get after  substitution (\ref{inverse-q})
$$
Z^{\mathbb{P}^1}_{D,y}=Z^{\mathbb{P}^1}(N+1)\, L(a',q_{crystal})\,
L(Q/a',q_{crystal}), \label{brane-P1-comp}
$$
which is the same result as (\ref{p1-brane}).

It is also possible to insert several branes on the external or internal axis. For example, for $M$ branes on the compact axis in $(-1)$ framing we take $V_i=a_i=q^{N_i+1/2}$ and then get analogous factors as above. The K\"ahler parameter gets modified to $N+M$, and brane positions $D_i$ get absorbed similarly as above into $N_i'$ and modified moduli $a_i'$. We also have to take the stretched strings 
between the branes into account (see (\ref{stretched})).
All these factors combine to
\begin{equation}
Z^{\mathbb{P}^1}_{M\,branes} =  {Z^{\mathbb{P}^1}(N')} \Big[\prod_{i<j} (1-\frac{a_i'}{a_j'}) 
\Big] \Big[ \prod_{i=1}^{M} L(a_i',q_{crystal})\,
L(Q/a'_i,q_{crystal}) \Big], \label{M-branes-P1-comp}
\end{equation}
which is the same as the crystal result found in \cite{okuda}.

\subsection{Brane and antibrane on two axes}

\begin{figure}
\begin{center}
\includegraphics[width=0.36\columnwidth]{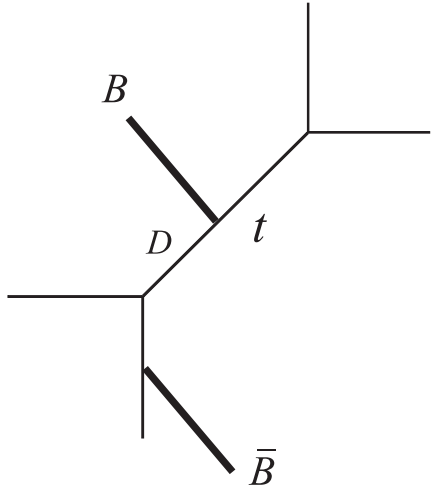}
\end{center}
\caption{\label{conifold1} The toric diagram of the conifold with K\"ahler parameter $t$, with a brane inserted at distance $D$ on the compact axis and an antibrane inserted on a non-compact axis.}
\end{figure}

Let us put one brane on the compact axis of the resolved conifold at distance $D$ with holonomy matrix $V_1$, and the second brane on non-compact axis with holonomy $V_2$ (Fig.~\ref{conifold1}).

 We also take one-dimensional holonomy matrices $V_i=q^{N_i+1/2}$, and absorb the position on the compact axis into $V_1$
\begin{equation}
a=q^{D+N_1+1/2}=q^{N_1'+1/2}, \qquad b=q^{N_2+1/2}.
\end{equation}
The partition function in $(-1,0)$ framing is
\begin{eqnarray}
Z^{\mathbb{P}^1}_{Dy;\bar{D}x} & = & \sum C_{R\otimes Q_L,P^t,\bullet} (-1)^{|P|} (-Q)^{|R|} C_{R^t\otimes Q_R\bullet\bullet} s_{Q_L}(a) s_{Q_R}(Q/a) s_{P}(b) \nonumber \\
& & \Big[(-1)^{|Q_L\otimes R|+|Q_R\otimes R^t|} q^{-\frac{\kappa_{Q_L\otimes R}+\kappa_{Q_R\otimes R^t}}{2}} \Big] = \nonumber  \\
& = & L\big(Q/a,q^{-1}\big) \sum s_{R^t}(-Qq^{\rho}) s_{Q_L}(-a) s_{P}(-b) \nonumber \\
& & c^{\alpha}_{R Q_L}  s_{\alpha^t/\eta}(q^{\rho}) s_{P^t/\eta}(q^{\rho}) q^{-\frac{\kappa_{R^t}}{2}}, \nonumber
\end{eqnarray}
where  the first dilog arises from $Q_R$ summation. Now summation over $P$ produces another dilogarithm, and we can also sum over $Q_L$ and use (\ref{schur-invert2})
to get
$$
Z^{\mathbb{P}^1}_{Dy;\bar{D}x}  =  L\big(Q/a,q^{-1}\big) L\big(b,q^{-1}\big) L(a,q^{-1}) \sum s_R(-Qq^{\rho}) s_{\eta /\alpha}(-a) s_{\eta}(-b) s_{R/ \alpha }(q^{\rho}),
$$
Performing the remaining sums over $R$, $\eta$ and finally $\alpha$ gives
the crystal result (\ref{p1-Dy-antiDx}) (after transformation (\ref{inverse-q})) 
\begin{equation}
Z^{\mathbb{P}^1}_{Dy;\bar{D}x} = \frac{Z^{\mathbb{P}^1}}{1-ab} \frac{L\big(Q/a,q_{crystal}\big) L\big(b,q_{crystal}\big) L\big(a,q_{crystal}\big)}{L\big(bQ,q_{crystal}\big)}.
\end{equation}
Here we contrast the simplicity of earlier crystal computations with the extensive use of summation formulae and Schur identities in the above vertex computation.

%***********************************

\section{Double $\mathbb{P}^1$} \label{subsec-P1P1}

In this section we rederive two-wall crystal amplitudes from the topological vertex perspective. At first
 we compute the partition function. Denoting the sizes of the right and the left axis by $t_i$
(and $Q_i=e^{-t_i}=q^{N_i}$), respectively for $i=1,2$, the vertex rules and some rearrangements give
\begin{eqnarray}
Z^{\mathbb{P}^1 \mathbb{P}^1} & = & \sum_{P,R} C_{P^t\bullet\bullet} (-Q_2)^{|R|} C_{PR\bullet}
(-Q_1)^{|P|} C_{R^t\bullet\bullet} = \nonumber \\
& = & \sum_{\eta} \Big[\sum_{\mu} s_{\mu}(q^{\rho})s_{\eta}(Q_1 q^{-\rho})
s_{\mu}(Q_1 q^{-\rho}) \Big] \nonumber \\
& & \Big[\sum_{\nu} s_{\nu}(q^{\rho})s_{\eta}(Q_2 q^{-\rho}) s_{\nu}(Q_2 q^{-\rho})
\Big] = \nonumber \\
& = & \exp \sum_{k>0} \frac{-Q_1^{k}-Q_2^{k}+ (Q_1 Q_2)^k}{k[k]^2}, \label{P1P1}
\end{eqnarray}
This result is the same as the crystal expression (\ref{crystal-P1P1}), up to
McMahon function invisible for the vertex. Since this is a partition 
function without any brane insertions, it is also unaffected by $q$ inversion. 

\begin{figure}
\begin{center}
\includegraphics[width=0.36\columnwidth]{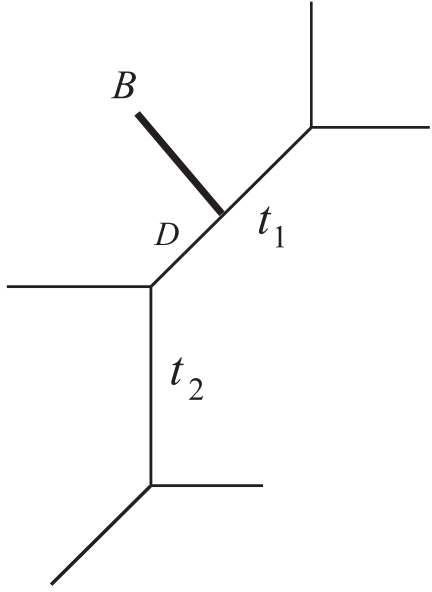}
\end{center}
\caption{\label{doublex} The toric diagram corresponding to a brane inserted in the double $\mathbb{P}^1$ geometry
 (with K\"ahler parameters $t_1$ and $t_2$) on the right compact axis at distance $D$
from the middle point.}
\end{figure}

The vertex computation with a brane on the right compact axis of double 
$\mathbb{P}^1$, in $-1$ framing also agrees with crystal result. Inserting 
this brane at position $D$ from the middle vertex, as shown in figure \ref{doublex}, the topological vertex 
rules lead to the amplitude

\begin{eqnarray}
Z^{\mathbb{P}^1 \mathbb{P}^1}_{D} & = & \sum C_{P^t\bullet\bullet} (-Q_2)^{|P|}  C_{R\otimes
Q_L,P,\bullet} (-Q_1)^{|R|}C_{R^t \otimes Q_R\bullet\bullet} \nonumber \\
& & q^{D|Q_L|} q^{(N_1-D)|Q_R|} \,Tr_{Q_L}V \,Tr_{Q_R}V^{-1} \nonumber \\
& &\Big[(-1)^{|Q_L\otimes R|+|Q_R\otimes R^t|}q^{-(\kappa_{Q_L\otimes R}+\kappa_{Q_R\otimes R^t})/2} \Big].
\end{eqnarray}
As before, we take one-dimensional $V=q^{N_0+1/2}$ and absorb the position into $V$ as
$$
a=q^{D+N_0+1/2}=q^{N_0'+1/2}.
$$
Performing sums over all representation in the appropriate order leads (after a little effort) to the result
\begin{equation}
Z^{\mathbb{P}^1 \mathbb{P}^1}_{D} = Z^{\mathbb{P}^1 \mathbb{P}^1} \frac{L\big(Q_1/a,q_{crystal} \big)
L\big(a,q_{crystal} \big)}{L\big(bQ_2,q_{crystal} \big)},
\end{equation}
which is the same as the crystal answer (\ref{crystal-D-P1P1}) after enlarging the
size of the right axis to $N_1+1$ due to the brane insertion. 

%***************************************

\section{$\mathbb{C}^3$ amplitudes}   \label{sec-c3-amplitudes}

The amplitude for several branes on one axis of $\mathbb{C}^3$ can be computed 
directly from the vertex rules, but since we already have the conifold result
it is easiest to take the  $N\to \infty$ limit in (\ref{M-branes-P1-comp}). This 
also gives result in $(-1)$ framing, and substituting (\ref{inverse-q}) we get
\begin{equation}
Z^{\C^3}_{M\,branes}=\Big[\prod_{i<j} (1-\frac{a_i}{a_j})  \Big] \prod_{i=1}^{M}
L(a_i,q_{crystal}), \label{M-branes-C3-comp}
\end{equation}
which is the result for the $\mathbb{C}^3$ crystal (\ref{nmbranes}). For one brane it reduces to a single dilogarithm.

For a brane on one axis at position $a$ and antibrane on the other at
position $b$, and in framing $(-1,0)$, the vertex gives
$$
Z^{\C^3\,vertex}_{D;\bar{D}}(a,b) = \frac{1}{1-ab} L(a,q_{crystal})  L(b,q_{crystal}).
$$
which reproduces the crystal answer (\ref{c3-antiDx-Dy}). In this case the vertex rules can be expressed in terms of Hopf link invariants (\ref{W-C})
$$
Z^{\C^3\,vertex}_{D;\bar{D}}(a,b) = \sum_{P,R} W_{PR} (-1)^{|P|+|R|}q^{-\frac{\kappa_P+\kappa_R}{2}} s_P(a)s_R(b),
$$
so that inverting $q$ (\ref{inverse-q}) according to our conventions and using (\ref{W-PtRt}) proves that this is the same Hopf link generating function as in the crystal case (\ref{c3-antiDx-Dy-HL}). 

The calculation for two branes, one in
each axis, is similar and also gives the crystal result in $(-1,0)$ framing\footnote{
This is also an example of a situation, which can be resummed in canonical framing, with the final result $L(a,q) L(b,q) \frac{1-a\sqrt{q}+ab}{1-a\sqrt{q}}$.
This result does not agree with the crystal one (in canonical crystal framing),
thus a proper choice of framing is indeed crucial.}
$$
Z^{\C^3\,vertex}_{D;\,D} = (1-ab)\frac{L(a,q_{crystal})}{L(b,q_{crystal})}.
$$

%*******************************

The configuration with  two branes on one axis and antibrane on the other is slightly more complicated.
The stretched string factors between the two branes on the same axis, 
at positions $a_i=q^{M_i+1/2}$ (for $i=1,2$) give an $(1-a_1/a_2)$ factor. The 
full amplitude, with antibrane at $b=q^{N_1+1/2}$, and in $(-1,0)$ framing 
can be written as
\begin{eqnarray}
&& Z^{vertex}_{2Dy,\, \bar{D}x}  =  (1-\frac{a_1}{a_2}) \sum C_{P_1\otimes P_2, R^t,\bullet} s_{P_1}(a_1) s_{P_2}(a_2) s_{R}(b) (-1)^{|R|}  \times  \nonumber \\
&\times & \Big[(-1)^{|P_1 \otimes P_2|} q^{-\frac{\kappa_{P_1\otimes P_2}}{2}}  \Big] \nonumber \\
& = & (1-{a_1 \over a_2}){L(b,q^{-1})} \sum  c^{\alpha}_{P_1 P_2} s_{\alpha^t/ \eta}(q^{\rho}) s_{P_1}(-a_1) s_{P_2}(-a_2) s_{\eta}(-b) .  \label{necklace-resum}
\end{eqnarray}
After performing summations in several steps and substitution (\ref{inverse-q}) we recover the crystal result 
(\ref{nmbranes})
\begin{equation}
Z^{vertex}_{2Dy,\, \bar{D}x} =\frac{1-\frac{a_1}{a_2}}{(1-a_1 b)(1-a_2 b)}L(a_1,q_{crystal})L(a_2,q_{crystal})L(b,q_{crystal}) . \label{ncry}
\end{equation}
Thus  another way to look at the crystal result (\ref{ncry}) is provided by the first line in the expansion of (\ref{necklace-resum}), which due to (\ref{W-C}) can be written in terms of Hopf link invariants (with all components in knot $(-1)$-framing) as
\begin{eqnarray}
Z^{vertex}_{2Dy,\, \bar{D}x} & = & (1-\frac{a_1}{a_2}) \sum W_{P_1\otimes P_2, R,\bullet} s_{P_1}(a_1) s_{P_2}(a_2) s_{R}(b)  \nonumber \\
& & \Big[(-1)^{|P_1 \otimes P_2|+|R|} q^{-\frac{\kappa_{P_1\otimes P_2}+\kappa_{R}}{2}}  \Big]. \nonumber 
\end{eqnarray}
Taking out the stretched string factors $(1-a_1/a_2)$ this result has a form of a generating function
 of invariants for a knot shown in figure \ref{fig-dbl-hopf}, arising from merging two Hopf-links along a common unknot. Because of a one-dimensional sources $V_i=a_i$, this is a generating function for representations with one row only. 

% \begin{figure}
% \begin{center}
% \includegraphics[width=0.4\columnwidth]{necklace2.eps}
% \end{center}
% \caption{\label{necklace2} The ``necklace" knot invariant generated by the insertion of
% 2+1 branes. The crystal only generates representations with a single row.}
% \end{figure}

\subsection{Two axes of $\mathbb{C}^3$ --- general situation}   \label{ssec-2axes}

Finally we consider $m$ branes on one axis at positions $a_i=q^{M_i+1/2}$, and $n$ antibranes on the next axis at $b_i = q^{N_i+1/2}$. 
As usual we take all branes in framing $(-1)$, which makes resummation doable. 
Using properties of tensor product, the part of the partition function without factors from strings stretching between branes on the same axis (\ref{stretched}) (which is denoted by ') can be written as
\begin{eqnarray}
Z'_{m,\bar{n}} & = & \sum C_{P_1\otimes \ldots \otimes P_m,\,R_1^t\otimes \ldots \otimes R_n^t,\bullet} (-1)^{\sum_i |R_i|} \Big[(-1)^{|\otimes_j P_j|} q^{-\frac{1}{2}\kappa_{\otimes_j P_j}} \Big]\cdot \nonumber \\
& & \cdot s_{P_1}(a_1)\ldots s_{P_m}(a_m)\cdot s_{R_1}(b_1)\ldots s_{R_n}(b_n) = \nonumber \\
& = & \sum s_{(P_1^t\otimes \ldots \otimes P_m^t)/\eta}(q^{\rho})  s_{P_1}(-a_1)\ldots s_{P_m}(-a_m)\cdot \nonumber \\
& & \cdot s_{(R_1^t\otimes \ldots \otimes R_m^t)/\eta} (q^{\rho})  s_{R_1}(-b_1)\ldots s_{R_n}(-b_n), \label{general-c3-vertex}
\end{eqnarray}
where it is understood that
$$
s_{(P_1^t\otimes \ldots \otimes P_m^t)/\eta} = \sum_{\alpha} c^{\alpha}_{P_1^t\ldots P_m^t} s_{\alpha/ \eta}.
$$
The antibrane part takes the form (here we write the partial result for
the $R$ summation only),
according to (\ref{multi-lit-rich}) 
$$
c^{\beta_{1}}_{R_1^t R_{2}^t} c^{\beta_{2}}_{\beta_{1} R_3^t}\ldots c^{\beta_{n-1}}_{\beta_{n-2} R_n^t} s_{\beta_{n-1}/\eta}(q^{\rho}) s_{R_1}(-b_1)\ldots s_{R_n}(-b_n) =
$$
$$
= s_{\beta_1^t/R_{1}}(-b_2) s_{\beta_{2}^t/\beta_{1}^t}(-b_3)\ldots  s_{\beta_{n-1}^t/\beta_{n-2}^t}(-b_n) s_{\beta_{n-1}/\eta}(q^{\rho}) s_{R_1}(-b_1) = 
$$
$$
= s_{\beta_{n-1}^t}(-b_1,\ldots,-b_n)s_{\beta_{n-1}/\eta}(q^{\rho}) = 
$$
\be
= L(b_1,q^{-1})\ldots L(b_n,q^{-1}) s_{\eta}(-b_1,\ldots,-b_n) \label{gen-c3-1}
\ee
In the same way, the brane part ($P$ summation separated) contributes
\be
L(a_1,q^{-1})\ldots L(a_m,q^{-1}) s_{\eta}(-a_1,\ldots,-a_m) \label{gen-c3-2}
\ee
The remaining summation over $\eta$ in (\ref{gen-c3-1}) and (\ref{gen-c3-2}) 
gives factors for strings stretched between all brane/antibrane pairs; also taking into account (\ref{stretched}) for each pair of branes (antibranes) on the same axis finally we get (after the q-inversion) 
\begin{eqnarray}
Z_{m,\bar{n}} & = & \Big[ \big(1-\frac{a_1}{a_2} \big)\ldots  \big(1-\frac{a_{m-1}}{a_m} \big) \Big]\Big[ \big(1-\frac{b_1}{b_2} \big)\ldots  \big(1-\frac{b_{n-1}}{b_n} \big)\Big] \nonumber \\
& & \frac{1}{1-a_1 b_1} \ldots  \frac{1}{1-a_{m} b_n} \prod_i L(a_i,q_{crystal})\ \prod_j L(b_j,q_{crystal}), \nonumber
\end{eqnarray}
and this is the same answer as we found from the crystal (\ref{nmbranes}). 

\begin{figure}
\begin{center}
\includegraphics[width=0.5\columnwidth]{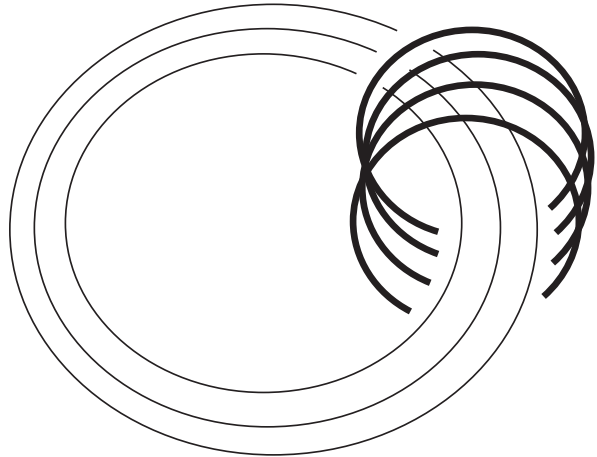}
\end{center}
\caption{\label{chain2} A ``multiple'' Hopf-link which corresponds to four branes on $y$ axis and three antibranes on $x$ axis in $\C^3$ geometry.}
\end{figure}

This more general case also can be understood as a generating function of Hopf-link invariants corresponding to tensor products of one-row representations, as the first line of (\ref{general-c3-vertex}) can be written using (\ref{W-C}) as
\begin{eqnarray}
Z'_{m,\bar{n}} & = & \sum W_{P_1\otimes \ldots \otimes P_m,\,R_1\otimes \ldots \otimes R_n}  \Big[(-1)^{|\otimes_j P_j|+|\otimes_k R_k|} q^{-\frac{1}{2}(\kappa_{\otimes_j P_j} + \kappa_{\otimes_k R_k})} \Big]\cdot \nonumber \\
& & \cdot s_{P_1}(a_1)\ldots s_{P_m}(a_m)\cdot s_{R_1}(b_1)\ldots s_{R_n}(b_n), \nonumber 
\end{eqnarray}
where factors from strings stretched between branes on the same axis (\ref{stretched}) are taken out. 
The corresponding knots are shown in figure \ref{chain2}  for the case of four branes and three antibranes inserted in the geometry.

To summarise, crystal amplitudes we considered can be
interpreted in two distinct ways: either as generating function of Hopf-link invariants for representations with several rows (as described in section \ref{crystal}) --- or (as shown here from the topological vertex point of view), when expanded without the stretched string factors, as generating multiple Hopf-link invariants (involving tensor products)  with a single row representations, in knot framing $(-1,-1)$.

%********************************

\section{Mirror Symmetry and B-model example} \label{sec-B-model}

In this thesis we focus mainly on the A-model topological string theory. This is equivalent to the B-model theory via powerful \emph{Mirror Symmetry} which has been mentioned at the end of section \ref{sec-calabi-yau}. Even though we do not discuss Mirror Symmetry in any more detail, its following manifestations in topological string theory will be important to us:
\begin{itemize}
\item A-model partition function on some Calabi-Yau manifold is equal to B-model partition function on a mirror manifold, 
\item under Mirror Symmetry A-branes on some Calabi-Yau manifold are mapped to B-branes on a mirror manifold.
\end{itemize}

We now discuss briefly how Mirror Symmetry is realised for toric manifolds and in the topological vertex theory \cite{B-vertex}. Let us consider a two-dimensional toric diagram for some particular toric three-fold. It can be shown a structure of a mirror manifold is encoded in a Riemann surface which is also related to this diagram. More precisely, if edges of the toric diagram are thickened, we obtain a Riemann surface which satisfies equation of the form
\be
H(u,v) = 0, \qquad\qquad \textrm{for}\quad (u,v)\in\C^2,    \label{B-surface}
\ee
and the mirror manifold is given as a subset of $\C^4$ described by the equation
\be
xy-H(u,v) =0,   \qquad\qquad \textrm{for}\quad (x,y,u,v)\in\C^4.   \label{B-cy}
\ee
For a generic point $(u_*,v_*)$ such that $H(u_*,v_*)\neq 0$ there is a cylinder over this point in remaining two complex dimensions, described by the equation $xy=H(u_*,v_*)=const$. A Riemann surface (\ref{B-surface}) is a locus where $xy=0$ and over this surface cylinders degenerate to pairs of intersecting lines $x=0$ or $y=0$ in $\C^2$. 

\begin{figure}
\begin{center}
\includegraphics[width=0.4\columnwidth]{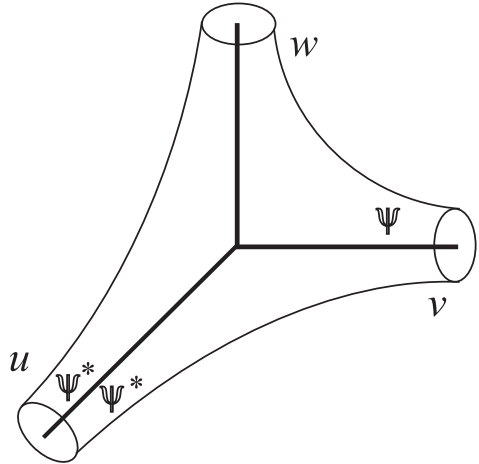}
\end{center}
\caption{\label{Bmodelpsi} Riemann surface representing B-model geometry mirror to $\C^3$ obtained from thickening edges of the trivalent A-model vertex, with two anti-branes and one brane inserted in two asymptotic patches.}
\end{figure}

As an example, a toric diagram for $\C^3$ is given by a trivalent vertex, and thickening its edges we obtain a Riemann surface shown in figure \ref{Bmodelpsi}, described by the equation
$$
H(u,v) = e^u + e^v + 1 = 0.
$$

%\begin{figure}
%\begin{center}
%\includegraphics[width=0.3\columnwidth]{B-vertex.eps}
%\end{center}
%\caption{Riemann surface representing B-model geometry mirror to $\C^3$ obtained from thickening edges of the trivalent A-model vertex.}  \label{fig-B-vertex}
%\end{figure}

Similarly as Chern-Simons theory is an effective theory for A-model open topological strings, there is also an effective theory for the B-model topological strings. This is called Kodaira-Spencer theory of gravity \cite{bcov}. Its excitations describe deformations of the complex structure of the target space manifold and can be encoded in variations of holomorphic three-form $\Omega$ given in (\ref{cy-omega}). For manifolds of the form (\ref{B-cy}) these excitations reduce to the Kodaira-Spencer scalar field $\phi(v)$ which lives on a Riemann surface (\ref{B-surface}) such that 
\be
\partial\phi(v) = u(v).   \label{KS-scalar}
\ee
Consistency of Kodaira-Spencer theory leads to the so-called \emph{W-constraints} which allow to find amplitudes for B-model theory in terms of expectation values of the scalar field $\phi$. In particular B-model partition functions found in this way agree with A-model partition functions on mirror manifolds.

As mentioned in section \ref{sec-open-top}, while A-branes in A-model theory wrap lagrangian cycles, the B-branes in B-model theory have to wrap complex submanifolds which have necessarily even real dimension. There is a special class of such branes in the geometry (\ref{B-cy}) which will be of particular interests to us. These branes are fixed to the Riemann surface $H(u,v)=0$ and they just wrap one of the complex lines $x=0$ or $y=0$ in remaining $\C^2$. Such a B-brane can move around the Riemann surface and its position on this surface is its modulus. These B-branes have a remarkable property: they are described by a two-dimensional free fermion field $\psi$ which is related to the scalar $\phi$ by usual bosonisation
\be
\psi = e^{\phi(v)/g_s},    \label{B-brane}
\ee
and similarly anti-branes are represented by a conjugated field $\psi^*$. However, these fermions are not standard fermions, which manifests in the fact how they transform between different patches on the Riemann surface. Instead of usual transformation rule as a half-differential, they transform according to some particular Fourier transformation. Nonetheless, W-constraints mentioned above also allow to find these transformation properties and compute expectation values of these fermions. 

Let us consider a configuration of branes in two asymptotic patches, at positions $e^{-u_i}=a_i$ in one patch and $b_j$ in the other patch. Fermionic B-brane operators (\ref{B-brane}) have a standard mode expansions (\ref{fermion-NS}) in these two patches
\bea
\psi(a) & = & \sum_k \psi_{k+1/2} \, e^{-(k+1)u_i} =  
\sum_k \psi_{k+1/2} a^{k+1}, \nn
\tilde{\psi}(b) & = & \sum_k \tilde{\psi}_{k+1/2} b^{k+1}, \nonumber
\eea
and only fermions from the same patch anticommute 
$$
\{ \psi_{-k-1/2},\,\, \psi^*_{l+1/2} \} = \delta_{k l},
$$
whereas the bare vacuum is annihilated by all positive modes (\ref{fermion-vac}) 
$$
\psi_{k+1/2}|0\rangle = \psi^{*}_{k+1/2}|0\rangle = \tilde{\psi}_{k+1/2}|0\rangle = \tilde{\psi}^{*}_{k+1/2}|0\rangle = 0 \qquad \textrm{for}\ k\geq 0.
$$

It can be shown that topological string partition functions with branes inserted in two patches at positions specified above are indeed reproduced by expectation values of these fermionic operators
\be
Z^{branes\ at\ a_i,\ b_i} = \langle 0 | \prod_i \psi(a_i)\, \prod_j \psi(b_j) | V\rangle,    \label{B-amplitude}
\ee
where $\langle 0 |$ is a standard vacuum state representing one patch of a Riemann surface, whereas $| V\rangle$ a non-trivial vacuum representing another patch. The state $| V\rangle$ encodes information how these patches are glued into entire surface, and it turns out to be given simply by a Bogoliubov transformation of the standard vacuum 
$$
|V\rangle = \exp \sum_{k,l\geq 0} \Big(a_{kl}\psi_{-k-1/2}\psi^{*}_{-l-1/2} + \tilde{a}_{kl}\psi_{-k-1/2}\tilde{\psi}^{*}_{-l-1/2}    \Big) |0\rangle,         \label{V}
$$
with all details of the surface encoded in coefficients $a_{kl}$ and $\tilde{a}_{kl}$ corresponding to two patches. 

In the case of $\mathbb{C}^3$ the state $|V\rangle$ is determined (up to $q^{1/6}$ factors) by
\begin{eqnarray}
a_{kl} & = & (-1)^l s_{hook(k+1,l+1)}(q^{\rho}), \nn
\tilde{a}_{kl} & = & (-1)^l q^{-\frac{\kappa_{(l+1)}}{2}} (W_{k+1,l+1} - W_{k+1}W_{l+1}), \nonumber
\end{eqnarray}
where $hook(m+1,n+1)$ is a hook representation with the relevant number of boxes in its row and column, and $W_{k+1,l+1}$ is Hopf-link invariant for two symmetric representations with relevant number of boxes. For symmetric representation, the value of Casimir is $\kappa_n = n^2 -n$.

Now we wish to analyse the configuration of B-branes shown in figure \ref{Bmodelpsi}. We insert two antibranes in one patch (these in framing $(-1)$) and a single brane in the other one. In fact, instead of $\langle 0|$, we should choose $\langle vac| = \langle 0| \tilde{\psi}_{1/2}$ vacuum (\ref{B-amplitude}) to ensure vanishing of the total fermion flux. In this case the only contribution comes from the third coefficient (with $1/2$ factor) in the exponent expansion of $|V\rangle$. Manipulations with fermion operators lead to \begin{eqnarray}
& & \langle vac|  \tilde{\psi}(b) \psi^*(a_1) \psi^*(a_2) |V^{(-1,0)}\rangle = \nonumber \\
& &  = \sum_{p,t,r \geq 0} \tilde{a}_{pt} \tilde{a}_{r0} \,b^{t+1}\big(a_1^{r+1}a_2^{p+1}- a_1^{p+1}a_2^{r+1} \big) (-1)^{-p-r}q^{-\frac{\kappa_{p+1}+\kappa_{r+1}}{2}}. \label{2Dbar-D-I} \nonumber
\end{eqnarray}
Performing the summation gives
$$
\langle vac|  \tilde{\psi}(b) \psi^*(a_1) \psi^*(a_2) |V^{(-1,0)}\rangle =  \frac{-a_1 a_2 b}{L(a_1,q)L(a_2,q)L(b,q)} \Big(\frac{1}{1-a_2 b} - \frac{1}{1-a_1 b}\Big)
$$
$$
= \frac{a_1 a_2^2 b^2}{L(a_1,q)L(a_2,q)L(b,q)} \frac{1-\frac{a_1}{a_2}}{(1-a_1 b)(1-a_2 b)}.
$$

We already know that inverting $q$ exchanges branes with antibranes. So if we started with two branes in the first patch and antibrane in the second, we would get dilogarithms in numerator. This agrees with the crystal result (\ref{nmbranes}) up to the irrelevant overall $a_1 a_2^2 b^2$ factor.

%*********************************************
%*********************************************

\section{Closed topological vertex and 'off-strip' geometries} \label{s-vertex}

In this section we reconsider the closed topological vertex geometry and prove its partition function is indeed equal to the generating function of the cube crystal (\ref{Z1Z2}). We derive this partition function using the topological vertex formalism. As we explain in section \ref{move-off} this derivation can be understood as 'moving off the strip' in the terminology of \cite{strip}, and  offers a possibility to simplify topological vertex techniques to a broader class of Calabi-Yau geometries. To support this claim we also consider a flop transition of the closed topological vertex and compute the partition function of the resulting geometry in section \ref{subsec-flop}.

\subsection{Derivation}

Now we derive the partition function (\ref{closed-ver}) for the closed topological vertex geometry from the topological vertex formalism, in a way which gives this result in the form explicitly related to (\ref{Z1Z2}). In computation we use various identities given in appendices \ref{app-schur} and \ref{app-vertex}. In fact the result can be obtained much faster, if certain sums in the amplitude are automatically performed using additional machinery of \cite{strip}, which we also review in appendix \ref{app-strip}. It turns out the derivation presented below can be understood as an extension of that machinery to a more general situation. Nonetheless, for completeness we first derive the closed topological vertex partition function from first principles. 

The basic topological vertex amplitude is given in (\ref{vertex}). It will turn out convenient to use cyclic symmetry to write the amplitude for the closed topological vertex from figure \ref{fig-closed-ver} as
\begin{eqnarray}
Z^{\mathcal{C}} & = & \sum_{R_1,R_2,R_3} C_{R_2 R_3 R_1} C_{R_1^t\bullet\bullet} C_{R_2^t\bullet\bullet} C_{R_3^t\bullet\bullet} (-Q_1)^{|R_1|}  (-Q_2)^{|R_2|} (-Q_3)^{|R_3|}  = \nonumber \\
%& = & \sum_{R_1,R_2,R_3,P} s_{R_3^t}(q^{\rho}) s_{R_2/P}(q^{R_3^t+\rho}) s_{R_1^t/P}(q^{R_3+\rho}) \nonumber \\
%& & \qquad \qquad s_{R_1}(-Q_1 q^{\rho}) s_{R_2^t}(-Q_2 q^{\rho}) s_{R_3}(-Q_3 q^{\rho}) = \nonumber \\
& = & \sum_{R_1,R_2,R_3,P} \Big[s_{P}(-Q_1 q^{\rho}) s_{P}(-Q_2 q^{\rho})\Big]\Big[ s_{R_3}(q^{\rho}) s_{R_3^t}(-Q_3 q^{\rho})  \Big] \nonumber \\ %\label{cl-vert-compute} 
& & \qquad \qquad \Big[ s_{R_2}(q^{R_3^t+\rho}) s_{R_2^t}(-Q_2 q^{\rho}) \Big] \Big[s_{R_1^t}(q^{R_3+\rho}) s_{R_1}(-Q_1 q^{\rho}) \Big].  \nonumber
\end{eqnarray}

%Let us now recall (as derived in more detail in \cite{strip}) that the relation (\ref{schur-sum-exp}) implies the following one
%\begin{equation}
%\sum_{P} s_{P}(q^{R+\rho}) s_{P^t}(-Q q^{\rho}) = \exp\Big[-\sum_n \frac{Q^n}{n[n]^2} \Big] \, \prod_k (1-Qq^k)^{C_k(R)},
%\end{equation}
%where coefficients $C_k(R)$ count boxes $(i,j)\in R$ with fixed $k=j-i$ and can be equivalently defined by
%\begin{equation}
%s_{[]}(q^{R+\rho}) s_{[]}(q^{\rho}) = \sum_k C_k(R) q^k + \frac{q}{(1-q)^2}.
%\end{equation}

This can be rewritten using (\ref{sum-schur-RRbis}) in the form
$$
Z^{\mathcal{C}} = \exp \Big[ \sum_n \frac{-Q_1^k -Q_2^k + Q_1^k Q_2^k}{n[n]^2} \Big] \times
$$
\begin{equation}
\times \sum_{R_3} s_{R_3}(q^{\rho}) s_{R_3^t}(q^{\rho}) (-Q_3)^{|R_3|} \prod_k (1-Q_1 q^k)^{C_k(R_3)} (1-Q_2 q^k)^{C_k(R_3^t)}. \label{cl-vert-compute-2}
\end{equation}
The exponent factor which arises above is equal to $Z_1$ (\ref{Z1-exp}), which is double-$\mathbb{P}^1$ partition function (up to the McMahon function). Thus we have to show the sum over $R_3$ above reproduces $Z_2$ (\ref{Z2-exp}). We use (\ref{schur-hooks})  to rewrite (\ref{cl-vert-compute-2}) as 
\begin{equation}
Z^{\mathcal{C}} = \frac{Z_1}{M(q)} \sum_{R_3} (-Q_3)^{|R_3|} X_{R_3} Y_{R_3},
\end{equation}
where
\begin{eqnarray}
X_{R_3} & = & s_{R_3^t}(q^{\rho}) \prod_k (1-Q_1 q^k)^{C_k(R_3)} =  \nonumber \\
& = & (-1)^{|R_3|} q^{|R_3|/2+n(R_3)} \prod_{(i,j)\in R_3} \frac{1-q^{M+j-i}}{1-q^{h(i,j)}}
\end{eqnarray}
and similarly
$$
Y_{R_3} = (-1)^{|R_3|} q^{|R_3^t|/2+n(R_3^t)} \prod_{(i,j)\in R_3^t} \frac{1-q^{L+j-i}}{1-q^{h(i,j)}}
$$
where we used identification (\ref{Q123-MLN}).

Finally, the crucial step is to rewrite $X_{R_3}$ and $Y_{R_3}$ using the identity (\ref{schur-finite}) for a Schur function with finite number of variables
\begin{eqnarray}
Z^{\mathcal{C}} & = & \frac{Z_1}{M(q)} \sum_{R_3} (-q Q_3)^{|R_3|} s_{R_3}(1,q,q^2,\ldots,q^{M-1}) s_{R_3^t}(1,q,q^2,\ldots,q^{L-1}) \nonumber \\
& = & \frac{Z_1}{M(q)} \prod_{i=1,\ldots L;\, j=1, \ldots, M} (1-Q_3 q^{j+i-1}) = \frac{Z_1 Z_2}{M(q)} \label{cl-vert-compute-3}
\end{eqnarray}
where the sum of Schur functions (\ref{schur-sum4}) was used in the last line. Because $Z_1 Z_2 = Z^{cube}$, we indeed obtain (\ref{claim})
$$
Z^{cube} = M(q) Z^{\mathcal{C}}.
$$

In the above derivation the same factors as in the crystal model arise, i.e. $Z_1$ associated with double-$\mathbb{P}^1$ partition function and $Z_2$ 'implementing' the third $\mathbb{P}^1$. In this form it also becomes a trivial observation that in $N \to \infty$ limit the proper result for double-$\mathbb{P}^1$ is recovered (\ref{crystal-P1P1}). 

Let us also remark on McMahon $M(q)$ factors. It is known the topological vertex computations do not produce this contribution, so to obtain the full topological string partition function (\ref{Z-top}) one should introduce one factor of $M(q)$ for each topological vertex $C_{PQR}$ used in the vertex computation of the amplitude \cite{foam}. On the other hand, in our crystal model just a single factor of $M(q)$ arises as the plane partitions it counts are anchored in one particular corner of the cube. This is why there is one factor of $M(q)$ in (\ref{claim}). 

%*************************************

\subsection{Moving off the strip}   \label{move-off}

It is interesting to relate this result to the formalism developed in \cite{strip} which allows to perform computations for toric geometries whose dual diagrams are given by a triangulation of a rectangle (or a 'strip' - hence the terminology). We briefly review this machinery in appendix \ref{app-strip}. A simple example of a geometry of this type is the double-$\mathbb{P}^1$ from figure \ref{fig-double-P1}, whose dual diagram is undoubtedly a triangulation of a strip.

\begin{figure}[htb]
\begin{center}
\includegraphics[width=0.4\textwidth]{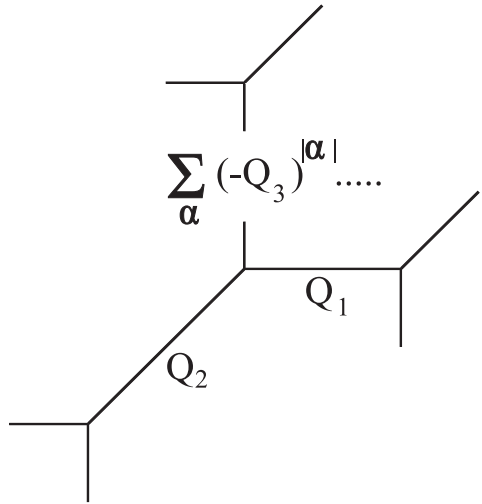}
\caption{The closed topological vertex as a strip with an additional $\mathbb{P}^1$ (of the size determined by $Q_3$) attached.} \label{fig-closed-ver-glue}
\end{center}
\end{figure}

Let us notice that the toric diagram for the closed topological vertex can be understood as a double-$\mathbb{P}^1$ with one additional sphere attached. This is shown in figure \ref{fig-closed-ver-glue}, with double-$\mathbb{P}^1$ given by spheres $Q_1-Q_2$ and the additional sphere denoted by $Q_3$. Even though the full diagram for the closed topological vertex cannot be drawn on a strip, a diagram for the double-$\mathbb{P}^1$ part can as presented above. Thus the partition function $Z^{\mathcal{C}}$ can be written as
$$
Z^{\mathcal{C}} = \sum_{\alpha} s_{\alpha^t}(q^{\rho}) (-Q_3)^{|\alpha|} Z_{\alpha},
$$
where $Z_{\alpha}$ is the factor for a double-$\mathbb{P}^1$ with one nontrivial representation, and is derived in appendix (\ref{Zalpha})
\begin{eqnarray}
Z_{\alpha} & = & s_{\alpha}\, \{\bullet \alpha \}_{Q_1} [\bullet \bullet ]_{Q_1 Q_2} \{\alpha^t \bullet \}_{Q_2} = \nonumber \\
& = & \frac{Z_1}{M(q)} \, s_{\alpha}(q^{\rho}) \prod_k (1-Q_1 q^k)^{C_k(\alpha)} (1-Q_2 q^k)^{C_k(\alpha^t)},
\end{eqnarray}
which altogether reproduces (\ref{cl-vert-compute-2}), and now the calculation continues as above and reproduces (\ref{cl-vert-compute-3}).

So, in this way we managed to 'move off the strip', which technically boils down to performing a sum of Schur functions with finite number of arguments (\ref{cl-vert-compute-3}). It thus seems likely this result might be generalised to a broader class of non-strip-like Calabi-Yau manifolds, and is interesting to pursue further. Below we consider an example of another 'off-strip' geometry related by a flop to the closed vertex, for which a partition function can also be derived using these methods.

%*******************************************************
%*******************************************************

\subsection{Flop transition} \label{subsec-flop}

The closed topological vertex consists of three $\mathbb{P}^1$'s with local bundles isomorphic to the resolved conifold. As is well known, such a bundle may undergo a flop transition. For the resolved conifold with K\"ahler parameter $t$ and $Q=e^{-t}$, a flop may be understood as a continuation to negative values of $t$. The partition function is invariant under this process and this should be seen order by order in genus expansion. To get a partition function of the flopped geometry, after the analytic continuation to negative $t$ one should expand the result again in positive powers of $Q$. For the resolved conifold the geometry before and after the transition is the same, which allows to fix the polynomial dependence of the free energy on $t$ \cite{G-V-transition}. In the case of the closed topological vertex the geometries before and after the flop are different, but it is possible to determine classical contribution to the free energy of $\mathcal{C}^{flop}$ in terms of those of $\mathcal{C}$ as we discuss below. Moreover, the invariance of the partition function under the flop implies in particular the Gopakumar-Vafa invariants should not change during the transition, providing the parameters on both sides of the transition are matched appropriately; such a behaviour indeed follows in general from the topological vertex rules as shown in \cite{flop}, and the calculation below proves the consistency of our method with these results.
 
\begin{figure}[htb]
\begin{center}
\includegraphics[width=\textwidth]{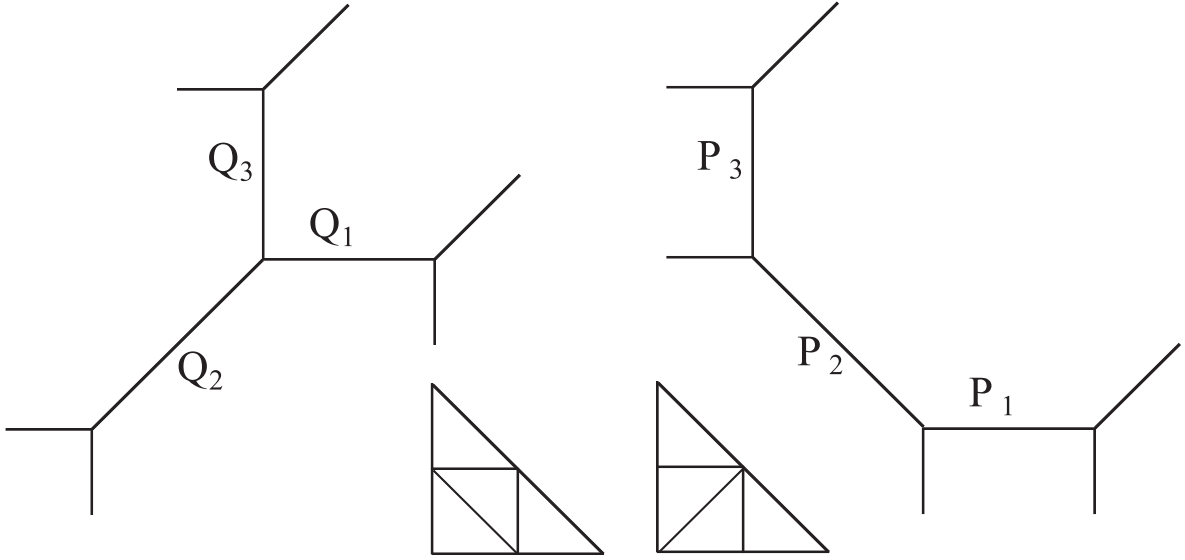}
\caption{The closed topological vertex $\mathcal{C}$ with its dual diagram (left) and the geometry after the flop $\mathcal{C}^{flop}$ (right).} \label{fig-flop-vertex}
\end{center}
\end{figure}

Let us focus on the closed topological vertex geometry, and the transition under which the conifold associated to $Q_2$ is flopped. We call the ensuing geometry $\mathcal{C}^{flop}$. The transition is presented in figure \ref{fig-flop-vertex}, and it is best understood in terms of a dual graph --- it is then represented by a tilt of a diagonal of a square corresponding to the conifold. The closed vertex on the left consists of three spheres meeting in one point, and after the flop it is replaced by a string of $\mathbb{P}^1$'s with two meeting points, and with a proper arrangement of bundles, as on the right side of the figure. We denote the parameters of the flopped geometry by $P_1, P_2, P_3$, and also express them in units of $g_s$ as
\begin{equation}
P_1 = q^{M_f}=e^{-s_1},\qquad P_2 = q^{L_f}=e^{-s_2}, \qquad P_3 = q^{N_f}=e^{-s_3}.  \label{P123-MLNf}
\end{equation}

How should the partition function for the flopped closed vertex look like after the transition? As mentioned above, it should be related to (\ref{closed-ver}) by a suitable identification of target space parameters. The geometry of the bundles in figure \ref{fig-flop-vertex} suggests the following relations
\begin{eqnarray}
Q_1 Q_2 & = &  P_1, \nonumber \\
Q_2  & = &  \frac{1}{P_2}, \label{t-s}  \\ 
Q_2 Q_3 & = &  P_3, \nonumber
\end{eqnarray}
from which it follows that $Q_1=P_1 P_2,\ Q_3=P_2 P_3,\ Q_1 Q_2 Q_3 =P_1 P_2 P_3, \ Q_1 Q_3 = P_1 P_2^2 P_3$. Substituting these into (\ref{closed-ver}) the only terms with negative powers we obtain are $P_2^{-n}$. Upon analytic continuation these should turn into $P_2^n$ together with appropriate change in the classical part of the free energy, as is the case for the conifold. Thus we expect the following instanton contribution to the partition function of $\mathcal{C}^{flop}$
\begin{equation}
Z^{\mathcal{C}^{flop}} = \exp\sum_{n>0}\frac{-P_1^n P_2^n - P_2^n - P_2^n P_3^n + P_1^n + P_1^n P_2^{2n} P_3^n + P_3^n - P_1^n P_2^n P_3^n}{n[n]^2}, \label{flop-ver}
\end{equation}
We first show by explicit calculation this is the correct result, and afterwords analyse classical contributions.

\begin{figure}[hbt]
\begin{center}
\includegraphics[width=0.4\textwidth]{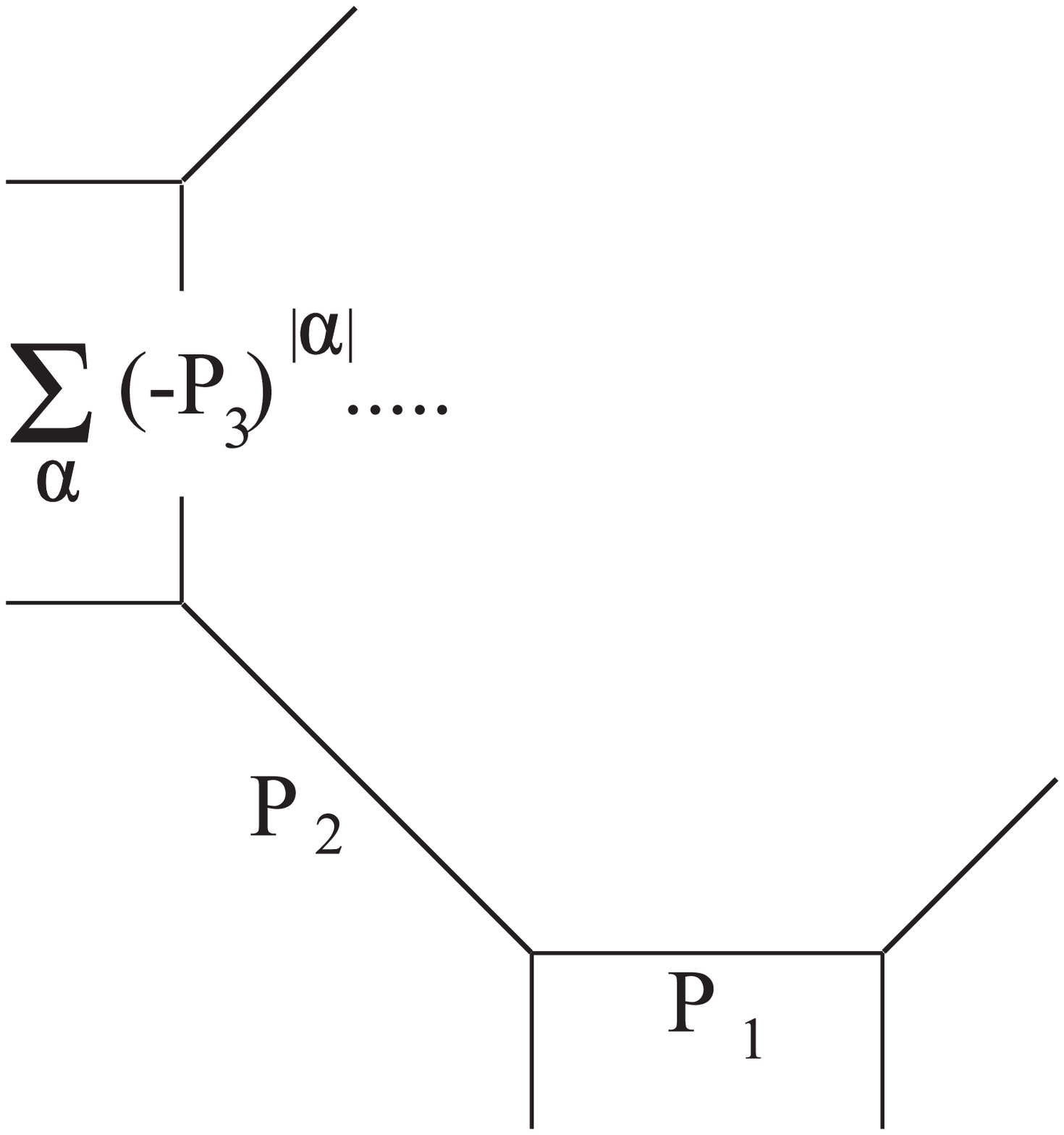}
\caption{The flopped closed topological vertex $\mathcal{C}^{flop}$ as a strip with attached $\mathbb{P}^1$ of the size determined by $P_3$.} \label{fig-flop-vertex-glue}
\end{center}
\end{figure}

To prove that (\ref{flop-ver}) is indeed correct we again use topological vertex rules. Similarly as for the closed vertex, we consider the flopped geometry as a strip with an additional $\mathbb{P}^1$ attached
$$
Z^{\mathcal{C}^{flop}} = \sum_{\alpha}  s_{\alpha^t}(q^{\rho}) (-P_3)^{|\alpha|} Z^{flop}_{\alpha} \, \Big[ (-1)^{|\alpha|} q^{-\kappa_{\alpha}/2} \Big],
$$
the factors in square brackets originating from a nontrivial framing of the additional sphere. This is presented in figure \ref{fig-flop-vertex-glue}, with $Z^{flop}_{\alpha}$ corresponding to the amplitude on the strip corresponding to a string of spheres $P_2-P_1$ 
\begin{eqnarray}
Z^{flop}_{\alpha} & = &  s_{\alpha}\, \{\alpha \bullet \}_{P_2} \{\alpha \bullet \}_{P_1 P_2} [\bullet \bullet ]_{P_1} = \nonumber \\
& = & e^{ \sum_n \frac{-P_2^n - P_1^n P_2^n + P_1^n}{n[n]^2} } \, s_{\alpha}(q^{\rho}) \, \prod_k (1-P_2 q^k)^{C_k(\alpha)} (1-P_1 P_2 q^k)^{C_k(\alpha)}, \nonumber
\end{eqnarray}
where we again used rules from the appendix \ref{app-strip} (now we read vertices from left to right, and they are of the types $A_{\alpha}-B-B$). Now we use (\ref{schur-hooks}) to write the full amplitude as
$$
Z^{\mathcal{C}^{flop}} = e^{ \sum_n \frac{-P_2^n - P_1^n P_2^n + P_1^n}{n[n]^2} } 
\sum_{\alpha} (q P_3)^{|\alpha|}  q^{2n(\alpha)}\prod_{(i,j)\in \alpha} \frac{1-P_2\, q^{j-i}}{1-q^{h(i,j)}}\, \frac{1-P_1 P_2\, q^{j-i}}{1-q^{h(i,j)}}.
$$
Finally, after identification (\ref{P123-MLNf}), using equality (\ref{schur-finite}) and summing over $\alpha$ we get
$$
Z^{\mathcal{C}^{flop}} = e^{ \sum_n \frac{-P_2^n - P_1^n P_2^n + P_1^n}{n[n]^2} } e^{ \sum_n \frac{P_3^n - P_2^n P_3^n - P_1^n P_2^n P_3^n + P_1^n P_2^{2n} P_3^n }{n[n]^2} }, 
$$
which is indeed the same as the expected result (\ref{flop-ver}) and consistent with \cite{flop}. 

\begin{figure}[htb]
\begin{center}
\includegraphics[width=0.8\textwidth]{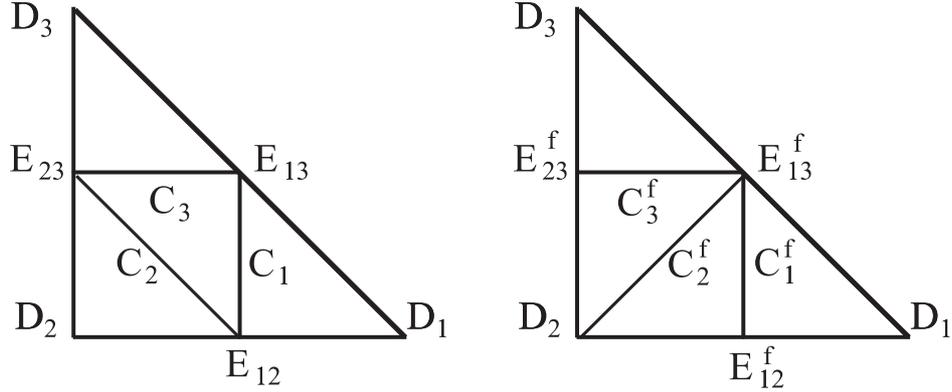}
\caption {Two geometries $\mathcal{C}$ and $\mathcal{C}^{flop}$ as a symmetric (left) and asymmetric (right) resolution of $\mathbb{C}^3/\mathbb{Z}_2 \times \mathbb{Z}_2$.} \label{fig-homologies}
\end{center}
\end{figure}

Let us finally turn to the issue of polynomial contributions. As mentioned above, those for $\mathcal{C}^{flop}$ and $\mathcal{C}$ are related to each other due to the invariance of the full partition function. We will show this relation is consistent with the values of cubic intersection numbers. These intersection numbers can be derived from the description of homology classes given in \cite{F-theory}, where both geometries are discussed as different resolutions of $\mathbb{C}^3/\mathbb{Z}_2\times\mathbb{Z}_2$ orbifold: the closed topological vertex is a symmetric resolution, and its flop an asymmetric resolution. The homology structure is in fact encoded in the dual diagram, with vertices corresponding to divisors and internal intervals to compact curves (each one arising as an intersection of two divisors at the end of the interval), as shown in figure \ref{fig-homologies}. There are three divisors $D_i$, $i=1,2,3$, in the singular orbifold, and additional three: $E_{ij}$ in its symmetric resolution, or $E_{ij}^{f}$ in asymmetric resolution, for $i\neq j$. The compact curves are the familiar by now three $\mathbb{P}^1$'s: $C_i$ with sizes given by $t_i$ for the closed vertex geometry and $C_i^f$ with sizes $s_i$ for its flop. For completeness, let us recall the intersection numbers derived in \cite{F-theory}. For the closed topological vertex these are
\begin{center}
\begin{tabular}{c|c c c c c c}
   & $D_1$ & $D_2$ & $D_3$   & $E_{23}$ & $E_{13}$ & $E_{12}$ \\
\hline
$C_1 = E_{12}\cap E_{13}$ & $1$ & $0$ & $0$ & $1$ & $-1$ & $-1$ \\
$C_2 = E_{12}\cap E_{23}$ & $0$ & $1$ & $0$ & $-1$ & $1$ & $-1$ \\
$C_3 = E_{13}\cap E_{23}$ & $0$ & $0$ & $1$ & $-1$ & $-1$ & $1$
\end{tabular}
\end{center}
whereas for its flop
\begin{center}
\begin{tabular}{c|c c c c c c}
   & $D_1$ & $D_2$ & $D_3$   & $E^f_{23}$ & $E^f_{13}$ & $E^f_{12}$ \\
\hline
$C^f_1 = E^f_{12}\cap E^f_{13}$ & $1$ & $1$ & $0$ & $0$ & $0$ & $-2$ \\
$C^f_2 = E^f_{12}\cap E^f_{23}$ & $0$ & $-1$ & $0$ & $1$ & $-1$ & $1$ \\
$C^f_3 = E^f_{13}\cap E^f_{23}$ & $0$ & $1$ & $1$ & $-2$ & $0$ & $0$
\end{tabular}
\end{center}

>From these intersection numbers we can deduce the form of genus zero prepotentials $F_0$ in terms of sizes of $\mathbb{P}^1$'s. In general
$$
F_0 = \frac{1}{6} J^3,
$$
with $J$ the Poincare dual to the K\"ahler form. For the closed vertex geometry $\mathcal{C}$ it can be parametrised by $e_i$ as
$$
J^{\mathcal{C}} = e_1 E_{12} + e_2 E_{13} + e_3 E_{23},
$$
so that the sizes of $\mathbb{P}^1$'s are given by
\begin{eqnarray}
t_1 & = & J^{\mathcal{C}} \cap C_1 = -e_1 -e_2 +e_3, \nonumber \\
t_2 & = & J^{\mathcal{C}} \cap C_2 = -e_1 +e_2 -e_3, \nonumber \\
t_3 & = & J^{\mathcal{C}} \cap C_3 =  e_1 -e_2 -e_3. \nonumber
\end{eqnarray}
Expressing $e_i$ in terms of $t_i$ and combining with genus one contributions (which depend on a single parameter $b$ due to symmetry in the geometry), the classical free energy (\ref{F-class}) for the closed topological vertex reads
\begin{equation}
F^{\mathcal{C}}_{class} = \Big(\frac{t_1^3 + t_2^3 + t_3^3}{6g_s^2} + \frac{t_1^2 t_2 + t_1 t_2^2 + t_1^2 t_3 + t_3 t_1^2 + t_2^2 t_3 + t_3 t_2^2}{4g_s^2} + \frac{t_1 t_2 t_3}{2g_s^2} \Big)  + \frac{b}{24}(t_1+t_2+t_3). \label{F-class-C}
\end{equation}
Similarly, for the flopped geometry the K\"ahler form has the Poincare dual
$$
J^{\mathcal{C}^{flop}} = e^f_1 E^f_{12} + e^f_2 E^f_{13} + e^f_3 E^f_{23},
$$
and the sizes of $\mathbb{P}^1$'s are
\begin{eqnarray}
s_1 & = & J^{\mathcal{C}^{flop}} \cap C^f_1 = -2e^f_1, \nonumber \\
s_2 & = & J^{\mathcal{C}^{flop}} \cap C^f_2 = e^f_1 -e^f_2 +e^f_3, \nonumber \\
s_3 & = & J^{\mathcal{C}^{flop}} \cap C^f_3 = -2e^f_3. \nonumber
\end{eqnarray} 
Including genus one contributions (for symmetry reasons now depending on two parameters $c,d$), we get classical free energy
\begin{eqnarray}
F^{\mathcal{C}^{flop}}_{class} & = & \Big(\frac{s_1^3 + 2s_2^3 + s_3^3}{6g_s^2} + \frac{2s_1^2 s_2 + 2s_1 s_2^2 + s_1^2 s_3 + s_3 s_1^2 + 2s_2^2 s_3 + 2s_3 s_2^2}{4g_s^2} + \frac{s_1 s_2 s_3}{2g_s^2} \Big)  + \nonumber \\
& + & \frac{c(s_1+s_3) + d s_2}{24} = \frac{t_1^3 + t_3^3}{6g_s^2} + \frac{t_1^2 t_2 + t_1 t_2^2 + t_1^2 t_3 + t_3 t_1^2 + t_2^2 t_3 + t_3 t_2^2}{4g_s^2} + \frac{t_1 t_2 t_3}{2g_s^2}  + \nonumber \\
& + & \frac{c(t_1+t_3)+t_2(2c-d)}{24}, \label{F-class-Cflop}
\end{eqnarray}
where we used the identification (\ref{t-s}) which equivalently reads $s_1=t_1+t_2,\ s_2=-t_2,\ s_3 = t_3+t_2$. Finally these classical terms can be combined with quantum ones $F^{\mathcal{C}}=\log Z^{\mathcal{C}}$ and $F^{\mathcal{C}^{flop}}=\log Z^{\mathcal{C}^{flop}}$ given in (\ref{closed-ver}) and (\ref{flop-ver}). Indeed, the full partition function is now explicitly seen to be invariant under the flop
$$
F^{\mathcal{C}}_{class}(t_1,t_2,t_3) + F^{\mathcal{C}}(t_1,t_2,t_3) = F^{\mathcal{C}^{flop}}_{class}(s_1,s_2,s_3) + F^{\mathcal{C}^{flop}}(s_1,s_2,s_3), %\label{flop-inv}
$$
under the identification between $t_i$ and $s_i$ (\ref{t-s}), and providing $c=d=b$. There are two important remarks to be made. Firstly, quantum genus zero contributions for $\mathcal{C}^{flop}$, given by trilogarithm (\ref{F-conifold}), are continued to negative $t_2$ using $Li_3 e^{t_2} - Li_3 e^{-t_2} \sim t_2^3/6$ (where we keep only a cubic term, the other being ambiguous for topological string). This continuation is precisely the origin of the well known shift in classical intersection numbers under the flop (in our case this is seen explicitly in expressions (\ref{F-class-C}) and (\ref{F-class-Cflop})). Secondly, the condition $c=d=b$ which must be enforced, is just a statement that the genus one classical part after the flop is entirely determined by the geometry before the flop.

%***********************************************************
%***********************************************************

\appendix

%\chapter*{Appendices}
%\addcontentsline{toc}{chapter}{Appendices}

\chapter{Starring} \label{app-star}

\section{Two-dimensional partitions}  \label{sec-partitions}

A \emph{two-dimensional partition} $R=(R_1,R_2,\ldots,R_l)$ is a set of non-increasing positive integers, $R_1\geq R_2\geq\ldots\geq R_l \geq 0$. $l=l(R)$ is called a \emph{length} of partition $R$. Sometimes it is convenient to understand a partition as an infinite sequence, such that $R_{l+1} = R_{l+2}=\ldots = 0$. A partition can be presented in a form of a Young diagram, which is a tableaux of $l$ rows of boxes, with $R_i$ boxes in $i$'th row. We often identify a partition with its diagram, and denote it usually by letters $P,Q,R$ or Greek ones $\alpha,\beta,\mu,\nu,\eta$. A \emph{dual} (or \emph{transposed}) partition is given by a Young diagram $R^t$ which arises from a transposition of the Young diagram corresponding to $R$.

A partition can also be presented in the so-called Frobenius notation. Let $d(R)$ denote the number of boxes on a diagonal of a Young diagram of $R$. Then
\be
R = \left( \begin{array}{cccc} a_1 & a_2 & \ldots & a_{d(R)} \\
b_1 & b_2 & \ldots & b_{d(R)} \\
\end{array} \right)  \label{frobenius}
\ee
where $a_i$ and $b_i$ have interpretation as distances from each diagonal element to the end of its row and column respectively and are given by
\be
a_i = R_i-i,\qquad b_i = R_i^t-i.  \label{frobenius-ab}
\ee
Sequences $(a_i)$ and $(b_i)$ are necessarily strictly decreasing. In Frobenius notation a transposition amounts to exchanging upper and lower elements in (\ref{frobenius}). Taking for example (in both standard and Frobenius notation)
$$
\Yvcentermath1 R=(5,4,2,1)=\left( \begin{array}{cc} 4 & 2 \\
3 & 1 \\
\end{array} \right) = \yng(5,4,2,1) 
$$
its transposition is
$$
\Yvcentermath1 R^t = (4,3,2,2,1)=\left( \begin{array}{cc} 3 & 1 \\
4 & 2 \\
\end{array} \right) = \yng(4,3,2,2,1)
$$

For a partition $R=(R_1,R_2,\ldots,R_l)$ we define its total number of boxes $|R|$ and other useful quantities as follows
\begin{eqnarray}
|R| & = &\sum_{i=1}^{l(R)} R_i = d + \sum_{i=1}^{d(R)} (a_i+b_i), \label{R-size}\\
||R||^2 & = & \sum_i R_i^2,\\ \label{R-double}
n(R) & = & \sum_{i} (i-1)R_i, \label{young-n}\\
{R \choose 2} & = & \sum_i {R_i \choose 2} = \frac{||R||^2}{2} - \frac{|R|}{2}, \label{young-choose}\\
\kappa_R & = & |R|+\sum_i R_i(R_i -2i) = 2{R \choose 2}- 2{R^t \choose 2} = -\kappa_{R^t}. \label{kappa}
\end{eqnarray}
One can also show
\be
\kappa_R = 2\sum_{\Box=(m,n)\in R} (n-m)
\ee
where $(m,n)$ denotes a position of a certain box $\Box$ in the diagram $R$, i.e. respectively its row and column. For such a single box, its hook length is defined as
\begin{equation}
h(\Box)=h(m,n)=R_m+R_{n}^{t}-m-n+1. \label{hook}
\end{equation}

Two partitions $P$ and $R$ \emph{interlace} one another $P\succ R$ if the following \emph{interlacing condition} is satisfied
\be
P\succ R \qquad \iff \qquad P_1\geq R_1 \geq P_2 \geq R_2 \geq P_3 \geq \ldots.  \label{interlace}
\ee

For any $n\in\mathbb{N}$ one can ask what is a number of partitions with such a number of boxes $p(n)$. These numbers are encoded in the generating function of all partitions which is equal to the inverse of the Euler function
\bea
\frac{1}{\varphi(q)} & = & \sum_{n=0}^{\infty} p(n)q^n = \sum_{R} q^{|R|} = \prod_{k=1}^{\infty} \frac{1}{1-q^k} = \frac{q^{1/24}}{\eta(q)} = \nonumber \\
& = & 1 + q + 2q^2 + 3q^3 + 5q^4 + 7q^5 + \ldots       \label{euler-fn}
\eea
The relation to $\eta(q)$ which has modular properties is an important factor in our considerations.

Sometimes we need to consider the following restricted sets of partitions. A set of Young diagrams with at most $l$ rows is denoted by $\cY_l$. A set of Young diagrams with at most $l$ rows and $k$ columns, i.e. those which fit into a rectangle of size $l\times k$, is denoted by $\cY_{l,k}$.

\subsection*{Coloured partitions}        \label{app-coloured}

A \emph{$N$-coloured partition} is a set of $N$ partitions $\vec{\bf{R}} = (R_{(1)},R_{(2)},\ldots,R_{(N)})$; each $R_{(i)}$ by itself is a usual partition: $R_{(i),1} \geq R_{(i),2} \geq \ldots \geq R_{(i),l(R_{i})} \geq 0$. A diagram of $N$-coloured partition is a set of $N$ diagrams corresponding to $R_{(i)}$'s. A size of $N$-coloured partition $\vec{\bf{R}} = \sum_{i,j} R_{(i),j} $ is equal to the total number of boxes in its diagram.

\section{Free fermion formalism}   \label{app-fermion}

Let us consider a complex fermion in the NS sector
\be
\psi(z) = \sum_{n\in\Z} \psi_{n+\hf} z^{-n-1},\qquad \psi^*(z) = \sum_{n\in\Z} \psi^*_{n+\hf} z^{-n-1},   \label{fermion-NS}
\ee
subject to anticommutation rules
$$
\{\psi_{n+\hf},\psi^*_{-m-\hf}\} = \delta_{m,n}.
$$
Particle annihilation and creation operators are $\psi^*_{n+\hf}$ with respectively $n\geq0$ and $n<0$. The vacuum state $|0\rangle$ is defined as
\be
\psi_{n+\hf} |0\rangle = 0,\qquad \psi^*_{n+\hf}|0\rangle = 0,\qquad \textrm{for}\ n\geq 0,  \label{fermion-vac}
\ee
and a basis of the total Fock space $\cF$ is given by states obtained by acting with creation operators on this vacuum. The space $\cF$ decomposes 
$$
\cF = \otimes_{p\in\Z}\, \cF_p
$$
into subspaces $\cF_p$ of fixed $U(1)$ charge with respect to the current
$$
J(z) = :\psi(z)\psi^*(z): = \sum_{k\in\Z} z^{-k-1} \sum_{n\in\Z} :\psi_{n+\hf}\psi^*_{k-n-\hf}: = \sum_{k\in\Z} \frac{J_k}{z^{k+1}}
$$
and vacua $|p\rangle$ with charge $p$ can be introduced
\bea
\psi_{n+\hf} |p\rangle & = & 0, \qquad \textrm{for}\ n\geq p, \nonumber \\
\psi^*_{m+\hf} |p\rangle & = & 0, \qquad \textrm{for} \ n\geq -p,
\eea
so that each subspace $\cF_p$ is generated from $|p\rangle$. 

There is a very interesting one-to-one correspondence between free fermion states and two-dimensional partitions. In $p=0$ sector the state
$$
|R\rangle = \prod_{i=1}^{d} \psi^*_{-a_i-\hf} \psi_{-b_i-\hf}|0\rangle
$$
corresponds to the partition $R=(R_1,\ldots,R_l)$ such that
$$
a_i = R_i-i,\qquad b_i = R_i^t-i,
$$
which are precisely the numbers which specify a partition in the Frobenius notation (\ref{frobenius}).

\begin{figure}[htb]
\begin{center}
\includegraphics[width=0.8\textwidth]{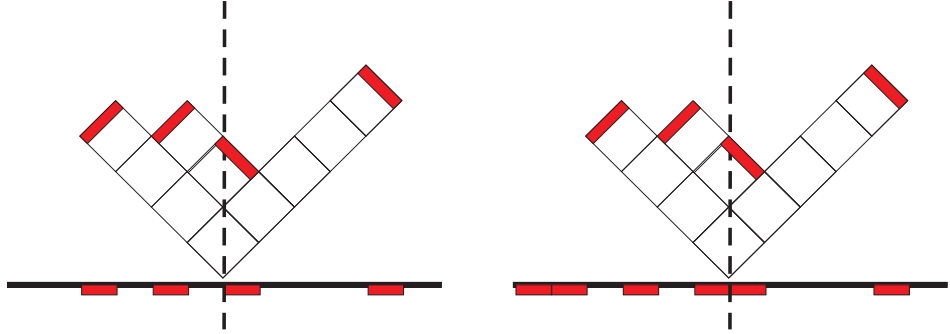}
\caption{A correspondence between partitions and states. Positions of particles and holes are given by projecting ends of rows and columns of a partition onto a Fermi sea. Particles are drawn in red, and according to one of the conventions used holes are drawn either in red (left), or just as holes (right). The partition drawn in the figure is $R=(5,2,2,1)$ and the corresponding state is $|R\rangle = \psi^*_{\hf} \psi_{-\frac{3}{2}} \psi^*_{-\frac{9}{2}}\psi_{-\frac{7}{2}} |0\rangle$.} \label{fig-fermi-part}
\end{center}
\end{figure}

It is easy to visualise this correspondence in terms of the Fermi sea. In particular the vacuum $|0\rangle$ is given by a Fermi sea with all negative states filled and it is mapped to the trivial partition $\bullet$. For a nontrivial partition one draws it with a corner fixed at the edge of the filled part of the Fermi sea. Then positions of particles and holes are read off by projecting ends of rows and columns of this partition onto the Fermi sea. There are then two conventions to draw such a state of the Fermi sea, as presented in figure \ref{fig-fermi-part}. 

For $p\neq 0$ one draws a partition with a corner fixed at position $p$ of the Fermi sea, and then reads off positions of particles and holes similarly as above, by projecting ends of rows and columns onto the Fermi sea. A state corresponding to partition $R$ of charge $p$ is denoted as $|p,R\rangle$.

A zero mode of Virasoro algebra is an important operator
$$
L_0 = \sum_{r\in\Z+\hf} r:\psi_{r}\psi^*_{-r}:
$$
States represented by partitions are its eigenstates with eigenvalues related to the number of boxes of a partition and a charge
\be
L_0 |p,R\rangle = \big(|R| + \frac{p^2}{2}\big)|p,R\rangle.   \label{L0}
\ee

%*******************************************************************

\subsection*{Bosonization}

Free fermion is related to a free boson system
\be
\phi(z) = \phi_0 -iJ_0\,\textrm{ln}\, z + i\sum_{k\neq 0} \frac{z^{-k}}{k}J_k,    \label{boson}
\ee
so that
\bea
i\partial \phi(z) & = & J(z), \nonumber \\
\psi(z) = :e^{i\phi(z)} :,& \qquad & \psi^*(z) = :e^{-i\phi(z)}:       \label{fermion}
\eea
The modes of this boson satisfy commutation relations
$$
[J_m,J_{-n}] = m \delta_{m,n}
$$
and a bosonic Fock space is built by applying creation operators $J_n$ with $n<0$ on the vacuum $|p\rangle$. The boson-fermion correspondence can also be presented as
$$
e^{\frac{1}{\hbar}J_{-1}}|p\rangle = \sum_R \frac{\mu_R}{\hbar^{|R|}}|p,R\rangle,
$$
where the so-called Plancherel measure on partitions is
$$
\mu_R = \prod_{i<j} \frac{R_i-R_j+j-i}{j-i} = \prod_{\Box\in R} \frac{1}{h(\Box)}.
$$

One can also introduce \emph{vertex operators} related to the positive and negative parts of the boson
\begin{equation}
\Gamma_{\pm}(z)=\exp \sum_{n>0} \frac{z^{\mp n}}{n} J_{\pm n}, \label{gamma}
\end{equation}
which satisfy a commutation relation
\begin{equation}
\Gamma_{+}(z)\Gamma_{-}(z')=\frac{1}{1-z/z'}\Gamma_{-}(z')\Gamma_{+}(z). \label{commute}
\end{equation}
Moreover
\be
\G_{\pm}(z) = z^{-L_0} \G_{\pm}(1) z^{L_0},   \label{gamma-z-1}
\ee
and 
\be
\G_-(1) |R\rangle = \sum_{P\succ R} |P\rangle,\qquad \G_+(1) |R\rangle = \sum_{P\prec R} |P\rangle. \label{gamma-1-build}
\ee
It can also be shown 
\be
\prod_{i=1}^N \G_-(1)(x_i) |R\rangle = \sum_{P\succ R} s_{P/R}(x^{-1}_1,\ldots,x^{-1}_N)|P\rangle  , \label{gamma-skew}
%,\qquad \G_+(1) |R\rangle = \sum_{P\prec R} |P\rangle. \label{gamma-1-build}
\ee
where $s_{P/R}$ are skew Schur functions defined in (\ref{def-skewSchur}).

\section{Affine Lie algebras and affine characters} \label{app-characters}

In this appendix we introduce a notation and collect various results concerning affine Lie algebras and affine characters, in particular for the $\widehat{su}(N)_k$ affine Lie algebra,which are used in several places in the thesis. This summary is necessarily very brief and incomplete, as any more detailed treatment would be beyond a scope of this thesis. Standard excellent monographs on these subjects are \cite{kac} and \cite{cft}.

%*******************************************************

\subsection*{$su(N)$ Lie algebra}  \label{app-suN}

Let us introduce first a notation for classical $su(N)$ Lie algebra, denoted also $A_{N-1}$ in Cartan classification. It is convenient to consider a particular realisation of $su(N)$ in terms of traceless $N\times N$ matrices. Let $E_{i,j}$ be $N\times N$ matrix with all elements equal to 0, apart from 1 at position $(i,j)$. They satisfy commutation relations
$$
[E_{i,j}, E_{k,l}] = \delta_{jk} E_{i,l} - \delta_{l,i} E_{k,j}.
$$
In terms of these matrices, Chevalley generators of $su(N)$ are given by
\bea
e_m &=& E_{m,m+1}, \nonumber \\
f_m &=& E_{m+1,m}, \nonumber \\
h_m &=& E_{m,m} - E_{m+1,m+1}, \nonumber \\
& & \textrm{for}\ m=1,\ldots,N-1.
\eea
The elements of Cartan subalgebra can be written as 
\be
u = \sum_{m=1}^{N-1} \tilde{u}_m h_m = \sum_{j=1}^N u_j E_{j,j}\qquad \textrm{such that}\ \sum u_j = 0.  \label{u-E}
\ee

Let $\epsilon_i$, for $i=1,\ldots,N$, denote the basis dual to $\{E_{j,j}\}$. We can think of both $E_{j,j}$'s and $\epsilon_i$'s as orthonormal basis of two dual $N$-dimensional spaces. We also introduce
$$
\overline{\epsilon}_i = \epsilon_i - \frac{1}{N} \sum_{j=1}^N \epsilon_j.
$$
Simple roots are 
$$
\alpha_m = \epsilon_m - \epsilon_{m+1},\qquad m=1,\ldots,N-1.
$$
The root lattice is given as
$$
Q = \{\sum_{m=1}^{N-1} k_m \alpha_m  \ | \ k_m\in \mathbb{Z}\}  = \{\sum_{i=1}^N n_i\epsilon_i\ | \ n_i\in \mathbb{Z} \ \textrm{and}\ \sum n_i =0 \}. 
$$
Weyl denominator formula is the following relation
\be
\sum_{w\in Weyl\, group} \epsilon(w) e^{w(\rho)} = \prod_{\alpha\in\Delta_+} 2\,\textrm{sinh}\frac{\alpha}{2},   \label{weyl}
\ee
where $w$ are elements of the Weyl group, $\Delta_+$ is a set of positive roots whose number is $|\Delta_+|$, and the Weyl vector is
$$
\rho = \frac{1}{2} \sum_{\alpha\in\Delta_+} \alpha.     
$$

\begin{figure}[htb]
\begin{center}
\includegraphics[width=0.4\textwidth]{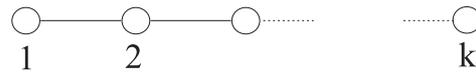}
\caption{Dynkin diagram for $A_k$ Lie algebra.} \label{fig-dynkin}
\end{center}
\end{figure}

The structure of Lie algebra can be encoded in a Dynkin diagram, with nodes associated to simple roots, joined by a number of lines related to angles between roots, as in example in figure \ref{fig-dynkin}.

Fundamental weights are
$$
\omega_m = \sum_{i=1}^m \overline{\epsilon}_i .
$$

Simple coroots $\alpha_m^{\vee}$ are elements of a basis dual to fundamental weights $(\omega_m | \alpha_n^{\vee})=\delta_{mn}$. Coroot lattice is
$$
Q^{\vee} = \{\sum_{m=1}^{N-1} k_m \alpha_m^{\vee}  \ | \ k_m\in \mathbb{Z}\}.
$$

Cartan matrix for $su(N)$ takes the form 
\be
A_{mn} = ( \alpha_m |  \alpha_n^{\vee}) = \left\{ \begin{array}{rl}
2  & \textrm{for}\quad m=n \\
-1 & \textrm{for}\quad |m-n|=1 \\
0 &  \textrm{otherwise} \end{array} \right.  \label{cartan}
\ee

There is a one-to-one correspondence between Young diagrams with at most $N-1$ rows and highest weights of $su(N)$:
\be
R=(R_1,\ldots,R_{N-1})\in \cY_{N-1} \to \lambda = \sum_{m=1}^{N-1} R_m \overline{\epsilon}_m = \sum_{m=1}^{N-1} \lambda_m \omega_m,   \label{R-lambda}
\ee
where $\lambda_m$ are usual Dynkin labels of a highest weight, related to parts of $R$ as
$$
R_m = \lambda_m + \lambda_{m+1} + \ldots + \lambda_{N-1}\ \iff\ \lambda_m = R_m-R_{m+1}.
$$

%*********************************************************

\subsection*{$\widehat{su}(N)_k$ affine Lie algebra}

Affine algebra is an extension of a finite Lie algebra by the central element $\widehat{k}$. This extension implies there is an extra (zeroth) simple root $\alpha_0$ and a corresponding coroot $\alpha_0^{\vee}$, as well as additional (zeroth) node in the Dynkin diagram.

\begin{figure}[htb]
\begin{center}
\includegraphics[width=0.4\textwidth]{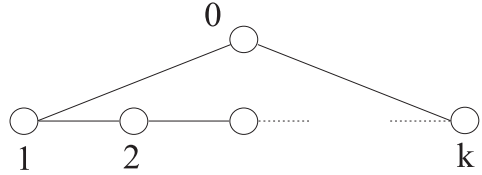}
\caption{Dynkin diagram for $\widehat{A}_k$ affine Lie algebra.} \label{fig-aff-dynkin}
\end{center}
\end{figure} 

For $\widehat{su}(N)_k$ there are $N$ affine fundamental weights, the basic (zeroth) one and $N-1$ corresponding to finite ones:
$$
\widehat{\omega}_0 \qquad \textrm{and} \qquad \widehat{\omega}_m = a_m^{\vee} \widehat{\omega}_0 + \omega_m.
$$
$a_m^{\vee}$ are eigenvalues of $\widehat{k}$ and are all equal to 1 for $\widehat{su}(N)_k$. An affine integrable weight $\widehat{\lambda}$ at level $k$ is an extension of its finite part $\lambda$, and is necessarily of the form
$$
\widehat{\lambda} =  k\widehat{\omega}_0 + \lambda = \sum_{m=0}^{N-1} \lambda_m \widehat{\omega}_m, \qquad \textrm{such that}\ k = \sum_{m=0}^{N-1} a_m^{\vee} \lambda_m.
$$
There are 
\be
\frac{(N+k-1)!}{(N-1)!\,k!}   \label{nr-weights}
\ee
$\widehat{su}(N)_k$ affine integrable weights, determined unambiguously by their finite parts. These finite parts, due to the level constraint, correspond in one-to-one way to Young diagrams with at most $N-1$ rows and at most $k$ columns $\cY_{N-1,k}$.

%Related to zertoh affine root there is an additional node in affine Dynkin diagram of $\widehat{su}(N)$. This affine Dynkin diagram  has an outer automorphism group $\mathbb{Z}_N$, permuting its nodes in the cyclic order. This induces also an action on affine integrable weights. The generator of outer automorphism group $\sigma$, the so-called basic outer automorphism, has a simple realisation in terms of a Young diagram $R=(R_1,\ldots,R_{N-1})$ corresponding to a given weight (\ref{R-lambda}), which amounts to adding a row of length $k$ as a first row of a diagram, and then reducing the diagram, i.e. removing $R_{N-1}$ columns which acquired a length $N$ (so that indeed $\sigma(R)\in \cY_{N-1,k}$),
%\be
%\sigma (R_1,\ldots,R_{N-1}) = (k-R_{N-1},R_1-R_{N-1},\ldots,R_{N-2}-R_{N-1}).   \label{cyclic}
%\ee
%Indeed, $\sigma^N(R)=R$, as should be expected for $\mathbb{Z}_N$ symmetry, and $N$ weights related by the action of $\sigma$ constitute an equivalence class, which is an orbit of $\mathbb{Z}_N$. We denote such an orbit as $[R] \subset \cY_{N-1,k}$. This is an example of $\mathbb{Z}_4$ orbit generated from $\widehat{su}(4)_3$ integrable weight $\widehat{\lambda} = \widehat{\omega}_0+ \widehat{\omega}_1+ \widehat{\omega}_2$ corresponding to a diagram $R=(2,1)\in \cY_{3,3}$
%$$
%\Yvcentermath1 \yng(2,1) \ \to \ \yng(3,2,1) \ \to \ \yng(2,2,1) \ \to \ \yng(2,1,1)
%$$

%*******************************************************
%*******************************************************

\subsection*{Affine characters}

In general, the character of an affine integrable weight $\widehat{\lambda}$ at a given level $l$ and finite part $\lambda$ can be written in terms of string functions $c^{\widehat{\lambda}}_{\widehat{\lambda}'}$ and $\Theta$-functions
\be
\chi_{\widehat{\lambda}}(q,\zeta) = \sum_{\widehat{\lambda}'} c^{\widehat{\lambda}}_{\widehat{\lambda}'} \Theta_{\widehat{\lambda}',\,l}, \qquad \quad  \Theta_{\widehat{\lambda}',\,l}(q,\zeta) = \sum_{\alpha^{\vee} \in Q^{\vee}} e^{2\pi i (l\alpha^{\vee}+\lambda'|\zeta)} q^{\frac{l}{2}|\alpha^{\vee}+\lambda'/l |^2}  \label{aff-char}
\ee
where $\zeta=\sum z_i \alpha_i^{\vee}$ is certain specialisation point and $q=e^{2\pi i\tau}$. 

Under the action of the modular group $SL(2,\Z)$ 
\be
\tau \to \frac{a\tau+b}{c\tau+d}, \qquad ad-bc=1,\qquad a,b,c,d\in\Z,    \label{SL-2Z}
\ee
affine characters have simple transformation properties. It is enough to specify these properties for generators of the modular group
\be
T:\ \tau\to \tau+1,\qquad\qquad S:\ \tau\to -\frac{1}{\tau}     \label{modular-TS}
\ee
for which respectively
\be
\chi_{\widehat{\lambda}}(\tau+1) = \sum_{\widehat{\mu}} \mathcal{T}_{\widehat{\lambda} \widehat{\mu}} \,  \chi_{\widehat{\mu}}(\tau), \qquad\quad \chi_{\widehat{\lambda}}(-\frac{1}{\tau}) = \sum_{\widehat{\mu}} \mathcal{S}_{\widehat{\lambda} \widehat{\mu}} \,  \chi_{\widehat{\mu}}(\tau),          \label{affine-TS}
\ee
where summation runs over all affine integrable weights $\widehat{\mu}$ and
\bea
\mathcal{T}_{\widehat{\lambda} \widehat{\mu}} & = & \delta_{\widehat{\lambda} \widehat{\mu}} e^{2 \pi i m_{\widehat{\lambda}}}, \nonumber \\
\mathcal{S}_{\widehat{\lambda} \widehat{\mu}} & = &  \frac{i^{|\Delta_{+}|} \, N^{-1/2}}{(k+N)^{\frac{N-1}{2}}} \sum_{w\in Weyl\, group} \epsilon(w) \exp\Big(-\frac{2\pi i}{k+N} \big(w(\rho+\lambda) \, | \, \rho+\mu \big) \Big), \label{S-modular} 
\eea
where the conformal anomaly is $m_{\widehat{\lambda}} = \frac{|\rho+\lambda|^2}{2N} - \frac{|\rho|^2}{2N}$.

%*********************************************************

\subsection*{$\chi^{\widehat{su}(k)_1}_r$ characters} \label{subsec-suN-char}

 For $\widehat{su}(k)_1$ there is a single string function in (\ref{aff-char}) which reads $c^{\widehat{\lambda}}_{\widehat{\lambda}}=\eta(q)^{-k+1}$, and there are $k$ integrable weights at level 1 with corresponding characters labelled by $r=0,\ldots,k-1$  , 
$$
\chi^{\widehat{su}(k)_1}_r(q,z_i) = \frac{\Theta_{\lambda_r}^{level\ 1}(q,\zeta)}{\eta(q)^{k-1}} = \frac{1}{\eta(q)^{k-1}} \sum_{\alpha^{\vee} \in Q^{\vee}} e^{2\pi i (\alpha^{\vee}+\lambda_r|\zeta)} q^{\frac{1}{2}|\alpha^{\vee}+\lambda_r |^2}.
$$
For $r$'th weight we can choose e.g. $\lambda_r = r\omega_1$. For an arbitrary element of coroot lattice $\alpha^{\vee} = \sum_{i=1}^{k-1} n_i \alpha^{\vee}_i$ we get
\begin{eqnarray}
\frac{1}{2}|\alpha^{\vee}+\lambda_r |^2 & = & \sum_i (n_i^2 - n_i n_{i+1})+n_1 r +\frac{r^2}{2}\frac{k-1}{k},  \nonumber \\
e^{2\pi i(\alpha^{\vee}+\lambda_r|\zeta)} & = & y_1^{2n_1-n_2+r} y_2^{2n_2-n_1-n_3}\ldots y_{k-1}^{2n_{k-1}-n_{k-2}} = \prod_{i=1}^{k-1} \tilde{y}_i ^{n_i + \frac{k-i}{k}r}, \nonumber
\end{eqnarray}
where we introduce
$$
y_i=e^{2\pi i z_i},
$$
\begin{equation}
\tilde{y}_1 = \frac{y_1^2}{y_2},\quad \tilde{y}_2 = \frac{y_2^2}{y_1 y_3},\ldots,\quad \tilde{y}_{k-1} = \frac{y_{k-1}^2}{y_{k-2}}.  \label{char-su-vars}
\end{equation}
With such a notation characters read
\bea
\chi^{\widehat{su}(k)_1}_r(q,y_i) & = & \frac{1}{\eta(q)^{k-1}}\sum_{n_1,\ldots,n_{k-1}} q^{\sum_i (n_i^2 - n_i n_{i+1})+n_1 r +\frac{r^2}{2}\frac{k-1}{k}} y_1^r \prod_{i=1}^{k-1} y_i ^{\sum_j A_{ij}n_j} = \nonumber \\
& = &  \frac{1}{\eta(q)^{k-1}}\sum_{n_1,\ldots,n_{k-1}} q^{\sum_i (n_i^2 - n_i n_{i+1})+n_1 r +\frac{r^2}{2}\frac{k-1}{k}} \prod_{i=1}^{k-1} \tilde{y}_i ^{n_i + \frac{k-i}{k}r}. \label{su-char}
\eea

%Sometimes it is more convenient to compute a character at a point $u$ specified in $h_m$ basis (\ref{u-E}), which we make use of for example in (\ref{Z-D4D6-char}).

%*********************************************************

\subsection*{$\chi^{\widehat{u}(1)_N}_j$ characters}

A notion of $\chi^{\widehat{u}(1)_N}_j$ characters is even more subtle. We just define them through the formula
\begin{equation}
\chi^{\widehat{u}(1)_N}_j(x) = \frac{1}{\eta(q)}\sum_{n\in\mathbb{Z}} q^{\frac{N}{2}(n+j/N)^2} x^{n+j/N}.  \label{u1-char}
\end{equation}

%**********************************************

\subsection*{$\chi^{\widehat{u}(k)_1}$ character and its decomposition}

\begin{figure}[htb]
\begin{center}
\includegraphics[width=0.4\textwidth]{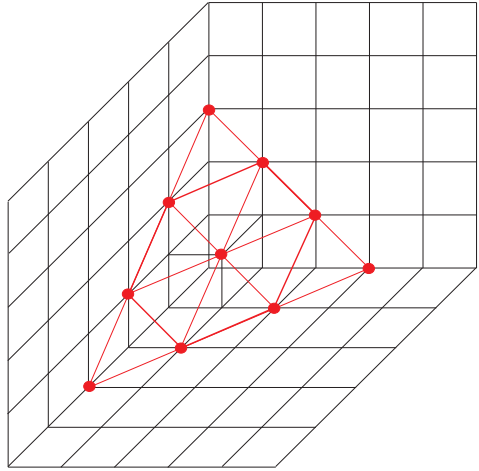}
\caption{Decomposition of $\mathbb{Z}^3$ lattice into $\mathbb{Z}$ (along the diagonal $x=y=z$) and $su(3)$ root lattice $Q_{su(3)}$ (along dotted plane).} \label{fig-decompose}
\end{center}
\end{figure}

$\widehat{u}(k)_1$ character is given by a trace over a Fock space of $k$ free complex fermions. It depends on variables $z_i,\ i=1,\ldots,k$ which couple to Cartan currents $J_i$
\begin{eqnarray}
\chi^{\widehat{u}(k)_1}(q, x_i) & = & Tr_{\mathcal{F}}\Big(e^{2\pi i \sum_i z_i J_i}\,q^{L_0 - \frac{k}{24}} \Big) = \nonumber \\
& = & q^{-\frac{k}{24}}\prod_{i=1}^{k} \prod_{p\in\mathbb{Z}_+ + \frac{1}{2}} (1+x_i q^p)(1+x_i^{-1}q^p) = \nonumber \\
& = & \frac{1}{\eta(q)^k}\sum_{\vec{p} = (p_1,\ldots,p_k)\in \mathbb{Z}} q^{\frac{1}{2}(p_1^2 + \ldots + p_k^2)}\, x_1^{p_1}\ldots x_k^{p_k} = \frac{\Theta_{\mathbb{Z}^k}(q;z_i)}{\eta(q)^k}, \label{u-k-1}
\end{eqnarray}
where $x_{i}=e^{2\pi i z_{i}}$. Here Jacobi triple product identity has been used, and in the last line $\Theta_{\mathbb{Z}^k}$ function is defined in terms of summation over $\mathbb{Z}^k$ lattice spanned by $\epsilon_j$ (for $j=1,\ldots,k$) defined is appendix \ref{app-suN}. We now show this lattice decomposes as a product of a one-dimensional lattice and $su(k)$ root lattice $Q_{su(k)}$
$$
\mathbb{Z}^k = \sum_{r=0}^{k-1} \mathbb{Z} \times Q_{su(k)},
$$
and rearrange summations appropriately. One-dimensional $\mathbb{Z}$ factor corresponds to $u(1)$ overall charge, so it is spanned by a diagonal $\vec{p}=(n,\ldots,n) \in \mathbb{Z}^k$. For a fixed point on this diagonal , $su(k)$ lattice is given by a perpendicular hyperplane which is spanned by $k-1$ vectors $\epsilon_j-\epsilon_{j+1}$. An example of this decomposition is shown in figure (\ref{fig-decompose}). To probe all points of the original $\mathbb{Z}^k$ in fact an additional shift $r=0,\ldots,k-1$ has to be introduced, and we have to sum over all values of $r$ each point in $\mathbb{Z}^k$ is uniquely specified by a set of numbers $(n;n_1,\ldots,n_{k-1};r)$:
\be
\vec{p}=(p_1,\ldots,p_k) = 
\left[\begin{array}{l}
n+n_1+r \\ 
n-n_1+n_2 \\
n-n_2+n_3 \\
\quad \vdots \\
n-n_{k-2}+n_{k-1} \\
n-n_{k-1}\end{array}\right] \in\mathbb{Z}^k. \label{p-decompose}
\ee

Rewriting the above character in terms of these variables we get
$$
\chi^{\widehat{u}(k)_1} = \frac{1}{\eta(q)^k}\sum_{r=0}^{k-1} \sum_{n;n_1,\ldots,n_{k-1}} q^{\frac{k n^2}{2} + \sum_i (n_i^2 - n_i n_{i+1}) +\frac{r^2}{2} + nr + n_1 r} x_1^{n+n_1+r}x_2^{n+n_2-n_1}\ldots x_k^{n-n_{k-1}} =
$$
\be
= \frac{1}{\eta(q)^k}\sum_{r=0}^{k-1} \sum_{n} q^{\frac{k}{2}(n+r/k)^2} \tilde{x}^{n+r/k} \sum_{n_1,\ldots,n_{k-1}} q^{\sum_i (n_i^2 - n_i n_{i+1})+n_1 r +\frac{r^2}{2}\frac{k-1}{k}} \prod_{i=1}^{k-1} \tilde{x}_i ^{n_i + \frac{k-i}{k}r} \label{uNchar}
\ee
where we introduced new variables
\begin{equation}
\tilde{x} = x_1 x_2\cdots x_k;\qquad \tilde{x}_i = \frac{x_i}{x_{i+1}},\ i=1,\ldots,k-1. \label{char-u-vars}
\end{equation}
Comparing with (\ref{u1-char}) and (\ref{su-char}), this can be written as 
\begin{equation}
\chi^{\widehat{u}(k)_1}(x_i) = \sum_{r=0}^{k-1} \chi^{\widehat{u}(1)_k}_r(\tilde{x})\, \chi^{\widehat{su}(k)_1}_r(\tilde{x}_i) \label{char-u-su}
\end{equation}
where $r$ runs over different $u(1)$ charges and $\widehat{su}(k)_1$ weights.

\section{Symmetric functions} \label{app-schur}

In this appendix we summarise properties of symmetric functions which often appear in formulae in this thesis.
%  Schur polynomials $s_{R}$, elementary $e_{R}$ and complete $h_{R}$ symmetric polynomials, Newton polynomials $P_{R}$.

A symmetric polynomial $S$ depends on a partition $R$, and its argument
is a string of variables $x=(x_1,x_2,\ldots)$, what we denote by
\be
S_R(x)=S_R(x_1,x_2,\ldots).
\ee
By $q^{R+\rho}$ we understand a string such that $x_i=q^{R_i-i+1/2}$ for
$i=1,2,\ldots$, thus
$$
S_R(q^{R+\rho})=S_R(q^{R_1-1/2},q^{R_2-3/2},\ldots).
$$
In particular
\be
S_R(q^{\rho})=S_R(q^{-1/2},q^{-3/2},\ldots).
\ee

One can concatenate two strings of variables, $x=(x_1,x_2,\ldots)$ and
$y=(y_1,y_2,\ldots)$, and then use it as an argument of a symmetric polynomial,
which is denoted by
$$
S_{Q}(x,y)=S_{Q}(x_1,x_2,\ldots, y_1,y_2,\ldots).
$$

One of the simplest examples of symmetric functions are \emph{Newton polynomials}
\begin{equation}
P_R(x)=\prod_{n} P_{R_i}(x), \quad \textrm{where} \quad  P_n(x)=\sum_{i=1}x_{i}^{n}.
\label{newton}
\end{equation}

Let us next introduce \emph{elementary} $e_n(x)$ and \emph{complete symmetric
functions} $h_n(x)$, for $n=0,1,2,\ldots$, in terms of a generating functions
\begin{eqnarray}
E(t) & = & \sum_{n=0}^{\infty} e_n t^n = \prod_{i} (1+x_i t), \label{e-n} \\
H(t) & = & \sum_{n=0}^{\infty} h_n t^n = \prod_{i} \frac{1}{1-x_i t}, \label{h-n}
\end{eqnarray}
and $h_{-1}=e_{-1}=h_{-2}=e_{-2}=\ldots=0.$
Then, for a partition $R=(R_1,R_2,\ldots)$,
\begin{eqnarray}
e_R & = & e_{R_1}e_{R_2}\cdots \nonumber \\ \label{e-R}
h_R & = & h_{R_1}h_{R_2}\cdots. \label{h-R} \nonumber
\end{eqnarray}

For a partition $R$, the \emph{Schur function} is defined as
\begin{equation}
s_R(x)=\det(h_{R_i -i+j})=\det(e_{R^{t}_i -i+j}). \label{s-det}
\end{equation}

Let us introduce Littlewood-Richardson coefficients $c^{P}_{QR}$ as
\begin{equation}
s_{Q\otimes R}=s_Q s_R=\sum_P c^{P}_{QR}s_P, \label{lit-rich}
\end{equation}
which have properties
\begin{equation}
c^{P}_{QR} =  c^{P^t}_{Q^t R^t} = c^{P}_{RQ}, \qquad c^{P}_{R\bullet} = \delta^P_R,  \label{lit-rich-prop} \\
\end{equation}
\begin{equation}
c^{P}_{QR} = 0 \ \textrm{for}\ |P| \neq |Q|+|R|. \label{lit-rich-nonzero}
\end{equation}
It is also convenient to define multiple coefficient
\begin{equation}
c^{P}_{R_1\ldots R_n} = \sum_{\alpha_i} c^{\alpha_1}_{R_1 R_2} c^{\alpha_2}_{\alpha_1 R_3} c^{\alpha_3}_{\alpha_2 R_4} \cdots c^{P}_{\alpha_{n-2} R_n}, \label{multi-lit-rich}
\end{equation}
in terms of which a multiple tensor product takes the form
\begin{equation}
R_1\otimes \ldots \otimes R_n = \sum_P c^{P}_{R_1\ldots R_n} P, \\
\end{equation}

Finally we define \emph{skew Schur functions}
\begin{equation}
s_{Q/R}=\sum_P c^{Q}_{RP}s_P. \label{def-skewSchur}
\end{equation}
For trivial representation $R=\bullet$, we have
$$
s_{Q/\bullet}=s_Q.
$$
and
$$
\textrm{If not}\ Q \subset  R \ \  \Leftrightarrow \ \  s_{R/Q}=0. \label{skew-zero}
$$

For Schur functions, we have the following identities
\bea
s_R(c x) &=& c^{|R|} s_{R} (x) \label{schur-c} \nn
s_R(q^{\rho}) &=& q^{\kappa_R /2} s_{R^t} (q^{\rho}) \label{schur-q} \nn
s_R(q^{\rho}) &=& (-1)^{|R|} s_{R^t} (q^{-\rho}) \label{schur-1} \nn
s_Q(q^{\rho})s_R(q^{Q+\rho}) &=& s_R(q^{\rho})s_Q(q^{R+\rho}) \label{schur-change}.
\eea
Skew Schur functions satisfy
\bea
s_{Q/R}(c x) &=& c^{|Q|-|R|} s_{Q/R} (x) \label{skew-c} \nn
s_{Q/R}(q^{\rho}) &=& (-1)^{|Q|-|R|} s_{Q^t/R^t} (q^{-\rho}) \label{skew-1}.
\eea
In addition, we have the summation formulae for Schur functions 
\bea
\sum_{R}  s_{R}(x) s_{R}(y) &=& \prod_{i,j} \frac{1}{1-x_i y_j}  \label{schur-sum3} \nn
\sum_{R}  s_{R}(x) s_{R^t}(y) &=& \prod_{i,j} (1+x_i y_j)  \label{schur-sum4}
\eea
and for skew Schur functions
\bea
\sum_{\eta}  s_{Q/ \eta}(x) s_{R/ \eta}(y) &=& \prod_{i,j} (1-x_i y_j) \ \sum_{\eta} 
s_{\eta/R}(x) s_{\eta/Q}(y) \label{schur-invert1} \nn
\sum_{\eta}  s_{Q/ \eta}(x) s_{R/ \eta}(y) &=&  \prod_{i,j} \frac{1}{1+x_i y_j} \
\sum_{\eta}  s_{\eta^{t}/R}(x) s_{\eta/Q}(y) \label{schur-invert2} \nn
\sum_{\eta}  s_{\eta/R}(x) s_{\eta}(y) &=& s_R(y) \sum_{\mu} s_{\mu}(x)  s_{\mu}(y) 
\label{schur-sum2} \nn
\sum_{\eta}  s_{\eta^t/R}(x) s_{\eta}(y) &=& s_R(y) \sum_{\mu} s_{\mu}(x)  s_{\mu^t}(y) 
\label{schur-sum6} \nn
\sum_{\eta}  s_{R/ \eta}(x) s_{\eta/Q}(y) &=& s_{R/Q}(x,y),  \label{schur-sum5} \nn
\sum_{\eta}  s_{R/ \eta}(x) s_{\eta}(y) &=& s_{R}(x,y)  \label{schur-sum1}.
\eea
the last two sums being  over partitions $\eta$ such that $Q\subset \eta \subset R$.

For the special case a partition with a single row $R=(R_1,0,0,\ldots)$,  the Schur function is
related to the quantum dilogarithm as
\be
s_{R=(R_1,0,\ldots)}(q^{\rho})  =  (-1)^{R_1}q^{R_1^2/2}\xi(q)L\Big((R_1+\frac{1}{2})g_s,q\Big)
\ee 
where 
$$ \xi(q) = \prod_{i=1}^{\infty} \frac{1}{1 - q^i} .$$

The following identity holds for Schur function with finite number of arguments
\begin{equation}
s_R(1,q,q^2,\ldots,q^{K-1}) = q^{n(R)} \prod_{(i,j)\in R} \frac{1-q^{K+j-i}}{1-q^{h(i,j)}}, \label{schur-finite}
\end{equation}
and for $K\to\infty$ this reduces to
\begin{equation}
s_R(q^{-\rho}) = q^{|R|/2+n(R)} \prod_{(i,j)\in R} \frac{1}{1-q^{h(i,j)}}.  \label{schur-hooks}
\end{equation}

A sum of Schur functions (\ref{schur-sum4}) can be rewritten as an exponent
\begin{eqnarray}
\sum_{P} s_{P}(x) s_{P^t}(y) = \exp\Big[-\sum_{n,i,j} \frac{(-1)^n}{n} x_i^n y_j^n \Big]. \label{schur-sum-exp}
\end{eqnarray}
As noticed in \cite{strip}, with $x=q^{R+\rho}$ and $y=q^{R'+\rho}$ this allows to write the sum as
\begin{equation}
\sum_{P} s_{P}(q^{R+\rho}) s_{P^t}(-Q q^{R'+\rho}) = \exp\Big[-\sum_n \frac{Q^n}{n[n]^2} \Big] \, \prod_k (1-Qq^k)^{C_k(R,R')}, \label{sum-schur-RRbis}
\end{equation}
where coefficients $C_k(R,R')$ are given by
\begin{equation}
\sum_{i,j} q^{R_i-i+1/2} q^{R'_j-j+1/2} = \sum_k C_k(R,R') q^k + \frac{q}{(1-q)^2}.
\end{equation}
From this statement the following properties are more or less easily deduced
\begin{eqnarray}
C_k(R,R') & = & C_k(R',R), \nonumber \\
\sum_k C_k(R,R')q^k & = & \sum_k C_k(R^t,R'^t)q^{-k}, \nonumber \\
\sum C_k(R,R') & = & |R|+|R'|, \nonumber \\
\sum k C_k(R,R') & = & \frac{\kappa_R + \kappa_{R'}}{2}. \nonumber \\
\end{eqnarray}
In particular, $C_k(R)=C_k(R,\bullet)$ counts the number of boxes $(i,j)\in R$ with fixed $k=j-i$, and so $\sum_k C_k(R)=|R|$ and $2\sum_k kC_k(R)= \kappa_R$.

%***********************************************************************
%***********************************************************************

\chapter{Properties of the topological vertex}  \label{app-topver}
%\addcontentsline{toc}{chapter}{Appendix B --- Properties of the topological vertex}

\section{Calculational framework} \label{app-vertex}

In this appendix we summarise topological vertex calculational framework. The most convenient form of the vertex is representation basis, in which vertex amplitudes can be expressed in terms of Schur functions. We recall the general formula for topological vertex in the canonical framing given in (\ref{vertex-chap})
\begin{equation}
C_{R_1 R_2 R_3} = q^{\frac{1}{2}(\kappa_{R_2}+\kappa_{R_3})} s_{R_{2}^{t}}(q^{\rho})
\, \sum_{P} s_{R_{1}/P}(q^{ R_{2}^{t}+\rho}) s_{R_{3}^{t}/ P}(q^{ R_{2}+\rho}).
\label{vertex}
\end{equation}

The crucial property of $C_{R_1 R_2 R_3}$ in the canonical framing is cyclicity
w.r.t. representations $R_i$. The above formula also immediately implies
\begin{equation}
C_{R_1 R_2 R_3} = q^{\frac{1}{2}\sum_{i} \kappa_{R_i}} C_{R_{1}^{t} R_{3}^{t}
R_{2}^{t}}. \label{C-transpose}
\end{equation}

The identities from appendix \ref{app-schur} lead to the following special cases, with some representations involved being trivial $\bullet$
\begin{eqnarray}
C_{R \bullet \bullet} & = & q^{\kappa_{R}/2} s_{R^{t}}(q^{\rho}) = s_{R}(q^{\rho}), \label{C-0R0} \\
C_{PR\bullet} & = & q^{\frac{1}{2}\kappa_R}s_{P}(q^{\rho})s_{R^t}(q^{\rho+P}) = \label{C-PR} \\
& = & q^{\frac{\kappa_P}{2}} \sum_{\eta} s_{R/ \eta}(q^{\rho}) s_{P^{t}/
\eta}(q^{\rho}). \label{C-PR-skew}
\end{eqnarray}

The vertex with one trivial representation is closely related to the leading term of the Hopf Link invariant $W_{PR}$, which also can be expressed in terms of Schur functions
\begin{eqnarray}
W_{PR}&=& q^{\kappa_{R}/2}C_{PR^t\bullet} = \label{W-C} \\
& = & s_{P}(q^{\rho})s_R(q^{\rho+P})= \label{W-schur} \\
& = & q^{\frac{1}{2}(\kappa_{P}+\kappa_{R})} \sum_{\eta} s_{R^t/ \eta}(q^{\rho})
s_{P^{t}/ \eta}(q^{\rho}),
\end{eqnarray}
and it is not difficult to show that
\be
W_{PR}(q) = (-1)^{|P|+|R|} W_{P^t R^t}(q^{-1}). \label{W-PtRt}
\ee

The important feature of the vertex is a framing ambiguity, which arises as a need to specify an
integer number for each stack of branes on a leg of $\mathbb{C}^3$. The vertex in a particular framing specified by numbers $f_1, f_2, f_3$ corresponding to representations $R_i$ on different axes is given as
\begin{equation}
C^{f_1,f_2,f_3}_{R_1 R_2 R_3} = (-1)^{\sum_i f_i |R_i|} q^{\sum_i f_i \kappa_{R_i}
/2} C_{R_1 R_2 R_3}, \label{framing}
\end{equation}
where $|R_i|$ denotes number of boxes in the Young diagram for a given representation. The canonical framing (\ref{vertex}) corresponds to $f_i=0$.

It is also possible to reverse orientation of the branes on one leg, what can 
be interpreted as changing branes to antibranes. To obtain vertex amplitude 
with an antibrane on the first axis one should substitute
\begin{equation}
C_{PQR}\ \to \ (-1)^{|P|} C_{P^t QR}, \label{vertex-orient}
\end{equation}
and similarly for any other leg.

To construct the full toric diagram, one has to glue together $\mathbb{C}^3$ 
patches. Gluing together just two patches gives a resolved conifold with K\"ahler parameter $Q=q^N=e^{-t}$, and the propagator is given by $(-Q)^{|R|}$. The orientations of two glued axes must be consistent, and sum over representations performed, what leads to
\begin{eqnarray}
Z^{\mathbb{P}^1} & = & \sum_R C_{\bullet \bullet R^t} (-Q)^{|R|} C_{R \bullet\bullet}  = \sum_R s_{R}(q^{-\rho}) s_{R}(Qq^{\rho}) = \nonumber \\
& = & \prod_{i,j=1}^{\infty} \frac{1}{1-Qq^{i-j}} =
\exp\Big(-\sum_{k=1}^{\infty}\frac{Q^{k}}{k[k]^2} \Big). \label{Z-P1}
\end{eqnarray}

\bigskip
It is also possible, though a bit more complicated, to consider branes on internal legs of toric diagram and configurations of several stacks of branes on one leg. In the case of one stack of branes on a compact leg one more parameter $d=g_s D$ should be introduced, which denotes the position of the brane along the compact leg, as measured from the left vertex. To properly glue two vertexes with additional brane between them, two additional summations must be introduced representing strings ending on the brane from the left and from the right, so the relevant vertex factor takes the form
\begin{equation}
\sum_{R,Q^L,Q^R} C_{\bullet\bullet R\otimes Q^L} (-1)^s q^f e^{-L} C_{R^t\otimes
Q^R\bullet\bullet}\, Tr_{Q^L }V\, Tr_{Q^R}V^{-1}  \label{comp-leg-brane}
\end{equation}
where a framing of the brane $p$ has also been taken into account, so that
\bea
L & = & |R|t + |Q^L|d + |Q^R|(t-d), \label{distances-L} \\
f & = & \frac{p}{2} \kappa_{R\otimes Q^L} + \frac{n+p}{2} \kappa_{R^t\otimes Q^R}, \nn
s & = & |R| + p|R\otimes Q^L| + (n+p)|R^t \otimes Q^L|.
\eea
The additional number $n=|v' \times v|$ is determined by planar directions of two axes $v$ and $v'$ of glued vertexes, and in all cases we consider it equals zero.

For more stacks of branes we need to specify a position of each stack as $d_i=g_s D_i$.  The holonomy
matrix corresponding to branes at $d_i$ is denoted $V_i$. In fact, to get agreement with crystal results we will need to absorb $d_i$ into $V_i$.  Moreover, in this case we have to choose different representations $R_i$ for each stack of branes, and for a given leg of the vertex consider the tensor product of representations $C_{P,Q,\otimes_i R_i}$. In addition, for each pair of branes at $d_i,d_j$ we have to introduce an additional factor from strings stretched between them
\begin{equation}
\sum_P (-1)^{|P|}\, Tr_P V_i\, Tr_{P^t} V_j^{-1}. \label{stretched}
\end{equation}

If there are several branes on an internal leg, also summations from the left and right vertexes for each brane must be introduced, as well as an overall summation over $|R|$ as in (\ref{comp-leg-brane}).

It is also important to make clear how summations over tensor products should 
be understood. The Hopf-link with a single factor of $|P_1 \otimes P_2|$ can be obtained from a fusion rule
\begin{equation}
W_{P_1\otimes P_2, R} = \sum_{\alpha} c^{\alpha}_{P_1 P_2} W_{\alpha R}  =  \sum_{\alpha} q^{\frac{\kappa_{\alpha}+\kappa_R}{2}} c^{\alpha}_{P_1 P_2} s_{\alpha^t/\eta}(q^{\rho}) s_{R^t/\eta}(q^{\rho}), \label{W-tensor}
\end{equation}
and then related to topological vertex by (\ref{W-C}). When a few factors of $|P_1 \otimes P_2|$ appear, the internal summation over $\alpha$ should also be introduced, with each such factor replaced by $\alpha$. For example, for two stacks of branes on one leg of $\mathbb{C}^3$ in $(-1)$ framing we have
\begin{displaymath}
C_{P_1\otimes P_2, R, \bullet} (-1)^{-|P_1\otimes P_2|} q^{-\frac{\kappa_{P_1\otimes P_2}}{2}} = \sum_{\alpha}  c^{\alpha}_{P_1 P_2} s_{\alpha^t/\eta}(q^{\rho}) s_{R/\eta}(q^{\rho}) (-1)^{|\alpha|} =
\end{displaymath}
\begin{equation}
= (-1)^{|P_1|+|P_2|} \sum_{\alpha} c^{\alpha}_{P_1 P_2} s_{\alpha^t/\eta}(q^{\rho}) s_{R/\eta}(q^{\rho}), \label{comp-brane-tensor-1}
\end{equation}
where two factors of $q^{\kappa_{\alpha} /2}$ (from vertex expression and $(-1)$ framing) cancelled each other, and formulae (\ref{lit-rich-prop}) and (\ref{lit-rich-nonzero}) have been used.

The internal summation arising in quantities with tensor product involved is a crucial and subtle issue. In particular, knot invariants in different framings but without tensor product differ just by an overall sign and factors of $q$. On 
the other hand, the summation implicit in tensor product formulae changes the structure of polynomials representing knot invariants. For example, in canonical framing we have
\begin{equation}
C_{1\otimes 1, 1, \bullet} = W_{1\otimes 1, 1, \bullet} = \frac{(q^2-q+1)^2}{(q-1)^3\sqrt{q}}.
\end{equation}
On the other hand, in framing $(-1,0)$
$$
C_{1\otimes 1, 1, \bullet} (-1)^{|1\otimes 1|}q^{-\frac{\kappa_{1\otimes 1}}{2}}=  W_{1\otimes 1, 1, \bullet} (-1)^{|1\otimes 1|}q^{-\frac{\kappa_{1\otimes 1}}{2}} =
$$
\begin{equation}
=  W_{1,2}\, q^{-1} + W_{1,2^t}\, q = \frac{(2q^2-3q+2)\sqrt{q}}{(q-1)^3},
\end{equation}
and this is also precisely the coefficient which we get from crystal expansion without the factor (\ref{stretched}) $(1-\frac{a_1}{a_2}) $ (and up to $q$ inversion) 
with two branes at $a_1, a_2$ on one slice of the crystal and antibrane on the other slice.

%************************************************************************
%************************************************************************

\section{Topological vertex on a strip} \label{app-strip}

There is a class of toric Calabi-Yau geometries whose dual diagrams are represented as a triangulation of a rectangle (or a \emph{strip}). Computation of their partition functions via topological vertex methods has been vastly simplified in \cite{strip}. As this simplification is quite convenient for a part of the calculations we perform, we briefly recall the \emph{rules on the strip}. A derivation and a more detailed discussion of these rules can be found in \cite{strip}.

\begin{figure}[htb]
\begin{center}
\includegraphics[width=0.8\textwidth]{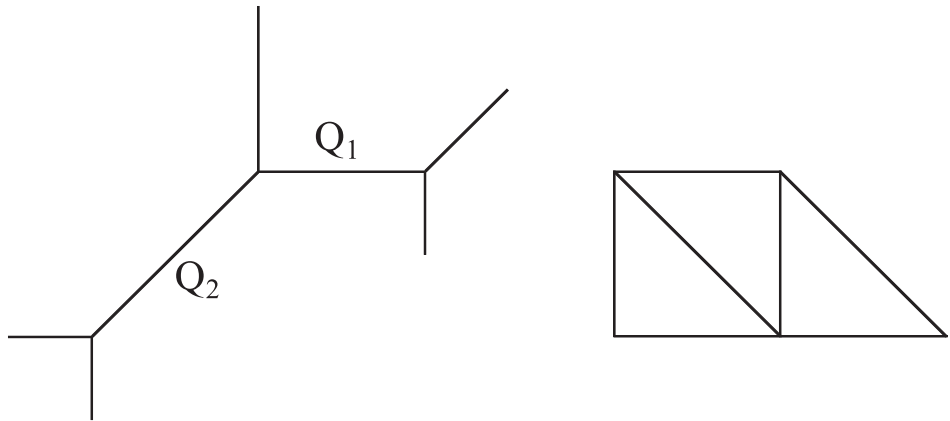}
\caption{Toric diagram (left) for the double-$\mathbb{P}^1$. The dual diagram (right) is a triangulation of a strip.} \label{fig-double-P1}
\end{center}
\end{figure}

A diagram which can be drawn on a strip is a string of $\mathbb{P}^1$'s with parameters $Q_i$, each represented by an interval between vertices $i$ and $i+1$. For every such interval we introduce representations $R_i$ which we sum over according to topological vertex rules. Additionally, with every external leg of each vertex we can associate one fixed representation $\alpha_i$ (for the first vertex in a string there are two external legs, but one of them must be associated with a trivial representation $\bullet$; the same statement must hold for the last vertex).

The partition function for such a system can be expressed in terms of the quantities
\begin{equation}
\{\alpha_i \alpha_j\}_{Q_{ij}} =  \exp \Big[-\sum_n \frac{Q^n_{ij}}{n[n]^2} \Big] \prod_k (1-Q_{ij}q^k)^{C_k(\alpha_i,\alpha_j)}, \label{amir-bracket}
\end{equation}
$$
[\alpha_i \alpha_{j}]_{Q_{ij}} = \{ \alpha_i \alpha_j \}^{-1}_{Q_{ij}},
$$
according to the following rules:
\begin{itemize}
\item determine the type of the first ($i=1$) vertex to be A or B if respectively its topological vertex factor is given by $C_{\bullet\alpha_1 R_1}$ or $C_{\bullet R_1 \alpha_1}$

\item determine recursively the type A or B of all other vertices: $(j+1)$'th vertex is of the same type as $j$'th if the local bundle of the sphere $Q_j$ is $\mathcal{O}(-2)\oplus\mathcal{O}$, and of a different type if this bundle is $\mathcal{O}(-1)\oplus\mathcal{O}(-1)$

\item for each pair of vertices $(i,j)$ in a diagram (with $i<j$) introduce a suitable factor according to their types $(A/B,A/B)$:
\begin{eqnarray}
(A,B) & \to & \{\alpha_i \alpha_j\} _{Q_{ij}}, \nonumber \\
(B,A) & \to & \{\alpha_i^t \alpha_j^t\} _{Q_{ij}}, \nonumber \\
(A,A) & \to & [\alpha_i \alpha_j^t]_{Q_{ij}}, \nonumber \\
(B,B) & \to & [\alpha_i^t \alpha_j]_{Q_{ij}}, \nonumber
\end{eqnarray}
where $Q_{ij} = Q_i Q_{i+1} \cdots Q_{j-1}$.

\item the full amplitude, with external representations $\alpha_i$ fixed, is given by a product of all factors above together with Schur functions for all external representations
$$
Z_{\prod \alpha_i} = \prod_i s_{\alpha_i}(q^{\rho})\ \prod_{i<j} [\{\alpha_i^{\dag} \alpha_j^{\dag} \}]_{Q_{ij}},
$$ 
with appropriate choice of the pairing $[\{\,\}]$, and with or without transposition $\dag=t,\cdot$.

\end{itemize} 

For example, in figure \ref{fig-closed-ver-glue} the strip-part consists of two intervals $Q_1$ and $Q_2$. We denote the right-most vertex by $i=1$, the middle one with external representation $\alpha$ by $i=2$ and the left-most by $i=3$. There is no external representation on the first vertex, so we are free to choose its type to be e.g. A; then we recursively determine types of all three vertices to be $A-B_{\alpha}-A$ (for convenience we explicitly write external representations associated with vertices, if they are nontrivial). Then according to the rules above the amplitude reads 
\begin{equation}
Z_{\alpha} = s_{\alpha}\, \{\bullet \alpha \}_{Q_1} [\bullet \bullet ]_{Q_1 Q_2} \{\alpha^t \bullet \}_{Q_2}. \label{Zalpha}
\end{equation}

%***********************************************************

\newpage

\addcontentsline{toc}{chapter}{Podziękowania --- Acknowledgements}

%\begin{footnotesize}

\selectlanguage{polish}

{\center \section*{Podziękowania}}

Z największą przyjemnością pragnę podziękować moim promotorom --- Jackowi Pawełczykowi z Uniwersytetu Warszawskiego oraz Robbertowi Dijkgraafowi z Uniwersytetu w Amsterdamie --- za nieocenione wsparcie oraz całą wiedzę którą mi przekazali. Współpraca z nimi była dla mnie niezwykłą i fascynującą przygodą.

Przez prawie trzy lata moich studiów doktoranckich miałem wielki przywilej prowadzić część badań w Amsterdamie, jako członek tamtejszej grupy strunowej. Współpracę tę umożliwili w szczególności profesorowie tej grupy: Jan de Boer, Robbert Dijkgraaf oraz Erik Verlinde, którym dziękuję zarówno za to, jak też za wiele inspirujących dyskusji i cennych wskazówek. Pod wieloma względami cenne było wsparcie, które okazali mi profesorowie Michał Baj, Marek Cieplak i Maria Krawczyk w Warszawie, oraz Kareljan Schoutens i Kostas Skenderis w Amsterdamie.

Jestem niezmiernie wdzięczny wszystkim osobom od których miałem przyjemność wiele się nauczyć oraz z którymi bliżej współpracowałem. Są to w szczególności Nick Halmagyi, Amir-Kian Kashani-Poor, Marcos Marino, Asad Naqvi oraz Ani Sinkovics, jak też: Xerxes Arsiwalla, Rutger Boels, Ruth Britto, Miranda Cheng, Sheer El-Showk, Bartomeu Fiol, Sebastian de Haro, Lotte Hollands, Paweł Jakubczyk, Nick Jones, Ingmar Kanitschnider, Liat Maoz, Jan Manschot, Ilies Messamah, Kyriakos Papadodimas, Thomas Quella, Krzysztof Sadlej, Jan Suffczyński, Rafał Suszek, Kuba Wagner, Brookie Williams i Rafał Wysocki. Dziękuję za wszystkie nasze dyskusje, Wasze tłumaczenia, pomoc w trudnych momentach, oraz cały wspólnie spędzony czas.

Nieocenione wsparcie dla mojej aktywności naukowej, a w szczególnosci licznych podróży, zapewniły Beata Czajkowska i Olga Mazur w Warszawie, oraz Yocklang Chong, Lotty Gilliéron i Paula Uijthoven w Amsterdamie.

Szczególne podziękowania składam mojej żonie Asi za ogromne wsparcie i wszelką pomoc na którą zawsze mogłem liczyć, oraz wiele cierpliwości i poświęcenia. Dziękuję całej mojej rodzinie, której tę pracę dedykuję. Słowa wdzięczności i podziękownia kieruję do wszystkich, z którymi w tych pięknych latach przedeptałem wiele ścieżek, zarówno naukowych, jak też leśnych, górskich i wszystkich innych.

Badania przeze mnie prowadzone w ramach niniejszej pracy były finansowane przez grant promotorski Ministerstwa Nauki i Szkolnictwa Wyższego N202-004-31/0060 ze strony polskiej oraz NWO Spinoza Grant ze strony holenderskiej.
 
\selectlanguage{english}

\newpage

{\center \section*{Acknowledgements}}

It is my utmost pleasure to thank my supervisors --- Jacek Pawełczyk from University of Warsaw and Robbert Dijkgraaf from University of Amsterdam --- for all the support, guidance and encouragement. Working under their supervision was a great and fascinating adventure.

For almost three years I shared my time between Warsaw and Amsterdam, conducting my research also as a member of Amsterdam String Theory Group. I would like to thank especially professors from the String Group: Jan de Boer, Robbert Dijkgraaf and Erik Verlinde, who made this collaboration possible and taught me a lot in many inspiring discussions over all this period. I also benefited a lot from interactions with professors Michał Baj, Marek Cieplak and Maria Krawczyk in Warsaw, as well as Kareljan Schoutens and Kostas Skenderis in Amsterdam.

I am very grateful to all the people I learnt from and worked with, in particular Nick Halmagyi, Amir-Kian Kashani-Poor, Marcos Marino, Asad Naqvi and Ani Sinkovics. I also owe many thanks to Xerxes Arsiwalla, Rutger Boels, Ruth Britto, Miranda Cheng, Sheer El-Showk, Bartomeu Fiol, Sebastian de Haro, Lotte Hollands, Paweł Jakubczyk, Nick Jones, Ingmar Kanitscheider, Liat Maoz, Jan Manschot, Ilies Messamah, Kyriakos Papadodimas, Thomas Quella, Krzysztof Sadlej, Jan Suffczyński, Rafał Suszek, Kuba Wagner, Brookie Williams and Rafał Wysocki. I appreciate a lot all our discussions, your insightful explanations, your help in difficult moments, or just sharing our time together.

My various scientific activities, in particular numerous trips, would not be possible without kind assistance of Beata Czajkowska and Olga Mazur in Warsaw, as well as Yocklang Chong, Lotty Gilliéron and Paula Uijthoven in Amsterdam.

My very special thanks go to my wife Asia for continuous support in many ways, encouragement, patience and devotion. I thank my entire family to whom this thesis is dedicated. My words of thanks go to all I walked with along many paths in those beautiful years, both through science, as well as forests, mountains, and all other surroundings.

My research related to this thesis was supported by MNiSW grant N202-004-31/0060 and the NWO Spinoza Grant.

%\end{footnotesize}

\end{document}